\documentclass[a4paper,12pt]{book}

\usepackage{amsmath,amssymb}
\usepackage{hyperref} % [breaklinks,bookmarks=true,hyperfigures=true]
\usepackage{graphicx}
\usepackage{lmodern}
\usepackage{subfig}
\usepackage{makeidx}
\usepackage[a4paper,text={155mm,220mm},centering]{geometry}
\makeindex

\graphicspath{{pics/}}

\newcommand{\op}{{\mathcal{O}}}
\newcommand{\nn}{\nonumber}
\newcommand{\tr}{{\rm Tr\,}}
\newcommand{\Tr}{{\rm Tr\,}}
\newcommand{\eps}{{\varepsilon}}
\newcommand{\sym}{{\mathcal{N}=4~ \text{SYM}}}
\newcommand{\syml}{\mathcal{N}=4~ \text{super Yang-Mills}}
\newcommand{\Syml}{\mathcal{N}=4~ \text{Super Yang-Mills}}
\newcommand{\scs}{\mathcal{N}=6~ \text{super Chern-Simons}}
\newcommand{\Scs}{\mathcal{N}=6~ \text{Super Chern-Simons}}
\newcommand{\Li}{{\text{Li}_2}}

\newcommand{\hi}{{\hat i}}   
\newcommand{\hj}{{\hat j}}   
\newcommand{\slsh}[1]{{#1}\!\!\!\! / \, }
\newcommand{\pslash}{p\!\! /}
\newcommand{\kslash}{k\!\! /}

\newcommand{\da}{{\dot\alpha}}
\newcommand{\db}{{\dot\beta}}
\renewcommand{\b}{\beta}
\renewcommand{\a}{\alpha}
\def \qqquad {\qquad\quad}

\newcommand{\ft}[2]{{\textstyle\frac{#1}{#2}}}

\newcommand{\notocsubsection}[1]{% 
    \refstepcounter{subsection}% 
    \subsection*{\thesubsection \quad #1}}% 

\begin{document}

\title{Superconformal Quantum Field Theory \\ in String -- Gauge Theory Dualities }
%\date{}
\author{Konstantin Wiegandt}

\thispagestyle{empty}

\begingroup
\centering
{\Large Perturbative Methods \\ for Superconformal Quantum Field Theories\\in String -- Gauge Theory  Dualities \\ \vspace{1.2cm}}
\endgroup

\begingroup
\centering
{Konstantin Wiegandt\footnote{k.wiegandt@qmul.ac.uk} \\ \vspace{.8cm}
{\small Centre for Research in String Theory\\
School of Physics and Astronomy\\
Queen Mary University of London\\
Mile End Road, London E1 4NS, UK\\
}
}
\endgroup

\begin{figure}[h]\centering
  \includegraphics[width=.8 \textwidth]{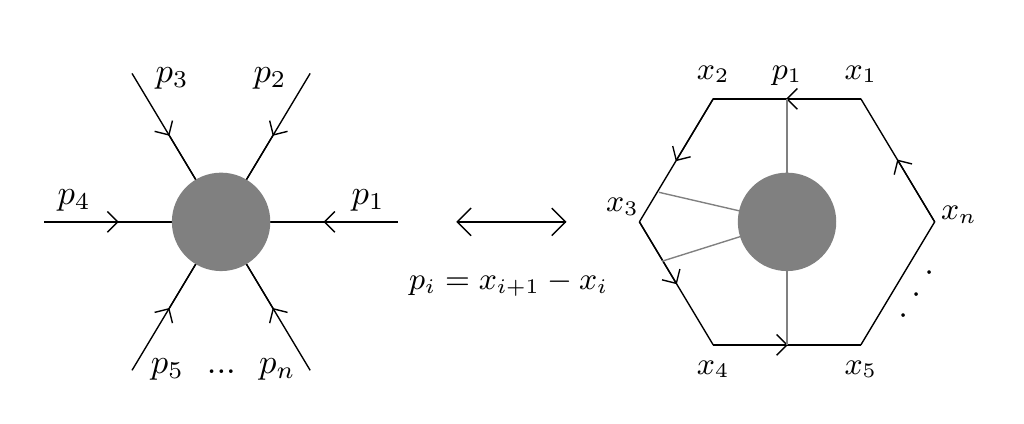}
\end{figure}

\begin{quotation}
\noindent This review covers a number of applications of conformal field theory methods for perturbative calculations in $\syml$ and ABJM theory.
After motivating the role of superconformal symmetry for elementary particle physics research, we review some basic facts about conformal symmetry and conformal transformations of fields in a pedagogical manner. The implications for perturbative quantum field theory calculations are discussed in terms of Ward identities. 
Recent developments and relations between amplitudes, correlators and Wilson loops are reviewed for $\syml$ as well as ABJM theory.
We give more details on previously published results on light-like polygonal Wilson loops in ABJM theory as well as the calculation of three-point functions of two BPS and one twist-two operator in $\sym$ theory.

This review is based on the author's PhD thesis and includes developments until April 2012.
\end{quotation}
\clearpage
\newpage

\thispagestyle{empty}

\begin{titlepage}

\begingroup
\centering
{\LARGE \textsc{Superconformal\\Quantum Field Theories\\in String -- Gauge Theory  Dualities} \\ \vspace{1.2cm}}
\endgroup

\begingroup
\centering
{{\large \textsc{Dissertation}}\\[.1cm] \normalsize zur Erlangung des akademischen Grades\\[.1cm] \large \textsc{doctor rerum naturalium}\\[.1cm] \normalsize (Dr. rer. nat.)\\[.1cm]im Fach Physik\\[1.2cm]}
\endgroup

\begingroup
\centering
{ eingereicht an der\\[.1cm]
Mathematisch-Naturwissenschaftlichen Fakult\"at I\\[.1cm]
der Humboldt-Universit\"at zu Berlin\\[1.2cm] }
\endgroup

\begingroup
\centering
{\normalsize von\\[.1cm] Herrn Dipl.-Phys. Konstantin Wiegandt\\[1.4cm]}
\endgroup

\begingroup
\centering
{ Pr\"asident der Humboldt-Universit\"at zu Berlin:\\[.1cm]Prof. Dr. Jan-Hendrik Olbertz\\[.2cm]
Dekan der Mathematisch-Naturwissenschaftlichen Fakult\"at I:\\[.1cm] Prof. Stefan Hecht, Ph.D.\\[2cm]}
\endgroup

Gutachter:\\[.25cm]
\begin{tabular}{ll}
~~~~~~1.& PD Dr. Harald Dorn
\\
~~~~~~2.& Prof. Dr. Gregory Korchemsky
\\
~~~~~~3.& Prof. Dr. Jan Plefka 
\\[.2cm] 
\end{tabular}

~\\
eingereicht am: 3.5.2012 \\
Tag der m\"undlichen Pr\"ufung: 14.8.2012

\end{titlepage}

\pagestyle{empty}

\cleardoublepage
  
~\\~\\~\\
\emph{"Every theoretical physicist who is any good\\ 
 knows six or seven different theoretical representations\\
  for exactly the same physics."}\footnote{R.~P. Feynman, 1965, p.168
\textit{``The character of physical law''}, MIT Press, Cambridge, MA.}

~\\[4cm]
\begingroup

\begin{figure}[h]\centering
  \includegraphics[width=.82 \textwidth]{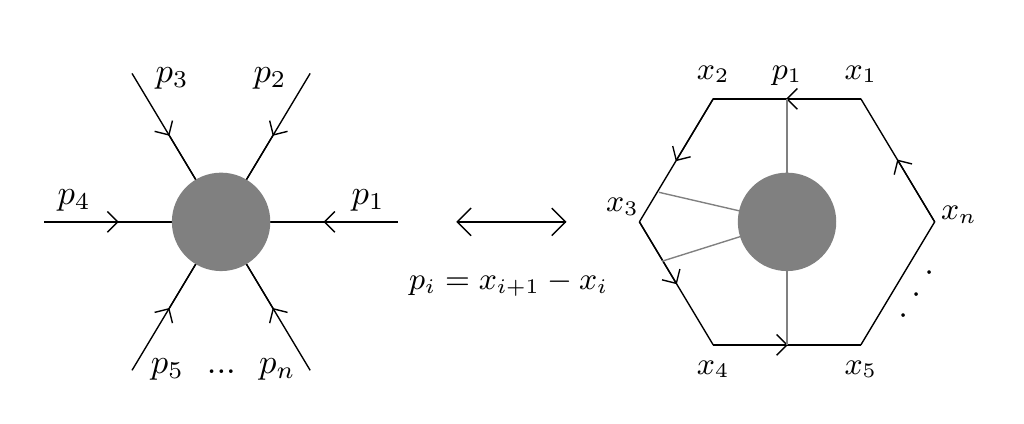}\\[-.15cm]
\end{figure}
\endgroup

\thispagestyle{empty}
  
 \tableofcontents
 \thispagestyle{empty}
 \addtocontents{toc}{\protect\thispagestyle{empty}}
\thispagestyle{empty}

\pagestyle{plain}
\chapter*{Executive Summary}
\renewcommand{\thepage}{\Roman{page}}
\setcounter{page}{1}

\addcontentsline{toc}{chapter}{Executive Summary}
The AdS/CFT correspondence \cite{Maldacena:1997re,Aharony:2008ug} states a duality between string theories living on  Anti-de Sitter space and supersymmetric conformal quantum field theories in Minkowski-space. Beyond the relations between gauge and string theories, the correspondence has lead to many new insights into the structure of supersymmetric quantum field theory itself.

In \cite{Alday:2007hr,Alday:2007he} a strong coupling calculation of $n$-gluon scattering amplitudes in terms of minimal surfaces in AdS was proposed, that formally resembles the calculation of a Wilson loop over a light-like polygonal contour made out of the gluon momenta $p_i$. Interestingly, the duality between gluon scattering amplitudes and light-like polygonal Wilson loops was found to be true also at weak coupling in $\syml$ theory \cite{Drummond:2007aua,Brandhuber:2007yx}. In \cite{Drummond:2007cf,Drummond:2007au} it was shown that the Wilson loop is governed by an anomalous conformal Ward identity that fixes the form of the expectation value for $n=4,5$ completely and allows for a function of conformally invariant cross ratios starting from $n=6$. The duality persists to higher loops and legs and also makes some properties of gluon scattering amplitudes more transparent, such as the dual (super-)conformal symmetry of the amplitudes found in \cite{Drummond:2006rz,Drummond:2008vq}. The dual superconformal symmetry combines with the superconformal symmetry to a Yangian symmetry \cite{Drummond:2009fd}, reflecting the integrability of $\syml$ theory in the planar amplitude sector. Another interesting development is the duality \cite{Eden:2010ce} between amplitudes and correlation functions in a limit where adjacent operators become light-like separated. 

Many of these developments have an analogue in ABJM theory \cite{Aharony:2008ug} and it is the subject of the first part of this thesis to investigate the Wilson loop side of a possible similar duality in this theory. The main result is, that the expectation value of light-like polygonal Wilson loops vanishes at one-loop order and has a non-vanishing form at two-loop order, which is identical in its functional form to the Wilson loop in $\syml$. We calculate the expectation value at two loops for the four-sided Wilson loop and show, that its form is governed by an anomalous conformal Ward identity. We generalise the results to $n$ points at two-loop order and find that the expectation value of the Wilson loop is given by
\begin{align}
\langle W_n \rangle^{\text{ABJM}}_{\text{2-loop}} &=  \left( \frac{N}{k}\right)^2 \Big [-\frac{1}{2}\sum_{i=1}^n  \frac{(-{\hat{\mu}}^2 \, x_{i,i+2}^2 )^{2\epsilon}}{(2\epsilon)^2} 
 + \mathcal{F}_n^{\text{WL}} + r_n + \mathcal{O}(\epsilon)
\Big ] \,,
\end{align}
where $r_n$ is a constant with linear dependence on $n$ and $\mathcal{F}_n^{\text{WL}}$ is the same finite part as that of the Wilson loop in $\syml$ theory at one-loop order.
These developments were presented in \cite{Wiegandt:2011uu,Wiegandt:2011zz,Henn:2010ps}. Most interestingly, it was found in \cite{Chen:2011vv,Bianchi:2011dg} that the four-point two-loop amplitude in ABJM theory with the tree-level part stripped off precisely agrees with this form, giving the first indication for a duality between amplitudes and Wilson loops in ABJM theory. Dual conformal and Yangian symmetry in the amplitude sector of ABJM theory \cite{Bargheer:2010hn,Huang:2010qy,Lipstein:2011ej} further hint at the existence of such a relation. Furthermore, non-trivial evidence for a duality between Wilson loops and correlators in ABJM theory was found  at one-loop level in \cite{Bianchi:2011rn}.
Recently, it was shown that, unlike the Wilson loop, the six-point amplitudes in ABJM theory do not vanish at one loop \cite{Bargheer:2012cp,Bianchi:2012cq} and we will discuss why this does not necessarily imply, that a relation between Wilson loops and amplitudes does not exist in ABJM theory.

In the second part of this thesis we calculate three-point functions of two protected operators and a twist-two operator $\hat{\op}_j$ with arbitrary even spin $j$ in $\syml$ theory. In order to carry out the calculations we project the indices of the spin $j$ operator to the light-cone and evaluate the correlator in a soft-limit where the momentum coming in at the spin $j$ operator becomes zero. This limit largely simplifies the perturbative calculation, since all three-point diagrams effectively reduce to two-point diagrams and the dependence on the one-loop mixing matrix drops out completely. We find that the correction to the structure constant has the simple form
\begin{equation}
C^\prime_{\op  \tilde{\op} j}(g^2) = {C^\prime}_{\op  \tilde{\op} j} ^{(0)} \left(1 +   \frac{g^2 N}{8 \pi^2}  \left(2H_j(H_j -  H_{2j})- H_{j,2}\right) + \op(g^4) \right)\,,
\end{equation} 
where $H_{j,m}$ are  generalised harmonic sums. The result is in agreement with the extraction of the structure constant from the analysis of the operator product expansion of four-point functions of half-BPS operators \cite{Dolan:2004iy}.

\section*{Overview}
\addcontentsline{toc}{chapter}{Outline of the Thesis}
The focus of this thesis lies on the gauge theory side of the AdS/CFT duality and we will in general refer to the literature instead of discussing the developments on the string theory side in order to keep the presentation more compact.

\begin{itemize}
\item  In {\bf chapter \ref{sec:general-intro}} we shortly comment on the role of supersymmetry in elementary particle physics, the question of quantum gravity and discuss some general features of the AdS/CFT correspondence  in order to embed the studies of this thesis into the the context of present research.
\item The following two chapters are of more general nature and may also be of interest independently of the research projects that are performed in this thesis.
In {\bf chapter \ref{sec:introduction-to-conformal-symmetry}} some well-known basic facts about conformal symmetry and conformal field theory are reviewed. We explicitly derive the transformation properties of fundamental fields under conformal transformations.
Furthermore we introduce $\syml$ and ABJM theory. 
 {\bf Chapter \ref{eqn:implications-of-conformal-symmetry}} is devoted to the study of the implications of conformal symmetry on correlation functions from the perspective of conformal Ward identities. Furthermore, we discuss some aspects of the renormalisation of composite operators, the operator product expansion and its relation to conformal three-point functions.

\item In {\bf chapter \ref{chapter:amplitude-wilson-loops-correlators}} we first discuss some techniques for the formulation and evaluation of scattering amplitudes, e.g. the spinor helicity formalism, BFCW recursions and unitarity methods. We then review dual superconformal and Yangian symmetry of scattering amplitudes in $\syml$ theory and the BDS ansatz for $n$-point gluon scattering amplitudes. In the remaining part of this chapter we discuss the duality between bosonic Wilson loops and MHV scattering amplitudes and shortly comment on more recent discoveries, such as the formulation of supersymmetric Wilson loops, the duality between superamplitudes and supercorrelators as well as the relation between form factors and periodic Wilson loops. 

\item In {\bf chapter \ref{sec:developments-in-ABJM}} we review the analogous developments in ABJM theory with a focus on a possible relation between scattering amplitudes and Wilson loops. In {\bf chapter \ref{sec:Wilson-loops-in-ABJM}} we present the calculation of the four-point light-like Wilson loop in ABJM theory, the calculation of the anomalous conformal Ward identity that constrains the Wilson loop as well as numerical studies for the hexagonal as well has higher-point Wilson loops and present the result for general $n$.
These calculations were partially presented in \cite{Wiegandt:2011uu,Wiegandt:2011zz,Henn:2010ps} and we give some more details, especially on the evaluation of the $n$-point Wilson loop.

\item {\bf Chapter \ref{sec:three-point-functions}} contains unpublished results\footnote{The results were published in \cite{Plefka:2012rd} after the completion of this thesis.}. We introduce different limiting procedures that can be employed for the calculation of three-point functions and present the general structure of the calculation that is necessary in order to read-off the structure constant of the three-point functions of two half-BPS operators and one twist-two operator. The remaining section contain some details of the calculation.

\item In {\bf chapter \ref{sec:conclusions}} we draw our conclusions and comment on possible future research directions.

\item We tried to delegate most explicit calculations to the Appendices and to present only the structure of calculations. Furthermore, {\bf the Appendices \ref{app:Conventions}, \ref{sec:details-conformal-symmetry}, \ref{sec:N-4-SYM-from-reduction}} contain our conventions, some technical details on conformal symmetry, an explicit derivation of the $\syml$ theory action from $\mathcal{N}=1$ super Yang-Mills theory in ten dimensions  and the Feynman rules. The {\bf Appendices  \ref{sec:introd-mellin-barnes}, \ref{sec:wilson-loop-op-and-gauge-invariance}}  contain an introduction to the Mellin-Barnes technique for the evaluation of integrals and some information on the path ordering and gauge invariance of Wilson loops. In {\bf Appendix \ref{sec:ABJM}} we summarise our conventions and Feynman rules for Chern-Simons and ABJM theory.
\end{itemize}

%\chapter*{Zusammenfassung}
\section*{{\Huge Zusammenfassung}}\sectionmark{Zusammenfassung}
\addcontentsline{toc}{chapter}{Zusammenfassung}~\\[.4cm]
Die AdS/CFT Korrespondenz  \cite{Maldacena:1997re,Aharony:2008ug} stellt eine Dualit\"at zwischen Stringtheorien im Anti-de Sitter Raum und supersymmetrischen konformen Quantenfeldtheorien im Min\-kowski\-raum dar. \"Uber die Relationen zwischen Eich- und Stringtheorien hinaus, hat die Korrespondenz zu vielen neuen Einsichten in die Struktur der Quantenfeldtheorie selbst gef\"uhrt.

In \cite{Alday:2007hr,Alday:2007he} wurde eine Berechnung von $n$-Gluon-Streuamplituden durch Minimalfl\"achen in AdS vorgeschlagen, welche formal der Berechnung einer Wilsonschleife \"uber eine lichtartige Kontur \"ahnelt, die aus den Impulsen $p_i$ der Gluonen zusammengesetzt ist. Interessanterweise wurde herausgefunden, dass die Dualit\"at zwischen Gluon-Streu\-amplituden und lichtartigen Wilsonschleifen in der $\Syml$ Theorie auch bei schwacher Kopplung gilt \cite{Drummond:2007aua,Brandhuber:2007yx}.  In \cite{Drummond:2007cf,Drummond:2007au} wurde gezeigt, dass die Wilsonschleife durch anomale konforme Ward-Identit\"aten bestimmt wird, die die Form des Erwartungswertes f\"ur $n=4,5$ vollst\"andig festlegen und von $n=6$ an eine beliebige Funktion von konform invarianten Kombinationen der kinematischen Variablen zulassen. Die Dualit\"at ist auch f\"ur h\"ohere Ordnungen der St\"orungstheorie und beliebige Anzahl $n$ von Punkten g\"ultig und macht einige Eigenschaften der Streuamplituden, wie z.B. die duale {(super-)konforme} Symmetrie, die in \cite{Drummond:2006rz,Drummond:2008vq} gefunden wurde, transparenter. Die duale superkonforme Symmetrie l\"asst sich mit der superkonformen Symmetrie zu einer Yangschen Symmetrie kombinieren \cite{Drummond:2009fd}, wodurch die Integrabilit\"at der $\Syml$ Theorie im planaren Amplitudensektor reflektiert wird. Eine weitere interessante Entwicklung in diesem Kontext ist die Dualit\"at \cite{Eden:2010ce}  zwischen Streuamplituden und Korrelationsfunktionen, die in einem Limes betrachtet werden, f\"ur den benachbarte Operatoren lichtartig entfernt sind.

Viele dieser Entwicklungen sind analog in der ABJM Theorie \cite{Aharony:2008ug} und es ist das Thema des ersten Teils dieser Dissertation, die Wilsonschleifenseite einer solchen m\"ogli\-chen Dualit\"at zu untersuchen. Das Hauptergebnis dieser Untersuchungen ist, dass der Erwartungswert der lichtartigen polygonalen Wilsonschleifen auf Einschleifenebene verschwindet und auf Zweischleifenebene dieselbe Form hat wie die Wilsonschleife in $\Syml$ Theorie auf Einschleifenebene. Es wird der Erwartungswert auf Zweischleifenebene f\"ur die vierseitige Wilsonschleife berechnet und gezeigt, dass deren Form durch eine anomale konforme Ward-Identit\"at bestimmt wird. Die Ergebnisse lassen sich auf den $n$-Punkt Fall auf Zweischleifenebene verallgemeinern, mit dem Ergebnis 
\begin{align}
\langle W_n \rangle^{\text{ABJM}}_{\text{2-loop}} &=  \left( \frac{N}{k}\right)^2 \Big [-\frac{1}{2}\sum_{i=1}^n  \frac{(-{\hat{\mu}}^2 \, x_{i,i+2}^2 )^{2\epsilon}}{(2\epsilon)^2} 
 + \mathcal{F}_n^{\text{WL}} + r_n + \mathcal{O}(\epsilon)
\Big ] \,,
\end{align}
wobei $r_n$ eine Konstante ist, die linear von $n$ abh\"angt und $\mathcal{F}_n^{\text{WL}}$ derselbe endliche Term wie der der 
Wilsonschleife in der $\syml$ Theorie auf Einschleifenebene ist.

Diese Ergebnisse wurden in \cite{Wiegandt:2011uu,Wiegandt:2011zz,Henn:2010ps} pr\"asentiert. Interessanterweise wurde in  \cite{Chen:2011vv,Bianchi:2011dg}  gefunden, dass die Zweischleifen-Streuamplitude mit vier externen Teilchen durch den Baumgraphenanteil geteilt, genau mit dieser Form \"ubereinstimmt und damit den ersten Hinweis auf eine m\"ogliche Dualit\"at zwischen Amplituden und Wilsonschleifen in der ABJM Theorie gibt. Duale konforme und Yangsche Symmetrie im Amplitudensektor  \cite{Bargheer:2010hn,Huang:2010qy,Lipstein:2011ej} weisen weiter auf eine solche Relation hin. Weiterhin wurden nicht-triviale Anhaltspunkte f\"ur eine Dualit\"at zwischen Korrelatoren und Wilsonschleifen  auf Einschleifenebene gefunden \cite{Bianchi:2011rn}. K\"urzlich wurde gezeigt, dass, entgegen der Wilsonschleife, die Streuamplituden mit sechs externen Teilchen auf Einschleifenebene nicht verschwinden \cite{Bargheer:2012cp,Bianchi:2012cq}. In der vorliegenden Dissertation wird auch kurz darauf eingegangen, warum dies nicht notwendiger Weise impliziert, dass in der ABJM Theorie keine Relation zwischen Wilsonschleifen und Amplituden existiert. 

Im zweiten Teil dieser Arbeit berechnen wir Dreipunktkorrelationsfunktionen von zwei gesch\"utzten Operatoren und einem Twist-Zwei Operator $\hat{\op}_j$  mit beliebigem geradzahligem Spin $j$ in der $\syml$ Theorie. Um diese Rechnungen durchzuf\"uhren, projizieren wir die Indizes des Spin $j$ Operators auf den Lichtkegel und werten den Korrelator in einem Infrarotlimes aus, in dem der Impuls, der bei dem Spin $j$ Operator einflie\ss t, null wird. Dieser Limes vereinfacht die perturbativen Rechnungen enorm, da alle Dreipunktdiagramme effektiv auf Zweipunktdiagramme reduziert werden und die Abh\"angigkeit der Einschleifenmischungsmatrix komplett herausf\"allt. Die Korrektur zur Strukturkonstanten hat die einfache Form
\begin{equation}
C^\prime_{\op  \tilde{\op} j}(g^2) = {C^\prime}_{\op  \tilde{\op} j} ^{(0)} \left(1 +   \frac{g^2 N}{8 \pi^2}  \left(2H_j(H_j -  H_{2j})- H_{j,2}\right) + \op(g^4) \right)\,,
\end{equation} 
hat, wobei  $H_{j,m}$ verallgemeinerte harmonische Summen sind. Dieses Ergebnis stimmt mit der Strukturkonstante, die sich aus der Operatorproduktentwicklung der Vierpunkt\-funktionen von 1/2-BPS Operatoren extrahieren l\"asst, \"uberein \cite{Dolan:2004iy}.

\chapter{Introduction -- New Directions in Quantum Field Theory}\chaptermark{Introduction -- New Directions in QFT}
\renewcommand{\thepage}{\arabic{page}}\label{sec:general-intro} \pagestyle{headings}
\setcounter{page}{1}
Einstein's theory of special relativity and the theory of quantum mechanics,  both developed at the beginning of the last century, induced major shifts of paradigm in physics and lead to the construction of a theory that incorporates them both, quantum field theory (QFT). 
The present \emph{standard model} of elementary particle physics makes use of the concepts
of QFT successfully and can be regarded as the experimentally best verified theory ever.
Feynman, one of the pioneers in the development of QFT, once said: \\

\emph{"Every theoretical physicist who is any good  
 knows six or seven different theoretical representations for exactly the same physics."}\footnote{R.~P. Feynman, 1965, p.168
\textit{``The character of physical law''}, MIT Press, Cambridge, MA.}
~\\~\\
Besides the debatable classification of good theoretical physicists, the second part of this sentence contains the important statement that there is not the \emph{one correct theory} that describes our world, but rather that any theory that \emph{correctly describes} the phenomena in this world is equally valid. The content of this sentence is possibly more up-to-date than ever before, as we will argue in the following.

The experiments in elementary particle physics are described in terms of scattering amplitudes.
As will be described in this thesis, there are many ways to calculate these scattering amplitudes, by at first sight completely different theoretical representations. These achievements are mainly due to the so-called \emph{AdS/CFT correspondence} \cite{Maldacena:1998re}, that will be described in section \ref{sec:ads-cft}. The correspondence relates a \emph{string theory} in a ten-dimensional space-time to a simplified theory of elementary particles physics, the so-called maximally supersymmetric $\syml$ theory in four-dimensional space-time, which is reasonably similar to the standard model of particle physics as we will argue later on. In the spirit of Feynman's words, if we could calculate all physical observables in either of the theories, there was no way to distinguish which of them is the \emph{correct one}. Furthermore, the duality triggered many relations between quantities purely on the field theory side of the duality. It turns out that scattering amplitudes are intimately related to a special type of so-called \emph{Wilson loops} as well as \emph{correlation functions} in a certain limit as will be described in detail in chapter \ref{chapter:amplitude-wilson-loops-correlators}. Using these relationships, it is even more convenient to analyse the quantum corrections to scattering amplitudes in terms of Wilson loops, providing new ways to interpret the quantum field theory itself. 

Possibly, these theoretical developments could only be achieved due to the lack of important new experimental input from collider physics in the last decades, which gave theoreticians time to reorganise the theory and to get a better understanding of it.
In light of the beginning \emph{"Era of the Large Hadron Collider (LHC)"}, these new theoretical developments are highly pleasing, since the discovery of new physics at the LHC requires a precise understanding of background processes of known physics. These processes are described by the theory of strong interactions called quantum chromodynamics (QCD), which is successfully described in the framework of quantum field theory.  Using the standard technique of Feynman diagrams, these processes are notoriously difficult to evaluate  and thus new methods are required. The status of available methods and tools can for instance be found in the introduction of  the review \cite{Bern:2007dw}.

Among the new physics that await its discovery at the LHC, there is the missing particle of the standard model, the \emph{Higgs particle}, which is ought to be responsible for the mechanism of mass generation. In this thesis we will deal a lot with symmetries and their consequences. In particular, we will deal with \emph{conformal symmetry}, which is introduced in chapters \ref{sec:introduction-to-conformal-symmetry}, \ref{eqn:implications-of-conformal-symmetry}
 and may be, loosely spoken, interpreted as an additional \emph{scale invariance} of the theory. 
One may note, that the only term in the standard model\footnote{assuming massless neutrinos} that violates this scale invariance and thus the conformal symmetry on the classical level, is the mass term of the Higgs particle. From a conceptual point of view one may prefer a theory which is fundamentally  scale invariant. Indeed, there are proposals for a classically conformal invariant standard model, with a dynamical mechanism of mass generation, see chapter \ref{sec:introduction-to-conformal-symmetry}.

\section{Supersymmetry and Particle Physics}
Another symmetry that we will encounter in this thesis is \emph{supersymmetry}. It relates particles of different spin to each other. Apart from the hope to find the superpartners of standard model particles at the LHC, supersymmetry can be considered as a calculational tool in non-supersymmetric theories such as QCD and has already proven as useful in this endeavour. 

At tree-level, gluon amplitudes in $\mathcal{N}=4$ super Yang-Mills and QCD are identical, since these amplitudes do not depend on the particle content of the theory. Furthermore,  using supersymmetric Ward identities, amplitudes with different external particles can be related to each other \cite{Bern:1993wt}. Also at loop-level, where the particle content in the loop is different in QCD and supersymmetric theories, supersymmetry leads to simplifications. For instance, one can decompose pure Yang-Mills amplitudes, i.e. amplitudes $A_g$ containing only gluon loops, into contributions of supersymmetric multiplets, see chapter 4 of \cite{Dixon:1996wi}
\begin{align}
 A_g &= (A_g + 4 A_f + 3 A_s) - 4(A_f+A_s) + A_s
= A^{\mathcal{N}=4} - 4 A^{\mathcal{N}=1} + A_s\,,
\end{align}
where the indices $s, f, g$ refer to scalars, fermions and gluons circulating in the loop. Taking the contributions of an $\mathcal{N}=4$ multiplet and an $\mathcal{N}=1$ (chiral) multiplet as granted, the task of calculating the gluon contribution is reduced to computing the contribution $A_s$ of a scalar in the loop. The supersymmetric amplitudes are simpler to analyse and, as we will see in section \ref{sec:amplitudes}, certain $\mathcal{N}=4$ SYM amplitudes are even tractable to all orders using methods of integrability.

\section{The Hydrogen Atom of Gauge Theories}\label{sec:hydrogen-atom-of-gauge-theories}
$\syml$ theory (SYM) is the maximally supersymmetric gauge theory. It resembles the standard model in that it is based on the same type of particles and interactions and both are renormalisable field theories in four-dimensional Minkowski-space. On the other hand, $\sym$ exhibits a few interesting features that make it simpler. The renormalisation group $\beta$-function of $\sym$ vanishes to all orders in the coupling constant and conformal symmetry is preserved at the quantum level, which is tightly linked to the supersymmetry of the theory. These properties make the model more tractable and in the \emph{planar limit} the \emph{integrability} of the gauge theory, see \ref{sec:integrability}, allows to extract certain  quantities to all orders in the coupling.

$\sym$ may therefore be regarded as a testing ground for the foundations of QFT and is sometimes referred to as the \emph{Hydrogen Atom of Gauge Theories}. Just as the understanding of the hydrogen atom was the key to the understanding of the spectra of other atoms, the full solution of $\sym$ is widely believed to have important consequences also for less symmetric theories such as QCD. \emph{Solving the theory} amounts to being able to calculate all physical quantities of interest to arbitrary order in the coupling constant or non-perturbatively. More details on
$\sym$ theory will be given in section \ref{sec:superconformal-algebra}.

\section{The Challenge of Quantum Gravity}
Another longstanding challenge in theoretical physics is the search for a viable theory of \emph{quantum gravity}. Naive quantisation of the linearised Einstein's equations turns out to fail due to perturbative non-renormalisability in the framework of standard quantum field theory \cite{'tHooft:1974bx,Deser:1974cz}. An interesting alternative is the asymptotic safety scenario \cite{Hawking:1979ig}, for a review see e.g. \cite{Lauscher:2005xz}.
An overview about different approaches towards the quantisation of gravity is given in the book  \cite{Kiefer:2004gr},  among which the \emph{covariant quantisation} of gravity and \emph{string theory} as well as \emph{canonical gravity} and \emph{loop quantum gravity} are reviewed.  

Another interesting direction is the research on \emph{composite gravity}.
In a famous paper by S. Weinberg and E. Witten \cite{Weinberg:1980kq} restrictions for these types of models have been obtained. An nice review on different models of emergent gravity can be found in \cite{Sindoni:2011ej}. Models of composite gravity in terms of {\itshape spinor gravity} can be found in \cite{Wetterich:2003wr, Hebecker:2003iw}. 

As mentioned above, a strong competitor in the realm of quantum gravity is string theory, which describes all fundamental constituents and interactions as one-dimensional objects with vibrational excitations. String theory is advocated to have the capability of unifying all matter and forces from one concept and naturally includes massless spin-two excitations that correspond to the graviton and in this sense naturally incorporates gravity. In contrast to other approaches, string theory has a well-defined low energy limit. Similarly as mentioned above for supersymmetry, even if string theory is not realised in nature, it justifies itself by the fact, that it has led to a wealth of new discoveries within QFT, especially through the AdS/CFT correspondence and has thus led to a better understanding of quantum field theory.

Even though the research communities of these different directions claim having constructed viable theories of quantum gravity, it is fair to say, that none of them can be regarded as generally accepted nor has any of them made a quantitative prediction that has been justified or falsified.

\section{String - Gauge Theory Dualities}\label{sec:ads-cft}
A fascinating new area of research in theoretical physics has been initiated by the so-called \emph{AdS/CFT correspondence} \cite{Maldacena:1997re,Aharony:2008ug}, that states a duality between string theories living on  \emph{Anti-de Sitter space} (AdS) and supersymmetric conformal quantum field theories (CFTs) in the planar limit \cite{'tHooft:1973jz}.
The striking feature of these dualitites is that they are \emph{strong-weak coupling dualities}, i.e. in a limit where one of the theories is strongly coupled, the other one is weakly coupled and one can thus access observables in the strongly coupled theory by calculating observables on the weakly coupled side with methods of perturbation theory. 
Another interesting aspect of this type of duality is, that it establishes a relation between a quantum theory that contains gravity (in terms of string theory) with a pure gauge theory without gravity. 
Prominent examples of such dualities are

\begin{enumerate}\centering
 \item $\mathcal{N}=4$ super Yang-Mills  theory $\leftrightarrow$ Type II\,B string theory on $AdS_5 \times S^5$
 \item $\mathcal{N}=6$ super Chern-Simons theory $\leftrightarrow$ Type II\,A string theory on $AdS_4 \times \mathbb{CP}^3$
\end{enumerate}
The gauge theory \emph{lives} on the $(d-1)$-dimensional boundary of the $AdS_d$ space, which is conformally equivalent to $\mathbb{R}^{1,d-2}$, i.e. flat Minkowski space. Since the information of the string theory in the higher-dimensional space is conjectured to be encoded by the theory on the boundary, one also uses the term \emph{holographic duality}.

In the following, we shall only mention a few key features, but not go into the details, since this thesis primarily deals with
applications of conformal field theory in general, even though the original motivation for most of this research is due to the correspondence.
 More information on the AdS/CFT correspondence can be found e.g. in the reviews \cite{Aharony:1999ti}, \cite{D'Hoker:2002aw}, \cite{Nastase:2007kj} and in the recent review of AdS/CFT integrability \cite{Beisert:2010jr}.\\

\notocsubsection{Symmetry and Duality Relations}
Having an exact duality means that all quantities of interest can be calculated in either of the theories. Therefore, it is necessary to set up a \emph{dictionary} between relevant observables. \\

Symmetry plays a central role in the correspondence and serves for the identification of obervables, i.e. for setting up the desired dictionary.  The $AdS_5 \times S^5$ space-time that the string theory lives in can be embedded into flat space as
\begin{equation}
AdS_{d+1} = \{ X \in \mathbb{R}^{d,2}| X \cdot X = -R^2 \}, \qquad S_{d+1} = \{ Y \in \mathbb{R}^{d+2}| Y \cdot Y = R^2 \}\,,
\end{equation}
where we have chosen equal \emph{radii} $R$ for both spaces. By construction, the isometry group of $AdS_{d+1}$ is $\text{SO}(d,2)$ and $SO(d+2)$ for $S_{d+1}$. Thus for $AdS_5 \times S^5$ the isometry groups are $SO(4,2) \simeq SU(2,2)$ and $SO(6) \simeq SU(4)$. Together with 32 fermionic directions the supergroup $PSU(2,2|4)$ is formed. The symmetry group of the $AdS_5 \times S^5$ superspace is reflected by the same symmetry group of the $\sym$ theory as we will discuss in section \ref{sec:superconformal-algebra}. More information on the matching of the symmetry groups can be found in \cite{Beisert:2010kp}.\\

String states with excitations in certain directions in the superspace are identified with composite gauge invariant operators in the QFT, which carry the corresponding charges of the global symmetry group of the field theory.
It turns out that the energies of the string states can be matched with the anomalous dimensions of the corresponding operators \cite{Gubser:1998bc,Witten:1998qj,Berenstein:2002jq,Gubser:2002tv,Metsaev:2001bj}. A nice review on the matching of string energies and scaling dimensions can be found in \cite{Plefka:2003nb}.

Another outcome of the duality is an exciting relation between gluon scattering amplitudes and Wilson loops that will be described in detail in chapter \ref{chapter:amplitude-wilson-loops-correlators}.\\

The coupling $g$ and the number of colours $N$ of the gauge theory are related to the radius $R$ of the $AdS_5 \times S^5$ spaces, the string coupling $g_s$ and the (inverse) string tension $\alpha^\prime$ via \cite{Maldacena:1997re}
\begin{equation}\label{eqn:matching-parameters}
\frac{R^4}{{\alpha^\prime}^2} = g^2 N, \qquad 4 \pi g_s = g^2\,.
\end{equation}
The so-called \emph{'t Hooft limit} consists in keeping the \emph{'t Hooft coupling} $\lambda = g^2 N$ fixed while taking $N\to \infty$ \cite{'tHooft:1973jz}, we are thus dealing with free ($g_s \to 0$) string theory. This limit is also called the \emph{planar limit} because in the gauge theory, Feynman diagrams that are non-planar\footnote{Diagrams which cannot be drawn on a plane without crossing of propagators.} are suppressed by factors of $1/N^2$ and can be neglected. The perturbative expansion on the gauge theory side is thus performed in powers of $\lambda$. In the \emph{weak coupling regime}  $\lambda<<1$ we can calculate observables in the quantum field theory with perturbation theory. However, due to the relation \eqref{eqn:matching-parameters}, this amounts to quantizing string theory in a highly curved space due to the fact that the radius $R$ is small in string units. On the other hand, in the regime, where $R$ is large in string units and the string energies can be calculated perturbatively in $1/ R$, the 't Hooft coupling $\lambda$  gets large and the regime where perturbation theory is valid on the gauge theory side is left. This is why the duality is called a \emph{weak-strong coupling} duality.
Luckily, there are overlapping regimes and due to integrability, see section \ref{sec:integrability}, it is possible to evaluate the gauge theory quantities at arbitrary coupling $\lambda$.\\

It should be noted, that the coincidence of the global symmetry groups of the theories constrains observables, it is however far from sufficient to prove a complete duality. The obvious symmetries do not uniquely fix the value of the observables on either side of the duality.

As an example, one may consider the equivalence of amplitudes and Wilson loops purely on the gauge theory side. As we will argue in chapter \ref{chapter:amplitude-wilson-loops-correlators}, the expectation values of the Wilson loop are constrained by the conformal symmetry. 
The amplitudes and Wilson loops do also agree for the so-called \emph{remainder function}, which is an arbitrary function of conformally invariant kinematical quantities, even though the conformal symmetry does not constrain this function. The fact that the amplitudes continue to agree with the Wilson loop shows that there is more to the amplitude Wilson loop duality than just  conformal symmetry.

\section{Solving the Theory -- Integrability}\label{sec:solving-the-theory}
As mentioned before, the strategy is to solve a simple theory, cf. section \ref{sec:hydrogen-atom-of-gauge-theories}, first and then proceed to the less symmetric cases. 
The theory can be considered as solved when one is able to calculate the expectation value of all relevant quantities such as scattering amplitudes, correlation functions, form factors, Wilson loops and so on, exactly or to all orders in perturbation theory.

As we will see in chapter \ref{eqn:implications-of-conformal-symmetry}, correlation functions of local gauge-invariant operators $\op_\alpha(x)$ of the type
\begin{equation}
\langle \op_{\alpha_1}(x_1)...\op_{\alpha_n} (x_n) \rangle
\end{equation}
are constrained by conformal symmetry.  The space-time structure of two-point functions is completely fixed by conformal symmetry and the only relevant parameter is the \emph{(anomalous) scaling dimension} $\Delta_\alpha(g)$ of the operator, that can be perturbatively calculated in orders of the coupling constant $g$.  The space-time structure of three-point functions of operators $\op_\alpha(x), \op_\beta(x), \op_\gamma(x)$ is also fixed and they only depend on the scaling dimensions of the operators as well as the so-called \emph{structure constants} $C_{\alpha\beta\gamma}(g)$.

As we will see in chapter \ref{eqn:implications-of-conformal-symmetry}, the restrictions from conformal symmetry do not uniquely fix the space-time structure of higher-point functions, but one can extract additional information from the so-called \emph{operator product expansion} (OPE). The OPE, together with the knowledge of the anomalous dimensions, in principle allows to determine any higher-point function and thus the scaling dimensions $\Delta_\alpha(g)$ and the structure constants $C_{\alpha\beta\gamma}(g)$ are the essential parameters that govern the behaviour of observables in the theory.

While two-point functions in $\mathcal{N}=4$ SYM are very well understood, largely due to the existence of integrability, three-point functions are explored to a lesser extent. Chapter \ref{sec:three-point-functions} is devoted to the investigation of the structure constants of so-called \emph{twist-two operators} of arbitrary spin.

Also on the amplitude side substantial progress has been made. As we will describe in chapter \ref{chapter:amplitude-wilson-loops-correlators}, certain four- and five-point scattering amplitudes in $\sym$ seem to be tractable to all orders in perturbation theory \cite{Bern:2005iz}.

Due to the several relations between amplitudes, correlation functions and Wilson loops that we will describe in detail in chapter  \ref{chapter:amplitude-wilson-loops-correlators}, one may expect that the exact solution of one of these objects has a strong impact on the others and it is thus certainly very reasonable to continue research on all of these objects.

\notocsubsection{Integrability}\label{sec:integrability}
Even though this thesis does not explicitly deal with integrability, we shall mention some of its key features, since the understanding of two-point functions is to a large extent related to the presence of integrability in the planar limit of the gauge theory. 

It was noted, that a very efficient tool for the field theory calculation of anomalous dimensions of composite operators is the \emph{dilatation operator} \cite{Beisert:2002ff,Beisert:2003tq}, which acts on the \emph{chain of fields} that the composite operator is built of. It was then remarked in \cite{Minahan:2002ve}, that this operator is equivalent to the operator of the \emph{Heisenberg spin chain}, that was solved by Bethe in 1931 \cite{Bethe:1931hc} with an ansatz that is now being referred to as the \emph{coordinate Bethe ansatz}. The integrability of the spin-chain is linked to the existence of higher charges that commute with the spin-chain Hamiltonian.

The generalization of the scalar sector analysed in \cite{Minahan:2002ve} to all local operators of planar $\sym$ was achieved in \cite{Beisert:2003yb,Beisert:2003jj}. 
In a certain limit, one obtains a set of algebraic equations, the so-called \emph{asymptotic Bethe equations} \cite{Beisert:2004hm,Beisert:2005tm,Beisert:2005fw}, allowing for the extraction of anomalous dimensions to arbitrary order in the coupling constant. For instance the cusp or soft anomalous dimension of the twist-two operators that we will encounter in chapter \ref{sec:three-point-functions}, can be extracted to all orders \cite{Staudacher:2004tk,Beisert:2006ez}. This would be practically impossible using usual Feynman-diagram techniques, because of the factorially growing complexity of the calculation with increasing perturbative order. 

There are many indications, that integrability is lost beyond the planar limit $N \to \infty$, but it may still help to make an expansion in powers of $1/ N$. It is notable, that integrable structures also appear in QCD \cite{Lipatov:1993yb,Faddeev:1994zg,Braun:1998id,Belitsky:1999bf}, for reviews see \cite{Braun:2003rp,Belitsky:2004cz,Korchemsky:2010kj}.
For general reviews of integrability see \cite{Tseytlin:2004cj,Belitsky:2004cz,Beisert:2004ry,Beisert:2004yq,Zarembo:2004hp,Plefka:2005bk,Minahan:2006sk,Arutyunov:2009ga} and especially the recent review of AdS/CFT integrability \cite{Beisert:2010jr}.

\cleardoublepage

%\part{The Power of Superconformal Symmetry}

\chapter{Introduction to Conformal Symmetry}\label{sec:introduction-to-conformal-symmetry}
A \emph{conformal field theory} is a theory which is invariant under conformal symmetry transformations at the classical level.  If the conformal symmetry persists at the quantum level, the theory is a \emph{conformal quantum field} theory. Examples for theories in which the conformal symmetry survives quantisation are the theories appearing in the AdS/CFT correspondence, i.e. $\syml$ theory and $\scs$ matter theory, see section \ref{sec:superconformal-symmetry}, but up to a certain perturbative order also scalar $\phi^3$-theory in six dimensions, which is introduced in section \eqref{eqn:classically conformal symmetry}. For these theories one can use the restrictions that conformal symmetry imposes on the expectation values of observables also at the quantum level. However, also in non-quantum-conformal theories as QCD, one can make use of conformal properties at a fixed point of the renormalisation group $\beta(g^*)=0$. Applications of conformal symmetry in QCD are reviewed in \cite{Braun:2003rp}.

In this and the following chapter we review some well-known facts about conformal field theory complemented by some explicit derivations in Appendix \ref{sec:details-conformal-symmetry}.
In section \ref{sec:conformal-transformations-of-coordinates}, we define conformal transformations and explicitly derive the form of infinitesimal and finite conformal transformations of the coordinates $x^\mu$. Section \ref{sec:conformal-field-theory-classical-super-trafos} is devoted to the transformation laws of fields under conformal transformations, as well as the definition of conformal primary operators.
In sections \ref{eqn:classically conformal symmetry} and  \ref{sec:superconformal-symmetry} we give some explicit examples of conformal and superconformal field theories.

\section{Conformal Transformations}\label{sec:conformal-transformations-of-coordinates}
By definition, \emph{conformal transformations} are coordinate transformations $x_\mu \to x_\mu^\prime(x)$ under which the metric is invariant up to a local scale factor, i.e. 
\begin{equation}\label{eqn:conformal-trafo-of-metric}
g^\prime_{\mu\nu}(x^\prime) %= \frac{\partial {x^\prime}^\rho}{\partial x^\mu} \frac{\partial {x^\prime}^\sigma}{\partial x^\nu} g_{\rho \sigma}(x^\prime) 
= \rho(x) g_{\mu\nu}(x)\,.
\end{equation}
Under conformal transformations angles are preserved as is easily seen in Appendix  \eqref{eqn:angles}. In this thesis we will be  interested in conformal transformations in flat $d$-dimensional Minkowski space with metric $\eta_{\mu\nu}=\text{diag}(+,-, ... ,-)$ and specialise to this case in the following.   
\subsection{Infinitesimal Conformal Transformations}
Writing an infinitesimal conformal transformation as
\begin{equation}\label{eqn:trafo-killing-vector}
{x^\prime}^\mu = x^\mu + k^\mu(x)\,,
\end{equation}
where every component of $k^\mu$ is infinitesimal, we would like to find the most general form of the \emph{conformal Killing vectors $k^\mu$} such that \eqref{eqn:conformal-trafo-of-metric} is satisfied. As reviewed in detail in Appendix \ref{sec:derivation-conformal-transformations}, the most general infinitesimal transformation can be written as
\begin{equation}\label{eqn:Killing-vector}
k^\mu = a^\mu + \omega_{\mu\nu} x^\nu + \lambda x^\mu + 2(x \cdot c) x^\mu - c^\mu x^2\,.
\end{equation}
The vector $a^\mu$ has $d$ independent components and parametrises translations, $\omega_{\mu\nu}=-\omega_{\nu\mu}$ is antisymmetric and has $d(d-1)/2$ independent components parametrising Lorentz transformations.
$\lambda$ parametrises dilatations, and $c^\mu$ has $d$ independent components and parametrises the so-called \emph{special conformal transformations}.

\subsection{Generators of Conformal Transformations}
The effect of infinitesimal conformal transformations\index{conformal transformation!infinitesimal} can be written in terms of generators G acting on the coordinates $x^\mu$ as
\begin{equation}\label{eqn:trafo-with-generators}
\delta x^\mu={x^\prime}^\mu - x^\mu  = i\,\epsilon \cdot G x^\mu  \,,
\end{equation}
where G is any of the differential operators\index{conformal algebra!generators}
\begin{align}\label{eqn:generators-conformal-algebra}
 &P_\mu &= -i&\partial_\mu & \text{(translations)} \\ \nn
  &M_{\mu\nu} &= \phantom{-}i &\left(x_\mu \partial_\nu-x_\nu \partial_\mu\right) & \text{(Lorentz transformations)} \\ \nn
 &D & = -i &( x \cdot \partial ) & \text{(dilatations)}\\ \nn
  &K_\mu &=-i& \left( 2 x_\mu (x \cdot \partial) - x^2 \partial_\mu \right) & \text{(special conformal transformations)}
\end{align}
and the expression $\epsilon \cdot G$ symbolically represents the appropriate index contraction of the infinitesimal transformation parameters with the generators, e.g. for translations we have $\epsilon \cdot G = a_\mu P^\mu$. Then \eqref{eqn:Killing-vector} is exactly\footnote{Due to the antisymmetry of $M_{\mu\nu}$ we define $\epsilon \cdot G = \frac{1}{2} \omega_{\mu\nu} M^{\mu\nu}$ for the Lorentztransformations.} reproduced
by \eqref{eqn:trafo-with-generators} and \eqref{eqn:generators-conformal-algebra}.
The first two generators in \eqref{eqn:generators-conformal-algebra} satisfy the Poincar\'e algebra
\begin{align}\label{eqn:poincare-algebra}\index{Poincar\'e algebra}
 [M_{\mu\nu},M_{\rho\sigma}] &= i \left(\eta_{\mu\sigma} M_{\nu\rho} + \eta_{\nu\rho} M_{\mu\sigma}  - \eta_{\mu\rho} M_{\nu\sigma}  - \eta_{\nu\sigma} M_{\mu\rho} \right)\,, \\ \nn
 [P_{\mu},M_{\rho\sigma}] &= i \left(\eta_{\mu\rho}P_\sigma -\eta_{\mu\sigma}P_\rho  \right)
%\hspace{5cm} [\mathbb{P}_{\mu},\mathbb{P}_\nu] = 0 
\end{align}
and together with the generators for dilatations and special conformal transformations satisfy the conformal algebra\footnote{These commutation relations can easily be derived by application of the differential operators to an arbitrary function $f(x)$ on the coordinates in the same way as the Poincar\'e algebra.}\index{conformal algebra}
\begin{align}\label{eqn:remaining-conformal-algebra}
 [D,P_{\mu}] &= + i\,P_{\mu}\,,& \qquad     [K_{\mu},M_{\rho\sigma}] &= i \left(\eta_{\mu\rho}K_\sigma -\eta_{\mu\sigma}K_\rho  \right)\,,& \\ \nn
 [D,K_{\mu}] &= -i\,K_{\mu}\,, &  \qquad    [K_{\mu},P_{\nu}] &= 2i \left(\eta_{\mu\nu}D -M_{\mu\nu} \right)& 
 %\\ \nn [\mathbb{D},\mathbb{M}_{\mu\nu}] &= 0
\end{align}
and all other commutators vanish. 

The antisymmetric operator $J_{ab}$ with $a,b=0,1,...d-1,d,d+1$ defined by
\begin{align}
J_{\mu\nu} = M_{\mu\nu},\qquad J_{d+1,d}=D,\qquad J_{\mu,d}=\frac{1}{2}(P_\mu-K_\mu),\qquad J_{\mu,d+1}=\frac{1}{2}(P_\mu+K_\mu)\,,
\end{align}
satisfies the commutation relations similar to the Lorentz algebra $so(1,3)$ in \eqref{eqn:poincare-algebra}.
 \begin{align}\label{eqn:so(2,d)}
 [J_{ab},J_{cd}] &= i \left(\eta_{ad} J_{bc} + \eta_{bc} J_{ad}  - \eta_{ac} J_{bd}  - \eta_{bd} J_{ac} \right)\,,
 \end{align}
where $\eta_{ab}=diag(+,-,-,...,-,+)$ if $\eta_{\mu\nu}$ is the Minkowski space metric. Therefore the conformal algebra in $d$-dimensional Minkowski-space is isomorphic to $so(2,d)$ and thus has $(d+1)(d+2)/2$ generators. In $d=4$ we therefore have 15 generators, corresponding to 4 translations, 6 Lorentz transformations (3 rotations and 3 boosts), 4 special conformal transformations and one dilatation. 
 
For conformal transformations acting on a set of points $x_i$ the generators can be generalised to
 \begin{align}\label{eqn:generators-acting-on-several-points}
D = \sum_i (x_i \cdot \partial_i), \qquad  K^\nu = \sum_i  \left( 2 x_i^\nu (x_i \cdot \partial_i) - x_i^2 \partial_i^\nu \right)
\end{align}
and analagously for the remaining generators. Note that for simplicity we have redefined the generators without the factor of $i$ in \eqref{eqn:trafo-with-generators} and \eqref{eqn:generators-conformal-algebra} here, thus $\delta_{\epsilon \cdot K} f(x_{ij}^2) = \epsilon_\mu K^\mu f(x_{ij}^2)$.

\subsection{Finite Transformations}
The finite conformal transformations\index{conformal transformation!finite}  can be obtained by exponentiating the infinitesimal transformations \eqref{eqn:generators-conformal-algebra}, \eqref{eqn:trafo-with-generators} and are
\begin{align}\label{eqn:finite-conformal-trafos}
 {x^\prime}^\mu &=  x^\mu + a^\mu & \text{(translations)} \\ \nn
  {x^\prime}^\mu &= \Lambda^\mu_{~\nu}  x^\nu & \text{(Lorentz transformations)} \\ \nn
  {x^\prime}^\mu & =\lambda x^\mu & \text{(dilatations)}
\end{align}
where $\Lambda_\mu^{~\nu}$ are the familiar Lorentz transformation matrices.
For the special conformal transformations it is less straightforward to obtain the finite transformation
\begin{equation}\label{eqn:finite-special-conformal-transformation}
   {x^\prime}^\mu = \frac{x^\mu-c^\mu x^2}{1-2 c \cdot x + c^2 x^2}  \qquad \text{(special conformal transformation)}
\end{equation}
it is however easy to see, that it coincides infinitesimally with the version given in \eqref{eqn:Killing-vector} by expanding \eqref{eqn:finite-special-conformal-transformation} to first order in $c^\mu$. The transformation can also be obtained by applying an inversion of the coordinates
\begin{equation}
 I(x^\mu) = \frac{x^\mu}{x^2}
\end{equation}
followed by a translation with vector $-c^\mu$ and another inversion, i.e. 
\begin{align}\label{eqn:I-P-I gives K}
I \circ P \circ I: \qquad
x^\mu \stackrel{I}{\longrightarrow} \frac{x^\mu}{x^2}  \stackrel{\mathbb{P}}{\longrightarrow}  \frac{x^\mu}{x^2} - c^\mu  \stackrel{I}{\longrightarrow} \frac{x^\mu-c^\mu x^2}{1-2 c \cdot x + c^2 x^2}\,.
\end{align}
Even though the inversion itself has no infinitesimal form and thus does not belong to the Lie algebra of conformal transformations, it is helpful to check the invariance of an expression under special conformal transformations. If an expression it is invariant under inversions and translations, it is consequently invariant under special conformal transformations. 

\subsection{Conformal Invariants}\label{sec:conformal-invariants}
For instance, we can easily check that the combination of four coordinates $x_i^\mu$
\begin{equation}\label{eqn:cross-ratios}
u_{ijkl} = \frac{x_{ij}^2 x_{kl}^2}{x_{ik}^2 x_{jl}^2}  
\end{equation}
form \emph{conformal invariants}. The $x_{ij}^2:=(x_i-x_j)^2$ are obviously invariant under translations and rotations and transform as
\begin{equation}\label{eqn:inversions-xij2}
x_{ij}^2 \stackrel{I}{\longrightarrow} \frac{x_{ij}^2}{x_i^2 x_j^2}, \qquad\qquad\qquad x_{ij}^2 \stackrel{D}{\longrightarrow} \lambda^2 x_{ij}^2
\end{equation}
under inversions respectively  dilatations. Therefore $u_{ijkl}$ are invariant under all conformal transformations. They can be built if we have at least four non-vanishing distances. These invariants are also called conformally invariant \emph{anharmonic ratios} or \emph{cross-ratios}.\index{cross ratios}\\

\subsection{Causality and Conformal Transformations}
Note that the transformation of $x^2$ under special conformal transformations \eqref{eqn:finite-special-conformal-transformation} is
\begin{equation}\label{eqn:x2 under conformal trafo}
 x^2 \rightarrow x^{\prime 2}=\frac{x^2}{1-2 c \cdot x + c^2 x^2}\,.%=\frac{1}{\left(\frac{x^{\mu}}{x^2}-c^{\mu} \right)^2}\,. ::Comment on singularity and 1point compactification
\end{equation}
Therefore, for every vector $x^\mu$ one can find a special conformal transformation which changes a spacelike into a timelike vector and vice versa, implying a possible clash with causality, see \cite{Rosen1968468} for a detailed discussion. Light-like distances are however preserved, i.e. $x^2=0 \to {x^\prime}^2=0$. We will make use of this property in the discussion of light-like Wilson loops in section \ref{sec:light-like-Wilson-loops}.

The problem of causality can be avoided by restricting to infinitesimal transformations\footnote{There is no continous deformation of light-like into spacelike distances.} or by discussing the conformal properties of Green's functions in the Euclidean region. The causality problem was further discussed in \cite{Swieca:1973gc,Schroer:1974ay} and to avoid the conflict  with local commutativity, i.e. causality, it was argued for a non-locality of the finite field transformations, reflected by the fact that instead of a true representation of the conformal group, one has in general a representation of its universal covering group \cite{Schroer:1974ay}.

\section{Conformal Field Theory}\label{sec:conformal-field-theory-classical-super-trafos}
 
In order to formulate a classical conformal field theory, we show how fields transform under conformal transformations following to a certain extent the original papers on  infinitesimal conformal transformations of fields \cite{Mack:1969rr,Callan:1970ze,Coleman:1970je}. A similar derivation can be found in \cite{DiFrancesco:1997nk}.

\subsection{Infinitesimal Conformal Transformations of Fields}\label{eqn:conformal-transformations-of-fields}
We are interested in the action of the generators of the conformal group on a field $\Phi_\alpha(x)$ of arbitrary spin-tensor structure, i.e. $\alpha$ represents a set of spinor and / or vector indices. 
By $\mathbb{G}$ we denote the generator of an infinitesimal conformal transformation $\delta \Phi$ of the fields\index{conformal transformation!of fields}
\begin{equation}
\delta \Phi(x) = \Phi^\prime(x)-\Phi(x) =  i \,\epsilon \cdot  \mathbb{G}_{\alpha\beta}\, \Phi_\beta(x)
\end{equation}
and the generators $\mathbb{G} \in \{\mathbb{P}_\mu, \mathbb{K}_\mu, \mathbb{D}, \mathbb{M}_{\mu\nu} \}$ form a representation of the conformal algebra \eqref{eqn:remaining-conformal-algebra}, \eqref{eqn:poincare-algebra} acting on the fields. The parameters $\epsilon$ are the same transformation parameters as in \eqref{eqn:trafo-with-generators}. In the following we will suppress the indices $\alpha,\beta$ but the reader should keep in mind the action of the generators on this index space. 

The total change of a field $d \Phi := \Phi^\prime(x^\prime)-\Phi(x)$ under an infinitesimal conformal transformation $x^\prime = x + \delta x$ can be split into a transformation of the field $\delta \Phi$ and a transformation induced by the change of coordinates as
\begin{equation}\label{eqn:change-field-and-coordinates}
d \Phi (x) = \delta \Phi (x) + \delta x^\mu \partial_\mu \Phi(x) + \op(\delta^2)\,,
\end{equation}
which follows from $\Phi^\prime(x+\delta x) - \Phi(x) =  \Phi^\prime(x) - \Phi(x) + \delta x^\mu \partial_\mu \Phi(x) + \op(\delta^2)$. To illustrate the purpose of $\delta \Phi$ imagine a scalar field $\phi(x)$ measuring the distribution of some quantity in spacetime with a local maximum at $x_0$. After a coordinate transformation in which $x_0 \to {x}^\prime_0$, the field must change in such a way, that it has a maximum at $x_0^\prime$, i.e. $\phi^\prime(x_0^\prime)=\phi(x_0)$ or  $d\phi=0$ and from \eqref{eqn:change-field-and-coordinates} we find that the infinitesimal change in the field is $\delta\phi(x)=-\delta x^\mu \partial_\mu \phi(x)$.

For a vector field $A_\mu(x)$ the situation is different, it does not only describe the absolute value of a distribution in space-time such as the scalar field, but additionally a direction in space-time. For instance under a rotation, the length of a vector is preserved just as the value of the scalar field, but its direction is changed. The change in the field $\delta A_\mu$ exactly measures the change of the field evaluated at the same position, i.e. $\delta A_\mu(x) = A^\prime_\mu(x)-A_\mu(x)$.  

As we will presently see, the change of the field $\delta \Phi(x)$ respectively the action of $\mathbb{G}$ on $\Phi(x)$ can be deduced from its action on $\Phi(0)$. It is thus reasonable to classify fields according to their transformation properties under transformations which leave the origin $x^\mu =0$ invariant, i.e. which form the \emph{stability or isotropy subgroup}  of $x=0$ (also called the \emph{little group}). It follows from  \eqref{eqn:finite-conformal-trafos}, \eqref{eqn:finite-special-conformal-transformation} that these are Lorentz transformations, dilatations and special conformal transformations. We define the action at $x=0$ as
\begin{align}\label{eqn:action-of-little-group-generators}
\mathbb{M}_{\mu\nu}\, \Phi(0) = \Sigma_{\mu\nu} \Phi(0)\,, \qquad
\mathbb{D}\, \Phi(0) = \tilde{\Delta} \Phi(0)\,, \qquad
\mathbb{K}_{\mu}\, \Phi(0) = \kappa_{\mu} \Phi(0)\,,
\end{align}
where $\Sigma_{\mu\nu}$ is the spin operator associated with the field $\Phi$ with explicit representations
\begin{align}\label{eqn:Lorentz-reps}
\Sigma_{\mu\nu}\, \phi = 0\,, \qquad 
(\Sigma_{\mu\nu}\, \psi)_\alpha = \frac{i}{4}[\gamma_\mu,\gamma_\nu]_{\alpha\beta}\, \psi_\beta\,,  \qquad
(\Sigma_{\mu\nu} A)_\rho = i\left( \eta_{\mu\rho}A_\nu - \eta_{\nu\rho} A_\mu \right) \,,
\end{align}
for scalar, Dirac and vector fields. The generators $\Sigma_{\mu\nu}, \kappa_\mu, \tilde{\Delta}$ satisfy the same commutation relations as $M_{\mu\nu}, K_\mu,  D$.
The action of $\Sigma_{\mu\nu}$ on tensors with more general indices can be obtained by taking a sum over terms like \eqref{eqn:Lorentz-reps} one for each of the indices, for more details see e.g. chapter 1.4 of \cite{ramond1982field}.

According to \cite{Mack:1969rr}, we choose a basis in index space in such a way that translations do not act on the indices\footnote{They act trivially in the field index space, i.e. $({\mathbb{P}_\mu}\Phi)_\beta  =  P_\mu \Phi_\beta$.}
\begin{equation}
\mathbb{P}_\mu \Phi(x) = -i \partial_\mu \Phi (x)\,.
\end{equation}
In order to evaluate the action of the other generators on $\Phi(x)$, one can use the translation operator $\Phi(x)=\exp(i \mathbb{P}_\mu x^\mu) \Phi(0)$ to write
\begin{equation}\label{eqn:transaltion-trick}
\mathbb{G} \,\Phi (x) = e^{i \mathbb{P}\cdot x}  e^{-i \mathbb{P}\cdot x} \, \mathbb{G} \, e^{i \mathbb{P}\cdot x} \Phi(0)\,.
\end{equation}
Note that the generators $\mathbb{G}$ only act on the fields, i.e. in particular $[\mathbb{P}_\mu, x_\nu]=0$ unlike $[P_\mu, x_\nu]= -i \eta_{\mu\nu}$. Therefore  we can use
\begin{equation}\label{eqn:commutators}
e^{-i \mathbb{P}\cdot x} \, \mathbb{G} \, e^{i \mathbb{P}\cdot x} = \sum_{n=0}^\infty \frac{i^n}{n!} x^{\mu_1}... x^{\mu_n} [...[\mathbb{G},\mathbb{P}_{\mu_1}],...],\mathbb{P}_{\mu_n}]\,.
\end{equation}
The sum on the right-hand side is finite and we need at most terms with two commutators, since all higher commutators vanish due to the commutation relations of the conformal algebra. Evaluating \eqref{eqn:commutators} for the different generators, applying the result to $\Phi(0)$ using \eqref{eqn:action-of-little-group-generators} and finally translating $\Phi(0)$ to $\Phi(x)$ using the remaining factor of $\exp(i \mathbb{P}\cdot x)$ in \eqref{eqn:transaltion-trick} we find
\begin{align}\label{eqn:infinitesimal-conformal-transformations-of-fields}
\mathbb{P}_\mu \Phi(x) &= - i \partial_\mu \Phi(x)\,, \\ \nn
\mathbb{D}\Phi(x) &=\left( i x^\mu\partial_\mu + \tilde\Delta \right) \Phi(x)\,,  \\ \nn
\mathbb{M}_{\mu\nu} \Phi(x) &= \left( -i(x_\mu \partial_\nu - x_\nu \partial_\mu) + \Sigma_{\mu\nu} \right) \Phi(x)\,, \\ \nn
\mathbb{K}_\mu \Phi(x) &=\left(\kappa_\mu- 2x_\mu \tilde\Delta + 2 x^\nu \Sigma_{\mu\nu} -i ( 2 x_\mu (x\cdot \partial)-x^2 \partial_\mu ) \right)\Phi(x)\,.
\end{align}
The operators that we consider here satisfy $\mathbb{K}_\mu \Phi(0)=0$, i.e. we choose $\kappa_\mu=0$ in \eqref{eqn:action-of-little-group-generators}. In principle $\kappa_\mu\neq 0$ and nilpotent \cite{Mack:1969rr} is possible  as well.

Since $[\mathbb{M}_{\mu\nu},\mathbb{D}]=0$ we have $[\Sigma_{\mu\nu},\tilde{\Delta}]=0$ and it follows by Schur's lemma, that any matrix that commutes with all generators $\Sigma_{\mu\nu}$ must be a multiple of the identity, and thus we write $\tilde{\Delta}=i\Delta$, where $\Delta$ is the  \emph{scaling dimension} of the field. \index{scaling dimension}

\subsection{Finite Conformal Transformations of Fields}
In \eqref{eqn:infinitesimal-conformal-transformations-of-fields} we have written the transformation of the fields under infinitesimal conformal transformations.  The transformation law for a scalar field $\phi$ under finite  conformal transformations $x \to x^\prime$ reads
\begin{align}\label{eqn:trafo-primary}
\phi^\prime(x^\prime) = \left| \frac{\partial x^\prime}{\partial x} \right|^{-\Delta/d}\phi(x)\,.
\end{align}
Under a finite scaling $x \to x^\prime = \lambda x$ \eqref{eqn:finite-conformal-trafos} the field thus transforms as $\phi^\prime = \lambda^{-\Delta}\phi$, which explains the name \emph{scaling dimension}.
The form of finite transformations for a general field with indices $\alpha$ reads
\begin{align}\label{eqn:finite-trafo-primary-general}
\Phi^\prime_\alpha(x^\prime) = \left| \frac{\partial x^\prime}{\partial x} \right|^{-\Delta/d} D_{\alpha}^{~\beta} (R)\, \Phi_\beta(x)\,,
\end{align}
where $D_{\alpha}^{~\beta}$ is the appropriate representation of the Lorentz transformation
\begin{equation}
R_\mu^{~\nu}= \rho^{-1/2}(x) \frac{\partial x^\prime_\mu}{\partial x_\nu}
\end{equation} 
and $\rho(x)$ is the scale factor of the conformal transformation \eqref{eqn:conformal-trafo-of-metric}. For more information see e.g. \cite{Osborn:1993cr,Erdmenger:1996yc,Jackiw:2011vz}. 

\subsection{Conformal Primary Fields and Operators}\index{conformal primary!field}\label{sec:conformal-primaries}
Fields transforming according to \eqref{eqn:trafo-primary}, \eqref{eqn:finite-trafo-primary-general} are called \emph{primary fields}\footnote{More precisely, in general dimension $d$ it is called a \emph{quasi-primary field}\index{primary!quasi-primary field}. The term \emph{primary field} is reserved for fields in $d=2$ which have a more specific transformation law, but are at the same time \emph{quasi-primary fields}, since they automatically satisfy the transformation laws of these.}\index{primary!field} or \emph{conformal primaries}. 
Correspondingly, a \emph{conformal primary operator}\index{primary!operator}\index{conformal primary!operator} is a local operator which transforms according to the above laws under conformal transformations. Recall, that we chose $\kappa_\mu=0$ to deduce these transformation laws, i.e. the definition of a conformal primary requires
\begin{equation}
\mathbb{K}_\mu \op(0) =0\,,
\end{equation}
where the scaling dimension of the primary is given by $\mathbb{D}\, \op(0) = i \Delta \op(0)$. From the commutation relation of the conformal algebra it follows that an operator $\op^\prime$ related to a primary operator via $\op^\prime(x)=\mathbb{P}\,\op(x)$ has the scaling dimension of $\op$ raised by one
\begin{equation}
\mathbb{D}\,\op^\prime (0)= [\mathbb{D},\mathbb{P}]\op(0) + \mathbb{P}(\mathbb{D}\,\op(0)) = i(1+\Delta)\mathbb{P}\,\op(0) = i \Delta^\prime \op^\prime(0)\,.
\end{equation} 
The operator $\op^\prime$ is called a \emph{descendant} of $\op$. Conversely, $\mathbb{K}$ lowers the dimension of an operator by 1 and thus the conformal primaries are the operators with lowest scaling dimensions.
In general it suffices to investigate the properties of conformal primaries, since relations for the descendants follow from the conformal algebra. \index{descendants}

\section{Classically Conformal Field Theories}\label{eqn:classically conformal symmetry}
A field theory is classically invariant if the action is invariant under the field transformations \eqref{eqn:infinitesimal-conformal-transformations-of-fields}. Due to the commutation relations  \eqref{eqn:remaining-conformal-algebra}  the mass operator  $M^2=P^2$ does not commute with conformal transformations and dilatations act as
\begin{equation}
e^{i\alpha D} P^2 e^{- i \alpha D} = e^{2\alpha} P^2\,.
\end{equation}
Therefore invariance under conformal transformations, either requires a continuous mass spectrum  or a massless theory. We will restrict to the latter ones in this thesis.
Furthermore, in  \cite{Mack:1969rr} it was derived that all non-derivative couplings with dimensionless coupling constant as well as all couplings of Yang-Mills type are conformally invariant. 

Therefore, at the classical level the only term in the standard model\footnote{assuming massless neutrinos} which is not invariant under conformal transformations is the term with the Higgs mass. An alternative for a classically conformal invariant standard model is discussed in \cite{Meissner:2006zh,Meissner:2007xv}.

\subsubsection{Energy momentum tensor}\index{energy momentum tensor}
The Noether theorem implies that the symmetries of a field theory are reflected by the existence of conserved currents $J^\mu$, e.g. the conserved current with respect to translations is the canoncical energy momentum tensor $T^{\mu\nu}_c$. The conserved currents associated with the invariance of the action are only defined up to total divergences and can thus be redefined. In particular, it is often possible \cite{Callan:1970ze} to construct a symmetric, divergenceless energy momentum tensor $\theta^{\mu\nu}$, which allows to write the associated currents of all conformal transformations as
\begin{equation}\label{eqn:redef-energy-mom}
J^\mu = k_\nu \theta^{\mu\nu},\qquad \theta^{\mu\nu}=\theta^{\nu\mu}, \qquad \partial_\mu \theta^{\mu\nu} = 0\,,
\end{equation}
where the conformal Killing vectors $k^\mu$ are defined in \eqref{eqn:Killing-vector}.
Using \eqref{eqn:conformal-killing-equation} the conservation of this current  can be written as
\begin{equation}
\partial_\mu J^\mu = \frac{1}{2}(\partial_\mu k_\nu+\partial_\nu k_\mu) \theta^{\mu\nu} + k_\nu \partial_\mu \theta^{\mu\nu} = \frac{1}{d} (\partial \cdot k)\,\theta^\mu_{~\mu}
\end{equation}
and thus the conservation of the current -- or the invariance of the action -- can be attributed to the tracelessness $\theta^{\mu}_{~\mu}=0$ of the energy momentum tensor. This symmetry may be broken in the quantum theory and the violation of the tracelessness is called the \emph{conformal anomaly}\index{conformal anomaly} or \emph{trace anomaly}\index{trace anomaly}.  

A detailed account on this subject can be found in section 2.4 of \cite{Braun:2003rp}. There, it is also explained that the violation of the dilatation symmetry in  massless QCD  is given by $\partial_\mu J^\mu_D=\beta(g)/(2g)G_{\mu\nu}^a G^{a,\mu\nu} $ where $\beta$ is the renormalisation group $\beta$-function and $G_{\mu\nu}^a$ the field strength.

\subsubsection{Simple classically conformal field theories}
As the simplest example, consider a scalar theory in $d$ dimensions, with interaction term $g\phi^n$ where $g$ is the coupling constant
\begin{equation}
S=\int d^dx\, \left(\partial_\mu \phi \partial^\mu \phi + g\phi^n \right)\,.
\end{equation}
The inverse length dimension $[d^dx]=-d$ of the measure must be compensated by the weight of the Lagrangian for the action to be dimensionless. Therefore, the dimension of the scalar field is $[\phi]= (d-2)/2$ and the requirement, that $g$ is dimensionless leads to the conclusion that $n=2d/(d-2)$. Therefore, in $d=4$ dimensions the only allowed interaction term is $\phi^4$ and in $d=6$ it is $\phi^3$.

\subsubsection{Scalar $\phi^3$-theory in six dimensions}
The latter  $(\phi^3)_{6}$ theory  is particularly interesting, because generalised to $N_f$ complex scalar fields $\phi_i$ with an interaction term with one real scalar $\chi$
\begin{equation}\label{eqn:phi3-model}
\mathcal{L}=\sum_{i=1}^{N_f} \partial_\mu \phi \partial^\mu \bar{\phi}_i + \frac{1}{2}\partial_\mu \chi \partial^\mu \chi + g \sum_{i=1}^{N_f} \phi_i \bar{\phi}_i \chi\,,
\end{equation}
it has a number of features in common with QCD in four dimensions. For example, its $\beta$-function has a structure similar to QCD and for a certain number $N_f$ of flavours  \cite{Mikhailov:1984cp}, it can be made to vanish up a certain order in the coupling, which allows to use conformal symmetry arguments up to this order. The similarity of this theory to $\sym$ will become clear in chapter \ref{sec:three-point-functions}.

\subsubsection{Chern-Simons theory}
Another example of a classically conformal field theory is Chern-Simons theory in three dimensions
\begin{equation}\label{eqn:Chern-Simons-theory-action}
\mathcal{S}_{\text{CS}}(A) = \frac{k}{4\pi} \int d^3x \epsilon^{\mu\nu\rho} \, \tr \left( A_{\mu}\partial_\nu A_{\rho} - i\frac{2}{3} A_{\mu}A_\nu A_{\rho}\right)
\end{equation}
and we will deal with the conformal properties of the theory in this thesis.

\subsubsection{Scale without Conformal Invariance}
In two dimensions scale invariance and unitarity of the theory imply conformal invariance \cite{Polchinski:1987dy}. However in $d>2$ it is possible to have scale invariant theories which are not invariant under conformal transformations  \cite{Polchinski:1987dy,Jackiw:2011vz,ElShowk:2011gz,Fortin:2011ks,Fortin:2011sz,Nakayama:2010zz} and thus do not allow for \eqref{eqn:redef-energy-mom}. The converse is however always true, conformal invariance implies scale invariance.

\section{Superconformal Symmetry}\label{sec:superconformal-symmetry}
The Poincar\'e algebra \eqref{eqn:poincare-algebra} can be extended to a \emph{super-Poincar\'e algebra}\index{Poincar\'e algebra!super-Poincar\'e algebra} by the introduction of fermionic supersymmetry\index{supersymmetry} generators $Q_\alpha^A$ and their conjugates $\bar{Q}_{\dot\alpha}^A$ which satisfy the anticommutation relations
\begin{equation}\label{eqn:poincare-super-charges}
\{Q_\alpha^A, \bar{Q}_{\dot\alpha}^B\} = 2 (\sigma^\mu)_{\alpha\dot\alpha} P_\mu \delta^{AB} \,,
\end{equation}
where $A=1,...,\mathcal{N}$ labels the number of supersymmetry generators $Q_\alpha^A$, $\sigma^\mu=({\bf 1},\vec{\sigma})$ are the Pauli matrices and $\alpha, \dot\alpha =1,2$ are the $su(2)$ Lorentz indices.   Additionally, we have a so-called \emph{R-symmetry}\index{R-symmetry}  group among which the supercharges transform. In the Lagrangian of the theory, this is reflected by a flavour symmetry of the fields.
Introductions to supersymmetry can be found e.g. in  \cite{Sohnius:1985qm,Wess:1992cp,West:1990tg,Bilal:2001nv}.

Analogously, a conformal symmetry algebra is extended to a \emph{superconformal algebra} by the introduction of \eqref{eqn:poincare-super-charges}, which due to the non-trivial commutation relations of the special conformal generators $K_\mu$ and the Poincar\'e supercharges $Q^A_{\dot\alpha}$ additionally contains the so-called \emph{conformal supercharges}  $S^A_{\alpha}$ and their conjugates  $\bar{S}^A_{\dot\alpha}$ satisfying
\begin{equation}\label{eqn:conformal-super-charges}
\{S_\alpha^A, \bar{S}_{\dot\alpha}^B\} = 2 (\sigma^\mu)_{\alpha\dot\alpha} K_\mu \delta^{AB} \,.
\end{equation}
Corresponding to the definition of conformal primaries in section \ref{sec:conformal-primaries}, \emph{superconformal primaries} are annihilated by the special conformal supercharges $S$.
More information on the superconformal algebra can be found e.g.  in chapter 3 of the review \cite{D'Hoker:2002aw}. 

%{\bf ::Comment: fixed anomalous dimensions in supermultiplet \cite{Belitsky:2003sh,Belitsky:2005gr}}

\subsection{$\Syml$ Theory}\label{sec:superconformal-algebra}
As implied by its name  the $\syml$ theory in $d=4$ dimensions has four supercharges  with associated \emph{$R$-symmetry} algebra $su(4)$. 
The Lagrangian can be derived by dimensionally reducing $\mathcal{N}=1$ super Yang-Mills theory in $D=10$ dimensions to $d=4$. We review this derivation in Appendix \ref{sec:N-4-SYM-from-reduction} and the resulting Lagrangian is given by\footnote{We omit gauge fixing and ghost terms here.}

\begin{align}\label{eqn:N=4SYM-Lagrangian} 
\mathcal{L}_{\text{SYM}} =   ~\tr \Big( &- \frac{1}{2} F_{\mu\nu}F^{\mu\nu} + \frac{1}{2} D_\mu \phi^{AB} D^\mu \bar{\phi}_{AB} + \frac{1}{8} g^2 [\phi^{AB},\phi^{CD}][\bar{\phi}_{AB},\bar{\phi}_{CD}] \\ \nn
\phantom{\tr} &+  2 i \bar{\lambda}_{\dot{\alpha}A} (\sigma_\mu)^{\dot{\alpha}\beta} D^\mu \lambda_\beta^A -\sqrt{2} g \lambda^{\alpha A} [\bar{\phi}_{AB},\lambda_\alpha^B] + \sqrt{2} g \bar{\lambda}_{\dot{\alpha}A}[\phi^{AB},\bar{\lambda}^{\dot{\alpha}}_B] \Big)\,,
\end{align}
where the field content comprises  four fermions $\lambda_{\alpha A}$, three independent complex scalars $\phi^{AB}$ which are related to their complex conjugates via $\bar{\phi}_{AB} = (\phi^{AB} )^* = \frac{1}{2}\epsilon_{ABCD} \phi^{CD}$ and a gauge field $A_\mu$ with field strength and covariant derivative
\begin{align}
F_{\mu\nu} &= \partial_\mu A_\nu-\partial_\nu A_\mu - i g  [ A_\mu , A_\nu],\qquad
D_\mu (\cdot) = \partial_\mu (\cdot) - i g [ A_\mu , (\cdot)]\,.
\end{align}
All fields are matrix valued $A_\mu=A_\mu^a T^a$ and $T^a$ are the generators of $SU(N)$ in the fundamental representation.
In this formulation the $SU(4)$ R-symmetry is manifest in the indices $A,B=1,...,4$. Very often, in the literature one encounters an equivalent formulation of $\sym$, with a different definition of the scalar fields, where the matrices $\Sigma$ in \eqref{eqn:definition-complex-scalar-fields} are not incorporated into the definition of the scalars but remain in the Lagrangian.
Then the $R$-symmetry is an $SO(6) \sim SU(4)$ rotation symmetry of 6 real scalars $\phi_i$. In this approach the real scalar fields are usually combined to complex fields as $Z=\phi_1+i\phi_2, W=\phi_3+i\phi_4, X=\phi_5+i\phi_6$. The approaches are of course completely equivalent, but the calculations slightly differ by the absorbed or non-absorbed $\Sigma$ matrices.

\subsubsection{Coupling Constant Renormalisation and $\beta$-Function}
The Lagrangian obviously satisfies the restrictions of section \ref{eqn:classically conformal symmetry} for classical conformal symmetry. Moreover, it turns out that the renormalisation group $\beta$-function of the theory vanishes in perturbation theory, which is believed to be true to all orders in the coupling constant \cite{Sohnius:1981sn,Mandelstam:1982cb,Brink:1982pd,Howe:1983sr,Sohnius:1985qm}. For an interesting discussion on the arguments of the all-loop vanishing of the $\beta$-function we refer the reader to chapter 13.2 of \cite{Sohnius:1985qm}.
Therefore, the theory remains exactly scale invariant at the quantum level. Thus also at loop-level conformal symmetry is not broken and restricts correlation functions as we will discuss in detail in section \ref{eqn:implications-of-conformal-symmetry}.

\subsubsection{Superalgebra $su(2,2|4)$}
In $\sym$ the bosonic symmetry algebra, which consists of the conformal algebra \eqref{eqn:so(2,d)} in four dimensions $so(2,4) \simeq su(2,2)$ and the $R$-symmetry algebra $su(4)$, is lifted to the superalgebra $psu(2,2|4)$.  The algebra $psu(2,2|4)$ plays an important role in the $\text{AdS}_5/\text{CFT}_4$ correspondence and a detailed account on the properties of the algebra as well as the isometries of the $AdS_5\times S^5$ superspace can be found in \cite{Beisert:2010kp}. 

For the formulation of the full superconformal symmetry algebra, it is useful to write all generators  in $su(2)$ notation, e.g. the Lorentz generators are split  as $\mathbb{M}_{\alpha\beta}$, $\overline{\mathbb{M}}_{\dot\alpha\dot\beta}$ acting on $su(2)_L$ respectively $su(2)_R$ indices.

For further reference, we adopt the formulation of the non-trivial commutation relations of the generators of the algebra\footnote{We omit the  discussion of the hypercharge $\mathcal{B}$. Including it, the resulting algebras were $u(2,2|4)$, $pu(2,2|4)$. A detailed discussion can be found in \cite{Beisert:2004ry}.} $su(2,2|4)$ from \cite{Drummond:2008vq}. The Lorentz generators
$\mathbb{M}_{\a \b}$, $\overline{\mathbb{M}}_{\da \db}$ and the $su(4)$ \emph{R-symmetry} generators
$\mathbb{R}^{A}_{~B}$ act canonically on the remaining generators carrying Lorentz or $su(4)$
indices, as explicitly given in Appendix D of \cite{Beisert:2004ry}. The dilatation $\mathbb{D}$ acts via 
\begin{equation}
[\mathbb{D},\mathbb{J}] = {\rm dim}(\mathbb{J})\mathbb{J}\,,
\end{equation}
where the non-zero dimensions of the various generators are
\begin{align} \notag
& {\rm dim}(\mathbb{P})=1, 
\quad &{\rm dim}(\mathbb{K})=-1, 
\quad {\rm dim}(\mathbb{Q}) = {\rm dim}(\overline{\mathbb{Q}}) =
\tfrac{1}{2}, 
\quad {\rm dim}(\mathbb{S}) = {\rm dim}(\overline{\mathbb{S}}) = -\tfrac{1}{2}\,.
\end{align}
The remaining non-trivial commutation relations are,
\begin{align} \label{eqn:superconformal-algebra-n4sym}
& \{\mathbb{Q}_{\a A},\overline{\mathbb{Q}}_{\da}^B\}  =  \delta_A^B \mathbb{P}_{\a \da},
   \qquad \{\mathbb{S}_{\a A},\overline{\mathbb{S}}_{\da}^B \} = \delta_A^B \mathbb{K}_{\a \da},
\\ \notag
& {}[\mathbb{P}_{\a \da},\mathbb{S}_{\b}^A] = \epsilon_{\da \db} \overline{\mathbb{Q}}_{\da}^A,
 \qqquad [\mathbb{K}_{\a \da},\mathbb{Q}_{\b A}] = \epsilon_{\a \b} \overline{\mathbb{S}}_{\da A},
\\ \notag
& {}[\mathbb{P}_{\a \da},\overline{\mathbb{S}}_{\db A}]  =  \epsilon_{\da \db} \mathbb{Q}_{\a A},
\qquad [\mathbb{K}_{\a \da}, \overline{\mathbb{Q}}_{\db A}]  =  \epsilon_{\da \db} \mathbb{S}_{\a
A},
\\ \notag
& [\mathbb{K}_{\a \da},\mathbb{P}^{\b \db}] = \delta_\a^\b \delta_\da^\db \mathbb{D} +
\mathbb{M}_{\a}{}^{\b}
 \delta_\da^\db + \overline{\mathbb{M}}_{\da}{}^{\db} \delta_\a^\b,
\\ \notag
& \{\mathbb{Q}_{\a A},\mathbb{S}_\b^B\} = \epsilon_{\a \b} \mathbb{R}^{B}_{~A} + \mathbb{M}_{\a \b}
\delta_A^B + \epsilon_{\a \b}\delta_A^B (\mathbb{D}+\mathbb{C}),
\\ \nn
& \{\overline{\mathbb{Q}}_{\da}^{A},\overline{\mathbb{S}}_{\db B}\} = \epsilon_{\da \db}
\mathbb{R}^{A}_{~B} + \overline{\mathbb{M}}_{\da \db} \delta_B^A + \epsilon_{\da \db}\delta_B^A
(\mathbb{D}-\mathbb{C})\,,
\end{align}
where $\mathbb{C}$ is the central charge. If the central charge vanishes, the resulting algebra is $psu(2,2|4)$
The commutations relations of the \emph{super Lie algebra} can be summarised as
\begin{equation}
[ j_a,j_b \} = f_{ab}^{~~c} j_c\,,
\end{equation}
where $j_a$ is any of the generators and $f_{ab}^{~~c}$ are the structure constants of the algebra and $[.\,,.\}$ is the graded commutator. We will use this in the discussion of the \emph{Yangian algebra} in section \ref{sec:Yangian-symmetry}.

\subsection{$\Scs$ Theory}\label{eqn:super-chern-simons}
The relevant field theory in the $\text{AdS}_4/\text{CFT}_3$ correspondence is a three-dimensional $\mathcal{N}=6$ superconformal Chern-Simons theory, called ABJM theory \cite{Aharony:2008ug}. Schematically, it reads
\begin{align}\nn
S_{\text{ABJM}} = S_{\text{CS}}(A)-S_{\text{CS}}(\hat{A})+ S_{\text{matter}}\,,
\end{align}
where we have two copies of the Chern-Simons action \eqref{eqn:Chern-Simons-theory-action} with two different $U(N)$ gauge fields $A_\mu, \hat{A}_\mu$ and the fields in the matter part 
\begin{equation}
S_{\text{matter}}= \int d^3x\, \Tr(D_\mu\, \phi_I\, D^\mu {\bar \phi}^{I}) + i\, \Tr(\bar\psi^I
\, \slsh{D}\, \psi_I) + \mathcal{L}_\text{int}
\end{equation}
are in the bifundamental representation and thus carry an $SU(N)$ index for each of the two gauge groups, i.e. $(\phi_I)_{i\hat{i}},(\psi_I)_{\hat{i}i}$. The interaction term $\mathcal{L}_{int}$ schematically consists of Yukawa type terms $\phi^2 \psi^2$ as well as terms sextic in the scalars  $(\phi)^6$. We give more details in Appendix \ref{sec:Lagrangian of ABJM theory}.
For Chern-Simons level\footnote{For $k=1,2$ the algebra is enhanced to $osp(8|4)$.} $k>2$ the superconformal symmetry group is the orthosimplectic supergroup $OSp(6|4)$ its bosonic part being the $R$-symmetry group $SU(4) \simeq  SO(6) $ and the conformal group \eqref{eqn:so(2,d)} in three dimensions $SO(2,3) \simeq Sp(4)$. A reformulation with manifest $SU(4)$ $R$-symmetry and a very clear presentation of the superconformal symmetry can be found in \cite{Bandres:2008ry}. A detailed account on $OSp(6|4)$ can be found in \cite{Papathanasiou:2009en}.

The generalization of ABJM theory to different gauge groups $SU(N)_k, SU(M)_{-k}$ for the Chern-Simons actions of level $k$ respectively $-k$ is referred to as ABJ theory \cite{Aharony:2008gk}.

\chapter{Implications of Conformal Symmetry}\label{eqn:implications-of-conformal-symmetry}
Conformal symmetry imposes restrictions on observables in quantum field theory. These are due to \emph{conformal Ward identities}, which we review in section \ref{sec:conformal-Ward-id}.
In section \ref{eqn:correlation-functions} we show how this fixes the form of two-, three- and higher point correlation functions of scalar operators and in \ref{eqn:correlation-functions-spin} we generalise this to correlation functions involving operators with spin.
 In section \ref{eqn:renormalisation-of-composite-operators} we comment on the renormalisation properties of composite operators and introduce a \emph{conformal renormalisation scheme}.

\section{Conformal Ward Identities}\label{sec:conformal-Ward-id}
The symmetries of the quantum field theory translate into constraints on expectation values through so-called \emph{Ward identities}\index{Ward identities}\index{Ward identities!conformal}. The Ward identities can be derived by considering the change of the Green's function under a transformation of the fields
\begin{equation}\label{eqn:some-field-trafo}
\Phi \to \Phi^\prime  =\Phi + \delta \Phi
\end{equation}
as given by \eqref{eqn:infinitesimal-conformal-transformations-of-fields}. Writing the Green's function as a path integral
\begin{align}
G_n= \langle \Phi_1(x_1)...\Phi_n (x_n) \rangle &= \int \mathcal{D}\Phi\, e^{i S[\Phi]}\Phi_1(x_1)...\Phi_n(x_n) 
\end{align}
and performing a change of variables $\Phi \to \Phi^\prime$ in the functional integral,  we find\footnote{Note that we do not take into account the Jacobi determinant in the functional integral, since it is a c-number and does not modify the Ward identity, for more information see \cite{Sarkar:1974xh,Mueller:1993hg,Braun:2003rp}. Furthermore, we supress a normalisation factor of the path integral here.}
\begin{align}
 \langle \Phi_1(x_1)...\Phi_n (x_n) \rangle &= \int \mathcal{D}\Phi^\prime\, e^{i S[\Phi^\prime]}\Phi^\prime_1(x_1)...\Phi^\prime_n(x_n) \\ \nn
 &=  \int \mathcal{D}\Phi\, e^{i S[\Phi+\delta \Phi]}(\Phi_1+\delta \Phi_1)...(\Phi_n+\delta \Phi_n) \,.
\end{align}
Considering a conformal transformation \eqref{eqn:some-field-trafo}, the action of a conformal field theory is classically invariant, $\delta S =0 $ and infinitesimally we get the Ward identity
\begin{align}\label{eqn:Ward-identity}
\sum_{i=1}^n  \langle \Phi_1(x_1)...\delta \Phi_i (x_i) ...\Phi_n(x_n) \rangle = 0\,.
\end{align}
This identity imposes restrictions on the the correlation function. Considering for instance translations and Lorentz transformations, we conclude, that the correlation function must be a function of the translation and Lorentz invariant combinations $x_{ij}^2:=(x_i-x_j)^2$ of the space-time points. Using  \eqref{eqn:infinitesimal-conformal-transformations-of-fields} the dilatation Ward identity reads
\begin{align}\label{eqn:dilatation-Ward-identity-general}
\sum_{i=1}^n (x_i \cdot \partial_i + \Delta_i)  \langle \Phi_1(x_1)...\Phi_i (x_i) ...\Phi_n(x_n) \rangle = 0\,,
\end{align}
where $\Delta_i$ is the scaling dimension of the field $\Phi_i$. This \emph{dilatation Ward identity} and the corresponding \emph{special conformal Ward identity} 
\begin{align}\label{eqn:special-conformal-Ward-identity-general}
\sum_{i=1}^n \left(2x_i^\mu (x_i \cdot \partial_i + \Delta_i) + 2i \Sigma^\mu_\nu x_i^\nu -x_i^2 \partial_i^\mu \right)\langle \Phi_1(x_1)... \Phi_i (x_i) ...\Phi_n(x_n) \rangle = 0
\end{align}
constrain the space-time structure of Green's functions as we will see in the sections \ref{eqn:correlation-functions}, \ref{eqn:correlation-functions-spin}.

\subsection{Anomalous Conformal Ward Identity}\index{anomalous Ward identities}
If the action $S$ is not invariant under conformal transformations, we get an \emph{anomalous conformal Ward identity}\index{conformal Ward identity!anomalous}\index{Ward identities!anomalous}
\begin{align}\label{eqn:general-anomalous-conformal-ward-identity}
\sum_{i=1}^n  \langle \Phi_1(x_1)...\delta \Phi_i (x_i) ...\Phi_n(x_n) \rangle = -i  \langle \delta S \Phi_1(x_1)...\Phi_n(x_n) \rangle \,,
\end{align}
where $\delta S$ is the variation of the action under the transformation \eqref{eqn:some-field-trafo}. This is for example the case, if the conformal symmetry of the action is broken through a regulator, e.g. in dimensional regularisation. Then, the weight of the measure $d^dx$ in $d=4-2\epsilon$ does not match the weight of the Lagrangian and thus the action $S=\int d^dx\, \mathcal{L} $ is only invariant up to order epsilon terms under dilatations and special conformal transformations. Note that dimensional regularization preserves Poincar\'e invariance and thus the statement that the expectation values must be functions of $x_{ij}^2$ remains true.

It turns out, that these anomalous conformal Ward identities are particularly useful for the evaluation of the Wilson loop as will be explained in sections \ref{sec:ward-identity-wilson-loops} and \ref{sec:anomalous-ward-identities}. More details on conformal Ward identities\index{conformal Ward identity} can be found in \cite{Sarkar:1974xh,Mueller:1993hg,Mueller:1997ak,Belitsky:1998gc} and the review \cite{Braun:2003rp}.

\subsection{Anomalous Dimensions}\label{sec:anomalous dimension}\index{anomalous dimension}\index{scaling dimension!anomalous}
Consider a Green's function of $n$ fields $\Phi(x)$ of classical scaling dimension $\Delta$. Due to divergences, we have to renormalise the fields at loop-level. In a theory with vanishing $\beta$-function, the relations \eqref{eqn:dilatation-Ward-identity-general}, \eqref{eqn:special-conformal-Ward-identity-general} remain valid for the renormalised Green's function up to a change in the scaling dimension: 
\begin{equation}\label{eqn:replacement-anomalous-dimension}
\Delta \to \Delta + \gamma(g)\,.
\end{equation}
It acquires an \emph{anomalous dimension $\gamma(g)$}\index{scaling dimension}\index{anomalous dimension} which has an expansion in the coupling constant $g$. This can be seen as follows.

For dimensional reasons, the renormalised Green's function may be written in a form where the canonical dimension is uniquely carried by the scale $\mu$ and a dimensionless function $G_n$, depending on the dimensionless quantities\footnote{and the dimensionless coupling.} $\mu^2x_{ij}^2$
\begin{equation}
\langle \Phi^R(x_1)...\Phi^R(x_n) \rangle =  \mu^{n \cdot \Delta} G_n(\mu^2 x_{ij}^2)\,.
\end{equation}
We can thus write
\begin{equation}\label{eqn:dilatation-ward-id-renormalised}
\sum_{i=1}^n \left(\Delta + x_i \cdot \partial_i \right)  \langle \Phi^R(x_1)...\Phi^R(x_n) \rangle = \mu \frac{\partial}{\partial \mu}  \langle \Phi^R(x_1)...\Phi^R(x_n) \rangle\,,
\end{equation}
which is the equivalent \cite{Mueller:1997ak,Belitsky:1998gc,Braun:2003rp} of the ordinary Callan-Symanzik equation 
\begin{equation}
\left[ \mu \frac{\partial}{\partial \mu}+\beta\frac{\partial}{\partial g} + n \gamma  \right]  \langle \Phi^R(x_1)...\Phi^R(x_n) \rangle = 0\,.
\end{equation}
For a theory with vanishing $\beta$-function or at a fixed point of the renormalisation group $\beta(g^*)=0$, we can thus rewrite \eqref{eqn:dilatation-ward-id-renormalised}
\begin{equation}
\sum_{i=1}^n \left(\Delta + \gamma + x_i \cdot \partial_i \right)  \langle \Phi^R(x_1)...\Phi^R(x_n) \rangle = 0\,.
\end{equation}
Therefore, using the replacement \eqref{eqn:replacement-anomalous-dimension}, the dilatation Ward identity can be brought to the simple form \eqref{eqn:dilatation-Ward-identity-general} that it has in a free theory. The special conformal Ward identity can also be brought to the form \eqref{eqn:special-conformal-Ward-identity-general}  under the same replacement, see e.g. \cite{Braun:2003rp} for details. 

This has the important consequence, that the restrictions on the Green's functions imposed by the free Ward-identities also apply in the interacting case, i.e. the space-time structure of the Green's functions is fixed in the same way as in the free case under the replacement \eqref{eqn:replacement-anomalous-dimension}. As we will see in the following sections, this uniquely fixes the structure of two and three-point functions.

%\subsubsection{Unitarity bounds for anomalous dimensions}
%Bound on anomalous dimension follows from the Wightman positivity condition or the unitarity condition for the conformal (or just Poincar\'e?) group, see I.2. in \cite{Fradkin:1996is}.
%\begin{equation}
% d \geq 2(h-1)+j
%\end{equation}

\subsubsection{Composite operators, mixing and conformal scheme}
The same statements are true for composite operators of local fields $\op(x)=\Phi(x)...\Phi(x)$. Even though the anomalous Ward-identities for the renormalised composite operators are not diagonal in the $\overline{\text{MS}}$-scheme, i.e. under renormalisation the operators mix with other operators, one can recover conformally covariant operators in the \emph{conformal scheme} as will be explained in section \ref{conformal scheme}.

\subsection{Protected Operators}\label{sec:protected-operators}
Operators that do not get quantum corrections and consequently do not acquire anomalous dimensions are called \emph{protected operators}\index{protected operators}.

\emph{Chiral primary operators} or \emph{BPS operators}\index{protected operators!chiral primary}\index{protected operators!BPS}\index{chiral primary operators}\index{BPS operators}   commute with a certain number of supercharges and are protected from quantum corrections. According to the amount of supercharges that they commute with, they are called $1/2$ BPS, $1/4$ BPS, $1/8$ BPS operators. We will make use of this type of operators  in chapter \ref{sec:three-point-functions}.
A detailed discussion of BPS operators can be found in section 3.4 of the review \cite{D'Hoker:2002aw}.

%\subsubsection{Current operators}
%Operators that are conserved currents are uniqeuly normalised by the charge integral. Thus the anomalous dimension of these must be zero. (see \cite{Peskin:1995ev}, page 430). The same argument applies to the energy-momentum tensor. too manu subtleties in gauge theories, omit this, discussion too long.

\section{Correlation Functions of Scalar Operators}\label{eqn:correlation-functions}
The form of two-, three- and four-point functions of conformal primary operators \eqref{eqn:trafo-primary} was originally found by Polyakov \cite{Polyakov:1970xd}. The form of the correlators can be deduced by requiring the invariance of the correlation functions under dilatations and inversions, which by virtue of \eqref{eqn:I-P-I gives K} guarantees the invariance under special conformal transformations. We give the results in the following sections and review a very simple way to derive the functional form of the correlators in detail in Appendix \ref{sec:correlators-in-conformal-field-theory}.

\subsection{Two-Point Functions}
Using the invariance under dilatations and inversions, one can deduce that the two-point function of two scalar conformal primary operators with scaling dimensions $\Delta_A$, $\Delta_B$ is
\begin{equation}\label{eqn:conformal-two-point-function}
\langle  \op_A(x_1)\op_B(x_2) \rangle =  \delta_{AB} \frac{C_A}{(x_{12}^2)^{\Delta_A}}\,,
\end{equation}
where $C_A$ is a constant. We derive this result in detail in Appendix \ref{sec:derivation-two-point-function}. It is easy to see, that this two-point function indeed satisfies the Ward-identities  \eqref{eqn:dilatation-Ward-identity-general}, \eqref{eqn:special-conformal-Ward-identity-general}. 

\subsection{Three-Point Functions}
We show in \ref{sec:derivation-three-point-function} that three-point functions are also uniquely fixed to the form
\begin{equation}\label{eqn:conformal-three-point-function}
 \langle \op_A (x_1) \op_B (x_2) \op_C (x_3) \rangle = \frac{C_{ABC}}{|x_{12}|^{\Delta_A+\Delta_B-\Delta_C} |x_{13}|^{\Delta_A+\Delta_C-\Delta_B} |x_{23}|^{\Delta_B+\Delta_C-\Delta_A}}\,,
\end{equation}
where $C_{ABC}$ are called the \emph{structure constants}\index{structure constant}.  As will be explained in section \ref{sec:OPE-in-CFT} these structure constants are same constants that appear in the operator product expansion. It is easy to see, that these three-point functions indeed satisfy the Ward-identities  \eqref{eqn:dilatation-Ward-identity-general}, \eqref{eqn:special-conformal-Ward-identity-general}. 

\subsection{Higher-Point Functions}
The space-time structure of four- and higher-point correlators is not uniquely fixed by conformal symmetry, since they may depend on arbitrary functions of the conformally invariant cross ratios $u_{ijkl}$ in \eqref{eqn:cross-ratios}.  

At four points we can build two of these cross ratios 
\begin{equation}
u= \frac{x_{12}^2 x_{34}^2}{x_{13}^2 x_{24}^2}, \qquad v= \frac{x_{14}^2 x_{23}^2}{x_{13}^2 x_{24}^2}
\end{equation} and for example the four-point function of  four scalar fields has the form
\begin{equation}\label{eqn:four-point-functions}
 \langle \phi(x_1) \phi(x_2) \phi(x_3) \phi(x_4) \rangle = \frac{1}{(x_{12}^2 x_{34}^2)^{\Delta_\phi}} \mathcal{G}(u,v)\,,
\end{equation}
where $\mathcal{G}(u,v)$ is an a priori arbitrary function of the conformal cross ratios $u$ and $v$. It is possible to extract information on the structure of four- and higher-point functions using a \emph{double operator product expansion} \cite{Ferrara:1973vz,Dolan:2000ut,Arutyunov:2000ku,Dolan:2001tt,Dolan:2004iy}, see section \ref{sec:OPE-in-CFT} for a short introduction to the operator product expansion\index{operator product expansion}.

\section{Correlation Functions of Operators with Spin}\label{eqn:correlation-functions-spin}
In this thesis we will also consider correlation functions with \emph{spin $j$} operators\footnote{See the comments in section \ref{eqn:spin-j-operators} for the notion of spin.} 
\begin{equation}\label{eqn:spin-j-operator}
\op_{\mu_1...\mu_j}(x)\,,
\end{equation} 
which are totally symmetric in the indices $\{\mu_i\}$ and traceless in the sense 
\begin{equation}
g^{\mu_1\mu_2}\op_{\mu_1...\mu_j}(x) =0\,.
\end{equation}
Due to the total symmetry of the operator, the tracelessness applies to every index pair. The \emph{twist}\index{twist} of these operators is defined as the difference of the scaling dimension $\Delta_j$ and the spin $j$ of the operator
\begin{equation}\label{eqn:twist}
\theta = \Delta_j - j\,,
\end{equation}
where $\Delta_j$ is the scaling dimension of the spin $j$ operator.

\subsection{Two-Point Functions of Spin $j$ Operators}
Using the transformation law under dilatations and inversions one can show, see Appendix \eqref{sec:transformations-of-scalar-conformal-primaries-inversions}, that two-point functions of spin $j$ operators must be of the form
\begin{equation}\label{eqn:two-point-function-spin-j}
\langle \op_{\mu_1...\mu_j}(x_1) \op_{\nu_1...\nu_j}(x_2)  \rangle = \frac{C_j}{(x_{12}^2)^{\Delta_j}} \left( I_{\{\mu_1\nu_1}...I_{\mu_j\}\nu_j}(x_{12}) - \text{traces} \right)\,,
\end{equation}
where $\{..\}$ indicates that the right-hand side is totally symmetrised in all indices $\mu_k, \nu_l$ and we subtract all traces in $\mu_i$ resp. $\nu_i$ in accord with the symmetries of the operators on the left-hand side and where $I_{\mu\nu}(x)$ is the inversion tensor
\begin{equation}\label{eqn:Imunu}
I_{\mu\nu}(x)=\eta_{\mu\nu} - 2\frac{x_\mu x_\nu}{x^2}\,.
\end{equation}
\subsection{Three-Point Functions with Spin $j$ Operators}
Consider a three-point function involving two scalar operators $\op_A(x),\op_B(x)$ and one spin $j$ operator $\op_{\mu_1...\mu_j}$. Conformal symmetry restricts these correlators, see Appendix \ref{sec:three-point-function-with-spin}, to the form
\begin{align}\label{eqn:structure-three-point-function-spin-j}
\langle \op_A(x_1) \op_B(x_2)  \op_{\mu_1...\mu_j}(x_3) \rangle = \frac{ C_{ABj} \left(Y_{\mu_1}...Y_{\mu_j}(x_{13},x_{23}) -\text{traces}\right)}{|x_{12}|^{\Delta_A+\Delta_B-\theta}|x_{13}|^{\Delta_A+\theta-\Delta_B}|x_{23}|^{\Delta_B+\theta-\Delta_A}}\,,
\end{align}
where $\theta=\Delta_j-j$ is the \emph{twist}\index{twist} of the operator $\op_{\mu_1..\mu_j}$ and
\begin{equation}
Y^\mu (x_{13},x_{23}) = \frac{x_{13}^\mu}{x_{13}^2} - \frac{x_{23}^\mu}{x_{23}^2} = \frac{1}{2}\partial_{x_3}^\mu \ln \left(\frac{x_{13}^2}{x_{23}^2}  \right)\,.
\end{equation}
In agreement with the operators on the left-hand side of \eqref{eqn:structure-three-point-function-spin-j}, the right-hand side is symmetric and traceless in the indices.

\subsubsection{Three-point functions involving two operators with spin}
The reason that conformal symmetry restricts \eqref{eqn:structure-three-point-function-spin-j} up to the structure constant $C_{ABj}$ is, that there is only one possible tensor structure with the correct behaviour under dilatations and inversions. When we have one spin $j$ field $\op_{\mu_1..\mu_j}$ a scalar operator $\op_C$ and a vector operator $\op_\mu$ we have two possible tensor structures, which are compatible with the constraints imposed by dilatation and inversion, see e.g. \cite{Fradkin:1996is}, and thus conformal symmetry restricts the correlator to
\begin{align}\label{eqn:three-point-two-spin-operators}
\langle \op_{\mu_1...\mu_j} (x_1) \op_C(x_2) \op_\mu(x_3) \rangle
&= \frac{1}{|x_{12}|^{\theta_j+\Delta_C -\theta_1}|x_{13}|^{\theta_j+\theta_1-\Delta_C}|x_{23}|^{\Delta_C +\theta_1-\theta_j}} \times \\ \nn
\Big( &~A \,Y_\mu(x_{13},x_{23}) \left(Y_{\mu_1}  ...Y_{\mu_j}  (x_{12},x_{13})\right) \\ \nn
 + ~&\frac{B}{x_{13}^2} \left[ \sum_{k=1}^j I_{\mu\mu_k} (x_{13}) Y_{\mu_1}..Y_{\mu_{k-1}}Y_{\mu_{k+1}}..Y_{\mu_j}(x_{12},x_{13}) -\text{traces} \right]\Big)\,,
\end{align}
where $\theta_j$ is the twist \eqref{eqn:twist} of the spin $j$ operator and $\theta_1=\Delta_{\op_\mu}-1$ is analogously defined as the twist of the operator $\op_\mu$ and where $A$ and $B$ are constants. If we take a spin 2 operator $\op_{\mu\nu}$ instead of $\op_\mu$, there are 3 possible tensor structures and thus dilatation and inversion symmetry determine the correlator only up to these three constants. 
In general, the higher the rank of the tensors involved, the more possible independent tensor structures that fulfill the invariance conditions arise and each of them can enter with an a priori arbitrary coefficient.
For a more detailed discussion also involving correlators with fermionic fields we refer the reader to chapter III.2 of \cite{Fradkin:1996is} or the review  \cite{Fradkin:1997df} as well as the article \cite{Sotkov:1976xe}.

%~\\Review \cite{Fradkin:1997df} contains information on three-point correlators with more than one spin $j$ field, see also formula 2.139 in this review. :: Two operators with spin: have several poissible tensro structures, i.e. more than one structure constant. Ord do we get more restrcitions if we do not just do it with inversions as Palchik, Fradkin? E.g. do it like for the Lorentztrafos of vector fields $\sim \partial_\mu \phi(x)$, we could just derive $\Sigma_{\mu\nu}$ in this way.::

\subsubsection{Correlators in Momentum Space}
In general, it can be quite complicated to  map expressions like \eqref{eqn:structure-three-point-function-spin-j} to momentum space via Fourier transformation.  In this thesis, will only make use of the Fourier transformed correlators in a simplifying limit. However, if necessary, also the full position space correlator structures can be mapped to momentum space, see e.g. \cite{Coriano:2012wp} for some useful tricks and more details. 

\subsection{Light-Cone Projection of Operators}\label{sec:light-cone-projection}
Correlation functions of operators with spin \eqref{eqn:two-point-function-spin-j}, \eqref{eqn:structure-three-point-function-spin-j} contain many terms with a repeated space-time structure and therefore in some sense redundant information.
Perturbative calculations can be largely simplified by contracting the indices with light-like vectors, i.e.
\begin{equation}
  \hat{\op}_j =  \op_{\mu_1..\mu_j}z^{\mu_1}...\,z^{\mu_j} \qquad  \text{where} \qquad z^2 = z_{\mu}z^{\mu} =0\,.
\end{equation}
The operator with the desired symmetry $\op_{\mu_1..\mu_j}$ can be recovered by applying a second order differential operator $\Delta^\mu$
\begin{equation}
\Delta^\mu= \left(\frac{d}{2}-1+z\cdot \partial_z\right)\partial_z^\mu -\frac{1}{2} z^\mu \partial_z \cdot \partial_z
\end{equation}
in the presence of the constraint $z^2=0$, see \cite{Dobrev:1975ru,Bargmann:1977gy} and an application in \cite{Belitsky:2007jp}.  Then,  the correct operator can be recovered by
\begin{align}
 \Delta^{\mu_1}...\,\Delta^{\mu_j} \hat{\op}_j \sim   \op^{\mu_1..\mu_j}\,.
\end{align}
Due to $[\Delta^\mu,\Delta^\nu]=0$ and $\Delta^\mu \Delta_\mu=0$, symmetrisation and tracelessness are automatically incorporated. For more information on the light-cone projection see e.g. \cite{Belitsky:2003sh,Belitsky:2004sc}. \\

We denote spacetime indices that are contracted with $z^\mu$ with a \emph{hat}, i.e.
\begin{equation}
\hat{x} =  z_\mu x^\mu , \qquad \hat{\partial} = z_\mu \frac{\partial}{ \partial x_\mu} \qquad \Rightarrow \qquad \hat{\partial}  \hat{x} = z^2=0\,.
\end{equation}
As an example, consider the two-point function \eqref{eqn:two-point-function-spin-j}. The first term in \eqref{eqn:Imunu} as well as all trace terms vanish\footnote{We take the same light-like vector for both operators $\hat\op_j$, $\bar{\hat{\op}}_j$, i.e. $z_1^\mu=z_2^\mu$.} and we get\footnote{Here we have redefined $C_j$ and absorbed the combinatorical factor from the permutations in \eqref{eqn:two-point-function-spin-j}. In general we have $j!$ terms of type $I_{\mu_i\nu_j}$ and additionally all traces that are subtracted, which do however vanish in the lightcone-projection.}
\begin{equation}\label{eqn:two-point-function-spin-j-lighcone-projection}
\langle \hat{\op}_j(x_1) \hat{\op}_j(x_2)  \rangle = C_j (-2)^j \frac{(\hat{x}_{12})^{2j}}{(x_{12}^2)^{\Delta_j+j}} \,.
\end{equation}
Therefore, we have only one term instead of the rapidly growing number (with increasing $j$) of terms in \eqref{eqn:two-point-function-spin-j}. In perturbative calculations, this simplicity is reflected by the fact that $\hat{\partial} \hat{x} = z^2=0 
$ and therefore multiple derivatives $\hat{\partial}$ acting on propagators produce only one term. This largely reduces the effort for the determination of the structure constants $C_j, C_{ABj}$ of two-point and three-point functions with spin as we will see in a practical example in chapter \ref{sec:three-point-functions}. Analogously, the three-point  functions \eqref{eqn:structure-three-point-function-spin-j} read
\begin{align}\label{eqn:structure-three-point-function-spin-j-light-cone-projection}
\langle \op_A(x_1) \op_B(x_2)  \op_{\mu_1...\mu_j}(x_3) \rangle = \frac{ C_{ABj} \left(\hat{Y}(x_{13},x_{23}) \right)^j}{|x_{12}|^{\Delta_A+\Delta_B-\theta}|x_{13}|^{\Delta_A+\theta-\Delta_B}|x_{23}|^{\Delta_B+\theta-\Delta_A}}\,,
\end{align}
where $|x_{ij}|=(x_{ij}^2)^{1/2}$.
The generalization of \eqref{eqn:structure-three-point-function-spin-j}, \eqref{eqn:three-point-two-spin-operators} to three operators $\hat{\op}_j, \hat{\op}_k, \hat{\op}_l$ of arbitrary spin can conveniently be written in the light-cone projection, see \cite{Sotkov:1976xe,Fradkin:1996is}.
% 
%\begin{align}
%\langle \hat{\op}_j(x_1) \hat{\op}_k(x_2) \hat{\op}_l(x_3) \rangle = g(x_i, \theta_i) \sum_{p,q,r}^{j,k,l}
%c_{pqr} \left(Y(x_{12},x_{13})\right)^{j-1-r}...
%\end{align}
%where $\theta_i$ is the twist of the respective operators at position $x_i$ and where
%\begin{equation}
%g(x_i,\theta_i) = \frac{1}{|x_{12}|^{\theta_j+\theta_k -\theta_l}|x_{13}|^{\theta_j+\theta_l -\theta_k}|x_{23}|^{\theta_k+\theta_l -\theta_j}}
%\end{equation}

\subsection{Comment on the Notion of Spin and Higher Spin Fields}\label{eqn:spin-j-operators}
The operators \eqref{eqn:spin-j-operator} are called spin $j$ fields due to their transformation properties under the Lorentz group and in analogy to the theory of \emph{massive} higher spin fields introduced by Fierz and Pauli \cite{Fierz:1939zz,Fierz:1939ix}, which are described by traceless and symmetric tensors $\phi_{\mu_1...\mu_j}$ that satisy the Klein-Gordon equation $(\partial^2+m^2)\phi_{\mu_1...\mu_j}=0$ and are divergence-free $\partial^{\mu_1}\phi_{\mu_1...\mu_j}=0$. 

\emph{Massless} higher spin fields were shown to be given by tensors that satisfy the weaker statement of double tracelessness of $\phi_{\mu_1...\mu_j}$ derived in \cite{Fronsdal:1978rb,Fang:1978wz}, i.e.  e.g. $\phi^{\mu~\nu}_{~\mu~\nu,...,\nu_j}=0$. This weakened a previously assumed condition on the source to be divergence-free, while still guaranteeing that only helicities $\pm s$ are transmitted between sources. This encouraged for the search towards a theory of interacting massless higher spin fields.
Higher spin theory\index{higher spin theory} is an active area of research, an older comprehensive review is \cite{Vasiliev:1995dn}, for a more recent overview see \cite{Sorokin:2004ie}. Massless \cite{Metsaev:2008ks} and massive \cite{Metsaev:2009hp} higher spin fields were also studied in $AdS$ space.

\section{Renormalisation of Composite Operators}\label{eqn:renormalisation-of-composite-operators}
\index{renormalisation!composite operators}
The Callan-Symanzik equation for Green's functions of elementary fields \cite{Symanzik:1970rt,Callan:1970yg,Wilson:1970wp} is derived in standard quantum field theory textbooks and we shortly review it here and comment on the generalization for operators and operator mixing.

\subsection{Renormalisation of Elementary Fields}
In quantum field theory calculations,  Green's functions 
\begin{equation}
G_n=\langle \Phi(x_1)...\Phi(x_n) \rangle
\end{equation}
are typically divergent at loop-level and the original (bare) fields (and couplings) are redefined as
\begin{align}
\Phi^R(x) = Z_\Phi \Phi(x)
\end{align}
in order to absorb the divergences of loop calculations. These \emph{renormalisation constants} depend on the cutoff and the renormalisation scale $\mu$ that preserves the canonical dimension of the fields and consequently the renormalised Green's function depends on the same parameters
\begin{equation}\label{eqn:some-greens-function}
G^R_n (x, \mu, g) =\langle \Phi^R(x_1)...\Phi^R(x_n) \rangle = (Z_\Phi)^n \langle \Phi(x_1)...\Phi(x_n) \rangle\,.
\end{equation}
Since the bare Greens function $G_n$ does not depend on the scale $\mu$, we have
\begin{equation}
0=\mu \frac{d}{d\mu}  \left(Z_\Phi^{-n} G_n^R \right)= Z_\Phi^{-n}\left(-n \mu \frac{d}{d\mu}\ln Z_\Phi  + \mu \frac{\partial}{\partial\mu}  + \mu \frac{\partial g}{\partial \mu} \frac{\partial}{\partial g}  \right) G_n^R\,.
\end{equation}
Defining
\begin{equation}
\beta(g) =  \mu \frac{\partial g}{\partial \mu}\,,\qquad \gamma = -  \mu \frac{d}{d\mu}\ln Z_\Phi\,,
\end{equation}
we get the standard Callan-Symanzik renormalisation group equation
\begin{equation}
\left(   \mu \frac{\partial}{\partial\mu}  +\beta(g) \frac{\partial}{\partial g} + n \gamma \right) G_n^R = 0\,,
\end{equation}
where $\beta(g)$ determines the dependence of the running coupling $g(\mu)$ on the scale $\mu$ and $\gamma$ is called the \emph{anomalous dimension}\index{anomalous dimension} of the field $\Phi$, see the discussion in section \ref{sec:anomalous dimension}.

%For $i$ different particle species $\phi_i$ the equation can be generalised to
%\begin{equation}
%\left[ \mu \frac{\partial}{\partial \mu}+\beta\frac{\partial}{\partial g} + n_1 \gamma_1 + ... + n_i \gamma_i  \right]  G_n(x_1...x_n;\mu) = 0\,.
%\end{equation} where $\sum_i n_i =n $ is the total number of fields in the correlator.

\subsection{Renormalisation of Composite Operators}
The limit where two points in the Green's functions \eqref{eqn:some-greens-function} approach each other  may lead to additional divergences and therefore, in general, local composite operators require a renormalisation constant $Z_\op \neq (Z_\Phi)^2$
\begin{align}
 \op^R(x) = Z_\op \op(x) = Z_\op\,  \Phi(x)\Phi(x)
\end{align}
in order to obtain finite correlation functions. The analogous renormalisation group equation with $m$ such operators then reads
\begin{equation}
\left[ \mu \frac{\partial}{\partial \mu}+\beta\frac{\partial}{\partial g} + n  \gamma + m \gamma_\op  \right]  \langle \Phi^R(x_1)...\Phi^R(x_n) \op^R(y_1)...\op^R(y_m) \rangle = 0\,,
\end{equation}
where the anomalous dimension $\gamma_{\op}$ of the composite operator is defined in the same way as for elementary fields by 
\begin{align}
\gamma(g) = - \mu \frac{d}{d \mu} \ln Z_\op\,.
\end{align}
Typically, \emph{operator mixing} is present, i.e. the renormalised operators are linear combinations of other operators and we need a \emph{renormalisation matrix} $Z_{jk}$\index{renormalisation!matrix}
\begin{equation}\label{eqn:renormalization-with-mixing}
 \op_i^R = \sum_j Z_{ij} \op_j
\end{equation}
in order to define finite Green's functions. The operators $\op_j$ on the right-hand  side of \eqref{eqn:renormalization-with-mixing} need to have the same quantum numbers and equal or smaller canonical dimensions as compared to the original operator. Correspondingly, we get an  \emph{anomalous dimension matrix}
\begin{equation}
 (\gamma_\op)_{ij} = - (Z_\op^{-1})_{ik}\, \mu \frac{d}{d \mu} (Z_\op)_{kj}\,.
\end{equation}
The renormalisation matrix can be expanded in the coupling constant as
\begin{align}\label{eqn:Z-matrix-expansion}
Z_{jk} = \delta_{jk}+\sum_{n=1}^\infty g^{2n} \sum_{m=1}^n \frac{Z_{jk}^{[m](n)}}{\epsilon^m}\,.
\end{align}

\subsection{Conformal Scheme}\label{conformal scheme}\index{renormalisation!conformal scheme}
From the perspective of the Ward identities, the mixing is reflected by the appearance of the anomalous dimension matrix $\gamma_{nm}$ and the so-called special conformal anomaly matrix $\gamma_{nm}^c$ in the anomalous version of the Ward identities \eqref{eqn:dilatation-Ward-identity-general}, \eqref{eqn:special-conformal-Ward-identity-general}, originating from the right-hand side of \eqref{eqn:general-anomalous-conformal-ward-identity}.
These anomalies spoil the conclusions drawn in section \ref{eqn:correlation-functions}, \ref{eqn:correlation-functions-spin} on the form of the correlation functions of conformal operators. Formulated differently, the renormalised operators \eqref{eqn:Z-matrix-expansion} yield finite correlation functions, but are not conformally covariant in the sense \eqref{eqn:finite-trafo-primary-general} and thus do not have conformal correlation functions. 

One can however define a \emph{conformal renormalisation scheme} by applying another finite scheme transformation \cite{Mueller:1993hg}
\begin{equation}
\mathbb{O}_i = \sum_{j=0}^i B_{ij}^{-1} \op_j^R\,,
\end{equation}
where the finite matrix $B$ diagonalises the anomalous dimension matrix
\begin{equation}
\gamma_{j}(g^2) \delta_{jk}= (B^{-1} \gamma B )_{jk}
\end{equation}
and has the expansion
\begin{align}
B_{jk} = \delta_{jk}+\sum_{n=1}^\infty g^{2n}B_{jk}^{(n)}\,.
\end{align}
Then the full mixing matrix $\mathbb{Z}_{jk}$ that defines conformal operators at loop-level reads
\begin{equation}
\mathbb{O}_j = \sum_k \mathbb{Z}_{jk}  \op_{k} = \sum_k^j \sum_{m=k}^j B_{jm}^{-1} Z_{mk} \op_{k}\,.
\end{equation}

\subsubsection{Renormalisation of Twist-Two Operators}
In chapter \ref{sec:three-point-functions} we will encounter the renormalisation of so-called twist-two operators and see 
that the renormalisation matrix is diagonal at one loop, i.e.
\begin{equation}
Z_{jk}^{[1](1)} = \delta_{jk} Z_j^{(1)}\,.
\end{equation}
At one-loop level there are however two sources of non-diagonality of the two-point functions. On the one hand, there are finite non-diagonal one-loop corrections
\begin{equation}
\langle \op_j \op_l \rangle^{(1)}= \text{finite} \neq 0 \qquad \text{for} \quad j \neq l
\end{equation}
and on the other hand the minimal subtraction procedure \eqref{eqn:renormalization-with-mixing}, \eqref{eqn:Z-matrix-expansion} leads to non-diagonal contributions, if the operators $\hat \op_j$ are not orthogonal at order $\epsilon$. Then we have
\begin{equation}
\langle \hat \op_j^{R} \hat \op_l^{R}  \rangle = \left (1 + g^2 \frac{Z_j^{(1)}+Z_l^{(1)}}{\epsilon} \right) \left( \delta_{jl}\langle \hat\op_j \hat\op_l \rangle^{(0)} + \epsilon \langle \hat\op_j \hat\op_l \rangle^{(\epsilon)}  \right) + \langle \hat\op_j \hat\op_l \rangle^{(1)} \,,
\end{equation}
where the index $(\epsilon)$ on the correlator denotes the $\op(\epsilon)$ part of the tree-level correlator.
In order to define orthogonal operators at loop-level, we need to perform another finite renormalisation of the operators and take into account the mixing with descendants of lower dimension $\hat{\op}_{jk}= \hat{\partial}^{j-k} \hat{\op}_k$, such that the full mixing matrix $\mathbb{Z}_{jk}$ that yields finite orthogonal operators is
\begin{equation}
\hat{\mathbb{O}}_j = \sum_k \mathbb{Z}_{jk}  \hat \op_{jk} = \sum_k^j \sum_{m=k}^j B_{jm}^{-1} Z_{mk} \hat{\op}_{jk},\qquad \hat{\op}_{jk}= \hat{\partial}^{j-k} \hat{\op}_k\,.
\end{equation}
The one-loop expansion of the renormalisation constant then reads
\begin{align}\label{eqn:one-loop-expansion-mixing-matrix}
\mathbb{Z}_{jk} =\delta_{jk} + g^2\underbrace{(-B_{jk}^{(1)}+\frac{1}{\epsilon}\delta_{jk} Z_{j}^{(1)})}_{= \mathbb{Z}_{jk}^{(1)}} + \op(g^4)\,.
\end{align}
Thus at one loop the renormalised two-point function has the expansion
\begin{align}\nn
\langle \hat{\mathbb{O}}_i \hat{\mathbb{O}}_j \rangle &= \sum_{l,k} \left(\delta_{il}  + g^2(-B_{il}^{(1)}+\frac{1}{\epsilon}\delta_{il} Z_{i}^{(1)})  \right)  \left(\delta_{jk}  + g^2(-B_{jk}^{(1)}+\frac{1}{\epsilon}\delta_{jk} Z_{j}^{(1)})  \right) \langle \hat{\op}_{il} \hat{\op}_{jk} \rangle  \,.
\end{align}
The matrix $B$ accounts for the non-diagonal terms that arise from the one-loop diagrams and the $\op(\epsilon)$ expansion of the correlator and can be determined by requiring these terms to vanish.

\section{Operator Product Expansion in Conformal Field Theory}\label{sec:OPE-in-CFT}
\sectionmark{Operator Product Expansion in CFT}
K.~G.~Wilson proposed%\footnote{In \cite{PhysRev.179.1499} this is  introduced as a hypothesis.} 
 \cite{PhysRev.179.1499} that the effects of an operator product $\op_\alpha(x)\op_\beta(0)$ in the limit $x\to 0$ could be computed by replacing the product by a linear combination of local operators
\begin{equation}\label{eqn:OPE}
 \op_\alpha(x)\op_\beta(0) \longrightarrow \sum_\gamma C_{\alpha\beta\gamma}(x) \op_\gamma(0)\,,
\end{equation}
where the local operators $\op_\gamma$ must have the global symmetry quantum numbers of the product $\op_\alpha \op_\beta$. The coefficient functions $C_{\alpha\beta\gamma}(x)$ are c-number valued functions and are often called \emph{Wilson coefficients}\index{Wilson coefficients} and the corresponding operators \emph{Wilson operators}\index{Wilson operators}.
On dimensional grounds,  in a conformal field theory, the coefficient functions $C_{\alpha\beta\gamma}$ must be of the form\footnote{We restrict to scalar operators here to illustrate the general idea.}
\begin{equation}\label{eqn:coefficient-functions}
C_{\alpha\beta\gamma}(x)= \frac{C_{\alpha\beta\gamma}}{|x|^{\Delta_\alpha+\Delta_\beta-\Delta_\gamma}}
\end{equation} 
and therefore we can write
\begin{equation}\label{eqn:conformal-OPE}
 \op_\alpha(x_1)\op_\beta(x_2) \stackrel{x_1 \to x_2}{\longrightarrow} \sum_\gamma \frac{C_{\alpha\beta\gamma}}{|x_{12}|^{\Delta_\alpha + \Delta_\beta - \Delta_\gamma}} \op_\gamma(x_2)\,.
\end{equation}
It is now easy to see why the coefficients $C_{\alpha\beta\gamma}$ are just the structure constants of the three-point functions in conformally invariant theories. Taking the limit $x_1 \to x_2$ in the conformal three-point function \eqref{eqn:conformal-three-point-function} yields
\begin{equation}\label{eqn:strcture-constant-in-limit}
 \langle \op_\alpha(x_1) \op_\beta(x_2) \op_\gamma(x_3) \rangle  \stackrel{x_1 \to x_2}{\longrightarrow} \frac{C_{\alpha\beta\gamma}}{|x_{12}|^{\Delta_A + \Delta_B - \Delta_C}|x_{23}|^{2\Delta_\gamma}} \,.
\end{equation}
On the other hand, in the same limit, we can can rewrite the first two operators in terms of the OPE and find
\begin{align}\label{eqn:show-ope}
 \langle \op_\alpha(x_1) \op_\beta(x_2) \op_\gamma(x_3) \rangle & \stackrel{x_1 \to x_2}{\longrightarrow}
\sum_\delta \frac{C_{\alpha\beta\delta}}{|x_{12}|^{\Delta_\alpha + \Delta_\beta - \Delta_\delta}} \langle \op_\delta(x_2) \op_\gamma(x_3) \rangle \\ \nn
& = \sum_\delta \frac{C_{\alpha\beta\delta}}{|x_{12}|^{\Delta_\alpha + \Delta_\beta - \Delta_\delta}} \frac{\delta_{\delta \gamma}}{|x_{23}|^{2\Delta_\delta}} \\ \nn
& = \frac{C_{\alpha\beta\gamma}}{|x_{12}|^{\Delta_\alpha + \Delta_\beta - \Delta_\delta}|x_{23}|^{2\Delta_\gamma}} \,,
\end{align}
where we used the conformal two-point function \eqref{eqn:conformal-two-point-function} in the second line and assumed that the operators are normalised as $C_\delta=1$. Thus the coefficient $C_{\alpha\beta\gamma}$ coincides with the structure constant in \eqref{eqn:strcture-constant-in-limit}.

Above, we have restricted to a basis of scalar operators, in general however the complete set of operators also involves operators with spin \eqref{eqn:spin-j-operator}. The complete operator product expansion can then be organised in terms of \emph{Wilson operators} with increasing \emph{twist}. The \emph{leading} or \emph{lowest twist operators} in $d=4$ are \emph{twist-two} operators which can schematically be written as
\begin{equation}\label{eqn:twist-two-schematically}
\op_{\mu_1...\mu_j}(x) = D_{\mu_1}...D_{\mu_i} \Phi(x) D_{\mu_{i+1}}...D_{\mu_j}\Phi(x) +...\,.
\end{equation}
These composite operators transform covariantly under conformal transformations only for specific distributions of the covariant derivatives and we will be more specific on their form in chapter \ref{sec:three-point-functions}.
The coefficient functions \eqref{eqn:coefficient-functions} of these operators in the OPE have to be generalised appropriately to be compatible with the form of the three-point function \eqref{eqn:structure-three-point-function-spin-j}, see \cite{Dolan:2004iy} for more information. 

As mentioned before, using the OPE \eqref{eqn:show-ope} twice, one can extract information on higher-point functions. Schematically, the four-point functions \eqref{eqn:four-point-functions} can be written
\begin{equation}
\langle \op_\alpha(x_1) \op_\beta(x_2) \op_\gamma(x_3) \op_\delta(x_4) \rangle \sim \sum_{\sigma} \frac{C_{\alpha\beta\sigma} C_{\gamma\delta\sigma}}{|x_{12}|^{\Delta_\alpha+\Delta_\beta- \Delta_\delta}|x_{34}|^{\Delta_\gamma+\Delta_\delta- \Delta_\delta}}\frac{C_\sigma}{|x_{23}|^{2\Delta_\sigma}}\,.
\end{equation}
In \cite{Arutyunov:2003ad} it is for example shown, that a four-point two-loop correlator of  $1/2$-BPS operators can be completely reconstructed on the basis of the OPE and the knowledge of the one-loop anomalous dimensions for certain operators of twist 2 and 4. 

The OPE, together with the knowledge of the anomalous dimensions, in principle allows to determine any higher-point function.  Chapter \ref{sec:three-point-functions} is devoted to the investigation of the coefficients of twist-two operators in the OPE.

\subsection{OPE and Twist-Operators in QCD}
Here, we would like to give an example for an application of the operator product expansion and show where \emph{leading twist operators}\index{twist!leading twist} in deep inelastic scattering show up, following to a large extent chapter 18 of \cite{Peskin:1995ev}. We only sketch the ideas and refer the reader to \cite{Peskin:1995ev} for more details.

\subsubsection{Deep Inelastic Electron Scattering}
Deep inelastic electron scattering is decribed by the parton model, in which the electron scatters from quarks carrying fractions of the total momentum of the proton, which are determined by parton distribution functions. 

The amplitude $\mathcal{M}(ep \to eX )$ for the scattering of an electron and a proton into an electron and a final state $X$ can be related to the forward matrix element of two quark electromagnetic currents $J^\mu(x)= \bar{q}(x) \gamma^\mu q(x)$ via the optical theorem. The total cross section averaged over initial and final electron spins then reads
\begin{align}\nn
\sigma(ep \to e X) = \frac{1}{2s} \int \frac{d^3k^\prime}{(2\pi)^3} \frac{1}{2 k^\prime} e^4 \frac{1}{2} \sum_{\text{spins}}\bar{u}(k) \gamma_\mu u(k^\prime) \bar{u}(k^\prime) \gamma_\nu u(k) \cdot \left(\frac{1}{Q^2}\right)^2  2  \text{Im}\left(W^{\mu\nu} \right)\,,
\end{align}
where the {\itshape forward Compton amplitude} is
\begin{equation}\label{eqn:compton-amplitude}
 W^{\mu\nu}= i \int d^4x \,e^{i q \cdot x} \langle P | T \{ J^\mu(x) J^\nu(0) \} | P \rangle\,,
\end{equation}
and $|P\rangle$ represents the proton state of momentum $P^\mu$ see \cite{Peskin:1995ev} for details. Here we will illustrate the use of the OPE to analyse this amplitude. As mentioned before, in a massless theory the operators appearing on the right-hand side of the operator product expansion \eqref{eqn:OPE} must have the same symmetry quantum numbers as the product of operators on the left-hand side. 

\subsubsection{Dominant contributions from leading twist operators}
In $d$ dimensions the scaling dimension of a quark (fermion) is $[q(x)]=(d-1)/2$ and thus the dimension of the quark electromagentic current is $[J^\mu]=d-1$. Thus, in the operator product expansion of the current product
\begin{equation}
 J^\mu(x) J^\nu(0) = \sum C_i(x) \op_i(0)
\end{equation}
an operator $\op_i$ with dimension $\Delta_i$ must have a coefficient function of dimension $[C_i]= 2(d-1)- \Delta_i$. In a massless field theory, the dimension is fully carried by the $x$-dependence and we thus have
\begin{equation}
C_i(x) = c_i \left(\frac{1}{-x^2} \right)^{d-1-\Delta_i/2}\,,
\end{equation}
where $c_i$ is a dimensionless coefficient. The Fourier transformation in \eqref{eqn:compton-amplitude} thus yields, that the coefficient function is suppressed by a factor of 
\begin{equation}
 \left(\frac{1}{Q^2} \right)^{\Delta_i/2-1},\qquad Q^2=-q^2\,,
\end{equation}
for $d=4$. If we consider an operator with spin $j$  of the form\footnote{Here $\{..\}$ means symmetrisation over all indices and all possible traces are subtracted.}
\begin{equation}
 \op_j^{\mu_1..\mu_j}(x) =  \bar{q}(x) \gamma^{\{\mu_1} (iD^{\mu_2}) ... (iD^{\mu_j\}}) q(x) - \text{traces}\,,
\end{equation}
we get additional kinematic factors from the matrix element of the operator
\begin{equation}
 \langle P | \op_j^{\mu_1..\mu_j }|P \rangle \sim P^{\mu_1} ... ~P^{\mu_j} - \text{traces}\,.
\end{equation}
Thus, the contribution of a spin $j$ operator will be of order
\begin{equation}
 \left( \frac{2 P \cdot q}{Q^2} \right)^j  \left( \frac{1}{Q^2} \right)^{(\Delta_j - j -1)/2}\,.
\end{equation}
The quantity $ Q^2 / 2 P \cdot q = x$ is held fixed in deep inelastic scattering. Thus, the relative size of the contribution from the OPE operators to deep inelastic scattering does not depend on the dimension but rather on the {\emph{twist}} $\theta$ of the operator
\begin{equation}
 \theta= \Delta_j -j \,.
\end{equation}
Therefore, the \emph{leading  twist operators (smallest twist)} give the most important contributions to the OPE for deep inelastic scattering. We will consider operators of this type in chapter \ref{sec:three-point-functions}.

\chapter{Amplitudes, Wilson Loops and Correlators in $\mathcal{N}=4$ SYM}\label{chapter:amplitude-wilson-loops-correlators}
\chaptermark{Amplitudes, Wilson Loops and Correlators}
In this chapter we give an overview about some recent developments in quantum field theory, which have mostly been inspired by the AdS/CFT correspondence. In the following chapter we will describe some of the analogous developments in ABJM theory. In particular, we will focus on the so-called amplitude / Wilson loop duality.

In order to state the duality, we will introduce the necessary background on amplitudes in $\sym$ in section \ref{sec:amplitudes} and shortly comment on the methods that are presently being used to calculate amplitudes, such as the spinor helicity formalism, BCFW recursions and unitarity cuts. In section \ref{sec:BDS} we will discuss the so-called \emph{BDS ansatz}, which constitutes an all-loop proposal for scattering amplitudes in $\sym$ based on an iterative structure in perturbative amplitude calculations and the previously known exponentiation of the IR divergences of gluon scattering amplitudes.

Unexpectedly, the scattering amplitudes exhibit a hidden symmetry called \emph{dual conformal symmetry} distinct from the ordinary conformal symmetry of the theory. This symmetry extends to a \emph{dual superconformal symmetry} and 
together with the superconformal symmetry forms a \emph{Yangian symmetry}.

In section \ref{sec:amplitude-wilson-loop-duality} we introduce the \emph{duality between amplitudes and light-like polygonal Wilson loops} and see that the dual conformal symmetry of the amplitudes can be understood in terms of the conformal symmetry of the Wilson loop  defined in a \emph{dual configuration space}. The expectation value of the Wilson loop is governed by an \emph{anomalous conformal Ward identity} which fixes the result of the Wilson loop and, via the duality, the result of the amplitudes for $n=4,5$ edges resp. gluons completely. For $n \geq 6$ particles the Ward identity admits for arbitrary functions of conformally invariant cross ratios. Interestingly, even though the BDS ansatz for scattering amplitudes breaks down starting from $n=6$ at two loops, the duality continues to hold, thus also the part, which is not constrained by the known symmetries of the two objects, continues to agree, showing that there is more to the amplitude / Wilson loop duality than just dual conformal symmetry. We will also have a glimpse on the generalization of the duality to superamplitudes and super Wilson loops in section \ref{sec:super-Wilons-loop} as well as a new \emph{triality} between correlators, amplitudes and Wilson loops and its generalization to supersymmetric objects in section \ref{sec:duality-correlators-amplitude-and-wilson-loops}. We also comment on relations between form factors and periodic Wilson loops in section \eqref{sec:form-factors-and-wilson-loops}.

The duality between scattering amplitudes and Wilson loops has its origin in the strong coupling calculation of gluon amplitudes \cite{Alday:2007hr,Alday:2007he} in terms of minimal surfaces in AdS space, which formally resembles the calculation of a Wilson loop. 
The analogue of the dual superconformal symmetry on the string theory side is an invariance of the the free  $AdS_{5}\times S^{5}$ superstring under a combination of bosonic and fermionic T-dualities \cite{Berkovits:2008ic,Beisert:2008iq,Beisert:2009cs}. The T-duality exchanges the original with the dual superconformal symmetries and thus explains the presences of the dual symmetry.\index{T-duality}

In the following we focus on the weak coupling side of the duality and refer the reader to the literature for more information on the strong coupling side of the duality.

\section{Scattering Amplitudes in $\sym$}\label{sec:amplitudes}
It is possible to show \cite{Berends:1987cv,Mangano:1987xk}, see also the review \cite{Dixon:1996wi}, that a $n$-gluon tree-level amplitude can be written as
\begin{equation}\label{eqn:color-ordering}
A_n\left(\{p_i\,h_i,a_i\}\right) = \sum_{\sigma \in S_n/Z_{n}} \tr\left( T^{a_{\sigma(1)}}, ..., T^{a_{\sigma(n)}} \right) A_n(\sigma(1)^{h_{\sigma(1)}},..., \sigma(n)^{h_{\sigma(n)}})\,,
\end{equation}
where $p_i$ denote the momenta, $h_i=\pm1$ the helicities and $a_i$ the colour index of particle $i$ and the sum goes over all non-cyclic permutations $\sigma$. The amplitudes $A_n$ are called \emph{colour ordered} or \emph{partial amplitudes}. In order to obtain the full result for the amplitude it is thus sufficient to evaluate all colour ordered amplitudes using the simpler \emph{colour ordered Feynman rules}, see \cite{Dixon:1996wi}. In the following we restrict to the discussion of colour ordered amplitudes and just call them amplitudes from now on.

\subsection{Spinor Helicity Formalism}\label{sec:spinor-helicity}
The \emph{spinor helicity formalism}\index{spinor helicity formalism} is reviewed e.g. in \cite{Bern:2007dw, Dixon:1996wi} and we will not give a detailed survey of this method, but just introduce a few terms and notations which are useful for the following discussion of amplitudes.  

The basic idea of the spinor helicity formalism is to reformulate $n$-point scattering amplitudes $A_n$ of $n$ massless  particles in terms of commuting, complex two-component Weyl spinors $\lambda_\alpha, \tilde{\lambda}_{\dot{\alpha}}$ which automatically satisfy the masslessness condition $p_i^2=0$. They can be defined by
\begin{equation}\label{eqn:definition-spinor-helicity}
p_i^{\alpha\dot{\alpha}} = p_i^\mu (\sigma_\mu)^{\alpha\dot{\alpha}} = \lambda^\alpha_i \tilde{\lambda}_i^{\dot{\alpha}}\,,
\end{equation}
where $\sigma^\mu=({\bf 1},\vec{\sigma})$ and $\vec{\sigma}$ are the three Pauli matrices and indices are lowered and raised using the antisymmetric  $\epsilon_{\alpha\beta}=-\epsilon^{\alpha\beta}$, $\epsilon_{\dot{\alpha}\dot{\beta}}=-\epsilon^{\dot{\alpha}\dot{\beta}}$, i.e. $\lambda^\alpha = \epsilon^{\alpha\beta}\lambda_\beta$. One can introduce the Lorentz invariant\footnote{With respect to $SU(2)_L$, $SU(2)_R$} antisymmetric contractions
\begin{align}\label{eqn:Lorentz-invariants-spinor-helicity}
\langle i j \rangle = (\lambda_i)^\alpha \epsilon_{\alpha\beta} (\lambda_j)^{\beta}=(\lambda_i)^\alpha  (\lambda_j)_{\alpha}\,, \qquad
\left[ i j \right] = (\tilde{\lambda}_i)_{\dot{\alpha}} \epsilon^{\dot{\alpha}\dot{\beta}} (\tilde{\lambda}_j)_{\dot{\beta}}=(\tilde{\lambda}_i)_{\dot{\alpha}}(\tilde{\lambda}_j)^{\dot{\alpha}}\,,
\end{align}
which satisfy $\langle i j \rangle \left[ i j \right] = - 2\, p_i \cdot p_j$. Introducing suitable polarisation vectors $\epsilon^+_{i,\mu}, \epsilon^-_{i,\mu}$ for gluons of positive and negative helicity, one can evaluate gluon scattering amplitudes in the spinor helicity formalism purely in terms of \eqref{eqn:Lorentz-invariants-spinor-helicity}.

\subsubsection{MHV and non-MHV Amplitudes}
Amplitudes with $n$-gluons vanish if all $n$ or $n-1$ gluons have the same helicity. The simplest non-trivial amplitudes are the ones where at most $n-2$ gluons have the same helicity. These amplitudes are called \emph{maximally helicity violating} (MHV) amplitudes\index{MHV amplitudes} for $2$ negative helicities ($\overline{\text{MHV}}$ for 2 positive helicities) and have the surprisingly 
simple form \cite{Parke:1986gb,Berends:1987me}
\begin{equation}\label{eqn:MHV-gluon-amplitude}
A^{\text{MHV,tree}}_n (i^-, j^-) = i \frac{\langle i j \rangle^4}{\langle 1 2\rangle \langle 2 3\rangle ... \langle n 1\rangle} \delta^{(4)}(\sum_i p_i)\,,
\end{equation}
where particles $i,j$ have negative helicity and all others have positive helicity. For real momenta\footnote{For complex momenta one can also define three-particle amplitudes. For real momenta masslessness and momentum conservation forces these three momenta to be collinear and thus all scalar products $p_i \cdot p_j$ are zero.} the first non-trivial MHV amplitude is the amplitude with four particles. Amplitudes with $n-3$ particles of positive helicity are called next-to-MHV (NMHV), amplitudes with $n-2-k$ positive helicities $\text{N}^k$MHV.

\subsection{BCFW Recursion}\index{BCFW recursion}\label{sec:BCFW recursion}
Formula \eqref{eqn:MHV-gluon-amplitude} can be justified % :: I say justified, because this is no real proof since it already uses the form of the amplitude at (n) AND (n-1) and only under this condition relates A_n and A_{n-1}. It is only consistent, and possibly the only consistent expression
recursively, see e.g.  \cite{Bern:2007dw}, by first calculating the three-point amplitude in complex kinematics and using \eqref{eqn:MHV-gluon-amplitude} for $n-1$ to show its correctness for $n$. The reason that this is possible is that one can factorise the $n$-point amplitude into a three-point and a $(n-1)$-point MHV amplitude by shifting the complex momenta \eqref{eqn:definition-spinor-helicity} of two particles in a way that preserves momentum conservation 
and light-likeness: $p_i = \lambda_i \tilde{\lambda}_i \to  (\lambda_i + z \lambda_j) \tilde{\lambda}_i,~p_j = \lambda_j \tilde{\lambda}_j \to  \lambda_j (\tilde{\lambda}_j - z \tilde{\lambda}_i)$, where $z$ is a complex number. Choosing e.g. $i=1,j=n$ the amplitude \eqref{eqn:MHV-gluon-amplitude} becomes: 
\begin{equation}
A_n(z) = i \frac{\langle i j \rangle^4}{(\langle 1 2\rangle + z \langle n 2 \rangle)\langle 2 3\rangle ... \langle n 1\rangle}\,.
\end{equation}
The expression $A(z)/z$ has poles at $z=0$ and $z = - \langle 1 2\rangle / \langle n 2 \rangle$. 
Therefore, by Cauchy's theorem, the vanishing\footnote{There are some conditions on the shifts such that the contour-integral at infinity vanishes.} contour integral
\begin{equation}
 \oint \frac{dz}{2\pi i} \frac{A(z)}{z} =0= \sum_{z_k} \text{Res}(A(z)/z, z=z_k)
\end{equation}
is equivalent to the sum over the residues of these poles. The residue at $z=0$ is just the original amplitude, which can thus be expressed through the value of the other residue 
\begin{equation}
A_n=A_n(z=0)=-\text{Res}(A_n(z)/z, z=- \langle 1 2\rangle / \langle n 2 \rangle )\,.
\end{equation}
Evaluating the right-hand side one finds, see  \cite{Bern:2007dw} for more details, that the amplitude factorises into a three-point and a $(n-1)$-point MHV amplitude, which can be used to confirm \eqref{eqn:MHV-gluon-amplitude}.

The same trick can be used to construct recursions for non-MHV amplitudes and is known as the \emph{BCFW recursion} \cite{Britto:2004ap,Britto:2005fq}\index{BCFW recursion}, however there will in general be more terms on the right-hand side, corresponding to different factorization channels. 
There are also BCFW type recursions for processes involving all sorts of massless and also massive fields, see e.g. the references in \cite{Bern:2007dw}. Furthermore, there is a supersymmetric version of the tree-level recursion \cite{Brandhuber:2008pf,ArkaniHamed:2008gz,Elvang:2008na} that was solved in \cite{Drummond:2008cr}. A nice review on these methods can be found in \cite{Brandhuber:2011ke}.

\subsubsection{Superamplitudes}
One can also find a compact formula for the supersymmetric amplitude which contains all possible tree-level MHV amplitudes involving gluons $G^\pm(p)$, gluinos $\Gamma(p),\bar\Gamma(p)$ and scalars $S_{AB}(p)$. It is convenient to combine the fields of different helicity in a superstate $\Phi(\eta,p)$
\begin{equation}\label{eqn:superfield}
\Phi = G^+ + \eta^A \Gamma_A + \frac{1}{2!}\eta^A \eta^B S_{AB} + \frac{1}{3!}\eta^A \eta^B \eta^C \epsilon_{ABCD} \bar{\Gamma}^D  +  \frac{1}{4!}\eta^A \eta^B \eta^C  \eta^D \epsilon_{ABCD} G^- \,,
\end{equation}
where $\eta^A$ are anti-commuting Gra\ss mann parameters and $A=1...4$ is the $SU(4)$ index. The superamplitude can then be written \cite{Nair:1988bq} as
\begin{equation}\label{eqn:N4SYM-superamplitude}
\mathcal{A}_n^{\text{MHV}}(\Phi_1,...,\Phi_n) = i \frac{\delta^{(4)}(p)\delta^{(8)}(q)}{\langle 1 2 \rangle \langle 2 3 \rangle ... \langle n 1 \rangle}\,,
\end{equation}
where  $\delta(p)=\delta(\sum_i \lambda_i \tilde{\lambda}_i)$ is the momentum conserving delta function that also appears in \eqref{eqn:MHV-gluon-amplitude} and $\delta(q)= \prod_{A,\alpha}(\sum_i  \eta_i^A \lambda_i^\alpha))$ can be thought of as a super-momentum conserving delta function. A particular component amplitude is then the coefficient of the term in \eqref{eqn:N4SYM-superamplitude} that contains the powers of the Gra\ss mann variables $\eta_i^A$ corresponding to the component fields in \eqref{eqn:superfield}. E.g. the MHV gluon amplitude \eqref{eqn:MHV-gluon-amplitude} is the coefficient\footnote{More correctly, we should write $(\eta_i)^4 = \eta_i^1 \eta_i^2 \eta_i^3 \eta_i^4$ since each Gra\ss mann variable can of course appear at most once.} of $(\eta_i)^4 (\eta_j)^4$, since gluons $i$ and $j$ have negative helicity.

As mentioned before, the recursions can also be used to construct all non-MHV amplitudes \cite{Drummond:2008cr}. The non-MHV amplitudes can be included in the expression by multiplying \eqref{eqn:N4SYM-superamplitude} with a factor  $\mathcal{P}_n(\lambda_i,\tilde{\lambda}_i,\eta_i)= 1 + \mathcal{P}_n^{\text{NMHV}}+...$  \cite{Drummond:2008cr} and one can extract the desired amplitude as the coefficient of the corresponding powers of $\eta_i$, exactly as in the MHV case. The recursions and their solutions  \cite{Drummond:2008cr} can conveniently be implemented into computer programs, e.g. the MATHEMATICA package GGT \cite{Dixon:2010ik} yields compact analytical formulas for all tree-level colour ordered gauge theory amplitudes (MHV and non-MHV) with $n$ gluons and up to four massless quark-anti-quark pairs.
~\\
At loop-level the most general form of the amplitude is
\begin{align}\label{eqn:N4SYM-superamplitude-non-MHV-loop}
\mathcal{A}_n(\Phi_1,...,\Phi_n) &= \mathcal{A}_n^{\text{MHV}}+ \mathcal{A}_n^{\text{NMHV}} +... + \mathcal{A}_n^{\text{N}^k\text{MHV}}\\ \nn
&=
%\mathcal{A}_n^{\text{N}^k\text{MHV}}(\Phi_1,...,\Phi_n) = 
i \frac{\delta^{(4)}(p)\delta^{(8)}(q)}{\langle 1 2 \rangle ... \langle n 1 \rangle} \left(A_{n;0}(\lambda, \tilde \lambda, \eta,a)+... + A_{n;n-4}(\lambda, \tilde \lambda, \eta,a)\right)
\end{align}
and $A_{n;k}$ contains the coefficients $\mathcal{P}_n^{\text{N}^k \text{MHV}}$ as well as their loop corrections as a series in the coupling constant $a$. The loop corrections to the MHV amplitude are usually designated by $M_n \equiv A_{n;0}$ in the literature, i.e.
\begin{equation}\label{eqn:N4SYM-superamplitude-MHV-loop}
\mathcal{A}_n^{\text{MHV}}(\Phi_1,...,\Phi_n) = i \frac{\delta^{(4)}(p)\delta^{(8)}(q)}{\langle 1 2 \rangle ... \langle n 1 \rangle} M_n\,.
\end{equation}

\subsubsection{Superconformal invariance of the amplitudes}
The tree-level amplitudes defined by \eqref{eqn:N4SYM-superamplitude} and their generalization to non-MHV amplitudes are invariant under the ordinary superconformal symmetry, i.e.
\begin{equation}\label{eqn:invariance-under-conformal-symmetry}
j_a \mathcal{A}_n= 0\,,
\end{equation}
where $j_a \in \{p^{\alpha\dot\alpha}, q^{\alpha A},\bar{q}^{\dot \alpha}_A, m_{\alpha\beta}, \bar{m}_{\dot\alpha \dot \beta}, r^A_B,d,s^\alpha_A, \bar{s}_{\dot\alpha}^A, k_{\alpha\dot\alpha} \}$ are the generators of the ordinary superconformal symmetry $psu(2,2|4)$, which satisfy the commutation relations given in section \ref{sec:superconformal-algebra}. An explicit representation of these generators in \emph{on-shell superspace} $(\lambda_i^\alpha, \tilde \lambda_i^{\dot\alpha},\eta_i^A)$  can be found in \cite{Drummond:2008vq}.

The ordinary conformal symmetry is broken at loop-level by infrared divergences and furthermore it is broken through the \emph{holomorphic anomaly} already at tree-level due to collinear singularities \cite{Cachazo:2004by,Cachazo:2004dr,Bargheer:2009qu,Korchemsky:2009hm}. The symmetry is superficially broken at singular particle configurations where massless particles become collinear, but can be deformed to account for the holomorphic anomaly \cite{Bargheer:2009qu}.
At loop-level the situation is more complicated because particles in the loop integration can become collinear with others and thus proper treatment is required \cite{Bargheer:2009qu,Sever:2009aa,Beisert:2010gn}.

\subsection{Generalised Unitarity}\label{sec:generalized-unitarity}\index{unitarity method}
The \emph{state-of-the-art technique}\footnote{Some applications and developments of the unitarity methods were listed in \cite{Bern:2011qt} and we recite them as further reading suggestions  \cite{Bern:1994cg,Bern:1994zx,Bern:1995db,Bern:1996je,Bern:2000dn,Bern:1997sc,Bern:2004ky,Britto:2004nc,Buchbinder:2005wp,Bern:2007ct,Britto:2006fc,Britto:2007tt,Ossola:2006us,Britto:2008vq,Forde:2007mi,Badger:2008cm,Bern:2008pv,Bern:2010tq,Bern:2010qa}. 
} for the evaluation of gauge theory and gravity amplitudes at loop-level is the method of \emph{generalized unitarity}\index{unitarity method!generalized}. 
\emph{Unitarity methods} have a long history, see e.g. \cite{eden2002analytic} for applications in the 1960s, and have become a well-developed powerful method for the calculation of loop amplitudes. 
Since we will not make use of this method in this thesis we refer the reader to the reviews \cite{Bern:2007dw,Bern:2011qt} and just mention some of the key concepts.

The unitarity method relies on the unitarity of the S-matrix $S= 1+i T$, which turns into an equation for the non-trivial part of the scattering matrix $T$
\begin{equation}\label{eqn:unitarity-scattering-matrix}
S S^\dagger = 1 \qquad \Rightarrow \qquad T T^\dagger = -i (T-T^\dagger) \,. %= 2 \, \text{Im}(T) 
\end{equation}
Since the scattering matrix has an expansion in the coupling constant, from the last equality it follows that information on the amplitude is encoded in the product of lower loop amplitudes. It is important to note, that the matrix multiplication $T T^\dagger$ implicitly contains a sum over all possible intermediate on-shell states, i.e. also a phase space integral\footnote{To see this, one can write explicit matrix indices $\sum_j T_{ij}(T^\dagger)_{jf}= -i (T - T^\dagger)_{if}$ where $i,f$ denote a physical initial resp. final state and the sum over $j$ is the sum over all possible physical on-shell states, i.e. all possible particles with all possible momentum configurations. This is the origin of the phase space integral.}.

%The imaginary part $\text{Im}(T)$ corresponds to a discontinuity in the scattering amplitude, i.e. a branch cut in complex momenta on the real axis, explicitly given by logarithms, dilogarithms etc. originating from the loop integrals. 
The discontinuity on the left-hand side of \eqref{eqn:unitarity-scattering-matrix} in a given channel\footnote{A momentum invariant, e.g. $s=(p_1+p_2)^2$.} of a particular loop-level amplitude can be obtained \cite{Cutkosky:1960sp} by replacing  two propagators that carry this invariant with a delta function
\begin{equation}\label{eqn:cutkosky-replacement-disconinuity}
\frac{i}{p^2 + i \epsilon} \to 2\pi \delta^{(+)}(p^2)\,,
\end{equation}
which reduces the loop integral to a phase space integral, and the amplitude is \emph{cut} into a product of two on-shell amplitudes exactly as stated by \eqref{eqn:unitarity-scattering-matrix}. One could now continue to solve the phase space integral, but the idea of the unitarity method is to turn the step \eqref{eqn:cutkosky-replacement-disconinuity} around:

For a given channel, one starts from the left-hand side of \eqref{eqn:unitarity-scattering-matrix}, i.e. one sews together all tree amplitudes by converting the phase space integral into a full loop integral using \eqref{eqn:cutkosky-replacement-disconinuity} backwards. The key point is, that the full amplitude is a linear combination of a restricted set\footnote{In $\sym$ the set is further reduced by requiring dual conformal invariance, see section \ref{sec:dual-conformal-invariance}.} of scalar basis integrals $\mathcal{I}_j$
\begin{equation}\label{eqn:linear-combination-of-basis-integrals}
A_n = \sum_j c_j \mathcal{I}_j\,,
\end{equation}
where $c_j$ are the coefficients of the integrals in the amplitude.  By the above procedure of \emph{sewing together} tree amplitudes to loop amplitudes, one can thus directly read off the coefficients $c_j$ of the corresponding integrals. By considering a certain cut, one can thus determine all coefficients of the integrals that have the corresponding cut. Of course, by this method only those contributions which have these cuts can be detected.
Amplitudes that fall into this class are called \emph{cut constructible}\index{cut constructability} and were identified in \cite{Bern:1994cg}, e.g. $\sym$ amplitudes belong to this class.

This method is more efficient than considering single Feynman diagrams, because at every step one is dealing with gauge invariant physical amplitudes and does not need to consider diagrams which cancel in the end. Furthermore, certain tree-level properties of on-shell methods can be carried over to loop-level, see e.g. \cite{Bern:2011qt}.

To understand the details of this method it is instructive to review the calculation of the one-loop $n$-gluon MHV amplitude in $\sym$ which can be found in the lectures \cite{Dixon:1996wi}.

As mentioned, by considering the cut in a certain channel, after some algebra one finds an expression $\sum_j c_j \mathcal{I}_j$ with all basis integrals $\mathcal{I}_j$ that contain the cut\footnote{It should be noted, that only integrals that contain a cut can be detected by the cutting procedure. In QCD there are rational terms, which are not detected by standard unitarity methods and require additional care.}. Several integrals may contribute to the same cut. One can restrict the number of appearing integrals, by considering \emph{multiple cuts} of an integral, i.e. one sews together a loop integral from more than two amplitudes. The advantage of this method of \emph{generalised unitarity}\index{unitarity method!generalised} is that one has less algebra to perform, since less integrals contribute. Cutting the maximal number of propagators, i.e. four propagators, selects a single box integral and no further algebra is required. This goes under the name \emph{maximal generalised unitarity}\index{unitarity method!maximal generalised}, see also the review \cite{Bern:2011qt}.

\subsection{One-Loop Four-Point Amplitude}
The simplest loop-level gluon amplitude that we can consider in $\sym$ is the one-loop four-point amplitude. Calculating all Feynman diagrams or using the unitarity method described in \ref{sec:generalized-unitarity} and evaluating the cuts in the $s$ and $t$-channel one finds
\begin{equation}\label{eqn:4-point-amplitude-1loop-full}
A_4 = A_4^{\text{tree}} M_4\,,
\end{equation}
where $A_4^{\text{tree}}$ is given by \eqref{eqn:MHV-gluon-amplitude}
\begin{equation}\label{eqn:4-point-amplitude-1loop}
M_4= 1 - \frac{a}{2} s t I^{(1)}_4(s,t) + \op(a^2)
\end{equation}
with $a=g^2 N/(8\pi^2)$, $s=(p_1+p_2)^2$, $t=(p_2+p_3)^2$ and the box-integral
\begin{equation}
I_4^{(1)}(s,t) = \int \frac{d^dl}{(i\pi^{d/2})} \frac{1}{l^2(p_1+l)^2 (p_1+p_2+l)^2 (p_4-l)^2}
\end{equation}
is the only integral in \eqref{eqn:linear-combination-of-basis-integrals} that contributes. The integral is infrared divergent and evaluation  in dimensional regularisation with $d=4-2\epsilon$ yields \cite{Bern:2005iz}
\begin{equation}\label{eqn:I4(s,t)}
s t I_4^{(1)}(s,t) = \frac{2}{\epsilon^2} \left(-s / \mu^{2}\right)^{-\epsilon}+ \frac{2}{\epsilon^2}\left(-t/\mu^2\right)^{-\epsilon}- \ln^2\left(\frac{s}{t} \right) - \frac{4}{3} \pi^2 + \op(\epsilon)\,.
\end{equation}

\subsection{BDS Ansatz for $n$-point All-Loop Amplitudes}\label{sec:BDS}
Similar to  \eqref{eqn:4-point-amplitude-1loop-full}, it turns out that also $n$-point planar MHV amplitudes can be factorised into a tree-level part and a scalar function $M_n(a)$ that carries the complete coupling constant dependence 
\begin{equation}\label{eqn:n-point-amplitude-1loop-full}
A_n= A_n^{\text{tree}} M_n(a)\,,
\end{equation}
where $M_n(a)=(1+ a M_n^{(1)} + a^2 M_n^{(2)}+...)$.
Based on an iterative structure \cite{Anastasiou:2003kj,Bern:2005iz} for $M^{(3)}, M^{(2)}$ in terms of the one-loop scalar factor $M^{(1)}$ and an $n$-point one-loop calculation \cite{Bern:1994zx},  Bern, Dixon and Smirnov (BDS) proposed the so-called \emph{BDS-ansatz}\index{BDS ansatz} for the all-loop $n$-point gluon amplitude \cite{Bern:2005iz}
\begin{equation}\label{eqn:BDS-ansatz}
M_n(a) = 1 + \sum_l a^l M_n^{(l)} = \exp \left(\sum_{l=1}^\infty a^l \left(f^{(l)}(\epsilon) M_n^{(1)}(l \epsilon) + C^{(l)}+ E_n^{(l)}(\epsilon)\right) \right)\,,
\end{equation}
where $M_n^{(1)}(l\epsilon)$ is the all-orders in $\epsilon$ function $M_n^{(1)}(\epsilon)$, that appeas in the one-loop amplitude calculation with $\epsilon$ replaced by $l\epsilon$,
$f^{(l)}(\epsilon) = f^{(l)}(\epsilon) + \epsilon f^{(l)}_1(\epsilon) + \epsilon^2 f^{(l)}_2(\epsilon) $, $C^{(l)}$ are constants independent of $n$ and $E_n^{(l)}(\epsilon)$  is a constant that vanishes as $\epsilon \to 0$. 

\subsubsection{Relation of the divergent part to Wilson loops}
It was known before, that the infrared divergences, which in the expression above are hidden in $M_n^{(1)}(l\epsilon)$, exponentiate in a form equivalent to \eqref{eqn:BDS-ansatz}, i.e. that the amplitude can be written as
\begin{equation}
M_n = \exp(Z_n) \times \text{finite}\,,
\end{equation}
where
\begin{equation}\label{eqn:BDS-div-part}
Z_n = -\frac{1}{4} \sum_{i=1}^n \left(\frac{\Gamma^{(l)}_{\text{cusp}}}{(l\epsilon)^2}+\frac{\Gamma^{(l)}}{l\epsilon} \right) \sum_{i=1}^n \left( \frac{-s_{i,i+1}}{\mu^2} \right)^{-l \epsilon}
\end{equation}
and $\Gamma_{\text{cusp}}= \sum_{l} a^l \Gamma_{\text{cusp}}^{(l)}=2a - 2 \zeta_2 a^2 + \op(a^3)$ is the \emph{cusp anomalous dimension}\index{cusp anomalous dimension}, $\Gamma^{(l)}$ is the collinear anomalous dimension and $s_{i,i+1}=(p_i+p_{i+1})^2$ are the kinematical invariants of the scattering process. The name \emph{cusp anomalous dimension} refers to the fact, that Wilson loops with cusps exhibit UV divergences, which cannot be removed by gauge field and coupling constant renormalisation, but need an additional renormalisation with associated anomalous dimension\footnote{We refer to the part of the anomalous dimension, which is proportional to the angle  $\theta$ of the cusp. $\Gamma_{\text{cusp}}(a,\theta)=\theta \,\Gamma_{\text{cusp}}(a)$ for $\theta>>1$ which is true to all orders in the coupling \cite{Korchemsky:1987wg}.}  $\Gamma_{\text{cusp}}$ \cite{Polyakov:1980ca,Dotsenko:1979wb,Brandt:1981kf,Dorn:1986dt,Korchemsky:1987wg}.

It turned out that the anomalous dimensions of the Wilson loops play an important role in perturbative QCD, since the same universal function governs the infrared behaviour of relevant quantities in QCD \cite{Ivanov:1985np,Korchemsky:1985xj,Korchemsky:1985ts,Korchemsky:1986fj,Korchemsky:1988hd,Korchemsky:1988si,Korchemsky:1991zp}.
It is known from studies of QCD, that in properly defined physical observables such as inclusive cross-sections all IR divergences cancel, however they cannot be removed from the S-matrix itself. 

\subsubsection{Finite part of the amplitudes}
One of the main results of \cite{Bern:2005iz} is that \eqref{eqn:BDS-ansatz} also implies an exponentiation of  the finite part. After subtracting the divergences from $M_n(l\epsilon)$ in a suitable form based on the known exponentiation of the divergent term, the amplitude can be written as $\mathcal{M}_n = \exp(Z_n) \exp(F_n)$ or its logarithm as 
\begin{equation}
\ln M_n = Z_n+ F_n\,,
\end{equation}
where \cite{Bern:2005iz} 
\begin{equation}\label{eqn:BDS-finite-part}
F_n^{\text{BDS}} =  \frac{1}{2} \Gamma_{\text{cusp}}(a)F_n^{(1)} + C
\end{equation}
and where $F_n^{(1)}$ is the finite part calculated at one-loop order and $C$ is a constant, e.g.
for $n=4$, compare \eqref{eqn:I4(s,t)}
\begin{align}\label{eqn:BDS-finite-part-4-5}
 F_4^{(1)} &= \frac{1}{2} \ln^2 \left(\frac{s}{t} \right) + 4 \zeta_2\,,\\ \nn
 F_5^{(1)} & = \frac{1}{4} \sum_{i=1}^5 \left( \ln \left(\frac{s_{i,i+1}}{s_{i+3,i+4}} \right) \ln \left(\frac{s_{i+2,i+3}}{s_{i+1,i+2}} \right)  + 3 \zeta_2 \right)\,.
\end{align} 
The remaining one-loop functions $F_n^{(1)}$ for $n \geq 6$ can be found in \cite{Bern:1994zx,Bern:2005iz}. 

\subsubsection{All-loop amplitudes}
The remarkable fact about \eqref{eqn:BDS-finite-part}  is that the coupling constant dependence is completely captured by the cusp anomalous dimension  $\Gamma_{\text{cusp}}(a)$ and the full kinematical dependence on the momenta $p_i$ is determined by the one-loop finite part $F_n^{(1)}$. Furthermore, the cusp anomalous dimension is thought to be known to all orders in the coupling constant through the BES equations \cite{Beisert:2006ez}.

Quite impressively, the BDS ansatz constitutes a proposal for an all-loop solution for $n$-point gluon amplitudes in $\sym$. However, it turns out, that this ansatz breaks down starting from $n=6$ at two loops and has to be corrected \cite{Drummond:2007bm}, \cite{Bern:2008ap} by a remainder function $R_n$ of conformally invariant cross ratios \eqref{eqn:cross-ratios}.

As will be discussed in section \eqref{sec:amplitude-wilson-loop-duality}, even though the all-loop $n$-point BDS ansatz breaks down, a relation between light-like polygonal Wilson loops and scattering amplitudes remains valid, thus opening a new avenue for the all-loop solution of scattering amplitudes for $n \geq 6$.

\subsection{Dual Conformal Symmetry of Amplitudes}\label{sec:dual-conformal-invariance}

In \cite{Drummond:2006rz} it was remarked that the integrals appearing in planar amplitude calculations in $\sym$ admit a \emph{dual conformal symmetry}\index{dual conformal symmetry}. The symmetry is also referred to as a \emph{hidden symmetry}, because it has no direct Lagrangian origin, it is distinct from the ordinary conformal symmetry of the amplitudes \eqref{eqn:invariance-under-conformal-symmetry}.

This can be seen by rewriting the momenta in terms of dual variables $x_1,...,x_n$ as
\begin{equation}\label{eqn:dual-variables}
p_i^\mu = x_i^\mu - x_{i+1}^\mu \equiv x_{i,i+1}^\mu,\qquad x_{i+n}^\mu\equiv x_i^\mu\,.
\end{equation}
As an example, consider the loop integral that appears in the 4-point one-loop amplitude calculation \eqref{eqn:4-point-amplitude-1loop}
\begin{equation}\label{eqn:1-loop-conformal-invariance}
s t I_4^{(1)}(s,t)=\int \frac{d^dl}{(2\pi)^d} \frac{s t}{l^2(p_1+l)^2 (p_1+p_2+l)^2 (p_4-l)^2} = \int \frac{d^d x_l}{(2\pi)^d} \frac{x_{13}^2 x_{24}^2}{x_{1l}^2 x_{2l}^2 x_{3l}^2 x_{4l}^2}\,,
\end{equation}
where we have performed the change of variable \eqref{eqn:dual-variables} and shifted the loop variable\footnote{Conventionally, the loop variable is designated by $x_5$ or $x_0$ in most of the literature. Here we denote it by $x_l$ in order to avoid confusion with the second relation in \eqref{eqn:dual-variables}.}
$x_l^\mu:=l^\mu-x_1^\mu$.  It is easy to see, that \eqref{eqn:1-loop-conformal-invariance} is invariant under inversions \eqref{eqn:inversions-xij2} 
\begin{equation}
I: \quad x_{ij}^2 \to \frac{x_{ij}^2}{x_i^2 x_j^2},\qquad d^dx_l \to  \frac{d^dx_l}{(x_l^2)^d}\,.
\end{equation}
for $d=4$.
Dilatation and translation invariance of \eqref{eqn:1-loop-conformal-invariance} are trivial in $x$-space and thus  by \eqref{eqn:I-P-I gives K}  equation \eqref{eqn:1-loop-conformal-invariance} is invariant under dual conformal symmetry for $d=4$.
However, since the integrals suffer from infrared divergences, a regularisation needs to be introduced. Setting $d=4-2\epsilon$ in dimensional regularisation explicitly breaks the dual conformal invariance, thus admitting only a \emph{broken dual conformal invariance}\index{dual conformal symmetry!broken} of the amplitudes. The integrals are therefore also called \emph{pseudo-conformal}\index{dual conformal symmetry!pseudo conformal}.

The integrals appearing at higher orders in perturbation theory are also dual conformal invariant in the above sense, e.g. in \cite{Bern:2006ew}  the appearing integrals were found to have this property in a four-loop calculation.
Assuming dual conformal invariance to be valid to all orders, %::I think at least this was not proven for INDIVIDUAL amplitudes, contrary to what was said in 1203.6362
together with unitarity methods, provides a powerful mechanism for higher loop calculations.

This property was used in \cite{Bern:2007ct} to construct the four-point five-loop amplitude, by listing all \emph{pseudo-conformal} integrals and determining the coefficients $0, \pm 1$ of the integrals\footnote{The coefficient of the integrals normalised by the tree-level amplitude.} using unitarity techniques as introduced in section \ref{sec:generalized-unitarity}.
As mentioned in \cite{Bern:2007ct,Bern:2010tq} it would be nice to have an analogous rule for the selection of integrals in the non-planar sector. One might expect such a relation between planar and non-planar integrals, since the planar and non-planar diagrams are linked by certain numerator identities  \cite{Bern:2008qj,Mafra:2009bz,Tye:2010dd,BjerrumBohr:2010zs,Bern:2010yg}.

In \cite{Alday:2009zm} a new way to regulate the IR divergences of gluon scattering amplitudes was introduced. The idea is to break the gauge group $U(N+M)$ to $U(N) \times U(1)^M$ by giving a vacuum expectation value $\sim m_i \delta_{ij}$ to the scalar fields with gauge group indices $i,j=N+1,..,N+M$ into a certain direction in the $SO(6)$ internal space. This introduces heavy fields $\op_{ia}$ with masses $m_i$ and light fields $\op_{ij}$ with masses $m_i-m_j$, while the fields $\op_{ab}$ corresponding to the unbroken part of the gauge group $(a,b=1,..,N)$ remain massless. 
The integrals appearing in an amplitude calculation thus depend on the masses $m_i$. In \cite{Alday:2009zm} it was shown, that the integral appearing in the one-loop four-point amplitude calculation
is invariant under an \emph{extended dual conformal symmetry}\index{dual conformal symmetry!extended}
\begin{equation}\label{eqn:extended-dual-conformal-symmetry}
\hat{K}^\mu I_4^{(1)} (x_{ij}^2, m_i) = 0\,, \quad \hat{K}^\mu = \sum_{i=1}^4 2 x_i^\mu (x_i \cdot \partial_i)- x_i^2 \partial_i^\mu + 2 x_i^\mu m_i \partial_{m_i} - m_i^2 \partial_i^\mu\,.
\end{equation}
The masses $m_i$ can be considered as additional components in an \emph{extended dual space} $\{x_i^\mu, m_i\}$. Since there is no need for an additional regularisation, the dual conformal symmetry is exact in the sense \eqref{eqn:extended-dual-conformal-symmetry} at loop-level and not anomalous as in the case of dimensional regularisation. More details on this regularisation can be found in the recent review \cite{Henn:2011xk}. The obvious conjecture is that any amplitude at any loop-level is a linear combination of integrals, which satisfy the extended dual conformal symmetry \eqref{eqn:extended-dual-conformal-symmetry}.

The appearance of this hidden dual conformal symmetry can be explained by the ordinary conformal symmetry of the Wilson loop, if one accepts the duality between amplitudes and Wilson loops, see section \ref{sec:amplitude-wilson-loop-duality} for more details. 

\subsection{Dual Superconformal Symmetry}\label{sec:dual-superconformal-symmetry}
In \cite{Drummond:2008vq} it was shown that the \emph{hidden} dual conformal symmetry of the amplitudes in $\sym$ can be extended to a \emph{dual superconformal symmetry}\index{dual conformal symmetry!superconformal symmetry}.  As mentioned in the introduction of this chapter, on the string side the same symmetry emerges  from fermionic T-duality.

The defining relation for the dual space coordinates \eqref{eqn:dual-variables} in spinor helicity variables reads
\begin{equation}\label{eqn:definition-dual-space-spinor-helicity}
(x_i)^{\alpha\dot\alpha}-(x_{i+1})^{\alpha\dot\alpha} = \lambda_i^\alpha \tilde{\lambda}_i^{\dot\alpha} = p_i^{\alpha\dot\alpha}
\end{equation}
and the dual variables automatically satisfy the momentum conservation constraint imposed by $\delta(p)$ in  \eqref{eqn:N4SYM-superamplitude}, i.e. $\sum_i \lambda^\alpha \tilde{\lambda}_i^{\dot \alpha}=0$. Similarly, one can introduce variables $\theta_i^{A\alpha}$ \cite{Drummond:2008vq}, which automatically satisfy the super-momentum conservation imposed by $\delta(q)$ in \eqref{eqn:N4SYM-superamplitude}
\begin{equation}\label{eqn:definition-dual-super-space}
(\theta_i)^{A\alpha}-(\theta_{i+1})^{A\alpha} = \lambda_i^\alpha \eta_i^{A} = q_i^{A\alpha}\,.
\end{equation}
The dual superconformal symmetry can be formulated by the action of generators $J^a$ in the \emph{full dual superspace} $(\lambda_i^\alpha, \tilde\lambda_i^{\dot\alpha}, \eta_i^A, x_i^{\alpha\dot\alpha},\theta_i^{A\alpha} )$. The generators $J^a$ satisfy the commutation relations of the $psu(2,2|4)$ algebra \ref{sec:superconformal-algebra} and their explicit form is given in \cite{Drummond:2008vq}. In \cite{Drummond:2009fd} some of these generators were slightly reformulated to a form $J_a^\prime$ which leaves the amplitudes invariant  
\begin{equation}\label{eqn:dual-superconformal-invariance-of-amplitudes}
J_a^\prime \mathcal{A}_n^{\text{tree}} = 0\,.
\end{equation}
When restricting the action of the generators $J_a^\prime$ to the \emph{on-shell} superspace $(\lambda_i^\alpha, \tilde{\lambda}_i^{\dot\alpha},\eta_i^A)$, the generators $P_{\alpha\dot\alpha},Q_{\alpha A} \in J_a^\prime$ become trivial, the remaining generators coincide up to signs with those of $j_a$  \eqref{eqn:invariance-under-conformal-symmetry}, except for ${K^\prime}^{\alpha\dot\alpha}, {S^\prime}_\alpha^A$.
For more detailed information on dual superconformal and Yangian symmetry, see e.g. the reviews  \cite{Drummond:2010km,Ferro:2011ph}. 
At loop-level the dual conformal symmetry is broken, albeit in exactly the same manner for MHV and NMHV amplitudes, such that their ratio $\mathcal{R}$ is an invariant \cite{Drummond:2008vq,Drummond:2008bq,Brandhuber:2009xz,Elvang:2009ya,Brandhuber:2009kh} under dual conformal symmetry.
We comment on the realization of the symmetry at loop-level in the following section.

\subsection{Yangian Symmetry}\label{sec:Yangian-symmetry}
In \cite{Drummond:2009fd} it was shown, that in \emph{on-shell superspace} the generators $j_a$ and the non-trivial dual superconformal generator\footnote{One could also take the other non-trivial generator $S^\prime$, since they are related by the commutation relations  $[K^\prime,\bar{Q}]=S^\prime$ of the algebra \ref{sec:superconformal-algebra}. As described in the previous section  the remaining generators are trivial or coincide up to signs with those of $j_a$.} ${K^\prime}^{\alpha\dot\alpha} \in J_a^\prime$ together form the \emph{Yangian}\index{Yangian!symmetry} $\text{Y}(psu(2,2|4))$ of the superconformal algebra $psu(2,2|4)$. 

The generators of the ordinary superconformal symmetry form the level-zero generators $j_a^{(0)}\equiv j_a$ of the Yangian algebra \index{Yangian!algebra} and satisfy the superconformal algebra
\begin{equation}
[ j_a^{(0)},j_b^{(0)} \} = f_{ab}^{~~c} j_c^{(0)}\,,
\end{equation}
where $f_{ab}^{~~c}$ are the structure constants of $psu(2,2|4)$, see  \ref{sec:superconformal-algebra}. The level-one generators can then be written in terms of the single particle level-zero generators $j^{(0)}_{i,a}$
\begin{equation}
j_a^{(1)} = f_{a}^{~bc}  \sum_{i>j}^n j_{i,b}^{(0)} j_{j,c}^{(0)},\qquad j_a =\sum_i^n  j^{(0)}_{i,a} 
\end{equation}
and satisfy  
\begin{equation}
[ j_a^{(1)},j_b^{(0)} \} = f_{ab}^{~~c} j_c^{(1)}\,,
\end{equation}
as well as the Serre relations, see \cite{Drummond:2009fd}. It was shown in \cite{Drummond:2009fd} that the dual superconformal generators are equivalent to these level-one generators $j_a^{(1)}$ of the Yangian. This is sufficient to construct all higher-level generators $j_a^{(l)}$ and therefore
the superconformal together with the dual superconformal generators form an infinite-dimensional Yangian algebra. This translates into the following symmetry of the amplitudes
\begin{equation}\label{eqn:dual-superconformal-invariance-of-amplitudes}
y  \mathcal{A}_n^{\text{tree}} = 0, \qquad y \in Y(psu(2,2|4))
\end{equation}
and thus constrains the form of the amplitudes.  It has been argued that the tree-level amplitudes are uniquely determined  by requiring analytic properties such as the right collinear behaviour together with Yangian symmetries \cite{Korchemsky:2009hm,Bargheer:2009qu}. 

At loop-level the symmetries are broken. In \cite{Beisert:2010gn} superconformal anomalies are computed and it is shown that they may be cancelled through a universal one-loop deformation of the tree-level symmetry generators. 
The symmetry generators can be deformed in such a fashion as to render the amplitudes, defined in a dimensional regularisation scheme, invariant to one-loop order. In \cite{Drummond:2010zv} Yangian symmetry of a certain class of  light-like Wilson loops at one loop  was proven and it is argued that it can be thought of as the effect of the original conformal symmetry at loop-level.

In further investigations it became clear, that dual superconformal symmetry is not a symmetry of individual amplitudes, but of the S-matrix  \cite{Sever:2009aa,Korchemsky:2009hm,Bargheer:2009qu,Beisert:2010gn}. Equations relating the different amplitudes have been presented in \cite{Bullimore:2011kg} and it is shown that the complete planar S-matrix is fully dual supersymmetric, but this is not true for individual  N$^k$MHV loop-amplitudes. 
 
In \cite{CaronHuot:2011kk} an all-loop equation for the $\bar Q$-symmetry acting on the BDS-subtracted planar S-matrix of $\sym$ is presented, which is argued to amount for exact Yangian symmetry. In principle, these equations can be used to determine the all-loop S-matrix uniquely and to all orders in the coupling constant.

\subsubsection{Twistor space formulation of amplitudes and Yangian symmetry}
In \cite{Drummond:2009fd} it was also shown, that the level-zero and level-one generators can be written nicely in twistor space variables \cite{Witten:2003nn} as first and second order differential operators. Twistor space is very useful in order to construct Yangian invariants \cite{Korchemsky:2009jv,Drummond:2010uq,Korchemsky:2010ut,Drummond:2010qh} and leads to the remarkable construction of the manifestly Yangian invariant all-loop integrand for scattering amplitudes in the planar limit of $\sym$ \cite{ArkaniHamed:2009dn, ArkaniHamed:2009dg,Mason:2009qx,ArkaniHamed:2009vw,ArkaniHamed:2010kv,ArkaniHamed:2010gh,ArkaniHamed:2010gg}. 
Since we will not make further use of the formulation in  twistor space, we refer the reader to the reviews \cite{Drummond:2010km,Ferro:2011ph} and the references therein for more information.

\section{Duality between Amplitudes and Wilson Loops}\label{sec:amplitude-wilson-loop-duality}
In \cite{Alday:2007hr,Alday:2007he} Alday and Maldacena proposed a strong coupling description of planar gluon scattering amplitudes in terms of  minimal surfaces in $AdS$ space. The calculation formally resembles the strong coupling calculation of  a Wilson loop's expectation value 
\begin{equation}\label{eqn:Wilson-loop}
 \langle W_n  \rangle= \langle W(C_n)\rangle = \frac{1}{N} \left\langle  \tr \mathcal{P}\exp \left( i \oint_{C_n} A_\mu dz^\mu \right) \right \rangle\,,
\end{equation}
where the light-like polygonal contour $\mathcal{C}_n$ is made of $n$ points $x_i$ related to the gluon momenta via $p_i= x_{i+1}-x_i$. 

\begin{figure}[t]\centering
  \includegraphics[width=.75 \textwidth]{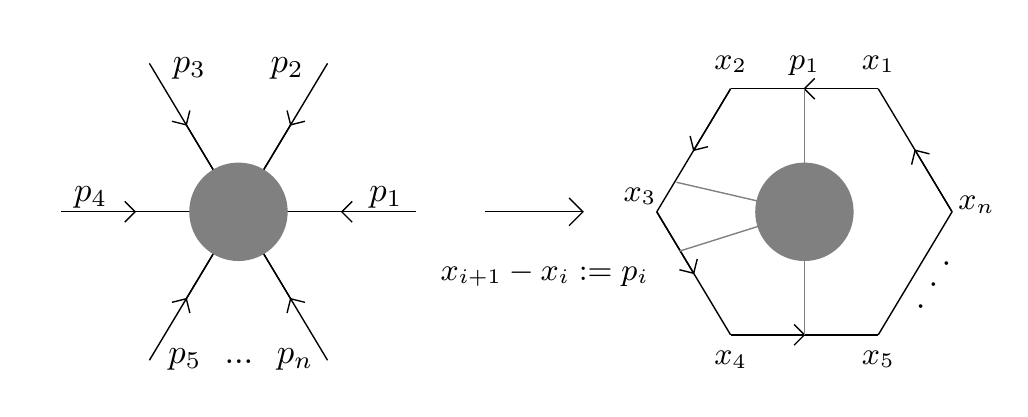}
\caption{Duality between MHV scattering amplitudes $A_n^{\text{MHV}}$ and the expectation value of Wilson loops $W_n$ in $\mathcal{N}=4$ super Yang-Mills.}
 \label{fig:duality}
\end{figure}

Interestingly, it was found in \cite{Drummond:2007aua} that the relation between gluon scattering amplitudes and Wilson loops is also valid in the weak coupling regime for $n=4$. The amplitude / Wilson loop duality states\footnote{Upon a specific identification of the renormalisation scales and regulators in the amplitude and the Wilson loop calculation, see \cite{Drummond:2007aua}. } that up to $\op(\epsilon)$ terms
\begin{equation}\label{eqn:duality-wilson-loop-amplitude}
\ln M_n = \ln  \langle W_n  \rangle + \text{const.} \,,
\end{equation}
where $M_n$ is the ratio of the planar MHV amplitude and its tree-level expression as defined in \eqref{eqn:N4SYM-superamplitude-MHV-loop}. The duality relation \eqref{eqn:duality-wilson-loop-amplitude} is illustrated in figure \ref{fig:duality}.
As explained before, $M_n$ was conjectured to be given to all-loop order by the BDS ansatz \eqref{eqn:BDS-ansatz}.

As mentioned in section \ref{sec:BDS}, relations between the divergent parts were already known before 
\cite{Korchemsky:1987wg,Korchemskaya:1992je,Korchemsky:1993uz},
but in particular the duality \eqref{eqn:duality-wilson-loop-amplitude} is also true for the finite parts\footnote{Strictly speaking $F_n^{\text{BDS}}$, $F_n^{\text{WL}}$ are the finite part of the logartihms of $M_n$ and $ \langle W_n \rangle$.} 
of the MHV amplitudes \eqref{eqn:BDS-finite-part} and the Wilson loops 
\begin{equation}
F_n^{\text{BDS}} = F_n^{\text{WL}}+ \text{const.}\,.
\end{equation} 
The duality was checked for n points at one loop \cite{Brandhuber:2007yx} as well as for 4, 5 and  6 points \cite{Drummond:2007cf}, \cite{Drummond:2007au}, \cite{Drummond:2008aq} at two loops. It was found that the expectation value of the Wilson loop is governed by an anomalous conformal Ward identity \cite{Drummond:2007cf,Drummond:2007bm} that completely fixes its form at 4 and 5 points and allows for an arbitrary function $R_n$ of conformal invariants starting from 6 points, as will be reviewed in  section \ref{sec:ward-identity-wilson-loops} . It was found that this so-called \emph{remainder function}\index{remainder function} $R_n$ is indeed present starting from 6 points at two-loop order and leads to a correction \cite{Drummond:2007bm}, \cite{Bern:2008ap} of the BDS ansatz \cite{Bern:2005iz} for planar gluon scattering amplitudes. 

The Wilson loop is numerically under control  at two loops  for arbitrary $n$  \cite{Anastasiou:2009kna} and the duality was also found to hold to order epsilon at two-loops  in \cite{Brandhuber:2010bj}. Much effort was put into the analytical evaluation of the remainder function at 6 points  \cite{DelDuca:2009au, DelDuca:2010zg,Zhang:2010tr} and it was possible to recast the result in a very compact form \cite{Goncharov:2010jf}  using the notion of the {\itshape symbol} of a function. This technique was applied to the three-loop hexagonal Wilson loop in \cite{Dixon:2011pw}. The remainder function was also analysed in the limit of self-crossing Wilson loops where it exhibits additional divergences \cite{Georgiou:2009mp,Dorn:2011gf,Dorn:2011ec}.

Constraints of an operator product expansion for Wilson loops were investigated \cite{Alday:2010ku,Gaiotto:2011dt,Belitsky:2011nn} and it was argued that this knowledge is enough to fix the full two-loop answer for arbitrary $n$. 
The duality has also been extended to tree-level amplitudes with arbitrary helicity states by introducing a suitable supersymmetric Wilson loop see section \ref{sec:super-Wilons-loop}. Reviews on the duality between scattering amplitudes and Wilson loops can be found in \cite{Henn:2009bd}, \cite{Alday:2008yw}.

\subsection{Light-Like Polygonal Wilson Loops}\label{sec:light-like-Wilson-loops}
In order to illustrate the simplest case of the duality relation \eqref{eqn:duality-wilson-loop-amplitude} we shortly review the one-loop calculation of the four-sided Wilson loop  \cite{Drummond:2007aua}. The contour of the $n$-sided polygon $\mathcal{C}$ is given by $n$ points $x_i$ $(i=1,...,n)$ and we can parametrise each edge $\mathcal{C}_i$ via
\begin{align}\label{eqn:zi-definition}
z_i^{\mu}(s_i)= x_i^{\mu} + p_i^{\mu} s_i,\qquad \qquad p_i^{\mu} = x_{i+1}^{\mu} -x_i^{\mu}\,,
\end{align}
where $s_i \in [\,0,1\,]$ and $\mathcal{C} = \mathcal{C}_1 \cup ... \cup\, \mathcal{C}_n$. The lowest order contribution to the expectation value of the Wilson loop \eqref{eqn:Wilson-loop}, see \eqref{eqn:Wilson-Loop-after-path-ordering}, is

\begin{align}\label{eqn:general-expression}
\langle W_n \rangle^{(1)} &= \frac{i^2}{N}\tr (T^aT^b) \int_{z_i>z_j} dz_i^\mu dz_j^\nu \langle A_\mu^a(z_i) A_\nu^b (z_j)\rangle = g^2 C_F \frac{\Gamma(1-\epsilon)}{4\pi^{2-\epsilon}}  \sum_{i>j} I_{ij}\,,
\end{align}
where $C_F=\delta^{aa}/(2N)$, we have used the gluon propagator given in \eqref{eqn:N-4-SYM-propagators} in position-space and using \eqref{eqn:products:p_ip_m}, \eqref{eqn:z_i-z_j^2} we defined
\begin{align}
I_{ij} &= \int ds_i ds_j\, p_i \cdot p_j (-(z_i-z_j)^2)^{1-d/2} \\ \nn
&= \frac{1}{2} \int_0^1 ds_i \int_0^1 ds_j 
\frac{ x^2_{i,j+1} +  x^2_{i+1,j} -  x^2_{i,j} -  x^2_{i+1,j+1}}{(-x^2_{ij}\bar{s}_i\bar{s}_j - x^2_{i+1,j} s_i \bar{s}_j - x^2_{i,j+1}\bar{s}_i s_j - x^2_{i+1,j+1}s_i s_j)^{d/2-1}}\,.
\end{align}
\begin{figure}[t]
\centering
\subfloat[]{\includegraphics[width=.3 \columnwidth]{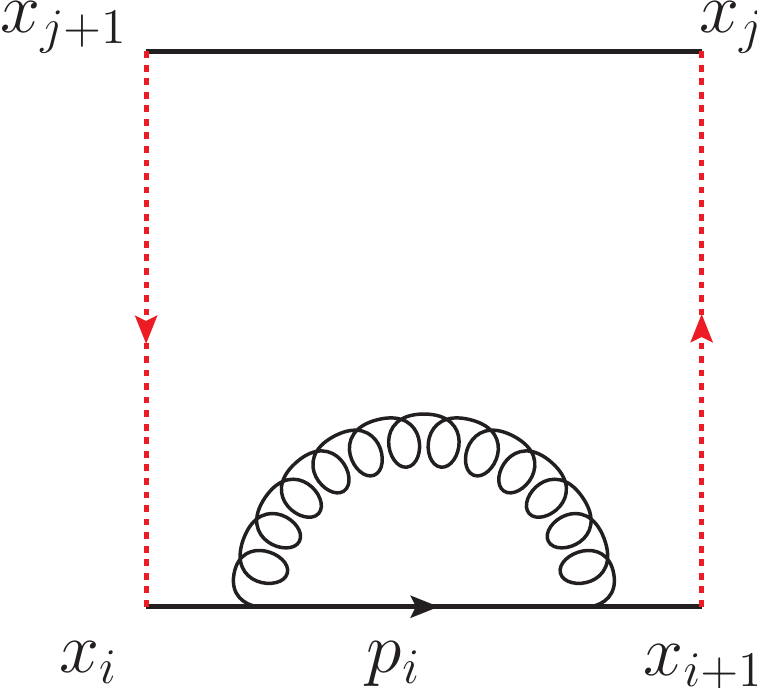}\label{fig:N4-general-diagram-same-edge}}~~~~~
\subfloat[]{\includegraphics[width=.325 \columnwidth]{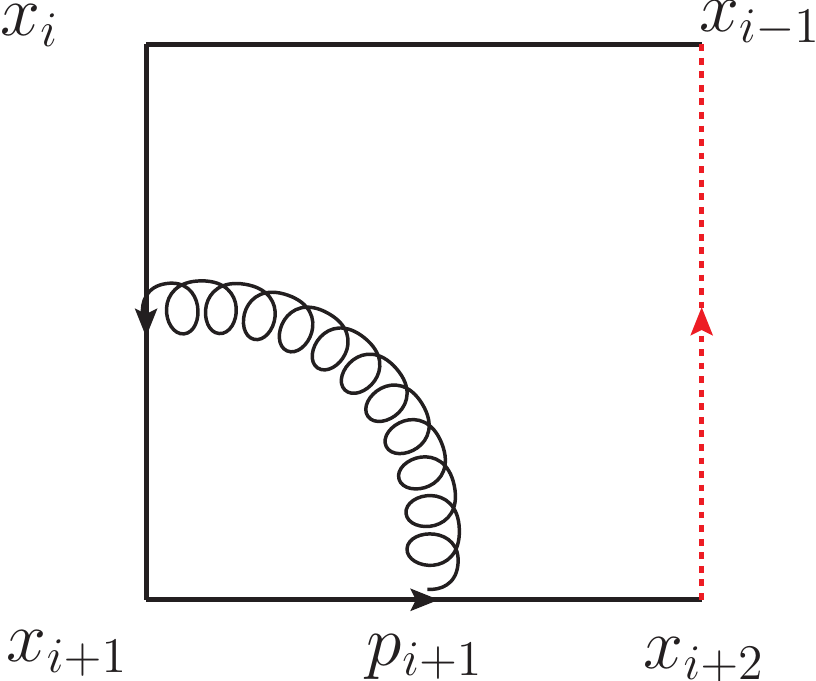}\label{fig:N-4general-diagram-adjacent}}~~~~~
\subfloat[]{\includegraphics[width=.3 \columnwidth]{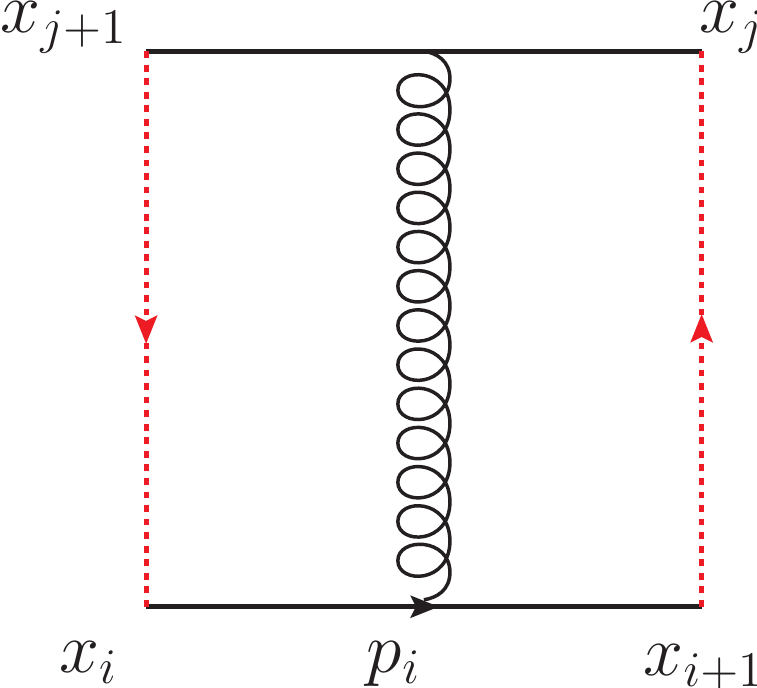}\label{fig:N-4general-diagram-non-adjacent}}~~~~~~~~~~
\caption{Examples for the three classes of one-loop diagrams with (a) a propagator connecting the same edge (vanishing), (b) propagator stretching between adjacent edges (divergent), (c) propagator stretching between non-adjacent edges (finite). Dotted lines represent one or more light-like distances.
}
 \label{fig:one-loop-diagrams-WL-N4}
\end{figure}

The sum is reduced to  $i\neq j$ since for $i=j$ we have $p_i^2=0$ and the contribution shown in figure \ref{fig:N4-general-diagram-same-edge} vanishes. 
There are $n$ divergent diagrams  $I_{i+1,i}$ of the type shown in fig. \ref{fig:N-4general-diagram-adjacent}
\begin{equation}
\nn
I_{i+1,i}  =
-\frac{1}{2} (-x_{i,i+2}^2)^{2-d/2} \int_0^1 ds_i ds_{i+1} \frac{1}{\left(\bar{s}_i  s_{i+2}\right)^{1-\epsilon}}
= -\frac{1}{2} \frac{(-x_{i,i+2}^2)^{\epsilon}}{\epsilon^2} 
\end{equation}
and one can see that the divergence arises from the cusps formed by two light-like edges, when the ends of the gluon propagator approach the cusp.
It is well-known \cite{Korchemskaya:1992je} that Wilson loops evaluated over light-like polygonal contours exhibit this type of UV-divergences.

Diagrams with $|i-j| \geq1$ shown in figure \ref{fig:N-4general-diagram-non-adjacent} yield finite contributions.
For $n=4$, the only diagrams of this type are $I_{31}= I_{42}$ and have the result
\begin{equation}
I_{31} = \frac{1}{2} \int ds_1 ds_3 \frac{x_{13}^2+x_{24}^2}{x_{13}^2 \bar{s}_1\bar{s}_3 + x_{24}^2 s_1 s_3} = \frac{1}{4} \left( \ln^2\left(\frac{x_{13}^2}{x_{24}^2} \right)+\pi^2\right)\,,
\end{equation}
such that the full one-loop contribution for $n=4$ is
\begin{align}\label{eqn:n-4sym-wilson-loop-one-loop}
\langle W_4 \rangle^{(1)}  &=  \frac{g^2 N}{8 \pi^2} \Bigg[-\frac{1}{2}\sum_{i=1}^4 \frac{(-x_{i,i+2}^2 {\mu}^2 \pi e^{\gamma_E})^{\epsilon}}{\epsilon^2}+ \frac{1}{2} \ln^2\left(\frac{x_{13}^2}{x_{24}^2} \right)+\frac{\pi^2}{3} \Bigg]\,,
\end{align}
where we have restored the regularisation scale $\mu$, let $C_F\to N/2$ in the planar limit and used $\Gamma(1-\epsilon)= \exp(\epsilon \gamma_E)+ \frac{1}{2}\zeta_2$. Indeed, upon a specific identification of the regularisation parameters $\epsilon$ and the scale $\mu^2$  \cite{Drummond:2007aua},  up to a constant, this coincides  with the one-loop correction to the planar scattering amplitude \eqref{eqn:4-point-amplitude-1loop}, \eqref{eqn:I4(s,t)} if we set $s=x_{13}^2, t=x_{24}^2$.
The one-loop result was generalised to $n$ points and reads \cite{Brandhuber:2007yx} 
\begin{align}\label{eqn:n-4sym-wilson-loop-one-loop}
\langle W_n \rangle^{(1)}  &=  \frac{g^2 N}{8 \pi^2} \Bigg[-\frac{1}{2}\sum_{i=1}^n \frac{(-x_{i,i+2}^2 {\mu}^2 \pi e^{\gamma_E})^{\epsilon}}{\epsilon^2}+ \mathcal{F}_n^{\text{WL}}  \Bigg]\,,
\end{align}
where the finite contribution $\mathcal{F}_n^{\text{WL}}$ is 
%\footnote{They are related via $\mathcal{F}_n^{\text{WL}}= F_n^{(1),\text{BDS}}-n/4 \zeta(2)$} 
equal to the finite part $F_n^{(1)}$ in the BDS conjecture \cite{Bern:2005iz} up to a constant, e.g. for $n=4$ and $n=6$
\begin{align}\label{eqn:BDS-WL-n4-n6}
\mathcal{F}_4^{\text{WL}} &= \frac{1}{2} \ln^2 \left( \frac{x_{13}^2}{x_{24}^2} \right) + \text{const.}\,, \\ \nn
\mathcal{F}_6^{\text{WL}} &= 
 \frac{1}{2} \sum_{i=1}^6\Bigg[ - \ln \left(\frac{x^2_{i,i+2}}{x^2_{i,i+3}} \right) \ln \left(\frac{x^2_{i+1,i+3}}{x^2_{i,i+3}} \right) \\ \nn
&\qquad\qquad+ \frac{1}{4} \ln^2 \left( \frac{x^2_{i,i+3}}{x^2_{i+1,i+4}}\right) 
-\frac{1}{2} \text{Li}_2 \left( 1-\frac{x_{i,i+2}^2 x_{i+3,i+5}^2}{ x_{i,i+3}^2 x_{i+2,i+5}^2} \right) \Bigg] + \text{const.}\,.
\end{align}

\subsection{Anomalous Conformal Ward Identities for Wilson Loops}\label{sec:ward-identity-wilson-loops}
In \cite{Drummond:2007cf,Drummond:2007au} it was shown that the finite part of the expectation value is governed by an \emph{anomalous conformal Ward identity}\index{conformal Ward identity!anomalous!for Wilson loops}, which we shall shortly review here.

As we have seen in the previous section, due to the cusp divergences of the light-like Wilson loops, a regularisation needs to be introduced in order to calculate the expectation value. As mentioned earlier, dimensional regularisation explicitly breaks conformal symmetry and thus conformal Ward identities \ref{sec:conformal-Ward-id} for expectation values become anomalous. This is due to the fact that the action  in $d=4-2\epsilon$ dimensions is not invariant under all generators of the conformal group. While dimensional regularisation preserves invariance under translations and Lorentz transformations, the action $S_\epsilon[\Phi]$ in $d=4-2\epsilon$ dimensions is not invariant under dilatations and special conformal transformations of the fields\footnote{$\Phi$ represents all fields that are integrated over in the path integral.} $\Phi \to \Phi + \delta \Phi$ as given in \eqref{eqn:infinitesimal-conformal-transformations-of-fields}. The action varies as \cite{Drummond:2007au}
\begin{align}\label{eqn:variation-action-under-conformal-trafos}
\delta_{D} S_\epsilon = \frac{2\epsilon}{\mu^{2\epsilon}} \int d^dx\, \mathcal{L}(x), \qquad
\delta_{K^\mu} S_\epsilon = \frac{4\epsilon}{\mu^{2\epsilon}} \int d^dx\,x^\mu \mathcal{L}(x) \,.
\end{align}
It is then easy to see how the generators of dilatations and special conformal transformations act on the expectation value of the Wilson loop
\begin{equation}
\langle W(\mathcal{C}_n) \rangle = \int \mathcal{D}\Phi\, e^{iS_{\epsilon}[\Phi]} \tr \left[\mathcal{P} \exp\left(i \oint_{C_n} dz^\mu 
A_\mu(z)\right) \right]\,.
\end{equation}
Consider a light-like polygon $\mathcal{C}_n$ defined by $n$ points $x_i^\mu$ satisfying $x_{i,i+1}^2=0$. The contour $\mathcal{C}^\prime_n$ that is obtained under an infinitesimal conformal transformations of the points $x_i$ as given by \eqref{eqn:generators-conformal-algebra}, \eqref{eqn:trafo-with-generators} is another light-like contour, because light-like distances trivially transform into light-like distances by \eqref{eqn:inversions-xij2}. By Poincar\'e symmetry, the expectation values $\langle W(\mathcal{C}_n) \rangle$, $\langle W(\mathcal{C}^\prime_n) \rangle$ are  functions of the coordinates $x_{ij}^2$, ${x^{\prime 2}_{ij}}$ and thus the variation of these expectation values under conformal transformations is generated by \eqref{eqn:generators-acting-on-several-points}, e.g. for special conformal transformation 
\begin{align}
\delta_{\epsilon \cdot K} \langle W(\mathcal{C}_n) \rangle &= \langle W(\mathcal{C}_n) \rangle - \langle W(\mathcal{C}^\prime_n) \rangle \\ \nn
&= \epsilon_\mu K^\mu \,\langle W(\mathcal{C}_n) \rangle \\ \nn
 &= \epsilon_\mu \sum_i  \left( 2 x_i^\mu (x_i \cdot \partial_i) - x_i^2 \partial_i^\mu \right) \, \langle W(\mathcal{C}_n) \rangle\,.
\end{align}
On the other hand, we can rewrite the expectation value of the Wilson loop over the transformed contour $\mathcal{C}_n^\prime$
\begin{equation}
\langle W(\mathcal{C}^\prime_n) \rangle = \int \mathcal{D}\Phi\, e^{iS_{\epsilon}[\Phi]} \tr \left[\mathcal{P} \exp\left(i \oint_{C_n^\prime} dz^\mu 
A_\mu(z)\right) \right]
\end{equation}
in terms of the original contour $\mathcal{C}_n$, if we evaluate it with the conformally transformed gauge field $A_\mu^\prime$. Rewriting $S_\epsilon[\Phi]=S_\epsilon[\Phi^\prime -\delta \Phi^\prime]=S_\epsilon[\Phi^\prime]-\delta S_\epsilon[\Phi^\prime]$ and using \eqref{eqn:variation-action-under-conformal-trafos} one can thus write
\begin{align}\label{eqn:Wilson-loop-Ward-identity}
D\,  \langle W(\mathcal{C}_n)  \rangle &= -\frac{2 i \epsilon}{ \mu^{2\epsilon}} \int d^dx \langle \mathcal{L}(x) W(\mathcal{C}_n) \rangle \,, \\ \label{eqn:special-conformal-Ward-identity}
K^\nu  \langle W(\mathcal{C}_n)  \rangle &= -\frac{4 i \epsilon}{\mu^{2\epsilon}} \int d^dx\, x^\nu \langle \mathcal{L}(x) W(\mathcal{C}_n) \rangle \,,
\end{align}
where $K^\nu, D$ are the differential operators given by \eqref{eqn:generators-acting-on-several-points}. It should be emphasised, that due to the factors of $\epsilon$ in \eqref{eqn:Wilson-loop-Ward-identity} and \eqref{eqn:special-conformal-Ward-identity}, it is sufficient to evaluate only the divergent term of the correlator on the right-hand side in order to get a differential equation for the finite part of the expectation value of the Wilson loop.

The dimensionally regularised  Wilson loop $\langle W(\mathcal{C}_n) \rangle$ is a dimensionless scalar function of 
the cusp points $x_i^\nu$, which appear paired with the regularisation scale as $x_{ij}^2 \mu^{2}$. 
As a consequence, the expectation values satisfies
\begin{equation}\label{eqn:wilson-loop-regularisation}
\left( \sum_{i=1}^n (x_i \cdot \partial_i) - \mu\, \frac{\partial}{\partial\mu} \right) \langle W(\mathcal{C}_n) \rangle =0 \,.
\end{equation}
This provides a consistency condition for the right-hand side of \eqref{eqn:Wilson-loop-Ward-identity}.

The all-loop anomalous conformal Ward identity of the light-like Wilson loop in $\sym$ was evaluated in \cite{Drummond:2007cf,Drummond:2007au}. The special conformal Ward identity for the finite part reads
\begin{align}
K^\mu\, F_n^{\text{WL}} &= \frac{1}{2} \Gamma_{\text{cusp}} (a) \sum_{i=1}^n \ln \left( \frac{x_{i,i+2}^2}{x_{i-1,i+1}^2}\right)  x_{i,i+1}^\mu \,.
\end{align}
Note that the dependence on the coupling constant $a$ only enters via the cusp anomalous dimension $\Gamma_{\text{cusp}}(a)$.
The solution\footnote{Using 
$\mathbb{K}^\nu x_{ij}^2 = 2 (x_i+x_j)^\nu x_{ij}^2\,, 
 \mathbb{K}^\nu x_{i}^\mu = 2 x_i^\nu x_i^\mu - x_i^2 \eta^{\mu\nu}$ it is easy to very that these expressions are indeed solutions.
} of this differential equation for $n=4,5$ is
\begin{align}
F_4^{\text{WL}} &= \frac{1}{4} \Gamma_{\text{cusp}}(a) \ln \left( \frac{x_{13}^2}{x_{24}^2} \right) + \text{const.}\,, \\ \nn
F_5^{\text{WL}} &= \frac{1}{8} \Gamma_{\text{cusp}}(a) \sum_{i=1}^5 \ln \left( \frac{x_{i,i+2}^2}{x_{i,i+3}^2} \right) \ln \left( \frac{x_{i+2,i+4}^2}{x_{i+1,i+3}^2} \right) + \text{const.}
\end{align}
and has exactly the form\footnote{It is easy to see the equivalence of the expressions, when rewriting $s_{i,i+1}=x^2_{i,i+2}$ and using the relation $x_{i+n}=x_i$.} of the finite part in the BDS conjecture \eqref{eqn:BDS-finite-part-4-5}, \eqref{eqn:BDS-finite-part}. Furthermore, one can check, that $F_n^{\text{BDS}}$  is a solution of the Ward identity also for $n\geq 6$.

However, starting\footnote{Due to the light-like separation of neighboring points $x_{i,i+1}^2$, non-trivial cross ratios can only be built starting from $n=6$, since we need at least 4 non-vanishing distances.}  from $n=6$, the functional form of $F_n$ is not uniquely fixed by the Ward identity, but the general solution of the anomalous conformal Ward identity contains an arbitrary function $f(u_{ijkl})$ of the conformal cross ratios \eqref{eqn:cross-ratios}. This is because the cross ratios \eqref{eqn:cross-ratios} are annihilated by all conformal generators, e.g.

\begin{equation}
K^\mu  \,u_{ijkl} =  K^\mu \left( \frac{x_{ij}^2 x_{kl}^2}{x_{ik}^2 x_{jl}^2} \right) =0 \,.
\end{equation}

\subsubsection{Remainder function}\index{remainder function}
Even though the BDS ansatz contains functions of these cross ratios, it was found in \cite{Drummond:2007bm}, \cite{Bern:2008ap}  that the BDS ansatz is not complete starting from two loops and  $n=6$, i.e.
\begin{equation}
F_6^{\text{BDS}} \neq F_6^{\text{MHV}}\,.
\end{equation}
It was first noticed in \cite{Drummond:2007bm} that the expectation value of the light-like Wilson loop at two loops for $n=6$ does not coincide with the BDS ansatz and in \cite{Bern:2008ap} it was confirmed, that the two-loop $n=6$ amplitude deviates in the same way from the BDS ansatz. Therefore, the duality between the Wilson loop and the amplitude continues to hold, i.e.
\begin{equation}\label{eqn:new-duality-relation}
F_6^{\text{MHV}} = F_6^{\text{WL}} + \text{const.}
\end{equation}
For this reason, the BDS ansatz \eqref{eqn:BDS-ansatz}, \eqref{eqn:BDS-finite-part}
only constitutes an all-loop conjecture for $n=4,5$ particles. The deviation of the amplitude respectively the Wilson loop expectation value from the BDS ansatz is called the \emph{remainder function}
\begin{equation}
R_n(u_{ijkl}) =F_n^{\text{BDS}} - F_n^{\text{WL}}\,,
\end{equation}
which is non-trivial starting from two loops and $n=6$. As mentioned above, much effort has been put into the numerical and analytical evaluation of this function. It is widely believed that the duality \eqref{eqn:new-duality-relation} continues to hold to all-loop order for arbitrary $n$. Since the evaluation of the Wilson loop is simpler -- also numerically faster -- than the amplitude calculation this provides a powerful tool for amplitude calculations.

\subsection{Supersymmetric Amplitudes and Wilson Loops}\label{sec:super-Wilons-loop}
The duality relation \eqref{eqn:duality-wilson-loop-amplitude}  is a relation between the expectation value of the bosonic Wilson loop \eqref{eqn:Wilson-loop} and the scalar function $M_n$ that contains the loop corrections to the MHV amplitudes \eqref{eqn:N4SYM-superamplitude-MHV-loop}.

A natural questions is, whether one can also construct a Wilson loop $\mathcal{W}_n$, that generates the other terms $A_{n;k}$ in \eqref{eqn:N4SYM-superamplitude-non-MHV-loop} corresponding to non-MHV amplitudes, i.e. a supersymmetric object that has an expansion in $\eta^A_i$ in the same way as the superamplitude \eqref{eqn:N4SYM-superamplitude-non-MHV-loop}.

Supersymmetric Wilson loops that reproduce the correct tree-level non-MHV amplitudes were proposed in  \cite{CaronHuot:2010ek,Mason:2010yk}. However, it was shown in \cite{Belitsky:2011zm} that the one-loop corrections to $\mathcal{W}_n$ do not match those of the non-MHV amplitudes. As explained in \cite{Belitsky:2011zm} the mismatch is due to singularities of the Wilson loop on a light-like contour in combination with the fact that the supersymmetry algebra of $\sym$ only closes \emph{on-shell}, i.e. modulo the equations of motion of the fields. Therefore, not only the conformal symmetry but also the chiral supersymmetry is broken at loop-level, while the \emph{dual} chiral supersymmetry of the amplitudes remains exact. The anomaly was further discussed in \cite{Belitsky:2012nu} and
in \cite{Beisert:2012xx} a regularization procedure that preserves super-Poincar\'e symmetry was proposed.
In the following section we shortly sketch another proposal for the dual description of scattering amplitudes, which does not suffer from this supersymmetric anomaly.

\section{Duality between Correlators and Wilson Loops}\label{sec:duality-correlators-amplitude-and-wilson-loops}
In addition to the duality between amplitudes and Wilson loops, a duality between light-like Wilson loops with $n$ cusps and $n$-point correlators $G_n=\langle \op(x_1)...\op(x_{n})\rangle $ of half-BPS protected operators was established in \cite{Alday:2010zy,Eden:2010zz,Eden:2010ce,Eden:2011yp,Eden:2011ku,Adamo:2011dq}. In the limit where the positions of adjacent operators become light-like, it is found that the one-loop $n$-point correlator divided by its tree-level expression coincides with a light-like $n$-polygon Wilson loop in the adjoint representation and thus in the planar limit via the amplitude / Wilson loop duality with the corresponding amplitude $A_n$ 
\begin{equation}
\lim_{x_{i,i+1}^2\to0} \ln (G_n/G_n^{\text{tree}} ) = \ln W^{\text{adj}}(\mathcal{C}_n)= \ln \left(A_n/A_n^{\text{tree}}  \right)^2\,.
\end{equation}
Therefore one often speaks about a \emph{triality} between correlators, amplitudes and Wilson loops. Unlike the Wilson loop / amplitude duality, see section \ref{sec:super-Wilons-loop}, the correlator / amplitude duality has a natural generalization to supercorrelators which are dual to superamplitudes \cite{Eden:2011yp,Eden:2011ku}.  An important feature of this duality is that objects are considered, which are well-defined  in $d = 4$ dimensions, thus avoiding the need of regularization and the associated anomalies.

\subsection{Hidden Permutation Symmetry in Correlation Functions}
Furthermore, a new hidden symmetry of four-point correlation functions and amplitudes in $\sym$  was discovered \cite{Eden:2011we,Eden:2012tu}. The all-loop integrand of the investigated correlation functions possesses an unexpected symmetry under the exchange of the four external and all internal (integration) points. Using this symmetry, the three-loop integrand can be fixed up to a few constants, which in turn can be determined using the duality between correlators and amplitudes with the knowledge of the three-loop amplitude result, impressively illustrating the power of these dualities. As a byproduct  the three-point function of two half-BPS operators and one Konishi operator at three-loop level is obtained \cite{Eden:2011we}. Furthermore, using the four-point correlation function of the stress-tensor multiplets \cite{Eden:2012tu,Eden:2012fe}, the five-loop anomalous dimension of the Konishi-operator was recently extracted.

\section{Form Factors and Wilson Loops}\label{sec:form-factors-and-wilson-loops}
Another duality with Wilson loops was proposed in \cite{Brandhuber:2010ad}. It relates \emph{form factors}\index{form factors} and periodic Wilson loops, see also \cite{Maldacena:2010kp} for a strong coupling description of form factors. Form factors are expectation values of the form
\begin{equation}
 \int d^4x e^{-iq x} \langle 1...n| \op(x) |0 \rangle\,,
\end{equation}
where $\langle 1...n|$ describes a multiparticle state with  momenta $p_1,...,p_n$ and $\op(x)$ is a gauge invariant operator. Form factors are therefore neither completely on-shell quantities such as scattering amplitudes nor completely off-shell quantities like correlation functions, but exhibit both features. Therefore, various on-shell techniques such as recursion relations and unitarity cuts can be applied.
The analysis of form factors was also extended to super form factors \cite{Brandhuber:2011tv} and
the form factors of half-BPS operators were evaluated analytically up to two loops in  \cite{Brandhuber:2012vm}.

\chapter{Amplitudes, Wilson Loops and Correlators in ABJM Theory}\label{sec:developments-in-ABJM}
\chaptermark{Amplitudes, Wilson Loops and Correlators}

As mentioned in the introductory chapter, there is another AdS/CFT correspondence, which involves a three-dimensional conformal field theory that ''lives'' on the conformal boundary of $AdS_4$ space and is therefore also called $AdS_4/CFT_3$ correspondence.

 This gauge theory is  the $\mathcal{N}=6$ ABJM theory \cite{Aharony:2008ug}, which is built upon an 
$SU(N)\times SU(N)$ gauge symmetry which allows for a planar $N\to\infty$ 
limit with $\lambda=N/k$ held 
fixed, where $k$ is the common absolute value of the Chern-Simons parameters 
of the two $SU(N)$ subgroups.
In this limit the ABJM theory is conjectured to be dual to type IIA string theory on $AdS_{4}\times\mathbb{CP}_{3}$ \cite{Aharony:2008ug,Benna:2008zy}, representing an exact
gauge-string duality pair very similar in nature to the well studied 4d $\mathcal{N}=4$ super 
Yang Mills/$AdS_{5}\times S^{5}$ string duality pair.

There are many structural similarities of
the 3d $\mathcal{N}=6$ superconformal ABJM theory to $\mathcal{N}=4$ super Yang-Mills, most
notably the emergence of hidden integrability in the planar limit
for the spectral problem of determining anomalous scaling dimensions of local operators
\cite{Minahan:2008hf,Gromov:2008qe,Bak:2008cp,Gaiotto:2008cg,Gromov:2008bz,Gromov:2008fy,Gromov:2009at}.  In the following we focus on relations between amplitudes and Wilson loops. More on the $AdS_4/CFT_3$ duality in general and the similarities and differences to the integrable structures in $\sym$ can be found in \cite{Klose:2010ki}.

Given these insights, the question arises whether there are also similar relations between
scattering amplitude, Wilson loops and correlators in ABJM theory as it is the case for $\sym$. We will shortly give an overview over developments in ABJM theory that parallel the ones in $\sym$  described in the previous chapter and elaborate in more detail on the Wilson loop side in chapter \ref{sec:Wilson-loops-in-ABJM}, which constitutes a part of the research performed in this thesis.

\section{Amplitudes in ABJM Theory}
Scattering
amplitudes in the ABJM theory have first been analysed in 
\cite{Agarwal:2008pu}, where in fact more general mass deformed superconformal 
Chern-Simons theories with extended supersymmetries were studied at the one-loop order. 
There, a vanishing result for the four-point one-loop amplitudes in the ABJM theory was found and
the authors speculated whether the two-loop scattering amplitudes in $\mathcal{N}=6$
Chern-Simons (ABJM theory) could be simply related to the one-loop $\sym$
amplitudes. 

In three dimensions there is no helicity degree of freedom for massless states. A spinor-helicity-like formalism for three dimensions was developed in \cite{Bargheer:2010hn,Huang:2010rn}. Similarly to the combination of all on-shell states into a supermultiplet \eqref{eqn:superfield} in $\sym$, one  can embed the matter fields of ABJM theory into two superstates using an $su(3)$ Gra\ss mann spinor $\eta^A$
\begin{align}
\Phi(\lambda,\eta) &=
	\phi^4(\lambda)+
	\eta^A\psi_A(\lambda)+
	\frac{1}{2!}\epsilon_{ABC}\eta^A\eta^B\phi^C(\lambda)+
	\frac{1}{3!}\epsilon_{ABC}\eta^A\eta^B\eta^C\psi_4(\lambda)\,, \\ \nn
\bar{\Phi}(\lambda,\eta) &=
	\bar{\psi}^4(\lambda)+
	\eta^A\bar{\phi}_A(\lambda)+
	\frac{1}{2!}\epsilon_{ABC}\eta^A\eta^B\bar{\psi}^C(\lambda)+
	\frac{1}{3!}\epsilon_{ABC}\eta^A\eta^B\eta^C\bar{\phi}_4(\lambda)\,.
\end{align}
Note, that the gauge fields do not appear in the superstates, since they have no on-shell degrees of freedom, Chern-Simons gauge fields in three dimensions have no freely propagating modes.

One can also define colour ordered superamplitudes and the colour decomposition requires to have $\Phi$ and $\bar{\Phi}$ fields to alternate, i.e. $A_n=A_n(\bar\Lambda_1,\Lambda_2,...,\bar{\Lambda}_{n-1},\Lambda_n)$, where $\Lambda_i=(\eta_i,\lambda_i)$.
In \cite{Agarwal:2008pu,Bargheer:2010hn} the four-point superamplitude was expressed as
\begin{equation}
A_4(\bar 1,2,\bar 3,4)=\frac{\delta^{(3)}(P)\,\delta^{(6)}(Q)}{\langle 12 \rangle \langle 23 \rangle}\,,
\end{equation}
similar to \eqref{eqn:N4SYM-superamplitude}. At tree-level, it was found that  amplitudes in ABJM theory exhibit Yangian and dual superconformal symmetry \cite{Bargheer:2010hn,Lee:2010du,Huang:2010qy,Gang:2010gy,Lipstein:2011ej}. A usual BCFW recursion for tree amplitudes does not work, since in three dimensions a linear shift vector as  in section \ref{sec:BCFW recursion},  which preserves momentum conservation and keeps the vectors light-like, is trivial. The solution is to introduce a non-linear shift \cite{Gang:2010gy}. Using these recursions, it was proven \cite{Gang:2010gy} that all tree-level amplitudes of the ABJM theory have dual superconformal symmetry.\\

\section{T-Duality of $AdS_{4}\times \mathbb{CP}_{3}$ Superstrings}

As mentioned before, from the string perspective the scattering amplitude/Wilson loop duality in the
$AdS_{5}/CFT_{4}$ system arises from
a combination of bosonic and fermionic T-dualities under which the free 
$AdS_{5}\times S^{5}$ superstring is self-dual \cite{Berkovits:2008ic,Beisert:2008iq}.
Hence, for the existence of an analogue duality in ABJM theory one would 
require a similar self-duality of the $AdS_{4}\times \mathbb{CP}_{3}$ superstring under the
combined T-dualities. The problem was analysed in \cite{Grassi:2009yj,Adam:2010hh,Adam:2009kt,Dekel:2011qw,Bakhmatov:2010fp,Bakhmatov:2011aa,Colgain:2012ca} but no T-self-duality could be established so far.

\section{Amplitudes and Wilson Loops in ABJM Theory}
In \cite{Henn:2010ps} we calculated the expectation value of the  Wilson loop operator \eqref{eqn:Wilson-loop} in the planar limit for light-like polygonal contours $\mathcal{C}_n$ in pure Chern-Simons and ABJM theory up to two loops. The details of this computation will be given in chapter \ref{sec:Wilson-loops-in-ABJM}.
As will be shown in section \ref{sec:anomalous-ward-identities}, conformal Ward identities force $\langle W(\mathcal{C}_{n})\rangle_{\text{1-loop}}$ 
to depend only on conformally invariant cross ratios \eqref{eqn:cross-ratios}.  At one-loop order in pure Chern-Simons and ABJM theory we find, see section \ref{sec:one-loop-WL}, that the Wilson loop with four and six cusps vanish, leading to the conclusion that the allowed function of conformal cross ratios is trivial at six points. The conjecture that the n-point Wilson loop vanishes
\begin{equation}\label{eqn:Wilson-loop-one-loop}
\langle W(\mathcal{C}_{n}) \rangle^{\text{CS}}_{\text{1-loop}} = 0 = \langle W(\mathcal{C}_{n}) \rangle^{\text{ABJM}}_{\text{1-loop}}
\end{equation}
was later  proven analytically in \cite{Bianchi:2011rn} for arbitrary $n$.

Furthermore, we computed the tetragonal Wilson loop $W_{4}$  at two-loop order in pure Chern-Simons and  ABJM theory. Remarkably, the result in dimensional
reduction regularisation with $d=3-2\epsilon$ for the Wilson loop in  ABJM theory is
% \begin{align}
% \label{eqn:here}
% & \langle W_4 \rangle^{\text{ABJM}}_{\text{2-loop}} \\ \nonumber &= \left( \frac{N}{k}\right)^2 \Big [-\frac{1}{2}\sum_{i=1}^4  \frac{(-{\mu^\prime}^2 \, x_{i,i+2}^2 )^{2\epsilon}}{(2\epsilon)^2}  \\ \nonumber
% & \qquad  \qquad \qquad  + \frac{1}{2}  \ln^2\left(\frac{x_{13}^2}{x_{24}^2}\right) +\text{const.}
% \Big ] 
% \end{align}
%where ${\mu^{\prime}}^2= \mu^2 8 \pi e^{\gamma_E}$.
%This is indeed 
of the same functional form as the one-loop result in $\mathcal{N}=4$ super
Yang-Mills theory.
% , where one has \cite{Drummond:2007aua} 
% \begin{align}
% & \langle W_4 \rangle^{\mathcal{N}=4\,\,\text{SYM}}_{\text{1-loop}} \\ \nonumber 
% &= \frac{g^{2}N}{8\pi^{2}} \Big [ - \frac{1}{2}\sum_{i=1}^4  \frac{(-{\hat{\mu}}^2 \, x_{i,i+2}^2 )^{\epsilon}}{\epsilon^2} \\ \nonumber
% & \qquad  \qquad  + \frac{1}{2}  \ln^2\left(\frac{x_{13}^2}{x_{24}^2}\right) +\text{const.}
% \Big ] \,.
% \end{align}
Most interestingly, it was then found by two independent approaches, using generalised unitarity methods in \cite{Chen:2011vv} and by a direct superspace Feynman diagram calculation in \cite{Bianchi:2011dg}, that the two-loop result for four-point scattering amplitudes in ABJM theory agrees with the Wilson loop %\eqref{eqn:here}
\begin{align}\label{eqn:4-point-ABJM} 
M_4^{(2)} = \frac{A_4^{(2)}}{A_4^{\text{tree}}} = \langle W_4 \rangle^{\text{ABJM}}_{\text{2-loop}}+ \text{const}. 
\end{align}
upon a specific identification of the regularisation scales \cite{Bianchi:2011dg,Chen:2011vv}.
This established the first non-trivial example of a Wilson loop / amplitude duality of the form \eqref{eqn:duality-wilson-loop-amplitude} in ABJM  theory.  In \cite{Chen:2011vv} the result \eqref{eqn:4-point-ABJM} has been obtained by assuming dual conformal invariance at quantum level, whereas in   \cite{Bianchi:2011dg} the same result is obtained by standard Feynman diagram calculations, thus giving a proof of the validity of on-shell dual conformal invariance at loop-level.

These result were extended to the more general case of ABJ theory in \cite{Bianchi:2011fc}. Furthermore, in  \cite{Bianchi:2011dg,Bianchi:2011fc} an analogue of the BDS ansatz for scattering amplitudes in ABJM theory is proposed and, using this, the form of the finite part at four loops is predicted. In order to verify this BDS-like ansatz, a four-loop computation of four-point scattering amplitudes in ABJM theory is required. Furthermore, in \cite{Bianchi:2011aa} an all-orders in $\epsilon$ identity between the four-point two-loop amplitude in ABJM and the four-point one-loop amplitude in $\sym$ is derived.

In light of these findings it is natural to ask, whether the duality between Wilson loops and amplitudes in ABJM theory continues to hold beyond $n=4$, as it does in $\syml$.  
  As a part of this thesis, in chapter \ref{sec:Wilson-loops-in-ABJM} we perform numerical computations to extend our findings of \cite{Henn:2010ps} to the $n$-sided Wilson loop at two-loop order. Interestingly, we find that the hexagonal Wilson loop at two loops also agrees with the corresponding Wilson loop in $\mathcal{N}=4$ super Yang-Mills at one-loop order.

We perform a detailed numerical analysis for the hexagonal Wilson loop leading to a guess for the $n$-point case, which we numerically check also for  $n>6$ in a limited set of kinematical points, see section \ref{sec:n-sided-wilson-loops-in-cs}. Again we find, that the result agrees with the result for the Wilson loop in $\mathcal{N}=4$ super Yang-Mills. It is thus natural to expect the result to hold for all $n$
\begin{align}\label{eqn:scs-two-loop-wl}
\langle W_n \rangle^{\text{ABJM}}_{\text{2-loop}} =  \left( \frac{N}{k}\right)^2 \Big [-\frac{1}{2}\sum_{i=1}^n  \frac{(-{\mu^\prime}^2 \, x_{i,i+2}^2 )^{2\epsilon}}{(2\epsilon)^2} 
+ \mathcal{F}_n^{\text{WL}} + r_n
\Big ] \,,
\end{align}
where ${\mu^{\prime}}^2= \mu^2 8 \pi e^{\gamma_E}$, $r_n$ is a constant that depends linearly on $n$ and is specified below \eqref{eqn:ABJM-result-Wn}. The details of the Wilson loop calculation in ABJM theory will be given in chapter \ref{sec:Wilson-loops-in-ABJM}.

Indeed, this is of the same form as the one-loop result \eqref{eqn:n-4sym-wilson-loop-one-loop} of the Wilson loop in $\mathcal{N}=4$ SYM  \cite{Brandhuber:2007yx} 
\begin{align}\label{eqn:n-4-one-loop-wl}
\langle W_n \rangle^{\mathcal{N}=4 \text{~SYM}}_{\text{1-loop}}  &=  \frac{g^2 N}{8 \pi^2} \Bigg[-\frac{1}{2}\sum_{i=1}^n \frac{(-x_{i,i+2}^2 {\mu}^2)^{\epsilon}}{\epsilon^2}+ \mathcal{F}_n^{\text{WL}}  \Bigg]\,,
\end{align}
where the contribution  $\mathcal{F}_n^{\text{WL}}$ in both formulas above is given by the finite part of the Wilson loop in $\mathcal{N}=4$ SYM, see  \eqref{eqn:BDS-WL-n4-n6} for the expressions with $n=4, n=6$. We comment on the differences between these expressions in section \ref{sec:equivalence-wilson-loops}.

Recently, also the six-point one-loop amplitudes became available \cite{Bargheer:2012cp,Bianchi:2012cq} and it was found, that they do not vanish in contrast to the Wilson loop  \eqref{eqn:Wilson-loop-one-loop} at one-loop order. In  \cite{Bargheer:2012cp} its is shown that scattering amplitudes in ABJM theory give rise to an anomaly of the superconformal symmetry, similarly as in $\sym$ theory. At tree- and one-loop level these anomalous terms can be used to determine the form of the amplitudes and it is shown that the symmetries predict a non-vanishing one-loop six-point amplitude. The prediction is confirmed by a calculation based on generalised unitarity methods. In \cite{Bianchi:2012cq} a superspace Feynman diagram calculation is used and it is argued, that the result, also for higher point functions, is in general non-vanishing.

This at least complicates a duality between Wilson loops and amplitudes in superconformal $\scs$ theory and it may very well be that there is no such duality in ABJM theory. If there was no self-T-duality in ABJM theory, there might be no reason to expect a duality between Wilson loops and scattering amplitudes.
On the other hand, the amplitude / Wilson loop duality in $\sym$ for the bosonic Wilson loops involved MHV amplitudes. The matching of non-MHV amplitudes in $\sym$ requires the introduction of a supersymmetric Wilson loop, see section \ref{sec:super-Wilons-loop}. The lack of helicity in three dimensions therefore complicates a map between these objects. Possibly, it is necessary to find an analog of the super-Wilson loop that matches with superamplitudes in ABJM theory.

\section{Duality between Wilson Loops and Correlators}
Furthermore, non-trivial evidence for a duality between Wilson loops and correlators in ABJM theory was found  at one-loop level in \cite{Bianchi:2011rn}.  Even though both quantities vanish at one loop, the
equality between correlators and Wilson loops in the light-like limit is already present at the level of integrands, before showing that these expressions actually vanish. It would be very desirable to perform a two-loop computation of the correlator in order to check whether the relation between Wilson loops and correlators continues to hold.

\chapter{Light-like Wilson Loops in Chern-Simons and ABJM Theory}\label{sec:Wilson-loops-in-ABJM}
\chaptermark{Light-like Wilson Loops in CS and ABJM Theory}
In this chapter we review the results for light-like polygonal Wilson loops in Chern-Simons and ABJM theory that were presented in \cite{Wiegandt:2011uu,Wiegandt:2011zz,Henn:2010ps}. We calculate the expectation value of the $n$-cusped Wilson loop operator 
\begin{align}\label{eqn:Wilson-loop-operator}
\langle W(\mathcal{C}_{n}) \rangle =  \frac{1}{N}
\, \langle 0 |\,
\tr \mathcal{P} \exp \left( i \oint_{\mathcal{C}_n} A_\mu dz^\mu \right)\, | 0\rangle
\end{align}
in the planar limit\footnote{I.e. we take the limit $N, k \rightarrow \infty$, $\frac{N}{k}$ finite.} for light-like polygonal contours $\mathcal{C}_n$ in pure Chern-Simons and ABJM theory. Note that due to the 
light-like contour, there is no difference in ABJM theory between the standard loop operator 
\eqref{eqn:Wilson-loop-operator} above and the 1/6 BPS supersymmetric loop operator of 
\cite{Drukker:2008zx}, as the terms in the exponential coupling to the scalars
drop out. 
 \\

\section{Wilson Loops in Chern-Simons Theory}

In 3d Chern-Simons theory, Wilson loops are the principal observables and topologically invariant
with exactly known correlation functions \cite{Witten:1988hf} in the Euclidean (or Wick rotated) theory.
This exact result is analytic in the inverse Chern-Simons parameter $k$ 
and perturbative studies in a loop-expansion of the effective coupling constant $1/k$
can reproduce the exact topological and finite 
result to the first orders \cite{Guadagnini:1989am,Alvarez:1991sx}, modulo regularisation
subtleties leading to or not leading to an integer shift of $k$ (for a 
review see \cite{Labastida:1998ud}).  
Wilson loops in Minkowski-space with cusps and 
light-like segments, however, display particularly strong divergences in 4d gauge theories and
seem to not have been considered in the 3d Chern-Simons literature before.\\

Wilson loops in Chern-Simons theory are usually defined with a framing procedure 
\cite{Witten:1988hf,Guadagnini:1989am,Alvarez:1991sx},
which may be thought of as a widening of the Wilson line to a ribbon.
This is necessary
in order to define an integer twisting number of the individual loop and acts
as a particular point-splitting regulator for collapsing gauge field propagators in perturbation theory,
while preserving the topological structure of the theory. 
Here, we refrain from framing our loops as we do not encounter the problem of 
collapsing gauge field propagators due to the piece-wise
linear structure of our loops. Moreover, the ABJM theory is not topological due to metric
dependent interactions in the matter sector, so that there is no need for framing from that
perspective either.
Instead, we regulate our correlators by the method of dimensional reduction,
which has been tested to the three-loop order in 
pure 3d Chern-Simons to yield a vanishing
$\beta$-function and to satisfy the Slavnov-Taylor identities \cite{Chen:1992ee}. 
Here, the tensor algebra is performed in 3 dimensions to obtain scalar integrands and then 
the dimension of the integrations are analytically continued.\\

\section{One Loop: Chern-Simons and ABJM Theory}\label{sec:one-loop-WL}
In this section we consider the one-loop expectation value of polygonal
Wilson loops with $n$ cusps. We would like to consider kinematical configurations for which all non-zero distances satisfy $-x_{ij}^2 >0 $, 
such that the result for the Wilson loops will be real (In particular, this allows us to
drop the $i \epsilon$ prescription of the propagators).
For $n$ odd, however, it is impossible to find vectors $p_{i}^{\mu}$
that lead to such configurations. For this reason, we will only discuss $n$ even.\\

We use the metric with signature $\eta_{\mu\nu}=\text{diag}(1,-1,-1)$ and employ a similar notation as in the $\sym$ case. An $n$-sided polygon can be defined by $n$ points $x_i$ ($i=1,..., n$), with the edge $i$ being the line connecting $x_i$ and $x_{i+1}$. Defining
\begin{equation}
p_i^{\mu}=x^{\mu}_{i+1}-x^{\mu}_i
\end{equation}
and parametrising the position $z^{\mu}_i$ on edge $i$ with the parameter $s_i \in [0,1]$ we have
\begin{equation} \label{eqn:z-parametrization-body}
z^{\mu}_i(s_i)= x^{\mu}_i + p^{\mu}_i s_i ,\quad z_{ij}^\mu := z_i^\mu - z_j^\mu\,.
\end{equation}
Furthermore, we use the notations
\begin{equation}
\epsilon(p,q,r) = \epsilon_{\mu\nu\rho} p^{\mu} q^{\nu} r^{\rho} \qquad 
 \text{and} \qquad \bar{s}_i= 1- s_i.
\end{equation}
We use the Lagrangian of ABJM theory given in \ref{sec:Lagrangian of ABJM theory}.

At one-loop level, we only need terms quadratic in the expansion of the
Wilson loop operator, and the free part of the action. 
Therefore,  at one loop order, the expectation value of (\ref{eqn:Wilson-loop-operator}) 
in ABJM theory coincides\footnote{For each of the gauge fields $A_\mu,\hat{A}_\mu$ present in ABJM theory.} with the one in pure Chern-Simons theory. 

\begin{figure}[t]
\centering
\subfloat[]{\begin{minipage}[c]{4cm}
		\includegraphics[width=0.85 \textwidth]{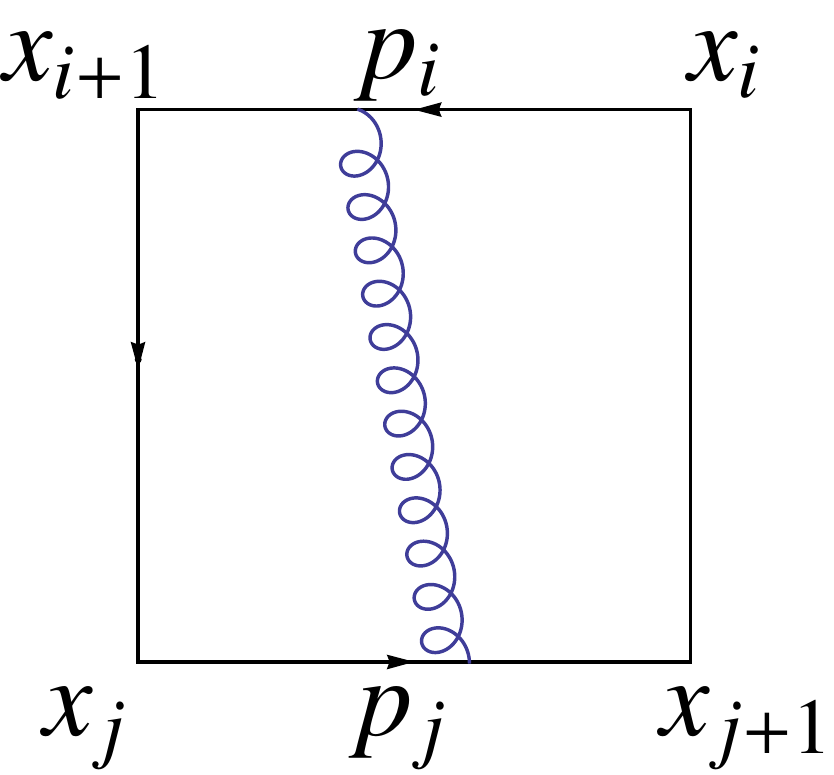}\label{fig:general-diagram}
		\end{minipage}
		
}
\subfloat[]{\begin{minipage}[c]{4cm}
		\includegraphics[width=0.6 \textwidth]{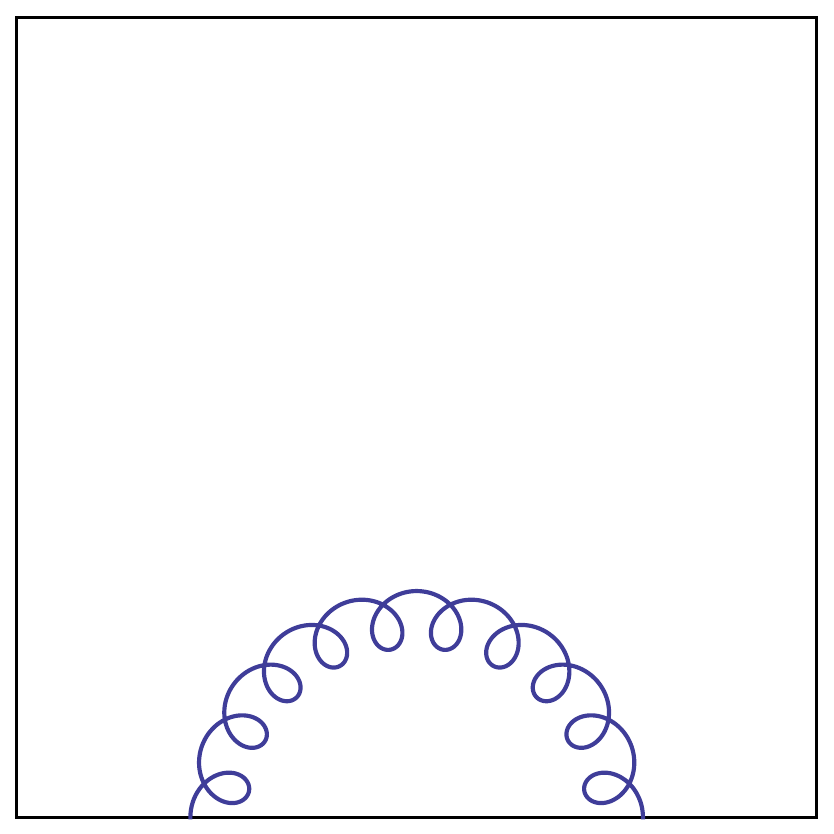}\label{fig: 1loopsameedge}
		\end{minipage}
		
}
\subfloat[]{\begin{minipage}[c]{4cm}
		\includegraphics[width=0.6 \textwidth]{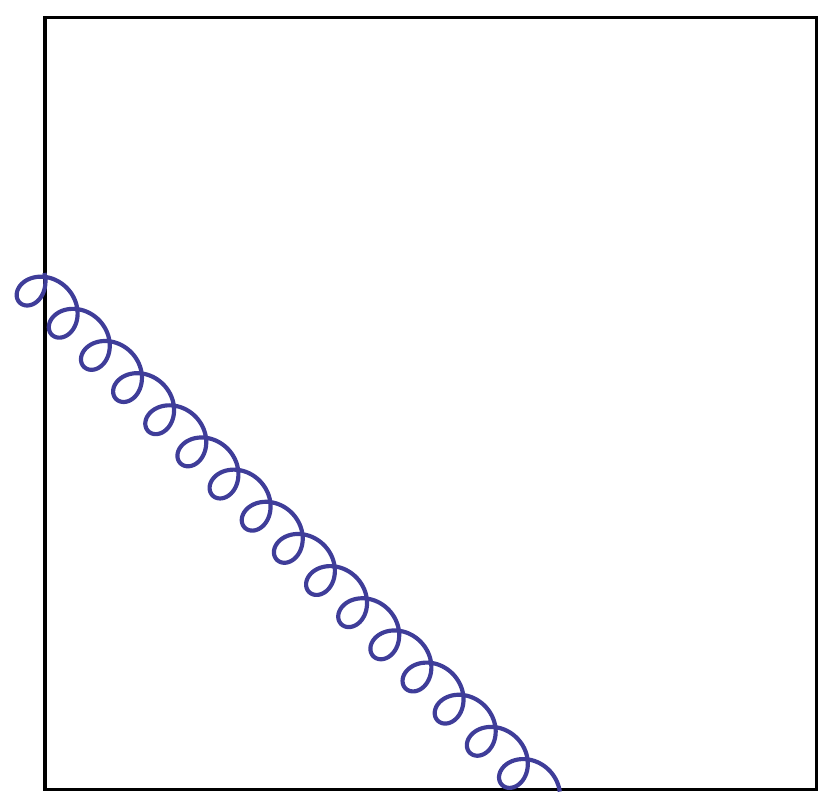}\label{fig: 1loopadjacentedge}
		\end{minipage}
		
}
\caption{
One-loop diagrams.
}
\label{fig:general-diagram-1-loop}
\end{figure}

The expectation value at one loop is a sum over all possible diagrams where the propagator stretches between edges $i$ and $j$,
\begin{align}\label{eqn:one-loop-general-expression1}
\langle W(\mathcal{C}_n) \rangle^{(1)}
&=  \frac{(i)^2}{N}\sum_{i \geq j}\int ds_i ds_j \dot{z}_{i}^{\mu} \dot{z}_{j}^{\nu} \langle \left(A_{\mu}\right)_{mn} (z_i)\left(A_{\nu}\right)_{nm}(z_j)\rangle \\ \nn
&=:  -\frac{N}{k}  \frac{\Gamma\left(\frac{d}{2}\right) }{\pi^{\frac{d-2}{2}}} \, \sum_{i \geq j} I_{ij} \,,
\end{align}
where the domain of integration is given by $\int_0^1 ds_i \int_0^1 ds_j$ for $i \neq j$,  $\int_0^1 ds_i \int_0^{s_i} ds_j$ for $i = j$ 
and $\dot{z}(s_i)={d z(s_i)}/{d s_i}=p_i$ and where we have introduced a normalisation factor for later convenience.
Here and throughout this chapter, we absorb the dimensional regularisation scale $(\mu^2 )^{\epsilon}$ into $k$ and only display it explicitly in our final results.
Using the Chern-Simons propagator  in the Landau gauge\footnote{We drop the $i \epsilon$ prescription in the propagator, since we consider kinematical configurations with $-x_{ij}^2>0$.}
\begin{align}\label{eqn:gluon-prop}
\langle \left(A_\mu\right)_{mn}(x) \left(A_\nu\right)_{kl}(y) \rangle &= \delta_{ml} \delta_{nk} \frac{1}{k}\left( \frac{\Gamma\left(\frac{d}{2}\right)}{\pi^{\frac{d-2}{2}}}\right) \epsilon_{\mu\nu\rho}\frac{(x-y)^\rho}{\left(-(x-y)^2\right)^{\frac{d}{2}}}\,,
\end{align}
and plugging in the expressions \eqref{eqn:z-parametrization-body} for $z_i$,
we obtain 
\begin{align}\label{eqn:one-loop-general-expression}
I_{ij}= \int ds_i ds_j \frac{\epsilon(p_i,p_j, p_i s_i - p_j s_j + \sum_{k=j}^{i-1}p_k)}{\left(-x^2_{ij}\bar{s}_i\bar{s}_j - x^2_{i+1,j} s_i \bar{s}_j - x^2_{i,j+1}\bar{s}_i s_j-x^2_{i+1,j+1}s_i s_j\right)^{\frac{d}{2}}}\,,
\end{align}
where $x_{i,j}^2 = (x_i - x_j)^2$, $\bar{s}_i= 1- s_i$ and $\epsilon(a,b,c) = \epsilon_{ijk} a^{i} b^{j} c^{k}$.
We can immediately see that in this gauge $I_{i,i}$ and $I_{i,i+1}$ vanish due to the antisymmetry of the $\epsilon$ tensor.
This corresponds to diagrams where the propagator ends on the same edge or on adjacent edges, as shown in Figures \ref{fig:general-diagram-1-loop}(b) 
and  \ref{fig:general-diagram-1-loop}(c), respectively.
Therefore we only need to keep diagrams of the type shown in Figure \ref{fig:general-diagram-1-loop}(a). 
The latter are manifestly finite in three dimensions and therefore we set $d=3$ in the remainder of this section.

\notocsubsection{Tetragon}
As explained above, in the Landau gauge, the 
only non-vanishing contributions to \eqref{eqn:one-loop-general-expression1} for the tetragon are $I_{31}$ and $I_{42}$.
Setting $d=3$, they are given by
\begin{align}
I_{31}=-\epsilon(p_1,p_2,p_3)\int ds_1 ds_3 \frac{1}{(-x^2_{13} \bar{s}_1\bar{s}_3 - x^2_{24} s_1 s_3)^{3/2}}\,, 
\end{align}
and
\begin{align}
I_{42}=-\epsilon(p_2,p_3,p_4)\int ds_2 ds_4 \frac{1}{(-x^2_{24} \bar{s}_2\bar{s}_4 - x^2_{13} s_2 s_4)^{3/2}}\,.
\end{align}
Taking into account that we have a closed contour, i.e. $\sum_i p_i =0$, we can write 
$\epsilon(p_2,p_3,p_4) = - \epsilon(p_2,p_3,p_1)= - \epsilon(p_1,p_2,p_3)$ and 
thus the contributions from the two diagrams cancel each other
\begin{align}
\langle W_n \rangle ^{(1)} & \propto \left( ~
\begin{minipage}[t]{50pt}
\vspace{-30pt}\includegraphics[height=50pt]{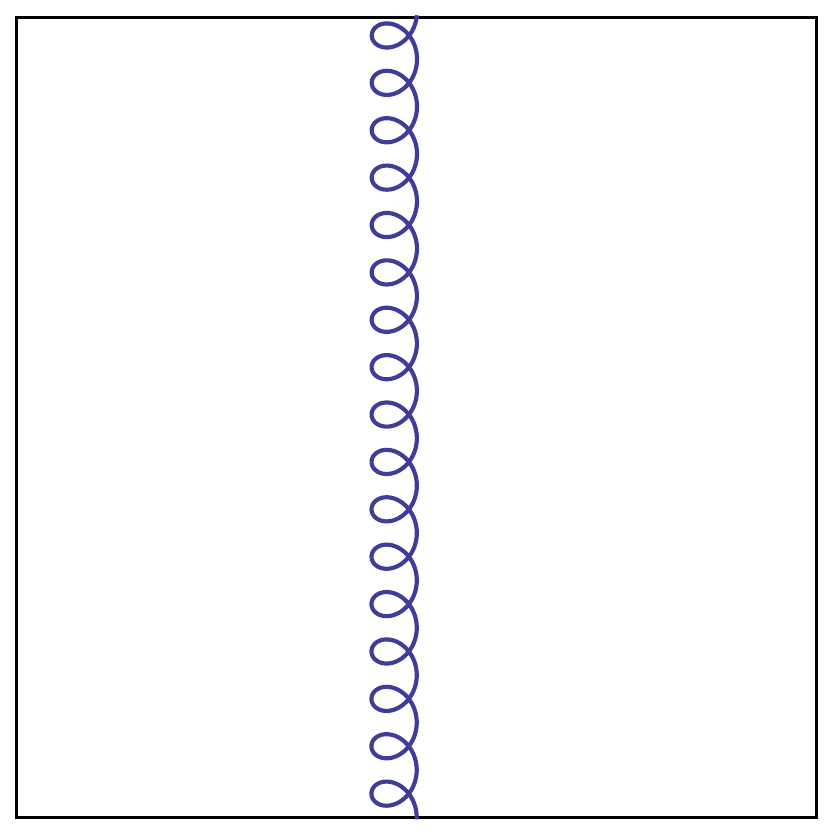} 
\end{minipage}
~~+~ 
\begin{minipage}[t]{50pt}
\vspace{-30pt}\includegraphics[angle=90,height=50pt]{i13} 
\end{minipage}~
\right) 
= I_{31} + I_{42} =0\,.
\end{align}
We will see in section \ref{sec:anomalous-ward-identities} that this result is compatible with the 
restrictions imposed by conformal symmetry.

\begin{figure}[t]
\centering
{\includegraphics[width=0.15 \textwidth]{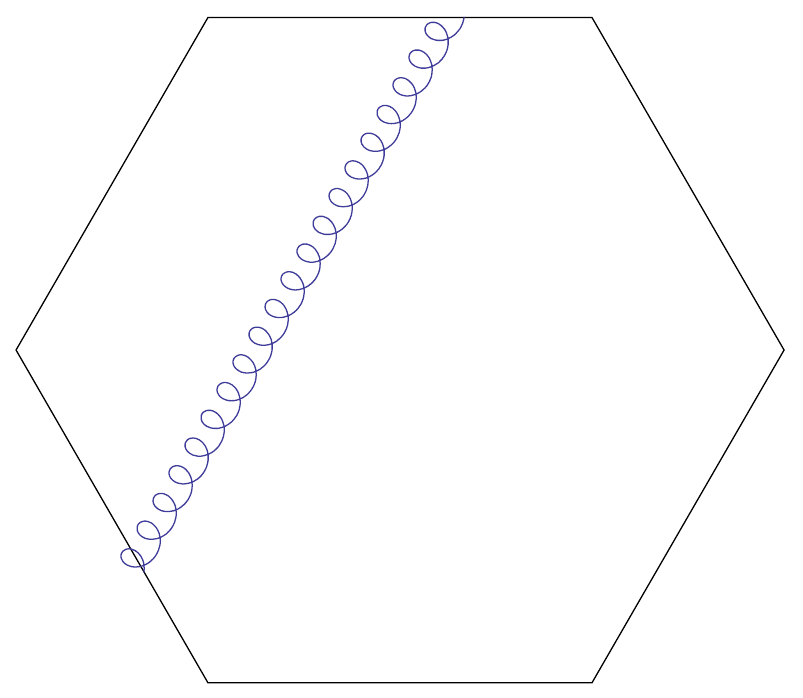}\label{fig:hexagon2}~~~~~~}
{\includegraphics[width=0.15 \textwidth]{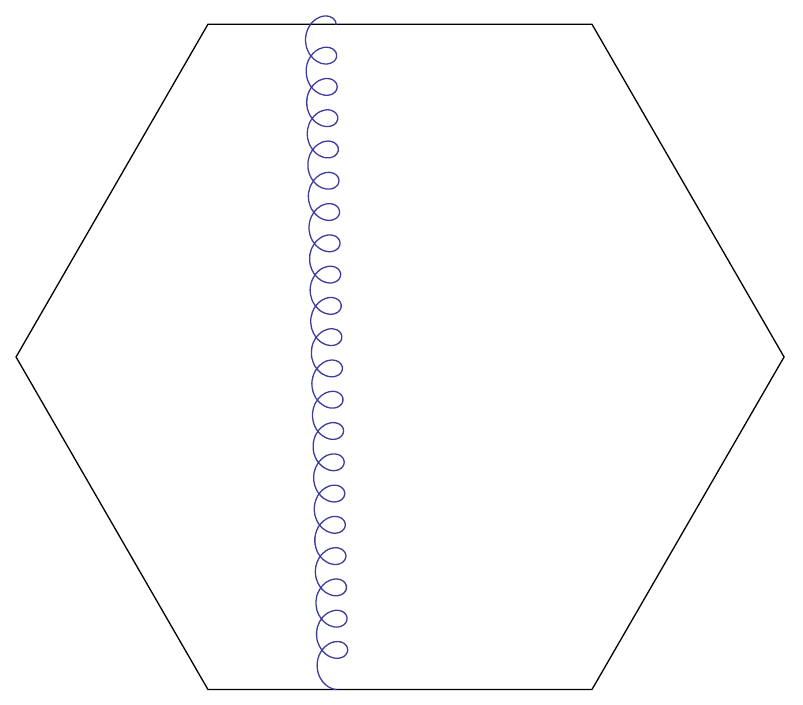}\label{fig:hexagon3}}
\caption{One-loop contributions $I_{i+2,i}$ and $I_{i+3,i}$ to the hexagonal Wilson loop.}
\label{fig:hexagon-one-loop}
\end{figure}

\notocsubsection{Hexagon and Higher Polygons}
For the hexagon there are two different non-vanishing types of contributions, 
$I_{i+2,i}$ and $I_{i+3,i}$, as shown in Figure \ref{fig:hexagon-one-loop}. 
The former appears in six orientations, $i=1\, \ldots 6$ (with the convention that $i + 6 \equiv i$), while the
latter appears in three orientations, $i=1, 2, 3$. 

Specialising the general formula (\ref{eqn:one-loop-general-expression}) to these cases we have
\begin{equation}
\label{eqn:I_ii+2}
I_{i+2,i}=   \int_0^1 ds_{i+2} ds_{i} \frac{\epsilon(p_{i+2},p_i,p_{i+1})}{(-\bar{s}_i\bar{s}_{i+2} x_{i,i+2}^2 - s_{i}\bar{s}_{i+2} x_{i,i+3}^2 - s_i s_{i+2}x_{i+1,i+3}^2 )^{3/2}}\,
\end{equation}
and 
\begin{equation}
\label{eqn:I_ii+3}
I_{i+3,i}=  \int_0^1 ds_{i+3}ds_i \frac{\epsilon(p_{i+3},p_i,p_{i+1}+p_{i+2}) }{(-\bar{s}_i\bar{s}_{i+3} x_{i,i+3}^2 - s_{i}\bar{s}_{i+3} x_{i+1,i+3}^2 - \bar{s}_i s_{i+3}x_{i,i+4}^2 - s_i s_{i+3} x^2_{i+1,i+4})^{3/2} }\,.
\end{equation}

We checked numerically for various non-symmetric hexagon 
configurations that the sum over all diagrams vanishes,
\begin{align}\label{eqn:six-vanishes}
\langle W(\mathcal{C}_6)  \rangle^{\text(1)} \propto \sum_{i>j}^6 I_{ij} = 0\,.
\end{align}
It is easy to see that \eqref{eqn:six-vanishes} is true for special configurations, as we will see presently.

Consider the configuration where opposite edges are anti-parallel, i.e. $p_i =- p_{i+3}$. 
{}From  \eqref{eqn:I_ii+3} we see that $I_{i,i+3}=0$ due to the antisymmetry of the $\epsilon$ tensor.
Furthermore, taking into account that for this configuration we have $x_{i,i+2}^2 = x_{i+3,i+5}^2$,
it is easy to see from equation \eqref{eqn:I_ii+2}  that the integrands of $I_{i,i+2}$ and $I_{i+3,i+5}$ are the same.
Finally, using $\sum_i p_i=0$ one can see that the Levi-Civita symbols produce a differing sign, such
that
\begin{align}
I_{i,i+2} + I_{i+3,i+5}&= 
\left(
\begin{minipage}{1.5cm}
{\includegraphics[width=1 \textwidth]{hexagon2}} 
\end{minipage} 
+ 
\begin{minipage}{1.5cm} 
{\includegraphics[width=1 \textwidth]{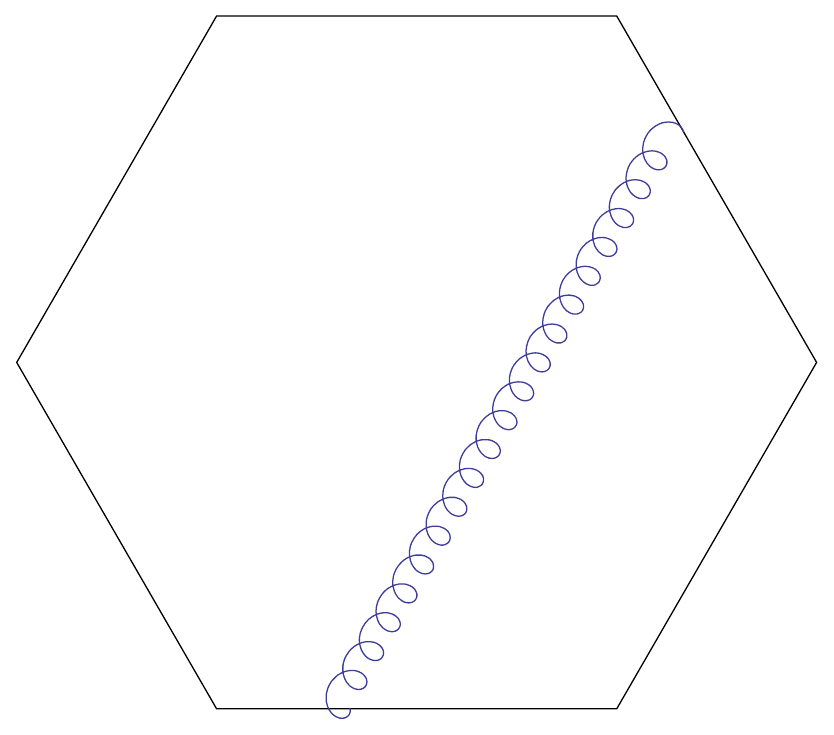}} 
\end{minipage}\right)_{p_i=-p_{i+3}} = 0 \,,
\end{align}
i.e. the contributions coming from those diagrams cancel pairwise,
and we arrive at equation  \eqref{eqn:six-vanishes}, in the specific anti-parallel kinematical configuration $p_i =- p_{i+3}$. 

The conjecture that all $n$-cusped Wilson loops vanish at one-loop order in Chern-Simons theory was indeed  proven in \cite{Bianchi:2011rn}.

\section{Four-Sided Wilson Loop in Chern-Simons Theory at Two Loops}\label{sec:two-loop}

\begin{figure}[t]
\centering
\subfloat[]{\begin{minipage}{4cm}\centering
		\includegraphics[width=.6 \textwidth]{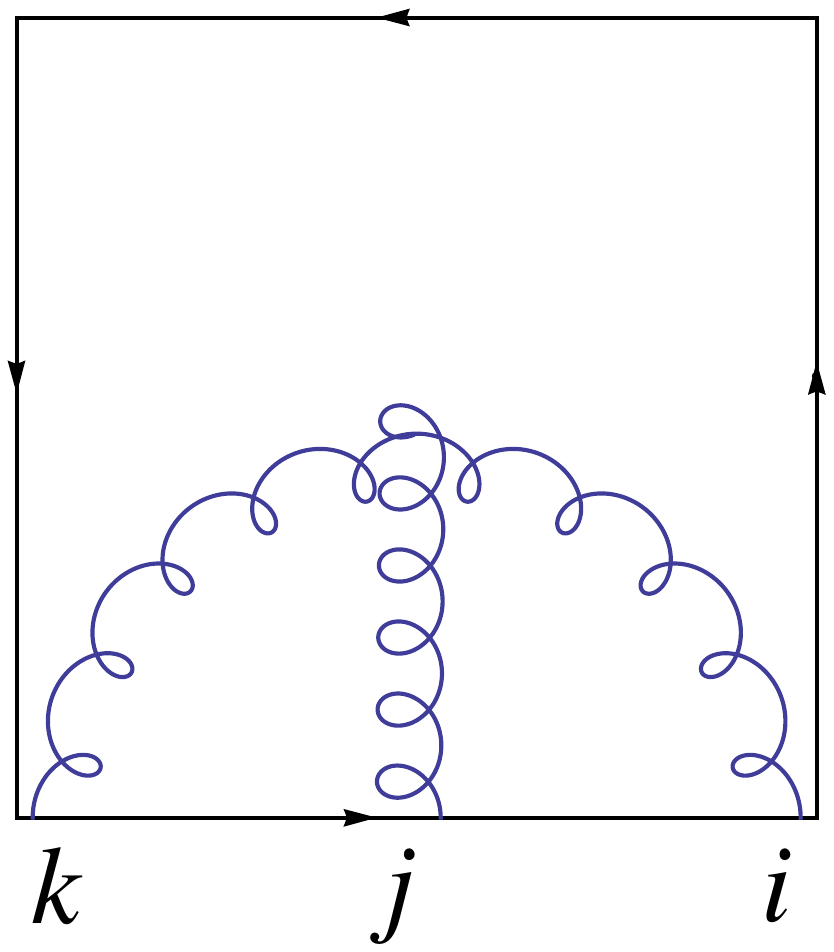}\label{fig:vertexI111}
		\end{minipage}
		
}
~~~~~
\subfloat[]{\begin{minipage}{4cm}\centering
		\includegraphics[width=.6 \textwidth]{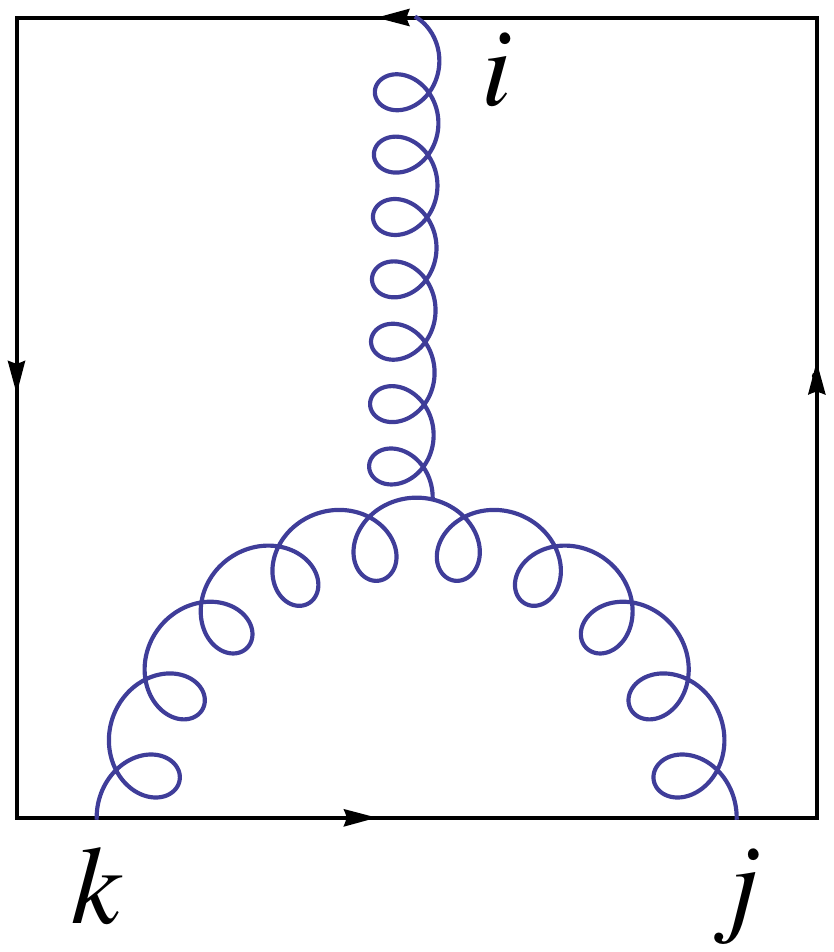}\label{fig:vertexI113}
		\end{minipage}
}
~~~~~
\subfloat[]{\begin{minipage}{4cm}\centering
		\includegraphics[width=.65 \textwidth]{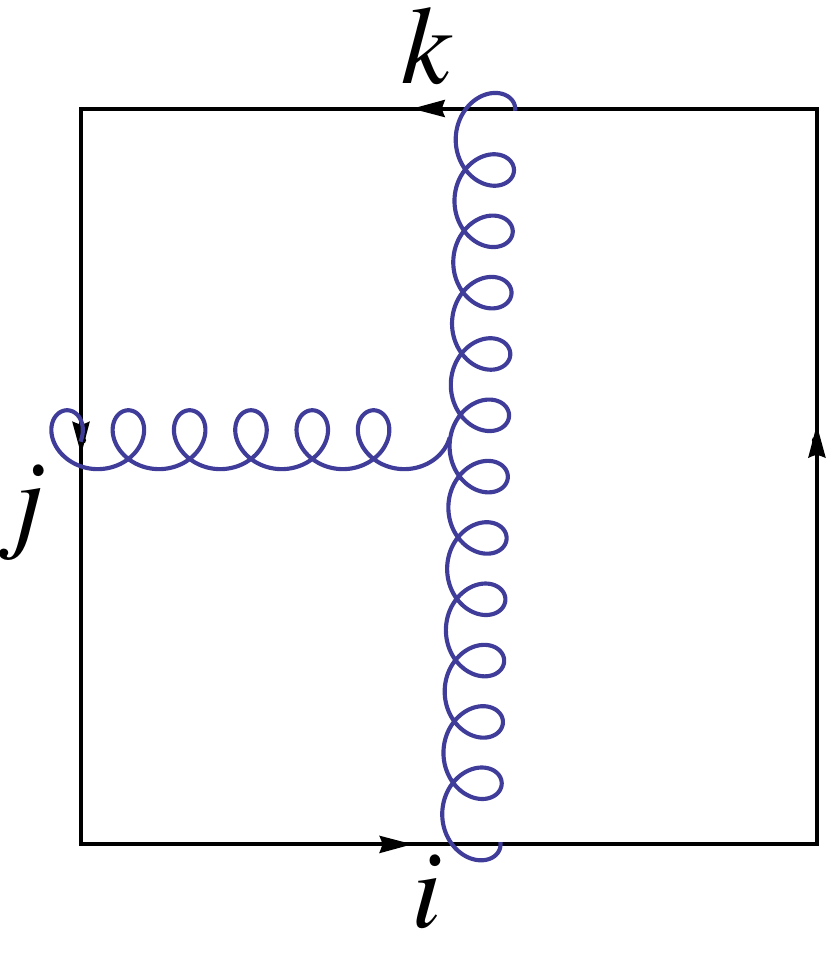}\label{fig:vertexI123}
		\end{minipage}
}\\

\subfloat[]{\begin{minipage}{4cm}\centering
		\includegraphics[width=.6 \textwidth]{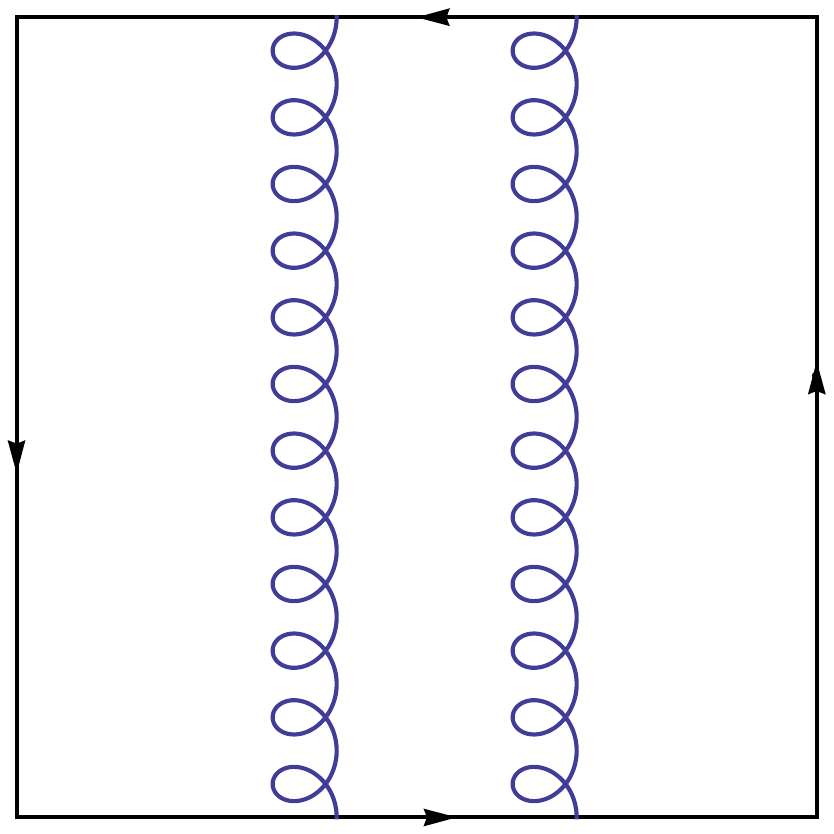}\label{fig: rect1331}
		\end{minipage}
}
\subfloat[]{\begin{minipage}{4cm}\centering
		\includegraphics[width=.6 \textwidth]{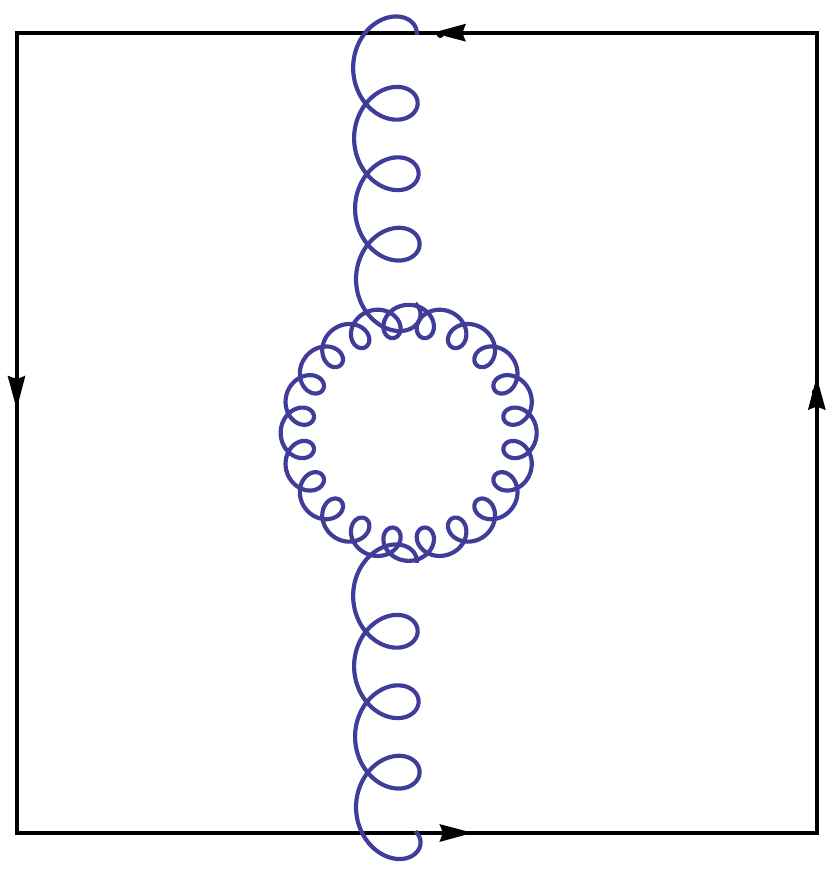}\label{fig:gluon-loop}
		\end{minipage}
}
\subfloat[]{\begin{minipage}{4cm}\centering
		\includegraphics[width=.6 \textwidth]{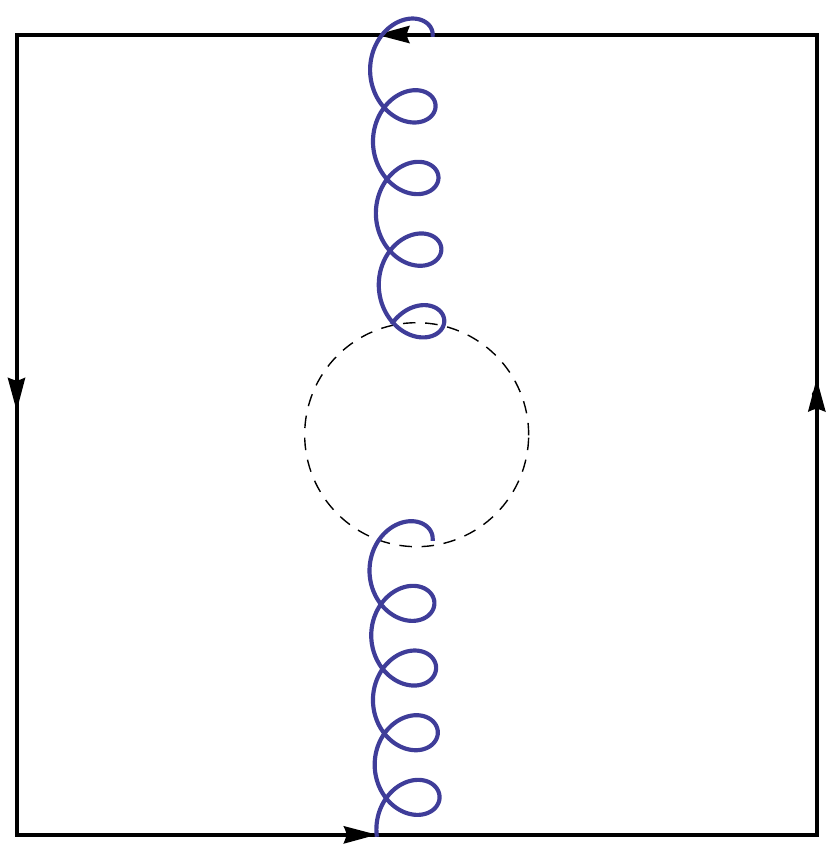}\label{fig:ghost-loop}
		\end{minipage}
}

\caption{Planar two-loop topologies appearing in the polygonal Wilson loop in CS theory.
Diagrams where one propagator is connected to a single edge or to two adjacent edges vanish in our gauge and are not displayed.}
\label{fig:3typesofdiagrams}
\end{figure}

In this section we calculate the two-loop contributions to the 
tetragonal light-like Wilson loop in pure Chern Simons theory. 
The results are 
consistent with
the anomalous conformal Ward identity to be discussed
in section \ref{sec:anomalous-ward-identities}.

Expanding the Wilson loop to quartic order, see \eqref{eqn:Wilson-Loop-after-path-ordering}, and 
performing Wick contractions leads to the topologies shown in Figure \ref{fig:3typesofdiagrams}.
We are taking the planar limit and therefore drop all non-planar graphs.
Moreover, all diagrams where one propagator is connected to a single edge or adjacent edges vanish
in our gauge for the same reason as at the one-loop order and are not displayed.

\notocsubsection{Ladder Diagrams}
Let us begin by computing diagrams of ladder topology as shown in Figure \ref{fig: rect1331}.
There are two different orientations of this diagram, and it is easy to see that they give the same
contribution. Taking into account this factor of $2$, the contribution of the ladder
diagrams is
\begin{align}
\langle W_n  \rangle^{(2)}_{\text{ladder}}
&= 2 \left( \frac{N}{k} \right)^2 \left(   \frac{\Gamma\left(\frac{d}{2}\right)}{\pi^{\frac{d-2}{2}}} \right)^2  \, I_{\text{ladder}}( x_{13}^2 , x_{24}^2 )\,,
\end{align}
where
\begin{align}\label{eqn:ladder-diagram-n-4}
I_{\text{ladder}}(x_{13}^2, x_{24}^2) 
% &=\frac{1}{N}(i)^4  \oint_{{z_i>z_j>z_k>z_l}} \hspace{-45pt} dz_{i,j,k,l}^{\mu,\nu,\rho,\sigma} \langle \tr A_\mu(z_i) A_\nu(z_j) A_\rho(z_k) A_\sigma(z_l) \rangle \\ \nonumber
&=   \int ds_{i,j,k,l} \frac{\epsilon(\dot{z}_i,\dot{z}_l,z_i-z_l)}{[-(z_i-z_l)^2]^{\frac{d}{2}}} \frac{\epsilon(\dot{z}_j,\dot{z}_k,z_j-z_k)}{[-(z_j-z_k)^2]^{\frac{d}{2}}} \,. 
\end{align}
The integral is finite and may be calculated for $d=3$
\begin{align}\nn
I_{\text{ladder}}(x_{13}^2,x_{24}^2) &= \frac{1}{4} \int_0^1 ds_i \int_0^{s_i} ds_j \int_0^1 ds_k \int_0^{s_k} ds_l \frac{ x_{13}^2 x_{24}^2 ( x_{13}^2 +x_{24}^2)}{[x_{13}^2 \bar{s}_i \bar{s}_l+ x_{24}^2 s_i s_l]^{\frac{3}{2}} [x_{13}^2 \bar{s}_j \bar{s}_k+ x_{24}^2 s_j s_k]^{\frac{3}{2}}}  + O(\epsilon) \,.
\end{align}
We computed this integral by first carrying out some of the parameter integrals and then deriving a differential equation for it, which could be solved.
The result is remarkably simple, 
\begin{align} 
I_{\text{ladder}}(x_{13}^2,x_{24}^2) &= \frac{1}{2} \left[ \ln^2 \left(\frac{x_{13}^2}{x_{24}^2}\right) + \pi^2\right] + O(\epsilon) \,.
\end{align}
Including the prefactors and dropping $O(\epsilon)$ terms, the contribution to the Wilson loop is
\begin{align}\label{eqn:result-ladder}
\langle W_4  \rangle^{(2)}_{\text{ladder}}
&=  \left( \frac{N}{k} \right)^2   ~\frac{1}{4} \left[ \ln^2 \left(\frac{x_{13}^2}{x_{24}^2}\right) + \pi^2\right]\,.
\end{align}

\notocsubsection{Vertex Diagrams}
\label{sect:vertex-diagram}

The diagrams with one three-gluon vertex shown in Figures \ref{fig:vertexI111} , \ref{fig:vertexI113} and \ref{fig:vertexI123}
are obtained by contracting the cubic term in the expansion of the 
Wilson loop in \eqref{eqn:Wilson-Loop-after-path-ordering} with the interaction term of the 
Lagrangian,
\begin{align}\label{eqn:vertex-diagram}
\langle W_n  \rangle^{(2)}_{\text{vertex}}
%&= \frac{1}{N} \langle (i)^3  \oint_{z_i>z_j>z_k} \hspace{-30pt} dz_{i,j,k}^{\mu,\nu,\rho}\, \tr \left( A_\mu A_\nu A_\rho \right) \left( i \int\, d^dw \mathcal{L}_{\text{int}}(w)\right)\rangle \\ \nonumber
%&= - \frac{1}{N} \frac{k}{4 \pi} \frac{2}{3}  (i)^5 \oint dz_{i,j,k}^{\mu,\nu,\rho}\int d^dw \langle \tr \left( A_\mu A_\nu A_\rho \right) \tr \left( A_\alpha A_\beta A_\gamma (w) \right) \epsilon^{\alpha\beta\gamma}\rangle \\ \nonumber
&=   \left(\frac{N}{k}\right)^2 \frac{i}{2 \pi}  \left( \frac{\Gamma\left(\frac{d}{2}\right)}{\pi^{\frac{d-2}{2}}}\right)^3 \sum_{i>j>k} I_{ijk} \,,
\end{align}
where 
\begin{equation}\label{eqn:Iijk}
I_{ijk}= -\int dz_i^\mu dz_j^\nu dz_k^{\rho}  \epsilon^{\alpha\beta\gamma}\epsilon_{\mu\alpha\sigma}\epsilon_{\nu\beta\lambda}\epsilon_{\rho\gamma\tau}\int d^dw \frac{(w-z_i)^{\sigma}(w-z_j)^{\lambda}(w-z_k)^{\tau}}{|w-z_i|^d|w-z_j|^d|w-z_k|^d}\,,
\end{equation}
and $|z_i|=(-z_i^2)^{\frac{1}{2}}$.
Here the indices of $I_{ijk}$ refer to the edges of the Wilson loop that the propagators attach to.
The expression can be shown to be antisymmetric under the exchange of any two indices,
and therefore the only non-vanishing contributions are the ones for $i\neq j\neq k$. 
As a consequence, topologies  \ref{fig:vertexI111} and \ref{fig:vertexI113} can be discarded.

Specialising to the tetragon, we have four contributions which are symmetric under $x_{13}^2 \leftrightarrow  x_{24}^2$ 
and thus it is sufficient to compute one of them
\begin{align}\label{eqn:vertex-Iijk}
I_{321}&%=-I_{123}
= \int d^dw \int_0^1 ds_{1,2,3} \frac{\epsilon(p_2,p_3,w)\epsilon(p_2,p_1,w)}{|w|^{d}|w-z_{12}|^{d}|w-z_{32}|^{d}} \\ \nonumber
%&\stackrel{\eqref{app:Vertex}}{=} 
&=
\frac{i \pi^{\frac{d}{2}}}{8}\frac{\Gamma(d-1)}{\Gamma\left(\frac{d}{2} \right)^3} x_{13}^2  x_{24}^2  \int_0^1 d^3 s_{1,2,3} d^3\beta_{1,2,3} \left(\beta_1\beta_2\beta_3 \right)^{\frac{d-2}{2}}\delta \left(\sum_{i=1}^{3} \beta_i-1\right)  \times \nonumber \\
& \qquad\qquad\qquad \qquad\qquad  \times \left( \frac{1}{\Delta^{d-1}}- 2 \frac{(d-1)}{\Delta^d} \beta_1 \beta_3 \bar{s}_1 s_3 (x_{13}^2+x_{24}^2)\right) \,, \nonumber
\end{align}
where the second and third line is obtained by introducing Feynman parameters in the standard way and  integrating over $w$. More details
may be found in Appendix \ref{app:Vertex}. $\Delta$ is given by
\begin{align}\label{delta-vertex}
\Delta &= -\beta_{1} \beta_{2} z_{12}^2 - \beta_{2} \beta_{3} z_{23}^2 - \beta_{1} \beta_{3} z_{13}^2  \nonumber \\
&= -x_{13}^2 \beta_1 \bar{s}_1 \left(\beta_3 \bar{s}_3+ \beta_2 s_2 \right) - x_{24}^2 \beta_3 s_3 \left(\beta_2 \bar{s}_2 + \beta_1 s_1  \right)\,.
\end{align}
\begin{figure}[t]
\centering
	\includegraphics[width=.15 \textwidth]{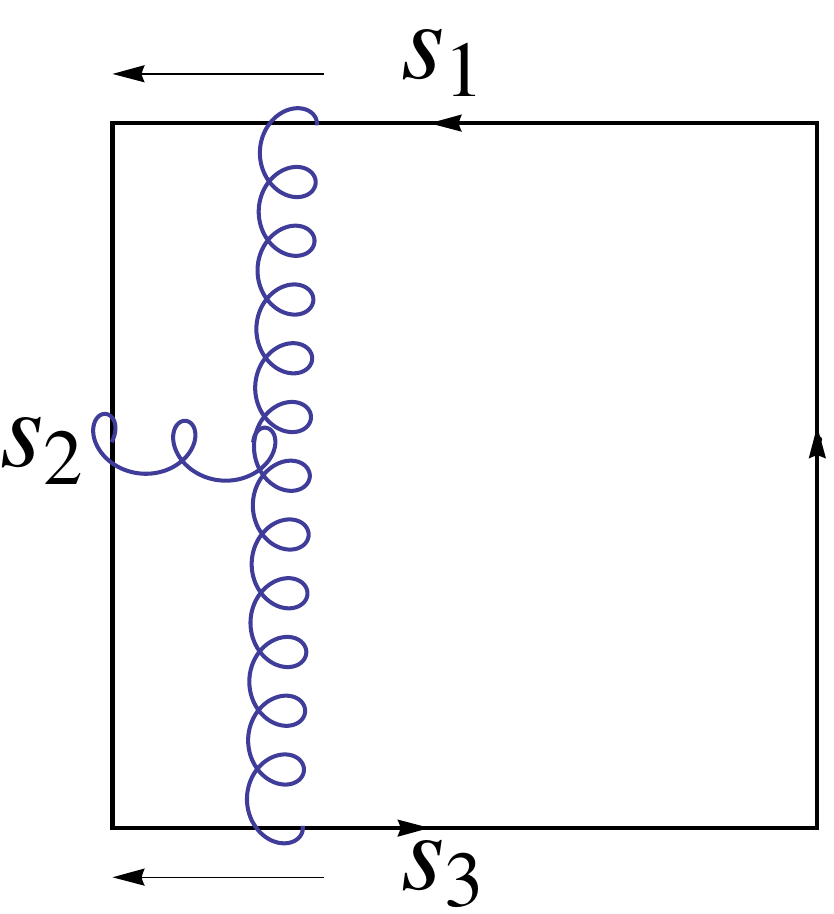}
\caption{The divergence in the vertex diagram arises from the integration region where 
$s_1 \rightarrow 1, {s}_3 \rightarrow 0$
(and $\beta_{1} \rightarrow 0, \beta_{3} \rightarrow 0 $), see equation (\ref{delta-vertex}).
%$s_1 , \bar{s}_3\rightarrow 0$ and $s_1, \bar{s}_3\rightarrow 1$.
}
\label{fig:divergence-vertex}
\end{figure}
~\\One might naively think that this diagram should give a finite answer due to the antisymmetry of the $\epsilon$ tensors.
The result would indeed be finite in the case of smooth contours \cite{Guadagnini:1989am} or contours with a single cusp. 
However, the presence of two cusps gives rise to a region in the integration space of Feynman parameters
where the first summand  in the third line of \eqref{eqn:vertex-Iijk} induces a divergent contribution.
The relevant region of Feynman parameters is 
%$s_1, \bar{s}_3\rightarrow 0$ and $s_1, \bar{s}_3\rightarrow 1$
$s_1 \rightarrow 1, {s}_3\rightarrow 0$
(and $\beta_{1} \rightarrow 0, \beta_{3} \rightarrow 0$), see equation \eqref{delta-vertex},
and is illustrated in Figure \ref{fig:divergence-vertex}. Due to the presence of three independent vectors
$p_{1}^{\mu}, p_{2}^{\mu}$ and $p_{3}^{\mu}$ the $\epsilon$ tensors do not suppress this region.
We find that this term produces a $1/\epsilon$ pole in dimensional reduction.
The second summand in the third line of \eqref{eqn:vertex-Iijk} is finite. 

The divergent and finite pieces can be seperated using Mellin-Barnes techniques. The details
can be found in Appendix \ref{app:Vertex}. We have not computed the coefficients of the
$\epsilon^{0}$ terms analytically (the coefficient of the pole-term is evaluated analytically in the Appendix \ref{sec:analytical-evaluation-of-pole-term}), but we have good numerical evidence that they give the following result:
\begin{equation}\label{eqn:I321}
I_{321}= \frac{i \pi^{\frac{d}{2}+1}}{8}\frac{\Gamma(d-1)}{\Gamma\left(\frac{d}{2} \right)^3}  \left[ 2 \ln(2)\frac{ (-x_{13}^2)^{2\epsilon}+(-x_{24}^2 )^{2\epsilon}}{\epsilon}  + \ln^2 \left(\frac{x_{13}^2}{x_{24}^2}\right) + a_6 + \mathcal{O}(\epsilon) \right]\,,
\end{equation}
where $a_6 =  8.354242685 \pm 2 \cdot 10^{{-9}} $, see  \eqref{eqn:a6} and below for an analytical guess of this constant. 
Taking into account all prefactors 
and restoring the regularisation scale, $k \rightarrow \mu^{-2\epsilon} k$, we can write
%\footnote{We have rewritten  $\frac{\Gamma(2-2 \epsilon)}{\epsilon}=
%\frac{(e^{\gamma_E})^{2\epsilon}}{\epsilon}-2+\mathcal{O}(\epsilon)$, 
% which appears in the prefactors and $c_2$.} 
the result, up to terms of order $\epsilon$,  as
\begin{align}\label{eqn:result-vertex}
\langle W_4  \rangle^{(2)}_{\text{vertex}}&= -\frac{1}{4}\left( \frac{N}{k}\right)^2  \left[ \ln(2) \sum_{i=1}^{4} \frac{(-x_{i,i+2}^2 \, \tilde \mu^2 )^{2\epsilon}}{\epsilon}+  \ln^2\left( \frac{x_{13}^2}{x_{24}^2} \right) +  a_6- 8\ln(2)  \right]\,,
\end{align}
where $\tilde \mu^2 = \mu^2 \pi e^{\gamma_E}$. 

\notocsubsection{Gauge Field and Ghost Loops}\label{sec:gauge-ghost-cancellation}
It is well known \cite{Chen:1992ee} that in the dimensional reduction (DRED) scheme
%i.e. performing algebraic operations with the epsilon-tensors in strictly 3 dimensions, 
the gauge field loop diagrams shown in Figure \ref{fig:gluon-loop} exactly cancel against the ghost loop diagrams 
shown in Figure \ref{fig:ghost-loop} :
%\footnote{Using the DREG scheme, the contributions cancel up to an order $\epsilon$ term}
\begin{align}\label{eqn:gluon-ghost-loops}
\langle W_n  \rangle^{(2)}_{\text{gluon loop}} = -  \langle W_n  \rangle^{(2)}_{\text{ghost loop}} \,.
\end{align}
Details of this cancellation can be found in Appendix \ref{app:gluon-and-ghost-loops}.

\notocsubsection{Result for the Two-Loop Tetragon in CS Theory}
\label{sect:CS-2loop-results}
Summing up the results \eqref{eqn:result-ladder}, \eqref{eqn:result-vertex} and \eqref{eqn:gluon-ghost-loops} for the tetragon, 
interestingly, the  $\ln^2(x_{13}^2/ x_{24}^2)$
terms in the two-gauge-field diagram and the vertex diagram 
exactly cancel and we obtain
\begin{align}\label{eqn:result-wilson-loop-two-loop}
\langle W_n  \rangle^{(2)} &=  -\left(\frac{N}{k}\right)^2  \frac{1}{4}\left[\ln(2) \sum_{i=1}^4\frac{(-x_{i,i+2}^2 \, \tilde \mu^2 )^{2\epsilon}}{\epsilon} + a_6 - 8 \ln(2) - \pi^2 \right]\,.
\end{align}
where $\tilde \mu^2 = \mu^2 \pi e^{\gamma_E}$ and we recall that $a_6 =  8.354242685 \pm 
2 \cdot 10^{{-9}} $.
As we will see in the next section, the cancellation observed here that led to the finite part of \eqref{eqn:result-wilson-loop-two-loop} being
a constant is in fact a consequence of the (broken) conformal symmetry of the Wilson loops under consideration.

\section{Anomalous Conformal Ward  Identities}\label{sec:anomalous-ward-identities}
The structure of the above results can be understood from conformal symmetry,
by deriving anomalous conformal Ward identities for the Wilson loops analogously to section \ref{sec:ward-identity-wilson-loops}.

If the Wilson loop $\langle W_n \rangle $ were well-defined in $d=3$ dimensional Minkowski space 
it would enjoy the conformal invariance of the underlying gauge theory 
and 
we would conclude that
$ \langle W(\mathcal{C}) \rangle = \langle W(\mathcal{C}^\prime) \rangle$.
This is indeed the case at one loop order, see section \ref{sec:one-loop-WL}.
However, as we have seen in section \ref{sec:two-loop}, starting from two loops, divergences force 
us to introduce a regularisation and calculate in $d=3-2\epsilon$ dimensions, 
thereby breaking the conformal invariance of the action.  The latter leads to an anomalous term in the
conformal Ward identities for the Wilson loops

\begin{align}\label{eqn:dilatation-Wardidentitiy}
D\,  \langle W_n \rangle &= -\frac{2 i \epsilon}{ \mu^{2\epsilon}} \int d^dx \langle \mathcal{L}(x) W_n \rangle \,, \\ \label{eqn:special-conformal-Wardidentitiy}
K^\nu  \langle W_n \rangle &= -\frac{4 i \epsilon}{\mu^{2\epsilon}} \int d^dx\, x^\nu \langle \mathcal{L}(x) W_n\rangle \,,
\end{align}
for {dilatations} and {special conformal} transformations. Here the operators on the left-hand sides act in
the canonical way on the coordinates of the cusp points,
\begin{align}\label{eqn:diff-equ-ward-id}
D    &= \sum_i (x_i \cdot \partial_i)  \,,\qquad  
K^\nu  = \sum_i \left(2 x_i^\nu (x_i\cdot \partial_i)- x_i^2 \partial_i^\nu   \right) \,.
\end{align}
We emphasise that thanks to the factor of $\epsilon$ on the r.h.s. of \eqref{eqn:dilatation-Wardidentitiy} and \eqref{eqn:special-conformal-Wardidentitiy} 
it is sufficient to know the divergent part of the integrals appearing on the r.h.s. of those equations in order to obtain information about
the finite part of $\langle W_n \rangle $.

\begin{figure}[h]
\centering
	\includegraphics[width=0.15 \textwidth]{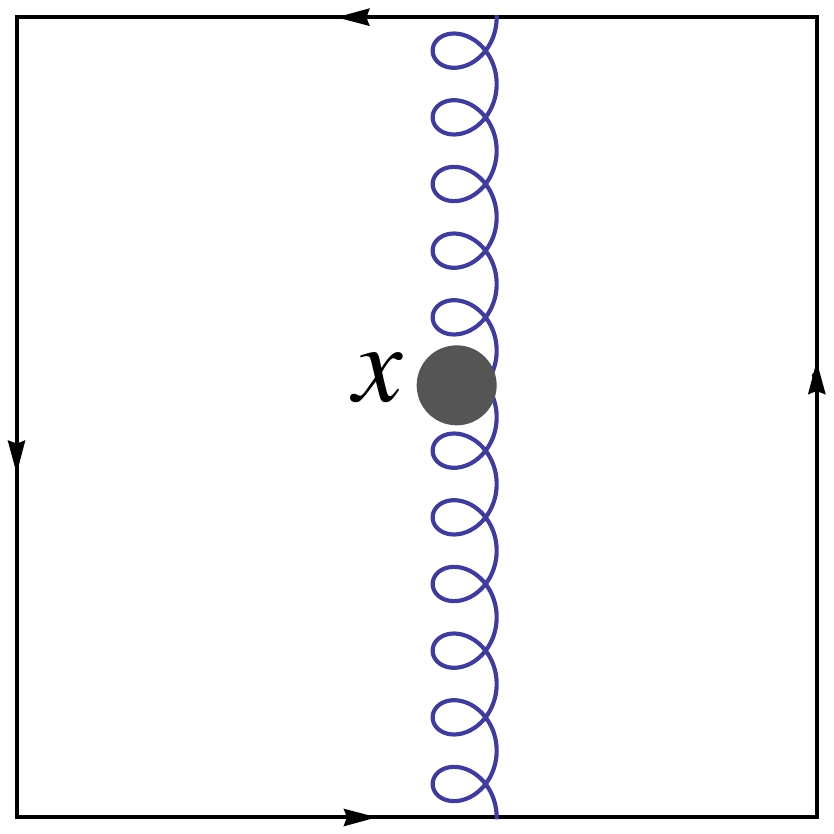}
	\caption{Lagrangian insertion contributing to the Ward identities at one loop. }
	\label{fig:lagrangian-insertion-1-loop}
\end{figure}

\notocsubsection{One-Loop Insertions}
At order  $N/k$ we have a contribution from the contraction of the kinetic part of the Lagrangian 
insertion with the second order expansion of the Wilson loop operator, shown in fig. \ref{fig:lagrangian-insertion-1-loop},
\begin{align}
\langle \mathcal{L}(x) W_n \rangle^{(1)}
= \langle \mathcal{L}_{\text{kin}}(x)W_n\rangle^{(1)} = \frac{(i)^2}{N}\int_{z_i>z_j} \hspace{-20pt} dz_{i,j}^{\mu,\nu} \epsilon^{\alpha\beta\gamma} \langle \tr ( A_\alpha \partial_\beta A_\gamma)(x) \tr (A_\mu A_\nu) \rangle^{(1)} \,.
\end{align}
~\\The direct calculation of the right-hand sides of
\eqref{eqn:dilatation-Wardidentitiy} and \eqref{eqn:special-conformal-Wardidentitiy}, 
yields a vanishing result as $\epsilon \rightarrow 0$. 
%For the dilatation Ward identity this is easy to see, since the integration 
%of 
%\begin{equation}
%\langle A_\mu(z_i) A_\alpha(x) \rangle \partial_\beta^{(x)} \langle A_\nu(z_j) A_\gamma(x) \rangle 
%\end{equation}
%over $d^dx$ 
%and contraction with the Levi-Civita symbols effectively yields\footnote{The same integral appears in \eqref{eqn:integration-yields-prop}. } a gluon propagator of type $\langle A_\mu(z_i) A_\nu(z_j) \rangle$. Thus, the dilatation Ward identity is equivalent to the one-loop diagrams which are all finite. Furthermore, one finds that there are no contributions to the special conformal Ward identity either.\\
Thus we have 
\begin{align}\label{eqn:ward-oneloop}
D\langle W_n  \rangle^{(1)} =  O(\epsilon)\,, \qquad {\rm and} \qquad K^\nu \langle W_n \rangle^{(1)}= O(\epsilon)\,,
\end{align} 
in other words the conformal symmetry is unbroken for $\epsilon = 0$.
As a consequence, the expectation value of the Wilson loop
is constrained to be a function of conformally invariant variables.
Starting from the Lorentz invariants $x_{ij}^2$ the most general
conformal invariants are the cross-ratios
\begin{equation}
u_{ijkl} := \frac{x_{ij}^2 x_{kl}^2}{x_{il}^2 x_{jk}^2}\,.
\end{equation}
In our case where neighbouring points are light-like separated, $x_{i,i+1}^2=0$,
non-vanishing cross-ratios can only be written down starting from $n=6$.
The special conformal Ward identities \eqref{eqn:ward-oneloop} then imply that $\langle W_n \rangle^{(1)}$
is given by a function of conformal cross-ratios,
\begin{equation}\label{eqn:sol-sc-WI-one-loop}
\langle W_n \rangle^{(1)} = g_{n}\left( u_{ijkl}\right)\,, \qquad \langle W_n \rangle^{(1)} = \text{const}. \,.
\end{equation}
Since there are no non-vanishing conformal cross-ratios at four points, $ \langle W_n \rangle^{(1)} $ must be a constant.

Let us now compare against the results of our one-loop computation of section \ref{sec:one-loop-WL}.
There, the constant on the r.h.s. of the second equation in \eqref{eqn:sol-sc-WI-one-loop} was 
found to be zero for the tetragon. 
Moreover, analytical investigations of certain symmetric contours and numerical investigations 
for non-symmetric contours show that $g_{6}(u_{ijkl})$ is zero for the hexagon. 
It was proven  in \cite{Bianchi:2011rn} for arbitrary $n$ that $g_n$ is zero.

\begin{figure}[t]
\centering
\begin{tabular}{ c c c c c c c}
% First ROW %%%%%%%%%%%%%
%\raisebox{35pt}{ $\langle \mathcal{L}(x)  W_n \rangle$ =} 
~& 
\subfloat[]{\label{fig:kinetic-insertion}
\includegraphics[width=.18 \textwidth]{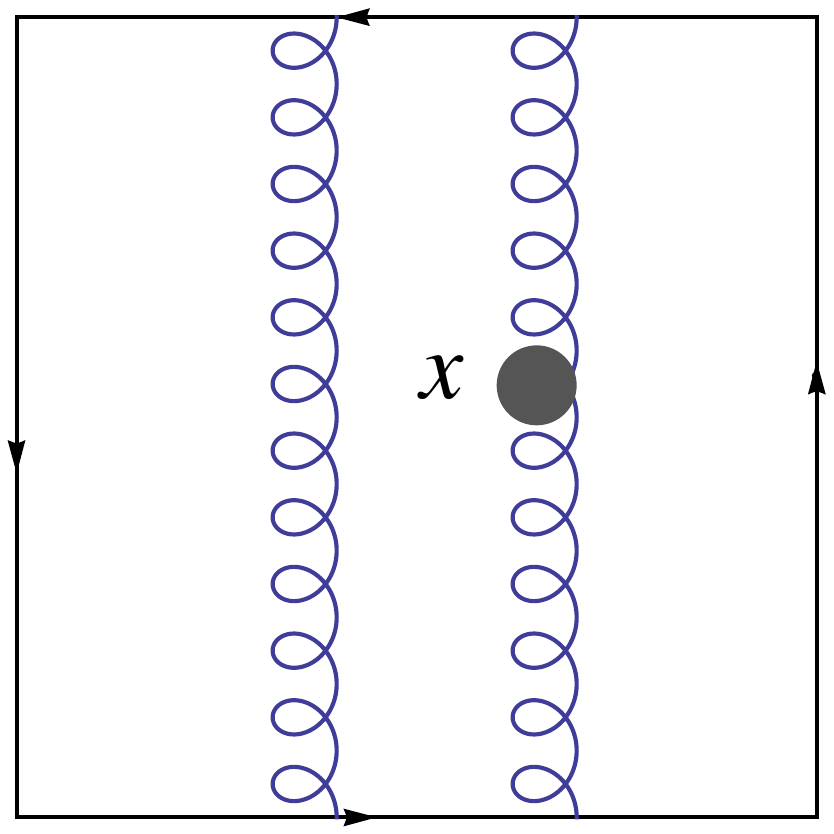}} & % \raisebox{35pt}{ + } 
&
\subfloat[]{\label{fig:vertex-insertion}
\includegraphics[width=.18 \textwidth]{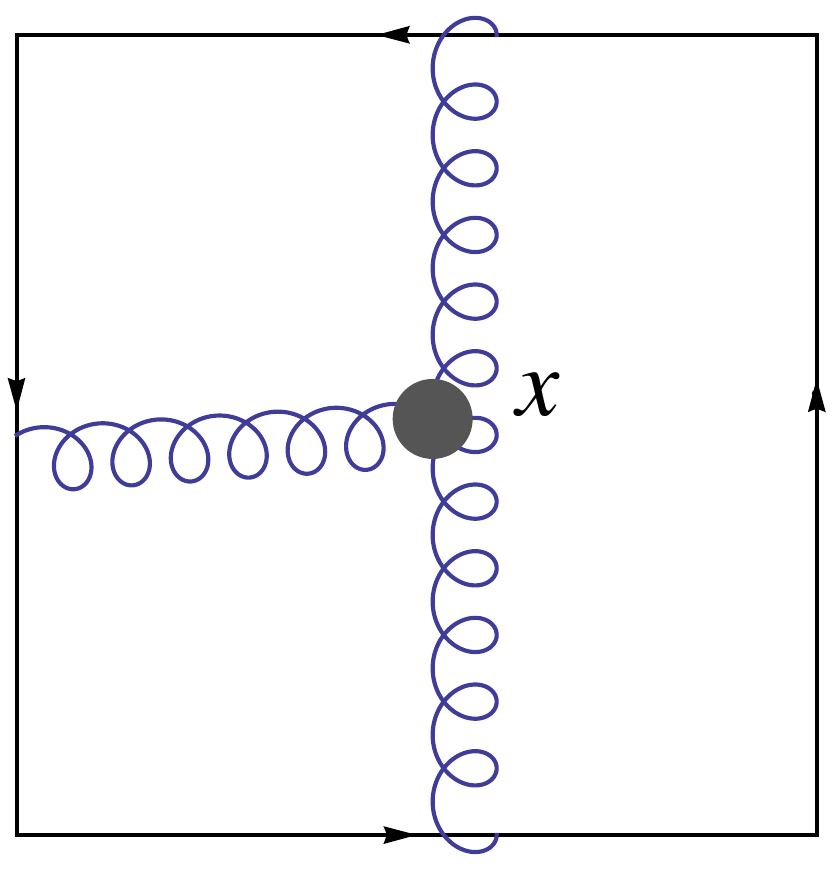}} & 
%\raisebox{35pt}{ + }
&
\subfloat[]{\label{fig:kinetic-insertion-in-vertex}
\includegraphics[width=.18 \textwidth]{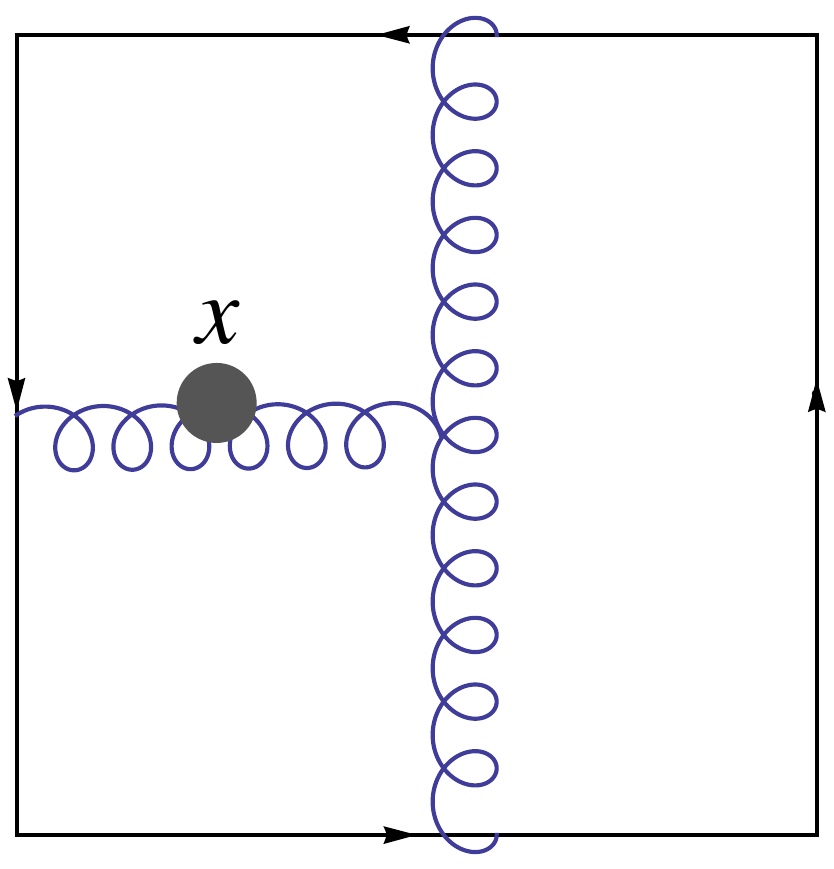}} &
%\raisebox{35pt}{ + }   
\\
% Second ROW %%%%%%%%%%%%
~& \subfloat[]{\label{fig:gluon-loop-vertex-insertion}
\includegraphics[width=.185 \textwidth]{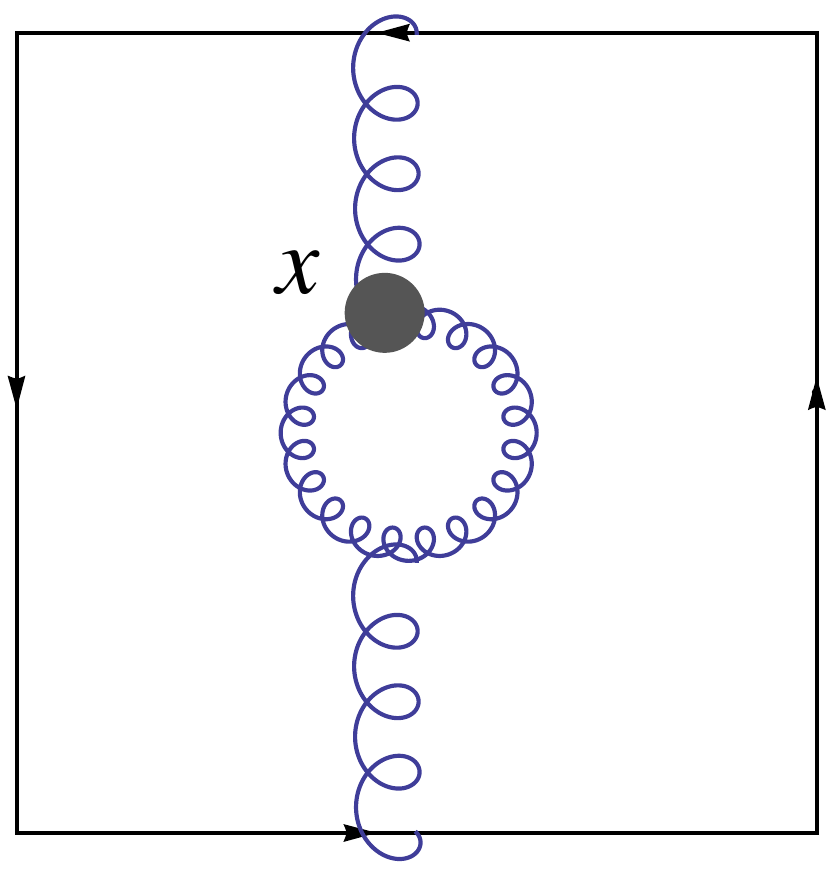}} & 
%\raisebox{35pt}{ + } 
&
\subfloat[]{\label{fig:gluon-loop-propagator-insertion}
\includegraphics[width=.185 \textwidth]{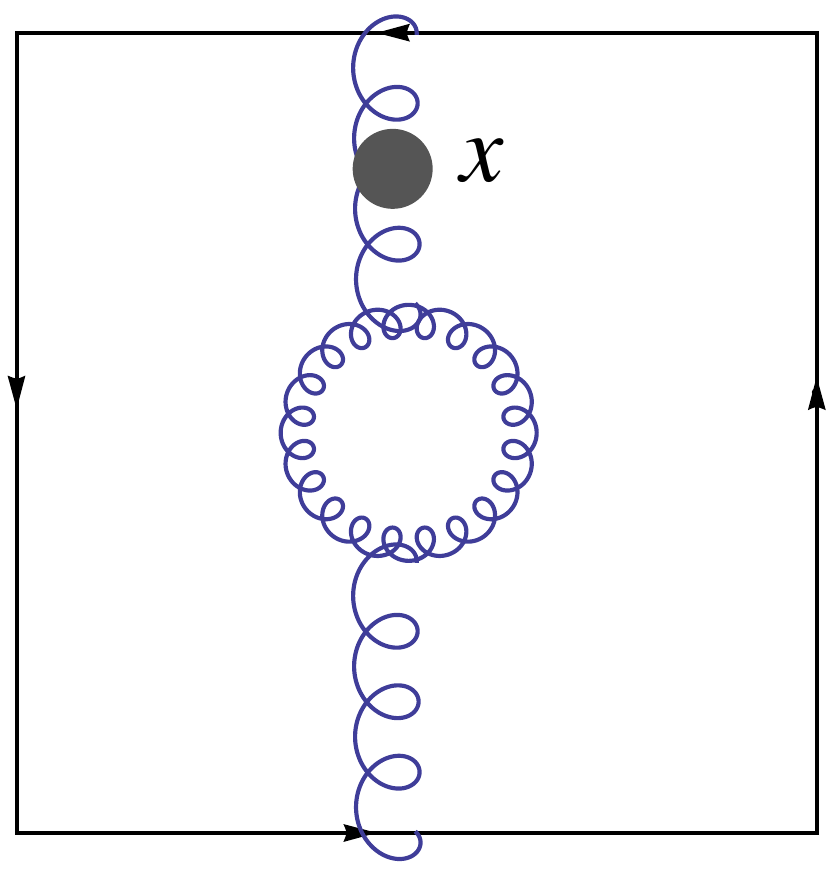}} & 
%\raisebox{35pt}{ + } 
&
\subfloat[]{\label{fig:gluon-loop-inloop-insertion}
\includegraphics[width=.185 \textwidth]{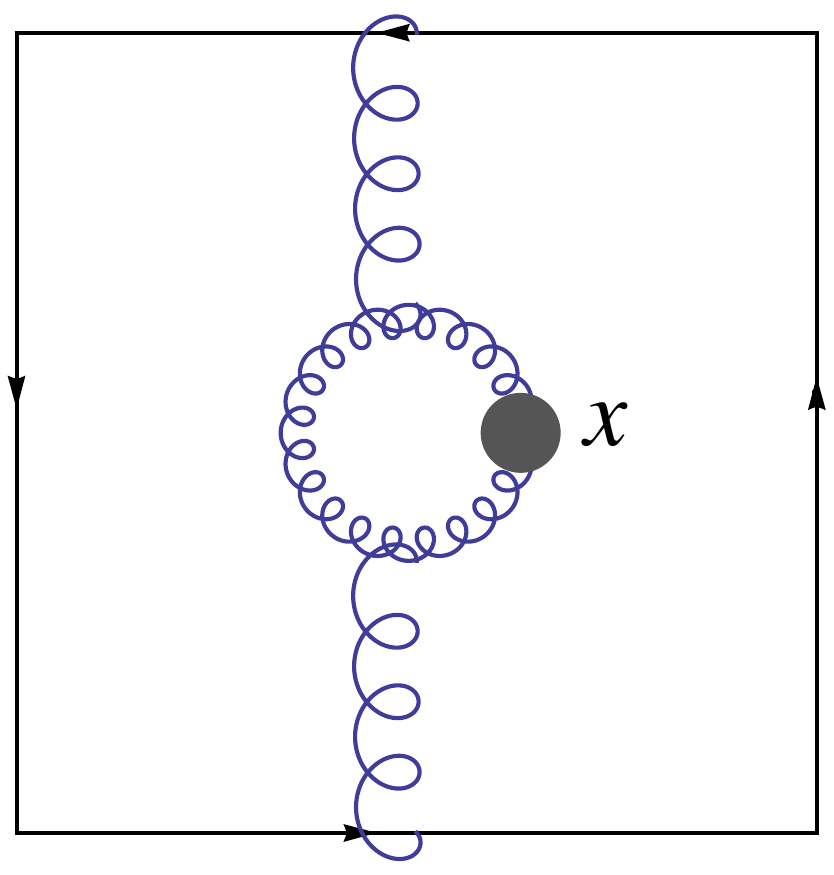} } &
%\raisebox{35pt}{ + }  
\\
% Third ROW %%%%%%%%%%%%%
~& \subfloat[]{\label{fig:ghost-loop-vertex-insertion}
\includegraphics[width=.185 \textwidth]{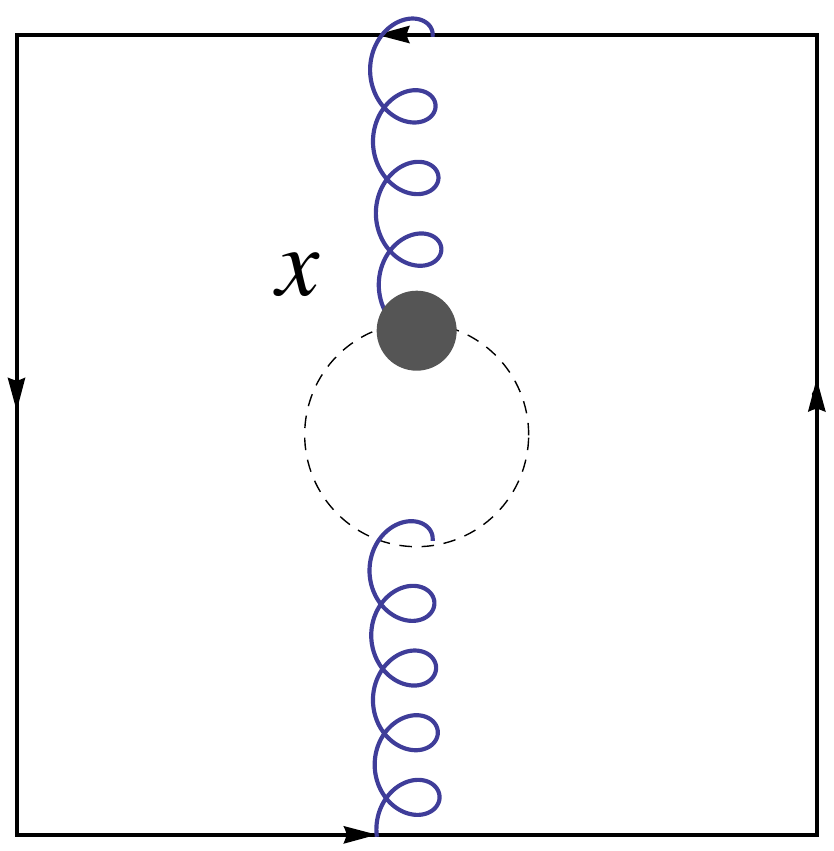}} & 
%\raisebox{35pt}{ + } 
&
\subfloat[]{\label{fig:ghost-loop-propagator-insertion}
\includegraphics[width=.185 \textwidth]{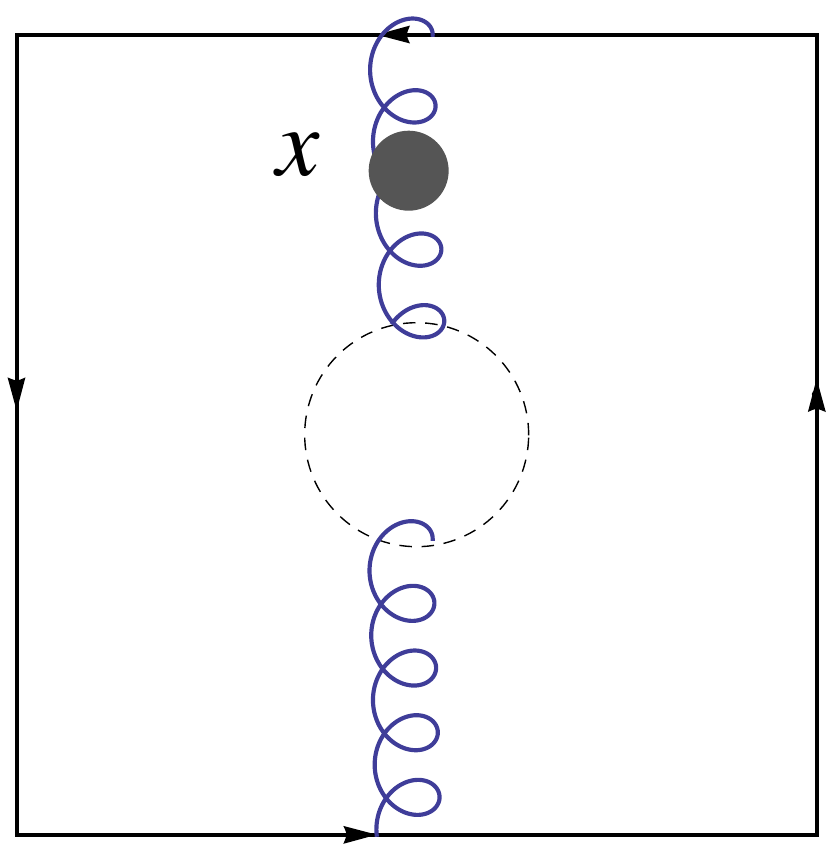}} & 
%\raisebox{35pt}{ + } 
&
\subfloat[]{\label{fig:ghost-loop-inloop-insertion}
\includegraphics[width=.185 \textwidth]{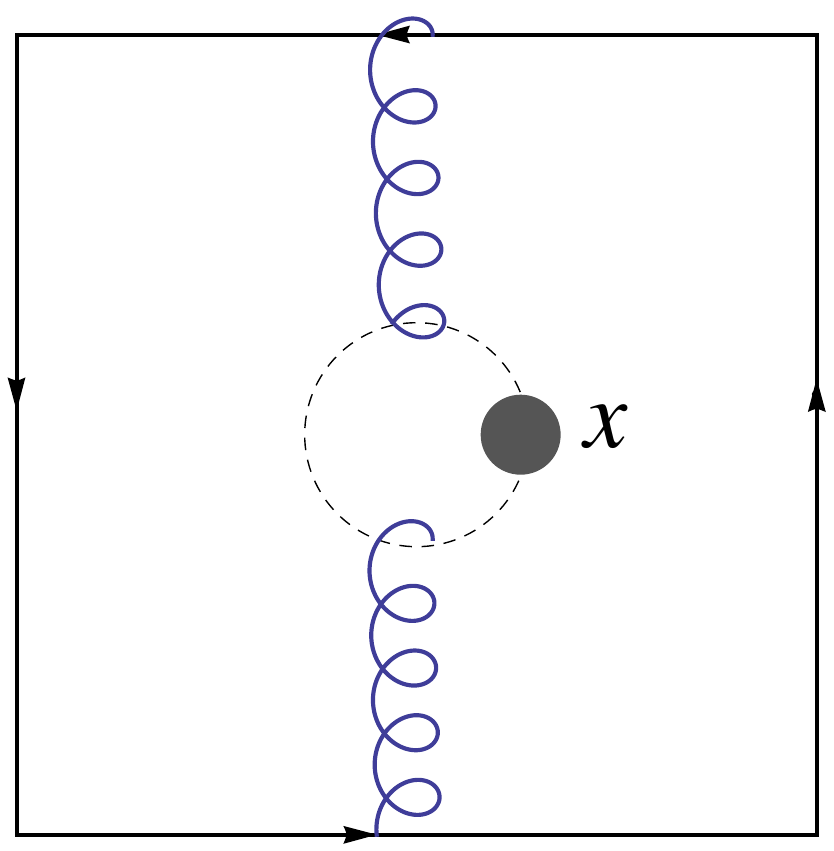}} & ~
\end{tabular}
\caption{Two-loop contributions to the anomalous conformal Ward identity. 
The grey point in the pictures denotes the point $x$ of the Lagrangian insertion 
in $\langle \mathcal{L}(x) W_n \rangle$. 
The diagrams of the second and third line cancel pairwise.
}
\label{fig:2-loop-diagrams-anomalous-conformal-ward-identity}
\end{figure}

\notocsubsection{Two-loop Insertions}
At two loops there are several diagrams that contribute to the insertion of the Lagrangian into the
Wilson loop,
$ \langle \mathcal{L}(x) W_n \rangle $,
that correspond to the kinetic term, the gauge field vertex, the ghost kinetic term and the ghost vertex in $\mathcal{L}(x)$. 
Those diagrams are shown in Figure \ref{fig:2-loop-diagrams-anomalous-conformal-ward-identity}. 
We do not display diagrams that vanish for kinematical reasons as at one-loop level.

Just as at one-loop level, only diagrams giving rise to divergent integrals will
contribute to the anomalous Ward identities.

\subsubsection{Insertion into the ladder diagram}
%First, we can contract the kinetic term of the insertion with the quartic order expansion of the Wilson loop, this corresponds to fig. \ref{fig:kinetic-insertion} 
%\begin{align}\label{eqn:insertion-kinetic-term}
%  \langle \mathcal{L}(x) W_4 \rangle^{(2)}_{\text{(a)}} = \langle \mathcal{L}_{\text{kin}}(x) \frac{1}{N}(i)^4 \int dz_{i,j,k,l}^{\mu,\nu,\rho,\sigma}\, \tr \left( A_\mu A_\nu A_\rho A_\sigma \right)  \rangle \,.
% \end{align}
Let us consider the insertion of the kinetic term of the action into the ladder diagram as shown in Figure \ref{fig:kinetic-insertion}.
For the dilatation Ward identity these are exactly the two-gluon diagrams calculated above, which are finite. 
For the special conformal Ward identity the integration is slightly more complicated, but the finiteness is easy to check for all contributions. 
%Thus we have
%\begin{align}
%\mathbb{D}:&\qquad \epsilon \int d^dx  \langle \mathcal{L}(x) W_4 \rangle^{(2)}_{\text{(a)}} = \mathcal{O}(\epsilon)\,, \\ \nonumber
%\mathbb{K}^\nu:& \qquad\epsilon \int d^dx\, x^\nu  \langle \mathcal{L}(x) W_4 \rangle^{(2)}_{\text{(a)}} = \mathcal{O}(\epsilon)\,.
%\end{align}
Thus, this diagram does not contribute to the anomalous Ward identities.

\subsubsection{Insertion of the interaction term}
Next, we can contract the cubic order expansion of the Wilson loop 
with the vertex term of the Lagrangian insertion, 
as shown in Figure \ref{fig:vertex-insertion},
\begin{align}
\langle \mathcal{L}(x) W_n\rangle^{(2)}_{\text{(b)}} 
&= \langle \mathcal{L}_{\text{int}}(x) \frac{1}{N}(i)^3 \int d^3z_{i,j,k}^{\mu,\nu\rho}\, \tr \left( A_\mu A_\nu A_\rho \right)  \rangle \,.
\end{align}
For the dilatation Ward identity, we trivially  have
\begin{align}\label{eqn:result-insertion-interaction-term}
\int d^dx  \langle \mathcal{L}(x) W_4 \rangle^{(2)}_{\text{(b)}} 
&= \frac{1}{i} \langle W_n \rangle^{(2)}_\text{vertex} = 
i \left( \frac{N}{k}\right)^2   \frac{\ln(2)}{\epsilon}  
+ O(\epsilon^0 )\,,
\end{align}
which is just the vertex diagram that was calculated in \eqref{eqn:vertex-diagram}, up to a factor of $i$. 

The contribution to the special conformal Ward identity is more complicated. 
We have
\begin{align}\label{eqn:special-conformal-wi-vertex-insertion}
\int d^dx\,  x^\nu \langle \mathcal{L}(x) W_n \rangle^{(2)}_{\text{(b)}} 
&= \left( \frac{N}{k} \right)^2 \frac{1}{2\pi}   \left( \frac{\Gamma\left(\frac{d}{2}\right)}{\pi^{\frac{d-2}{2}}}\right)^3 \sum_{i>j>k} I^\prime_{ijk} 
\end{align}
with
\begin{align}\label{eqn:I321prime}
I^\prime_{321} =& \int_0^1 ds_{1,2,3} \int d^dx (x + z_2)^\nu   \frac{\epsilon(p_2,p_3,x)\epsilon(p_2,p_1,x)}{|x|^{d}|x-z_{12}|^{d}|x-z_{32}|^{d}}
\\
=&  \frac{2 \pi i \ln (2)}{\epsilon} (x_2 + x_3)^\nu + \mathcal{O}(\epsilon^0)\,,
\end{align}
where the coefficient $\ln(2)$ was computed numerically to $10$ relevant digits. The reason the pole arises was discussed in section \ref{sect:vertex-diagram}.

Details of this calculation can be found in Appendix \ref{app:conf-wi-insertion-interaction-term}. Summing up all four contributions $I^\prime_{321}, I^\prime_{421}, I^\prime_{432}, I^\prime_{431}$ we arrive at
\begin{align}\label{eqn:special-conformal-wi-vertex-insertion}
\int d^dx\,  x^\nu \langle \mathcal{L}(x) W_n \rangle^{(2)}_{\text{(b)}} 
&=  \frac{i}{\epsilon}  \left( \frac{N}{k} \right)^2  \frac{\ln(2)}{4} \, \sum_{i=1}^{4} x_i^\nu \,.
\end{align}

\subsubsection{Insertion of the kinetic term into the vertex diagram}
Furthermore, we can contract one gauge field of the kinetic term of the insertion with the 
Wilson loop and the other one with the 3-gauge-field vertex, leading to a diagram of the type 
displayed in Figure \ref{fig:kinetic-insertion-in-vertex},
\begin{align}\label{eqn:kinetic-insertion-in-vertex}
\langle \mathcal{L}(x) W_n \rangle^{(2)}_{\text{(c)}} 
&=  \langle \mathcal{L}_{\text{kin}}(x) \frac{1}{N} \oint_{z_i > z_j > z_k} 
\hspace{-35pt} dz_{i,j,k}^{\mu,\nu,\rho}\, \tr \left( A_\mu A_\nu A_\rho \right) 
\left( i\int d^dw \mathcal{L}_{\text{int}}  \right)  \rangle \\ \nonumber
=& \left(\frac{k}{4 \pi}\right)^2   (i)^3  \frac{2}{3}\frac{1}{N} \int d^dx \int d^dw \oint dz_{i,j,k}^{\mu,\nu,\rho}\\ \nonumber
&\phantom{=}\qquad \epsilon^{\alpha\beta\gamma}\epsilon^{\delta\sigma\tau} \langle \tr \left(A_\alpha \partial_\beta A_\gamma \right)(x) \tr \left(A_\mu A_\nu A_\rho \right) \tr \left(A_\delta A_\sigma A_\tau \right)(w)  \rangle \,.
\end{align}
Let us Wick-contract
the kinetic term with $A_\nu(z_j)$ (the two other contractions are discussed below.)
We obtain
\begin{align}
\left( \frac{N}{k}\right)^2 \frac{i}{8 \pi^2}  \left( \frac{\Gamma\left(\frac{d}{2}\right)}{\pi^{\frac{d-2}{2}}}\right)^4 \int d^dw  \oint dz_{i,j,k}^{\mu\nu\rho} \epsilon^{\delta\sigma\tau} I_{\nu\sigma} G_{\mu\tau}(z_i-w) G_{\rho\delta}(z_k-w) \nonumber \,,
\end{align}
where $G_{\mu\nu}= \epsilon_{\mu\nu\rho}{(x-y)^\rho}/{\left(-(x-y)^2\right)^{\frac{d}{2}}}$ 
and where
\begin{align}
I_{\nu\sigma}(x- z_j,x-w) &=\epsilon^{\alpha\beta\gamma}\left[ G_{\alpha\nu}(x-z_j) \partial_\beta^{(x)} G_{\gamma\sigma}(x-w) + G_{\alpha\sigma}(x-w) \partial_\beta^{(x)} G_{\gamma\nu}(x-z_j) \right] \,.
\end{align} 
The only dependence on the insertion point $x$ is in $I_{\nu\sigma}$. For the dilatation Ward 
identity the integral $\int d^dx\, I_{\nu\sigma}$ can easily be computed (for details see Appendix \ref{app:conf-wi-insertion-kinetic-term-in-vertex})  
and effectively gives a propagator such that we have
\begin{align}\label{eqn:result-insertion-kinetic-term-vertex}
\int d^dx\, \langle \mathcal{L}(x) W_n \rangle^{(2)}_{\text{(c)}} 
&= %- 3 i \left( \frac{N}{k}\right)^2 \frac{1}{2 \pi}  \left( \frac{\Gamma\left(\frac{d}{2}\right)}{\pi^{\frac{d-2}{2}}}\right)^3 \sum_{i>j>k} I_{ijk}=
3 i \langle W_n \rangle^\text{vertex}
=- 3 i \left( \frac{N}{k}\right)^2   \frac{\ln(2)}{\epsilon}   + O(\epsilon^0 )\,, 
\end{align}
where a factor of 3 was included since the insertion can be in any of the 3 propagators of the vertex diagram and thus 
we get three times the same contribution.

The contribution to the special conformal Ward identity is more complicated, 
since the integration $\int d^dx\, x^\nu I_{\nu\sigma}$ does not just yield a propagator. 
Performing the calculation, we find 
\begin{align}\label{eqn:special-conformal-wi-vertex-insertion-in-propagator}
\int d^dx  \,x^\nu\langle \mathcal{L}(x) W_n \rangle^{(2)}_{\text{(c)}} 
&= - \frac{i}{\epsilon}  \left( \frac{N}{k} \right)^2 \frac{3}{4}   \ln(2) \, \sum_{i=1}^{4} x_i^\nu + O(\epsilon^0 ) \,,
\end{align}
where the coefficient $\ln(2)$ was computed numerically (details can be found in Appendix \ref{app:conf-wi-insertion-kinetic-term-in-vertex}).

\subsubsection{Insertions with gauge field and ghost loops}
The gauge-field-ghost-insertions in Figure \ref{fig:2-loop-diagrams-anomalous-conformal-ward-identity} cancel pairwise. 
For the dilatation Ward identity the insertions of the three gauge-field and the gauge-field-ghost vertices as shown in diagrams \ref{fig:gluon-loop-vertex-insertion}, \ref{fig:ghost-loop-vertex-insertion} are identical to the gauge field and ghost loop diagrams  \eqref{eqn:gluon-ghost-loops} and thus cancel. It is not necessary to perform the integration over the insertion point to see how the cancellation occurs and thus the contributions to the special conformal Ward identity cancel as well (for details on the cancellation see \ref{app:gluon-and-ghost-loops}).

Insertions of the kinetic term into the gauge field propagator as shown in diagrams \ref{fig:gluon-loop-propagator-insertion}, \ref{fig:ghost-loop-propagator-insertion} cancel as well, since the insertion is the same for both diagrams and thus the algebraic relations responsible for the cancellation in \eqref{eqn:gluon-ghost-loops} remain unchanged. 

Inserting gauge field respectively ghost kinetic terms into the propagators inside the loop  
in diagrams \ref{fig:gluon-loop-inloop-insertion}, \ref{fig:ghost-loop-inloop-insertion} 
produce slightly more complicated expressions.  Nevertheless they cancel as well as can 
be seen in a straightforward calculation. For the dilatation Ward identity the 
cancellation can be seen in an even simpler way by noticing that the integration 
over the insertion point $x$ effectively yields a gauge field respectively ghost propagator. 
Thus the diagrams are identical to the ones in \eqref{eqn:gluon-ghost-loops} and cancel.

\notocsubsection{Anomalous Ward Identities and Generalisation to Higher Polygons}
Summing up the divergent contributions of \eqref{eqn:result-insertion-interaction-term} \eqref{eqn:result-insertion-kinetic-term-vertex}, and inserting
them into the dilatation Ward identity \eqref{eqn:dilatation-Wardidentitiy},
we obtain
\begin{equation}\label{eqn:dilatation-wi}
D\, \langle W_n \rangle^{(2)} = 
-\left(\frac{N}{k}\right)^2  \ln(2) \left( \sum_{i=1}^{4} 1 \right)  + \mathcal{O}(\epsilon)\,,
\end{equation}
where we have written the factor $4$ as $(\sum_{i=1}^{4} 1)$ to emphasise its origin
from the sum of four vertex-type diagrams.
Note that only the divergent part of the vertex-diagram was required here.
Summing up \eqref{eqn:special-conformal-wi-vertex-insertion}  and 
\eqref{eqn:special-conformal-wi-vertex-insertion-in-propagator}, and inserting them into the 
special conformal Ward identity \eqref{eqn:special-conformal-Wardidentitiy}, we obtain
\begin{equation}\label{eqn:special-conf-wi}
K^\nu \langle W_n \rangle^{(2)}
= -2 \left( \frac{N}{k} \right)^2  \ln(2) \, \sum_{i=1}^{4} x_i^\nu + \mathcal{O}(\epsilon)\,.
\end{equation}
Let us now explain how these equations can be generalised from $n=4$ to arbitrary $n$.
In our two-loop computation, we found that the only diagrams contributing 
to \eqref{eqn:dilatation-wi} and \eqref{eqn:special-conf-wi}
are those producing poles in $\epsilon$.
The mechanism for how these poles are generated was described in section \ref{sect:vertex-diagram}, see in
particular Figure \ref{fig:divergence-vertex}. It is clear that for $n>4$ cusps, the same type of vertex diagram
will produce the divergent terms. Although those diagrams will depend on one further
kinematical variable w.r.t. the four-point case, this dependence cannot change the (leading) 
UV pole $\epsilon^{-1}$ of the diagrams. Since there are $n$ diagrams of this type at
$n$ points, we expect
\begin{equation}\label{eqn:dilatation-wi-n}
D\, \langle W_n \rangle^{(2)} = 
-\left(\frac{N}{k}\right)^2  \ln(2) \left( \sum_{i=1}^{n} 1 \right)  + \mathcal{O}(\epsilon)\,,
\end{equation}
and
\begin{equation}\label{eqn:special-conf-wi-n}
K^\nu \langle W_n \rangle^{(2)}
= -2 \left( \frac{N}{k} \right)^2  \ln(2) \, \sum_{i=1}^{n} x_i^\nu + \mathcal{O}(\epsilon)\,.
\end{equation}
We will now proceed to discuss the solution of these Ward identities and
compare them to the result of the two-loop computation of the tetragon Wilson loop in section \ref{sect:CS-2loop-results}.

\notocsubsection{Solution to the Anomalous Conformal Ward Identities}
Using
$ D\left( x_{ij}^2\right) = 2 x_{ij}^2 $
it is clear that the most general solution to the dilatation Ward identity \eqref{eqn:dilatation-wi-n} is 
\begin{align}\label{eqn:solution-dilatation-ward-identity-n}
\langle W_n \rangle^{(2)} = -\left(\frac{N}{k}\right)^2  \left[  \frac{\ln(2)}{4}\sum_{i=1}^n\frac{(-x_{i,i+2}^2 \tilde \mu^2 )^{2\epsilon}}{\epsilon} + f_{n}\left(\frac{x_{ij}^{2}}{x_{kl}^2}\right)\right] + O(\epsilon) \,,
\end{align}
where $f$ is an arbitrary function of dimensionless variables and we recall that  
$\tilde{\mu}^2 = \mu^2 \pi e^{\gamma_{E}}$. 
Of course, this is exactly what we expect from \eqref{eqn:wilson-loop-regularisation}. 

The result for the special conformal Ward identity is more interesting. 
Plugging \eqref{eqn:solution-dilatation-ward-identity-n} into the special
conformal Ward identity \eqref{eqn:special-conf-wi}
and using  $\mathbb{K}^\nu \ln (x_{kl}^2) = 2(x_k + x_l)^\nu $,
it is easy to see that the function $f_{n}$ is allowed to
depend on conformally invariant cross-ratios only, i.e. $f_{n}(x_{ij}^2 / x_{kl}^2 )$ = $g(u_{abcd})$.
Therefore, we finally have
\begin{align}\label{eqn:solution-special-conformal-ward-identity-xij}
\langle W_n \rangle^{(\text{CS})}_{\text{2-loop}} &= -\frac{1}{4}\left(\frac{N}{k}\right)^2  \Big[ 2\ln(2)\sum_{i=1}^n\frac{(-x_{i,i+2}^2 \tilde \mu^2 )^{2\epsilon}}{2\epsilon} + g_{n}(u_{abcd}) + \mathcal{O}(\epsilon) \Big]\,.
\end{align}
In the four-point case, there are no non-vanishing cross-ratios,
and therefore in particular $g_{4}$ must be a constant.
This is in agreement with \eqref{eqn:result-wilson-loop-two-loop} and thus 
represents an independent check of the direct perturbative computation,
including its finite part (recall that deriving the Ward identity does not rely 
on the finite parts of the direct perturbative computation). 
So, even though the result for the vertex diagram \eqref{eqn:result-vertex} was 
obtained numerically, its functional form is an analytical result, since we know the 
analytical expression for the ladder diagram and the sum of vertex and ladder diagram 
through the solution to the anomalous conformal Ward identity.

\section{N-Sided Wilson Loops in CS Theory at Two Loops}\label{sec:n-sided-wilson-loops-in-cs}
As mentioned in the previous section, the solution \eqref{eqn:solution-special-conformal-ward-identity-xij} of the Ward identity for the light-like polygonal Wilson loop in pure Chern-Simons theory contains a function $g_n$ that depends only on conformally invariant cross ratios $u_{abcd}=(x_{ab}^2x_{cd}^2)/(x_{ad}^2x_{cb}^2)$ which can be constructed starting from $n=6$. In this section we review the investigations \cite{Wiegandt:2011uu}, that lead to the conclusion that $g_n$ is a trivial function at two loops.

\begin{figure}[t]
\centering
\subfloat[]{\begin{minipage}{4.5cm}\centering
		~\includegraphics[width=.9 \textwidth]{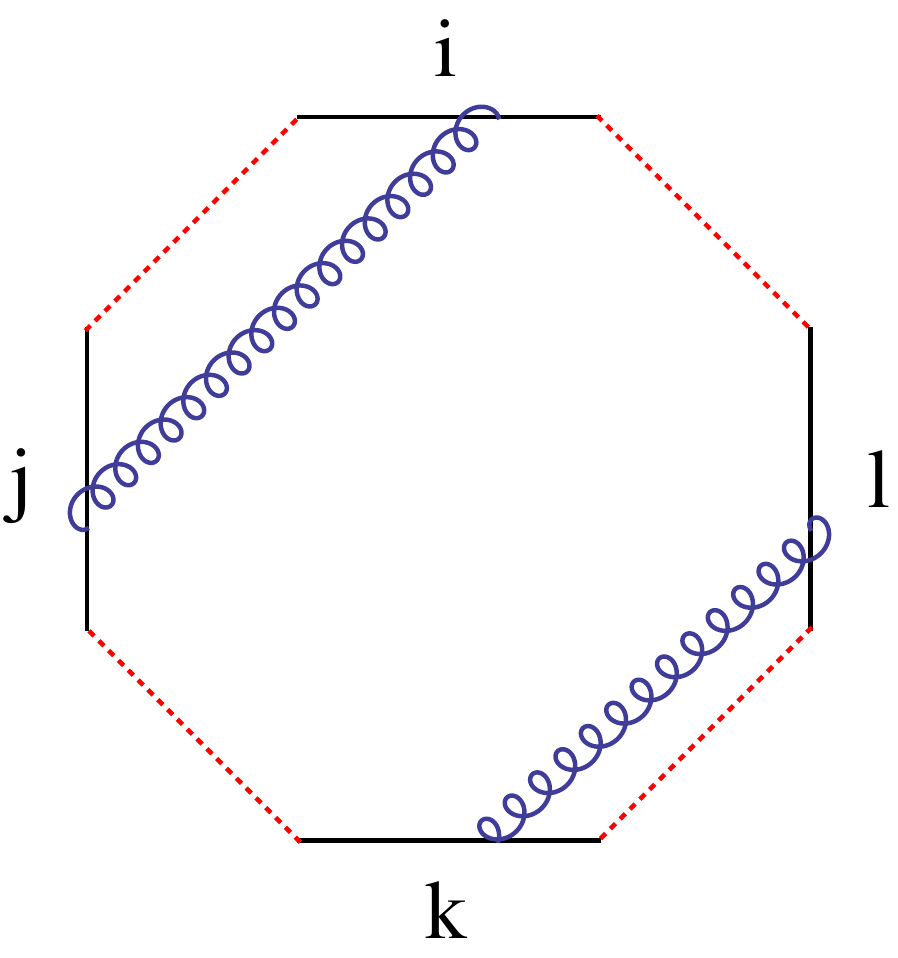}\label{fig: rect1331}\\
		\end{minipage}
}
~~~~~~~~~~~
\subfloat[]{\begin{minipage}{4.5cm}\phantom{a}~\\
\centering
		~\includegraphics[width=.85 \textwidth]{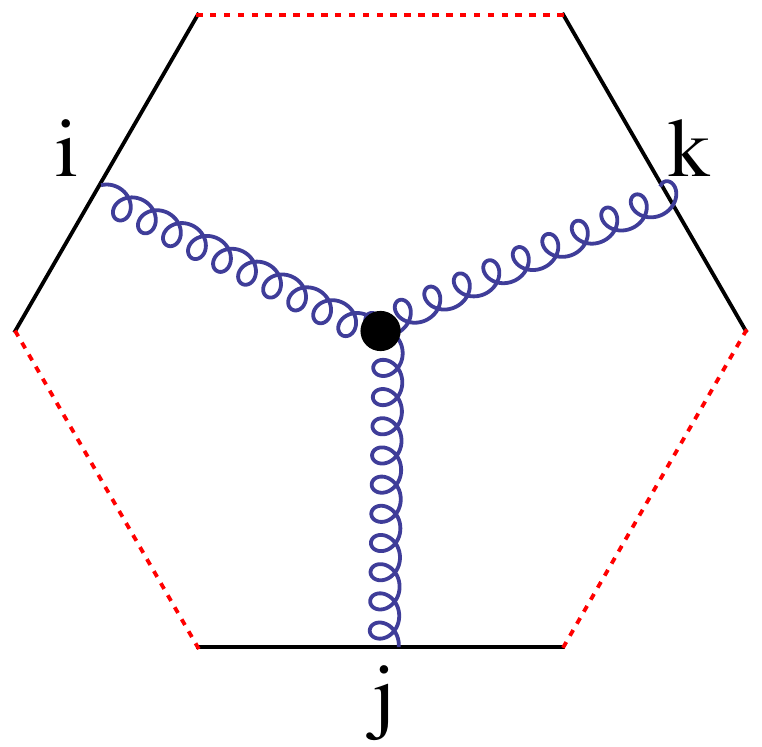}\label{fig:vertexI123}
		\end{minipage}
}
\caption{\ref{fig: rect1331} Two-gluon diagrams $I_{ijkl}$ given in \eqref{eqn:two-gluon-expr},  \ref{fig:vertexI123} vertex diagrams $I_{ijk}$ given in \eqref{eqn:vertexIaandIb}. Dashed lines represent one or more light-like distances.}
\label{fig:2typesofdiagrams}
\end{figure}

As in the four-point case, at two loops there are two types of contributions to the Wilson loop in pure Chern-Simons theory, one from two-gluon diagrams $\langle W_n\rangle^{\text{two-gluon}}=\sum  I_{ijkl}$ and another one from diagrams involving a three-gluon vertex $\langle W_n\rangle^{\text{vertex}} = \sum I_{ijk}$ such that
\begin{equation}
\langle W_n \rangle^{\text{CS}}_{\text{2-loop}} = \langle W_n\rangle^{\text{two-gluon}} + \langle W_n\rangle^{\text{vertex}} \,.
\end{equation}
The indices $i,j,k,l$ denote the edges that the propagators attach to, see figure \ref{fig:2typesofdiagrams} and their expressions are given below in \eqref{eqn:two-gluon-expr}, \eqref{eqn:vertex-diagrams}. Contributions from gauge- and ghost-loops cancel in dimensional reduction regularisation \cite{Chen:1992ee} as explained in \ref{sec:gauge-ghost-cancellation}.

\subsubsection{Two-Gluon Diagrams}\label{app:two-gluon}
The contributions from the two-gluon diagrams are all finite and can be written as
\begin{equation}\label{eqn:two-gluon-diagrams}
\langle W_n\rangle^{\text{two-gluon}}= \frac{1}{4} \left(\frac{N}{k}\right)^2 \sum_{i \geq j \geq k \geq l} \left( I_{ij,kl}+ I_{il,jk} \right)\,,
\end{equation}
where
\begin{equation}\label{eqn:two-gluon-expr}
I_{ij,kl} = \int ds_{i,j,k,l} \frac{\epsilon(\dot{z}_i,\dot{z}_j,z_i-z_j)}{\left(-(z_i-z_j)^2\right)^{3/2}} \frac{\epsilon(\dot{z}_k,\dot{z}_l,z_k-z_l)}{\left(-(z_k-z_l)^2\right)^{3/2}}
\end{equation}
and the integration boundaries haven to be chosen according to the path ordering,  such that $z(s_i)>z(s_j)>z(s_k)>z(s_l)$. The diagram vanishes due to the epsilon tensor contractions if the gluon propagator of at least one of the index pairs connects the same or adjacent edges for the same reason as in the four-point case. For $n=4$ only the diagram \eqref{eqn:ladder-diagram-n-4} appears.

%As a check of the numerics one can use the factorizing diagrams $I_{i+5,i+3,i+2,i}$, which are just a product of the analytically known one-loop diagrams \cite{Bianchi:2011rn}.

\subsubsection{Vertex Diagrams}\label{app:vertex}
We write the generalization of the vertex diagrams \eqref{eqn:Iijk}
as
\begin{align}\label{eqn:vertex-diagrams}
\langle W_n \rangle^{\text{vertex}} 
&=  -\frac{1}{4} \left(\frac{N}{k}\right)^2  \pi^{3-d} \Gamma(d-1) \frac{1}{4 \pi}  \sum_{i>j>k} \left( I^{(a)}_{ijk} + I^{(b)}_{ijk}\right) \\ \nn
&=  - \frac{1}{4} \left(\frac{N}{k}\right)^2 \left( \pi e^{\gamma_E}  \right)^{2 \epsilon} (1-2\epsilon)  \frac{1}{4\pi} \sum_{i>j>k} \left(I_{ijk}^{(a)}+I_{ijk}^{(b)} \right)+ \mathcal{O}(\epsilon)\,.
\end{align}
The indices $i,j,k$ indicate the edges the gluon-propagators connect to, see figure \ref{fig:vertexI123}, and
\begin{align}\label{eqn:vertexIaandIb}\nn
I^{(a)}_{ijk} &= \int ds_{i,j,k},d[\beta]_3  \epsilon^{\alpha\beta\gamma}\epsilon_{\mu\alpha\sigma}\epsilon_{\nu\beta\lambda}\epsilon_{\rho\gamma\tau}  p_i^\mu p_j^\nu p_k^\rho \\ \nn 
& \quad \left[  \partial_j^\lambda  \Delta  \partial_i^\sigma  \partial_k^\tau \Delta+   \partial_i^\sigma \Delta \partial_j^\lambda  \partial_k^\tau \Delta+  \partial_k^\tau \Delta  \partial_i^\sigma  \partial_j^\lambda \Delta    \right] \Delta^{1-d}\,, \\ \nn
I^{(b)}_{ijk} &=\int ds_{i,j,k},d[\beta]_3\epsilon^{\alpha\beta\gamma}\epsilon_{\mu\alpha\sigma}\epsilon_{\nu\beta\lambda}\epsilon_{\rho\gamma\tau}   p_i^\mu p_j^\nu p_k^\rho   \\ 
& \quad \left[\partial_i^\sigma \Delta \partial_j^\lambda \Delta \partial_k^\tau  \Delta \right] \Delta^{-d} (1-d)\,,
\end{align}
where $\partial_i^\mu = \partial/\partial z_{i,\mu}$ and 
\begin{align}\nn
 \int d[\beta]_3 & = \int_0^1 d\beta_{i,j,k}\delta(\sum_m \beta_m-1)(\beta_i \beta_j \beta_k)^{d/2-2}, \\ \nn
\Delta & = - z_{ij}^2 \beta_i \beta_j- z_{ik}^2 \beta_i {\beta}_k- z_{kj}^2 \beta_k {\beta}_j\,.
\end{align}
The above expressions are obtained by introducing Feynman parameters in the standard way and  integrating over $w$, exactly as in the four-point case in section \ref{sect:vertex-diagram}. The details are given in Appendix \ref{app:Vertex-n-gon}.

As in the four-point case, the vertex diagrams are divergent in the region of integration where all three propagators approach the same edge (all diagrams with more than one propagator on the same edge vanish identically due to the antisymmetry of the Levi-Civita symbol), and we split them up as in \eqref{eqn:split-up-fin-div} 
\begin{equation}
 \langle W_n \rangle^{\text{vertex}} =  \langle W_n \rangle^{\text{div}} + \langle W_n \rangle^{\text{finite}}\,.
\end{equation}
The divergent part is calculated analytically in Appendix \ref{sec:analytical-evaluation-of-pole-term} and reads
\begin{equation}\nn
\langle W_n \rangle^{\text{div}} = - \left( \frac{N}{k} \right)^2 \left( \frac{\ln (2)}{2}  \sum_{i=1}^n \frac{(-x_{i,i+2}^2 \tilde{\mu}^2)^{2\epsilon}}{2 \epsilon} \right)\,.
\end{equation}
Thus, the function $g_n$ in \eqref{eqn:solution-special-conformal-ward-identity-xij} is given by
\begin{equation}\label{eqn:contributions-to-gn}
 g_n(u_{abcd}) = \langle W_n\rangle^{\text{two-gluon}} + \langle W_n \rangle^{\text{finite}}\,.
\end{equation}
We evaluate these contributions using a Mathematica program that generates all $n$-point diagrams, performs the index-contractions and numerically integrates the diagrams for randomly generated kinematical configurations, see Appendix \ref{app:generation-kinematics}.\\

~\\
{\bf Hexagonal Wilson loop}\\~\\
For the hexagonal Wilson loop we evaluated the contributions in \eqref{eqn:contributions-to-gn} for a large set of conformally equivalent and conformally non-equivalent kinematical configurations. 

Conformally equivalent configurations must yield the same result, since, by the anomalous conformal Ward identity, the expectation value is constrained  to the form \eqref{eqn:solution-special-conformal-ward-identity-xij}, and thus the function $g_n$ depends only on conformally invariant quantities.

\begin{figure}[t]
\centering
 \includegraphics[width=.9 \columnwidth]{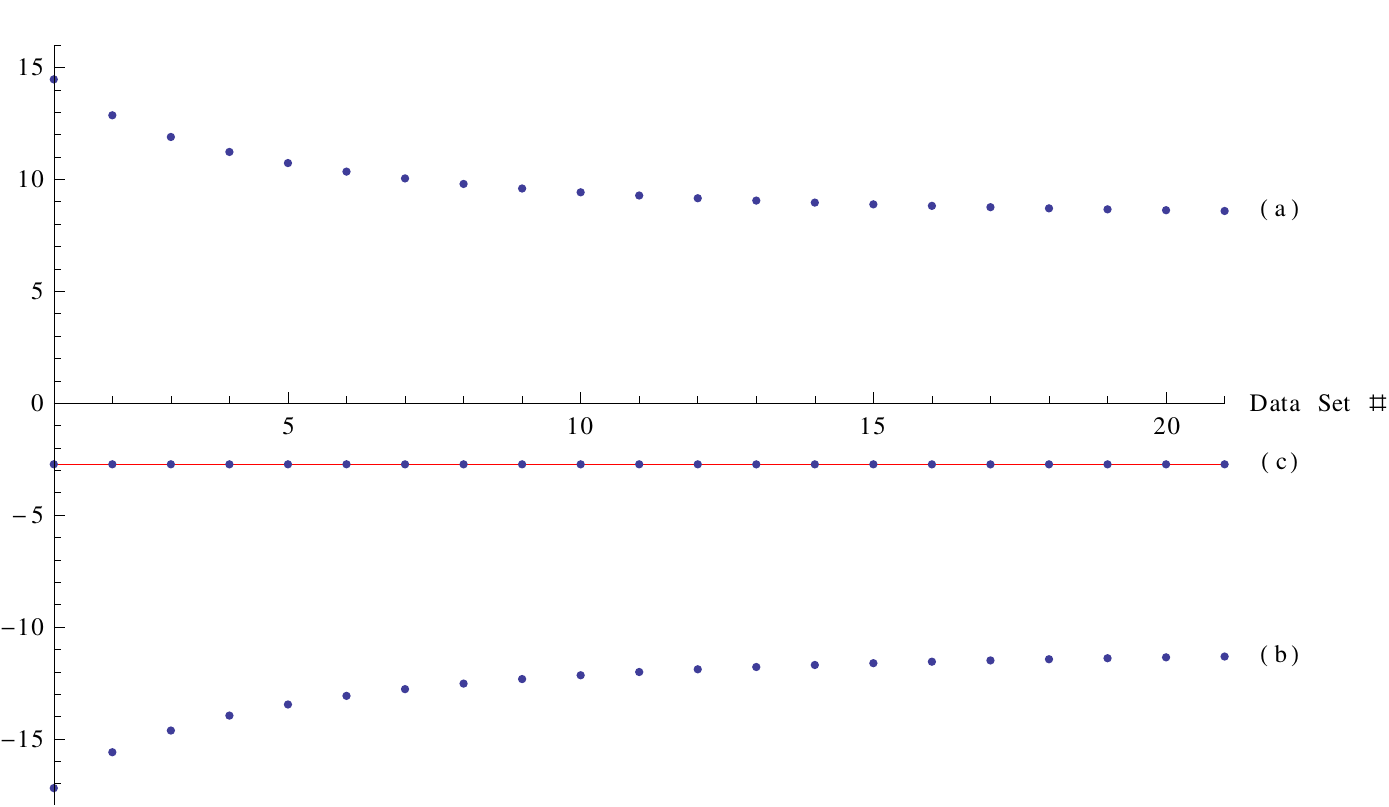}
\caption{Numerical results for the hexagon $n=6$:  $(a)=\langle W_n\rangle^{\text{two-gluon}}$, $(b)=\langle W_n \rangle^{\text{finite}}$ and their sum  $(c) = g_n(u_{abcd})$, see \eqref{eqn:contributions-to-gn}, for the kinematical points given in \ref{app:generation-kinematics}. We omit the common factor $\frac{1}{4}\left(\frac{N}{k}\right)^2$.}
\label{fig:Wvertextwogluonplot}
\end{figure}

It turns out, that even for kinematical configurations which are not conformally equivalent, the unknown function yields the same constant\footnote{The analytical term with $\ln(2)$ in \eqref{eqn:g6},\eqref{eqn:results-g_n} arises from the multiplication of the analytically known divergent term with an $\mathcal{O}(\epsilon)$ expansion of the prefactor, see \eqref{eqn:vertex-diagrams}.}
\begin{equation}\label{eqn:g6}
  g_6 (u_{abcd}) = c_6 - 12\ln(2),\quad  c_6 = 5.57 \pm 0.05\,.
\end{equation}
In figure \ref{fig:Wvertextwogluonplot} we show the results for the two-gluon and vertex contributions for a continuously deformed kinematical configuration, generated as explained in app. \ref{app:generation-kinematics}, in order to illustrate, how the different contributions vary while their sum remains constant.\\

\begin{figure}[h]
\centering
 \includegraphics[width=.9 \columnwidth]{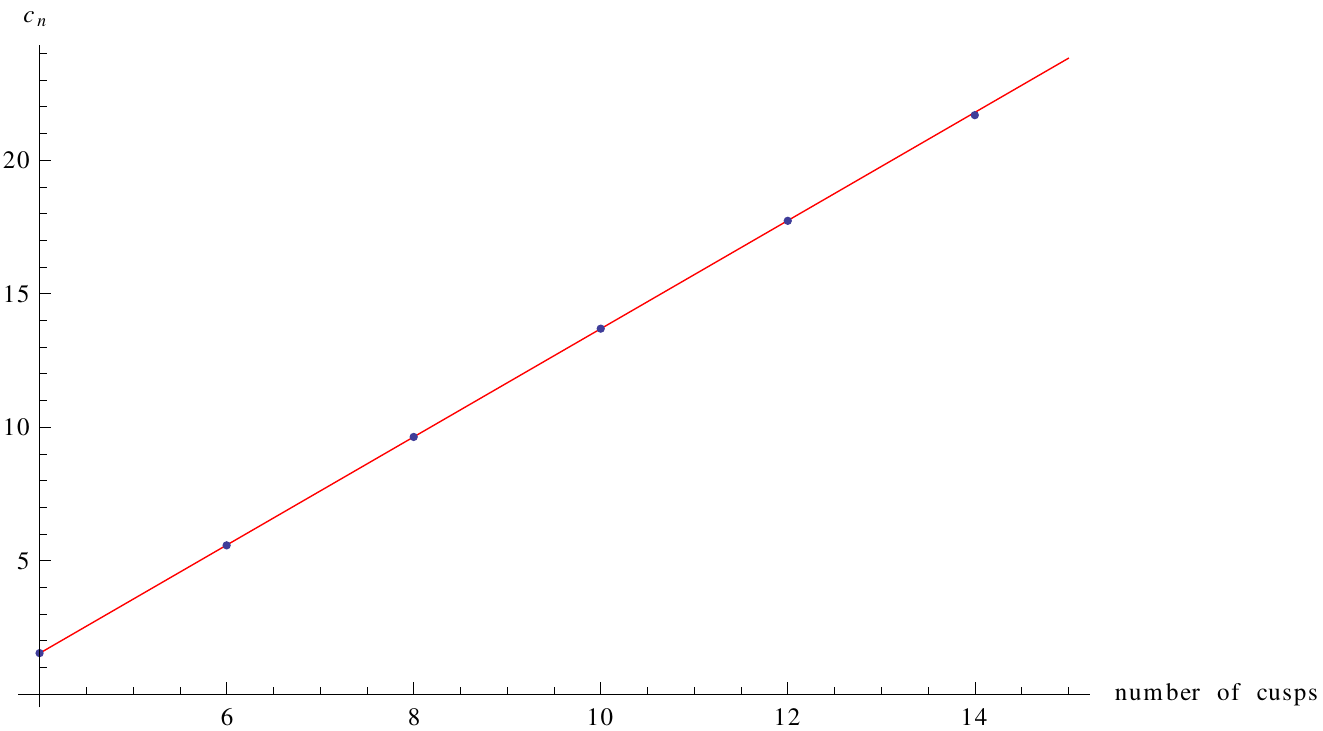}
\caption{This figure shows the dependence of the constant $c_n$ \eqref{eqn:n-dependence-constant} on the number of cusps $n$ of the Wilson loop.}
\label{fig:constant-dependence-on-number-of-cusps}
\end{figure}

~\\~\\
{\bf Generalization to $n$ Cusps}\\~\\
It  turns out that also for $n>6$ the function of conformal cross ratios is just a constant, i.e.
\begin{equation}\label{eqn:results-g_n}
 g_n  (u_{abcd})  = c_n -  2 n \ln(2)\,.
\end{equation}
In figure \ref{fig:constant-dependence-on-number-of-cusps} we show the dependence of the numerical constant $c_n$ on the number of cusps $n$ up to $n=14$. Clearly, the constant depends linearly on the number of cusps $n$. It seems reasonable to assume that this dependence holds for all $n$, i.e.  
\begin{equation}\label{eqn:n-dependence-constant}
 c_n = a + b \cdot n\,,
\end{equation}
which is the line shown in fig. \ref{fig:constant-dependence-on-number-of-cusps} with the parameters\footnote{We determine the constants in \eqref{eqn:n-dependence-constant} from the results at $n=4$ and $n=6$, since here we have the smallest number of integrals and thus the best numerical result. At n-points we have to evaluate $2 {n \choose 3} + 2 {n \choose 4}$  two-gluon  \eqref{eqn:two-gluon-diagrams} and vertex integrals \eqref{eqn:vertex-finite-pieces}. 
}  %$c_0=6.59884$  $d_0 = -2.02855$. 
$a=6.6\pm0.1$,  $b = -2.028\pm 0.025$. 
Thus, we expect the Chern-Simons contribution to the $n$-point Wilson loop to be
\begin{align}\label{eqn:n-point-conjecture-CS} 
\langle W_n \rangle^{\text{CS}} &= - \frac{1}{4} \left( \frac{N}{k} \right)^2 \Bigg[2 \ln (2)\sum_{i=1}^n \frac{(-x_{i,i+2}^2 \tilde{\mu}^2  )^{2\epsilon}}{2 \epsilon} + g_n \Bigg]\,,
\end{align}
where $g_n$ is the constant given by \eqref{eqn:results-g_n} and \eqref{eqn:n-dependence-constant}.

%%%%%%%%%%%%%%%%%%%%%%%%%%%%%%%%%%%%%%%%%%%%%%%%%%%%%%%%%%%%%%%%%%%%%%%%%%%%%%%%
\section{N-Sided Wilson Loops in ABJM Theory at Two Loops}
Here we explain how the results are modified in ABJM theory. We use the Wilson loop operator proposed in \cite{Drukker:2008zx}
\begin{align}\label{eqn:Wilson-loop-ABJM}
\langle W(A,\hat A) \rangle =  \frac{1}{2N} \left\langle   \tr \mathcal{P}\exp \left(i \oint_\mathcal{C} A_\mu dz^\mu \right) + \hat{~\tr} \mathcal{P}\exp \left(i \oint_\mathcal{C} \hat{A}_\mu dz^\mu \right)  \right\rangle \,.
\end{align}
Note that the sign(s) in the exponent(s) in \eqref{eqn:Wilson-loop-ABJM} are correlated to corresponding signs in the Lagrangian by the requirement of
gauge invariance, see Appendix \ref{sec:wilson-loop-op-and-gauge-invariance}.

\notocsubsection{Gauge Field Contributions}
In ABJM theory there is a second copy of the gauge field $\hat A_\mu$ with opposite sign in the Lagrangian \eqref{eqn:ABJM-Lagrangian}. 
Up to a sign, the gauge field contributions for both gauge groups are identical at one loop,
\begin{align}
\langle W \rangle^{(1)}_A = - \langle W \rangle^{(1)}_{\hat A}\,,
\end{align}
due to the different sign of the propagator for the second gauge field, $\langle A_\mu A_\nu \rangle = - \langle \hat A_\mu \hat A_\nu \rangle $. 
Thus, at one loop the diagrams cancel. This does not differ from the result of pure Chern-Simons theory, since the expectation value at one loop vanishes, 
as we found in section \ref{sec:one-loop-WL} for $n=4,6$.

At two loops, however, the sign has no effect, since the two-gluon diagram contains an even number of propagators 
and in the vertex diagram we have to take into account the sign of the interaction term as well. 
Therefore, the two-loop diagrams are identical
\begin{equation}
\langle W \rangle^{(2)}_A = \langle W \rangle^{(2)}_{\hat A}\,.
\end{equation}
Thus, up to two loops, the expectation value for pure gauge field contributions is the same in ABJM theory and Chern-Simons theory
\begin{align}
\langle W(A,\hat A) \rangle_{\text{gauge fields}}= \langle W(A) \rangle_{\text{CS}}\,.
\end{align}

\notocsubsection{Matter Contributions}\label{sec:matter-contributions}
In pure Chern-Simons theory the one-loop correction to the gauge field propagator is zero, 
since the contributions of gauge fields and ghosts exactly cancel against each other, see \eqref{eqn:gluon-ghost-loops}. 

In ABJM theory we have to take into account fermionic and bosonic matter in the loop. 
This gauge field self energy has been calculated in 
\cite{Gaiotto:2007qi,Bak:2008cp,Drukker:2008zx} and the corrected 
propagator reads\footnote{Recall that we absorbed the regularisation scale into the coupling constant: $k \rightarrow \mu^{-2 \epsilon} k $}

\begin{align}\label{eqn:one-loop-correctd-gluon-prop-tot-der}
G_{\mu\nu}^{(1)}(x) 
&= \left(\frac{2\pi}{k}\right)^2  \frac{N \delta^I_I}{8} \frac{\Gamma(1-\frac{d}{2}) \Gamma(\frac{d}{2})^2}{\Gamma(d-1)\,\pi^d}   \\  \nn
&\qquad \left(  \frac{\Gamma(d-2)}{\Gamma(2-\frac{d}{2})} \frac{\eta_{\mu\nu}}{(-x^2)^{d-2}} -\partial_\mu \partial_\nu\left( \frac{\Gamma(d-3)}{\Gamma(3-\frac{d}{2})} \frac{1}{4}  \frac{1}{(-x^2)^{d-3}}  \right) \right)\,,
\end{align}
for details see Appendix \ref{sec:one-loop-correction-propagator}.
%Evaluating the derivatives, we get
%\begin{align}
% G_{\mu\nu}^{(1)}(x) 
% &= \left(\frac{2\pi}{k}\right)^2  \frac{N \delta^I_I}{8} \frac{\Gamma(1-\frac{d}{2}) \Gamma(\frac{d}{2})^2}{\Gamma(d-1)\,\pi^d}    \left( \frac{\Gamma(d-1)}{\Gamma(3-\frac{d}{2})}  \frac{x_\mu x_\nu}{(x^2)^{d-1}}   + \underbrace{\frac{\Gamma(d-2)}{\Gamma(3-\frac{d}{2})} \frac{\eta_{\mu\nu}\left(-\frac{1}{2} + (2-\frac{d}{2}) \right)}{(x^2)^{d-2}}}_{\mathcal{O}(\epsilon)} \right)\,.
%\end{align}
%One can either use this form of the propagator or take \eqref{eqn:one-loop-correctd-gluon-prop-tot-der} and 
%drop the derivative term to calculate the expectation value.  Both calculations yield the same divergent and finite terms.
%
%We illustrate the calculation for the latter case, i.e. we drop the derivative term and use the propagator
We can drop the derivative term in \eqref{eqn:one-loop-correctd-gluon-prop-tot-der}  (it
would not contribute to the gauge-invariant Wilson loop) and instead use the propagator
\begin{align}
G_{\mu\nu}^{(1)}(x) 
&= -\frac{1}{N}\left(\frac{N}{k} \right)^2 \pi^{2-d} \Gamma \left( \frac{d}{2}-1 \right)^2   \frac{\eta_{\mu\nu}}{(-x^2)^{d-2}} 
%&= -\frac{1}{N}\left(\frac{N}{k} \right)^2 \left( (4 \pi e^{\gamma_E} )^{2 \epsilon} + \frac{\pi^2}{2}\epsilon^2 + \mathcal{O}(\epsilon^3) \right) \frac{\eta_{\mu\nu}}{(x^2)^{d-2}}
\,,
\end{align}
which up to two small differences is  the tree-level $\mathcal{N}=4$ SYM gluon propagator. 
The first difference is a trivial prefactor, and the second is that since we are at two loops, 
the power of $1/x^2$ is $1-2 \epsilon$ here,
as opposed to $1-\epsilon$ in the one-loop computation in $\mathcal{N}=4$ SYM.
Thus it is clear that the results will be very similar to the expectation value of the Wilson loop in $\mathcal{N}=4$ SYM.
The corresponding calculation in $\mathcal{N}=4$ SYM was carried out in \cite{Drummond:2007aua,Brandhuber:2007yx}
and we briefly summarise the results.

\begin{figure}[t]
\centering
\subfloat[]{\includegraphics[width=.31 \columnwidth]{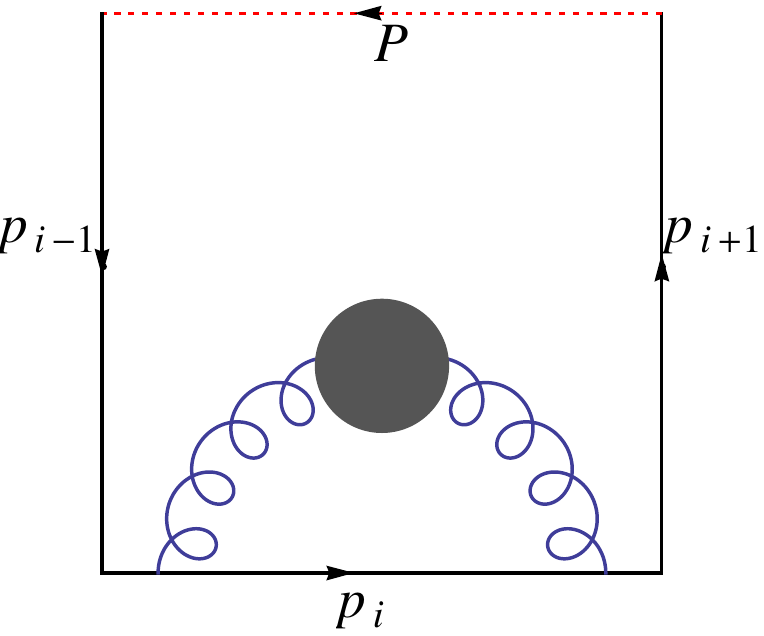}\label{fig:general-diagram-propagator-corrections-same-edge}}~~~~~
\subfloat[]{\includegraphics[width=.28 \columnwidth]{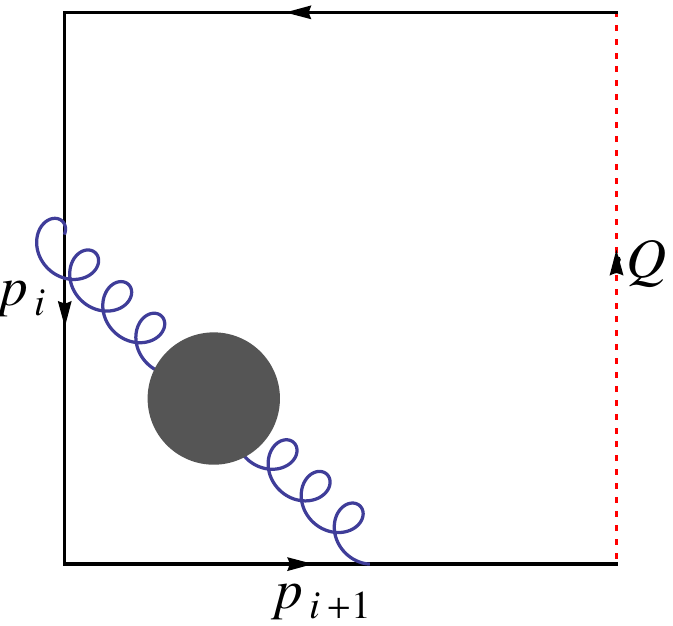}\label{fig:general-diagram-propagator-corrections-adjacent}}~~~~~
\subfloat[]{\includegraphics[width=.275 \columnwidth]{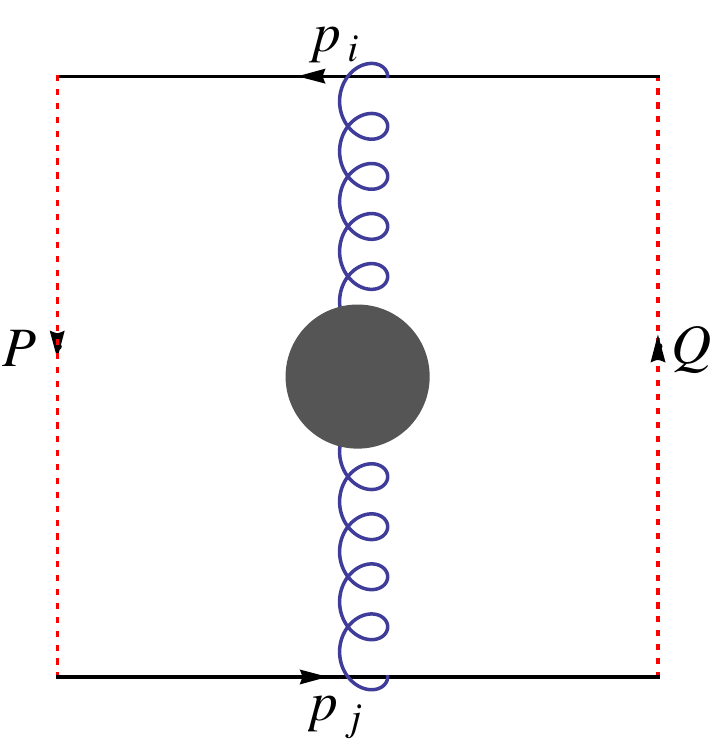}\label{fig:general-diagram-propagator-corrections-non-adjacent}}~~~~~~~~~~
\caption{Examples for the three classes of diagrams involving the gauge field self energy: diagrams with (a) a propagator connecting the same edge (vanishing), (b) propagator stretching between adjacent edges (divergent), (c) propagator stretching between non-adjacent edges (finite). These diagrams have the same structure as the 1-loop diagrams in $\mathcal{N}=4$ SYM. Dashed lines labelled with $P$ and $Q$ represent one or more light-like distances.
}
\label{fig:self-energy-diagrams}
\end{figure}

As in $\mathcal{N}=4$ SYM we have three classes of diagrams  shown in figure \ref{fig:self-energy-diagrams}. 
Diagram \ref{fig:general-diagram-propagator-corrections-same-edge} vanishes due to the light-likeness of the edges, 
whereas \ref{fig:general-diagram-propagator-corrections-adjacent} yields a divergent, and 
\ref{fig:general-diagram-propagator-corrections-non-adjacent} yields a finite contribution. We have
\begin{align}\label{eqn:general-expression}
\langle W_n \rangle^{(2)}_{\text{matter}} &= \frac{i^2}{N}\tr \int_{z_i>z_j} dz_i^\mu dz_j^\nu \langle A_\mu A_\nu \rangle^{(1)}\\ \nonumber
&=- N \sum_{i>j} \int ds_i ds_j p_i^\mu  p_j^\nu G_{\mu\nu}^{(1)}(z_i-z_j) \\ \nonumber
&= \left( \frac{N}{k}\right)^2 \left( (4 \pi e^{\gamma_E} )^{2 \epsilon} + \frac{\pi^2}{2}\epsilon^2 + \mathcal{O}(\epsilon^3) \right) \sum_{i>j} I_{ij}\,,
\end{align}

~\\where 
\begin{align}
I_{ij} &= \int ds_i ds_j\, p_i \cdot p_j (-(z_i-z_j)^2)^{2-d} \\ \nn
&= \frac{1}{2} \int_0^1 ds_i \int_0^1 ds_j 
\frac{ x^2_{i,j+1} +  x^2_{i+1,j} -  x^2_{i,j} -  x^2_{i+1,j+1}}{(-x^2_{ij}\bar{s}_i\bar{s}_j - x^2_{i+1,j} s_i \bar{s}_j - x^2_{i,j+1}\bar{s}_i s_j - x^2_{i+1,j+1}s_i s_j)^{d-2}}\,.
\end{align}
There are $n$ divergent diagrams  $I_{i+1,i}$ of the type shown in fig. \ref{fig:general-diagram-propagator-corrections-adjacent}
\begin{equation}
\nn
I_{i+1,i} % =
%-\frac{1}{2} (-x_{i+2,i}^2)^{3-d} \int_0^1 ds_i ds_{i+1} \frac{1}{\left(\bar{s}_i  s_{i+2}\right)^{1-2\epsilon}}
= -\frac{1}{8} \frac{(-x_{i,i+2}^2)^{2\epsilon}}{\epsilon^2} 
\end{equation}
and the finite diagrams $I_{ij}$ with $|i-j| \geq1$, see fig. \ref{fig:general-diagram-propagator-corrections-non-adjacent}, were solved in \cite{Brandhuber:2007yx} 
\begin{align}\nn
I_{ij} %&=  \frac{1}{2} \mathcal{F} (x^2_{i,j},x^2_{i+1,j+1},x^2_{i,j+1},x^2_{i+1,j}) \\ \nn
 = \frac{1}{2}\Big(&-\Li \left(1- a x_{i,j}^2 \right)-\Li \left(1- a x_{i+1,j+1}^2 \right)
& +\Li \left(1- a x_{i,j+1}^2 \right)+\Li \left(1- a x_{i+1,j}^2 \right)\Big)\,,
\end{align}
where 
\begin{equation}\nn
a= \frac{x_{ij}^2 + x_{i+1,j+1}^2 -x_{i,j+1}^2-x_{i+1,j}^2}{x_{ij}^2 x_{i+1,j+1}^2 -x_{i,j+1}^2 x_{i+1,j}^2} \,.
\end{equation}
The sum over all finite diagrams $\sum_{i>j+1}^n I_{ij}= \mathcal{F}_n^{\text{WL}}$ is related to the well-known finite part of the BDS conjecture \cite{Bern:2005iz} via $\mathcal{F}_n^{\text{WL}}=\mathcal{F}_n^{\text{BDS}}+ \text{const.}$, explicitly spelled out in \eqref{eqn:BDS-WL-n4-n6} for $n=4$ and $n=6$. Then, the full matter part reads 
\begin{align} \label{eqn:full-matter}
\langle W_n \rangle^{\text{matter}}_{\text{2-loop}} 
&= -\frac{1}{4} \left( \frac{N}{k}\right)^2 \Bigg[\sum_{i=1}^n \frac{(-x_{i,i+2}^2 \mu^2 4 \pi e^{\gamma_E})^{2\epsilon}}{2\epsilon^2}  -4 \mathcal{F}_n^{\text{WL}} + \frac{n}{4} \pi^2 \Bigg]\,.
\end{align}
where we have restored the regularisation scale $\mu$.

\section{Equivalence of Wilson Loops in $\sym$ and ABJM Theory}\label{sec:equivalence-wilson-loops}
Taking into account the Chern-Simons result \eqref{eqn:n-point-conjecture-CS},  the full result in ABJM theory can be written as
\begin{align}\label{eqn:ABJM-result-Wn}\nn
\langle W_n \rangle^{\text{ABJM}}_{\text{2-loop}} &=  \left(\frac{N}{k}\right)^2 \Bigg[-\frac{1}{2}\sum_{i=1}^n \frac{(-x_{i,i+2}^2 {\mu^\prime}^2 )^{2\epsilon}}{(2\epsilon)^2} + \mathcal{F}_n^{\text{WL}} + r_n \Bigg]\,,
\end{align}
where ${\mu^{\prime}}^2= \mu^2 8 \pi e^{\gamma_E}$,  $r_n= -(\frac{n}{4} \pi^2 + c_n -2 n\ln (2) - 5 n  \ln^2(2))/4$ and $c_n$ is the numerical constant
given by \eqref{eqn:n-dependence-constant}.

Indeed, this is of the same form as the one-loop result of the Wilson loop in $\mathcal{N}=4$ SYM  \cite{Brandhuber:2007yx} 
\begin{align}
\langle W_n \rangle^{\mathcal{N}=4 \text{~SYM}}_{\text{1-loop}}  &=  \frac{g^2 N}{8 \pi^2} \Bigg[-\frac{1}{2}\sum_{i=1}^n \frac{(-x_{i,i+2}^2 {\mu}^2)^{\epsilon}}{\epsilon^2}+ \mathcal{F}_n^{\text{WL}}  \Bigg]\,,
\end{align}
up to a few small differences. The first difference is a trivial prefactor, which is due to the different definitions of the coupling constant in the two theories. Secondly, since we are at two loops in ABJM theory, in the divergent part we have $2\epsilon$ instead of $\epsilon$ in the one-loop calculation in $\mathcal{N}=4$ SYM. The finite pieces are exactly the same up to the $n$-dependent constant $r_n$.

\subsubsection{Remark on the analytical guess for the four-point result}
We would like to remark, that in \cite{Henn:2010ps} for $n=4$ we found an analytical guess for the numerical constant $a_6$ in the vertex diagram, see \eqref{eqn:a6}, which leads to the final result  
\begin{align}\nn
\langle W_4 \rangle^{\text{ABJM}}_{\text{2-loop}} &=  \left(\frac{N}{k}\right)^2 \Bigg[-\frac{1}{2}\sum_{i=1}^4 \frac{(-x_{i,i+2}^2 {\mu^\prime}^2 )^{2\epsilon}}{(2\epsilon)^2} + \frac{1}{2} \ln^2 \left(\frac{x_{13}^2}{x_{24}^2}\right) + \frac{2}{3} \pi^2 + 3 \ln^2(2) -2 \ln(2)\Bigg]\,.
\end{align}
Note that the $\ln(2)$ term has a different transcendentality than the other terms. It stems from the vertex diagrams of the pure CS part \eqref{eqn:result-wilson-loop-two-loop} of the Wilson loop. We carefully rechecked the appearance of this term, since from the experience in $\sym$ one might expect only terms of homogenous transcendentality, but conclude that -- given our regularization procedure (dimensional reduction) -- it is correct. Furthermore, it is interesting to note, that up to this $\ln(2)$ term, the constant exactly agrees with the constant determined in the four-point amplitude calculation in \cite{Bianchi:2011dg}.

\chapter{Three-Point Functions of Twist-Two Operators in $\sym$}\label{sec:three-point-functions}
\chaptermark{Three-Point Functions  in $\sym$}
As mentioned in chapter \ref{sec:solving-the-theory}, two-point functions in $\mathcal{N}=4$ SYM are very well understood, largely due to the existence of integrability,  while three-point functions are explored to a lesser extent. Progress for three-point functions, comparable to the one that was made for the two-point functions with the help of integrability, would have important implications for all higher-point functions, via the operator product expansion as explained in section \ref{sec:OPE-in-CFT}. The investigation of structure constants can thus be seen as the next step towards the full solution of the theory in the sense of section \ref{sec:solving-the-theory}. 
In general, the structure constants of three-point functions, see section \ref{eqn:correlation-functions}, receive radiative corrections
\begin{equation}
C_{\alpha\beta\gamma}(\lambda) = C_{\alpha\beta\gamma}^{(0)} + \lambda C_{\alpha\beta\gamma}^{(1)} + ...
\end{equation}
and like for two-point functions, there are non-renormalisation theorems such that the three-point correlators of half-BPS operators do not get quantum corrections \cite{Lee:1998bxa,Eden:1999gh,Arutyunov:2001qw,Heslop:2001gp,Basu:2004nt}. We will explicitly see such cancellations in a special case of the three-point functions that we consider in this chapter.

Direct computations of three-point functions in $\sym$ theory have been performed in \cite{Bianchi:2001cm,Beisert:2002bb,Chu:2002pd,Roiban:2004va,Okuyama:2004bd,Alday:2005nd,Alday:2005kq,Georgiou:2008vk,Georgiou:2009tp}. In \cite{Georgiou:2012zj} a large number of three-point functions involving scalar primary operators of up to and including length five is considered.

The role of integrability for three-point function calculations was first addressed in \cite{Okuyama:2004bd,Roiban:2004va,Alday:2005nd}.  and applications of integrability methods  can be found in \cite{Escobedo:2010xs,Escobedo:2011xw,Gromov:2011jh,Kostov:2012jr} at tree-level and in \cite{Gromov:2012vu,Kostov:2012jr} for loop-level three-point functions of scalar single trace operators. It does indeed turn out, that three-point functions can be studied efficiently using integrability.

The question of a weak-strong coupling matching was addressed in \cite{Escobedo:2011xw,Bissi:2011dc,Bissi:2011ha,Grignani:2012yu,Georgiou:2011qk}. In \cite{Grignani:2012yu} for instance it is found that the correlator of three BMN states, each with two impurities and with a non-zero momentum,  completely agrees with the string-theoretical calculation of the 3-string vertex matrix elements.

Here, we calculate the structure constants of three-point correlators with two protected operators $\op, \tilde\op$ and a twist-two operator $\hat{\op}_j$ of arbitrary even spin $j$ in $\mathcal{N}=4$ supersymmetric Yang-Mills theory at one loop.
\begin{figure}[t]
\centering
 \includegraphics[width=.25\textwidth]{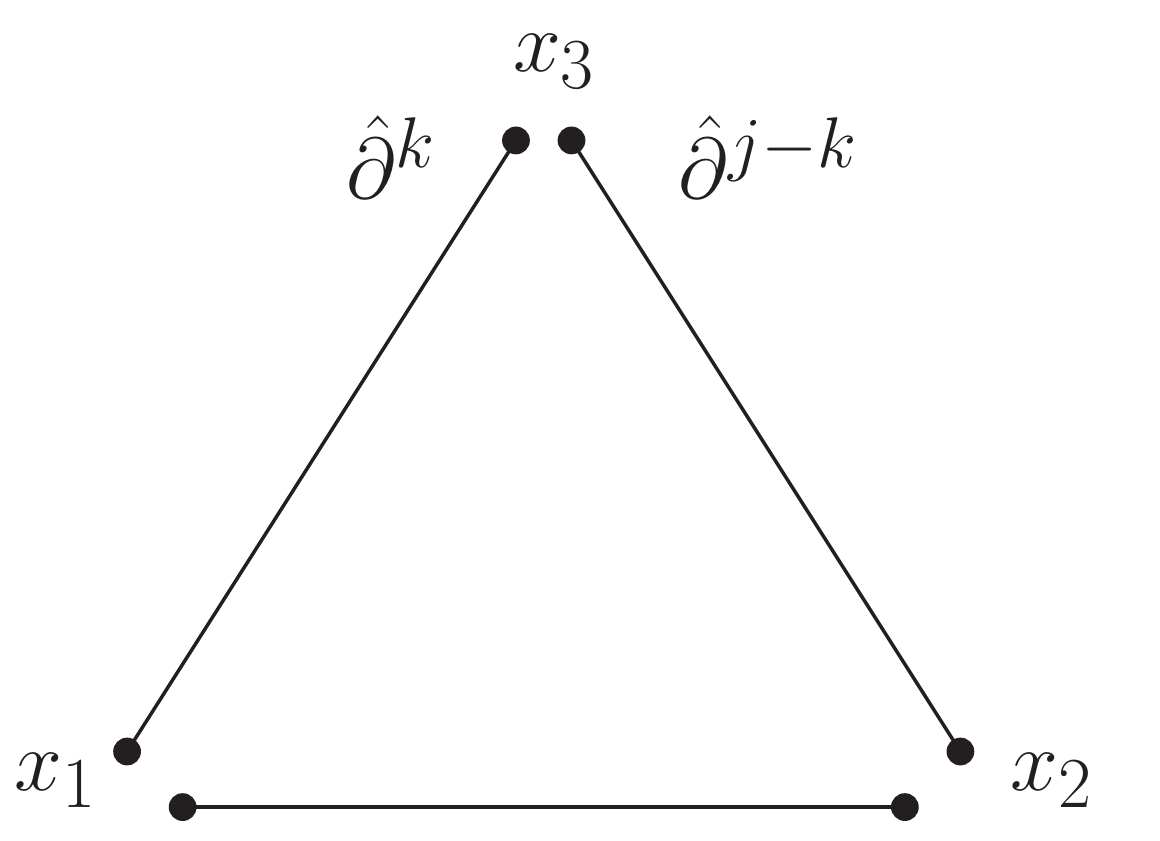}
\caption{Three-point function of two BPS and one twist-two spin $j$ operator at tree-level. The derivatives acting on $x_3$, indicated with $\hat \partial^k$, $\hat \partial^{j-k}$, are distributed in a certain way as explained in section \ref{sec:definition-operators}.}
\label{fig:three-point function at tree-main}
\end{figure} 
We perform a special infrared limit on the correlator in order to simplify the calculations. In this limit we do not rely on any knowledge of the one-loop mixing matrix, since it drops out completely. Furthermore, in this limit the three-point calculation is effectively reduced to calculating two-point functions.
We find that the normalisation invariant structure constant has the simple form \eqref{eqn:result}
\begin{equation}\label{eqn:result-normalization-invariant-structure-constants}
C^\prime_{\op  \tilde{\op} j}(g^2) = {C^\prime}_{\op  \tilde{\op} j} ^{(0)} \left(1 +   \frac{g^2 N}{8 \pi^2}  \left(2H_j(H_j -  H_{2j})- H_{j,2}\right) + \op(g^4) \right)\,,
\end{equation} 
where  $H_{j,m}$ are generalised harmonic sums
\begin{equation}
H_{j,m}=\sum_{r=1}^j \frac{1}{r^m},\qquad H_j=\sum_{r=1}^j\frac{1}{r},\qquad H_{j,2}=\sum_{r=1}^j\frac{1}{r^2}\,.
\end{equation} 
The structure constant can also be deduced from the double OPE of four-point functions of half-BPS operators and can be found in equation (5.27) of  \cite{Dolan:2004iy}, which agrees with our direct calculation and thus confirms the validity of  the operator product expansion.
The method that we use here however, can  also be applied to more general operators than the ones considered in   \cite{Dolan:2004iy}.

The explicit calculations are rather technical and we shift most of them to the appendix. In the following sections we  give an outline of the calculation. In section \ref{sec:renormalized-three-point-function} we give a plug-in formula for the structure constant and specify the location of the corresponding contributions in the appendix.

\section{Limits on Three-Point Functions}\label{sec:limits-of-three-point-correlators}
The calculation of the structure constant $C_{\op  \tilde{\op} j}$ can be largely simplified by considering appropriate limits of the correlator. In contrast to two-point functions, we can take several limits or integrate out parts of the space-time structure of the three-point correlator without trivializing its structure. Taking the limits corresponds to a projection onto a certain component of the $2^j$ terms in the correlator structure \eqref{eqn:structure-three-point-function-spin-j-light-cone-projection}.

\notocsubsection{Limit $x_2\to \infty$}\label{sec:Limit-x2-to-infty}
Multiplying by appropriate factors of $x_{12}^2 ,x_{23}^2$ and taking the limit  $x_2 \rightarrow \infty$ in \eqref{eqn:structure-three-point-function-spin-j-light-cone-projection} we obtain
\begin{align}\label{eqn:3-point limit}
 \lim_{x_2 \rightarrow \infty} |x_{12}|^{\Delta_1+\Delta_2-\theta} |x_{23}|^{\theta+\Delta_2-\Delta_1} \langle \op  \tilde{\op} \hat{\op}_j \rangle = C_{\op  \tilde{\op} \op_j} \frac{(\hat{x}_{13})^j}{|x_{13}|^{j + \theta + \Delta_1 -\Delta_2}}\,.
\end{align}
Taking the same limit in the evaluation of the correlator on the left-hand side, only very few terms in the Feynman-diagram calculation remain and the structure constant can easily be read off. We use this limit in order to calculate the tree-level structure constant in Appendix \ref{sec:three-point-limit-x2-to-infinity} for general dimension. For the one-loop calculation it is however more appropriate to calculate the diagrams in the limit described in the next section.

\begin{figure}[t]
\begin{align}
 \int d^dx_3
\begin{minipage}[c]{2.6cm}
	 \includegraphics[width=1 \textwidth]{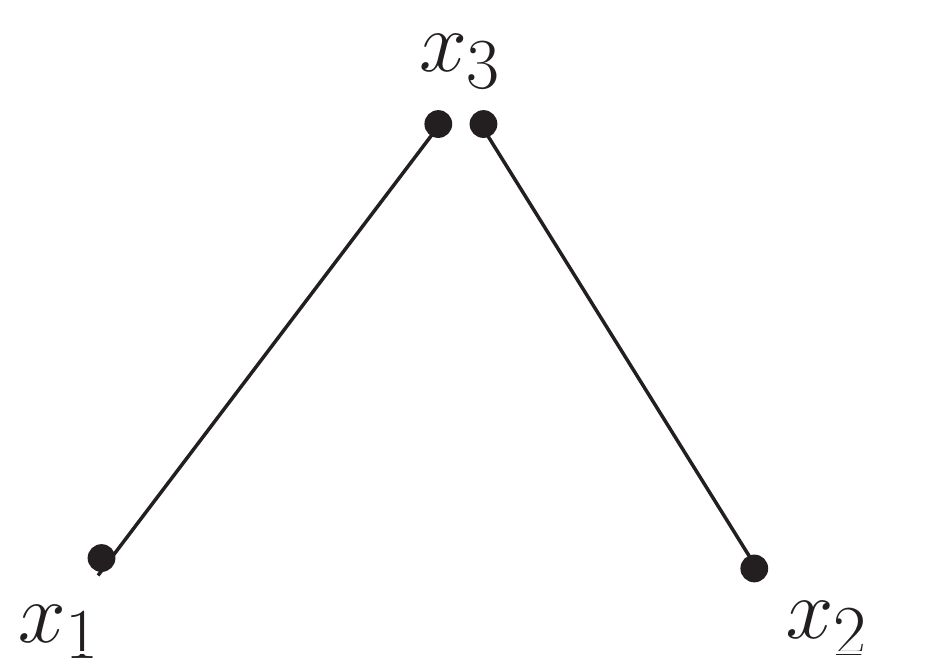}
\end{minipage}
= \int \frac{d^dp_1}{(2\pi)^d }~ e^{i p_1 x_{12}}
\begin{minipage}[c]{3cm}
	 \includegraphics[width=1 \textwidth]{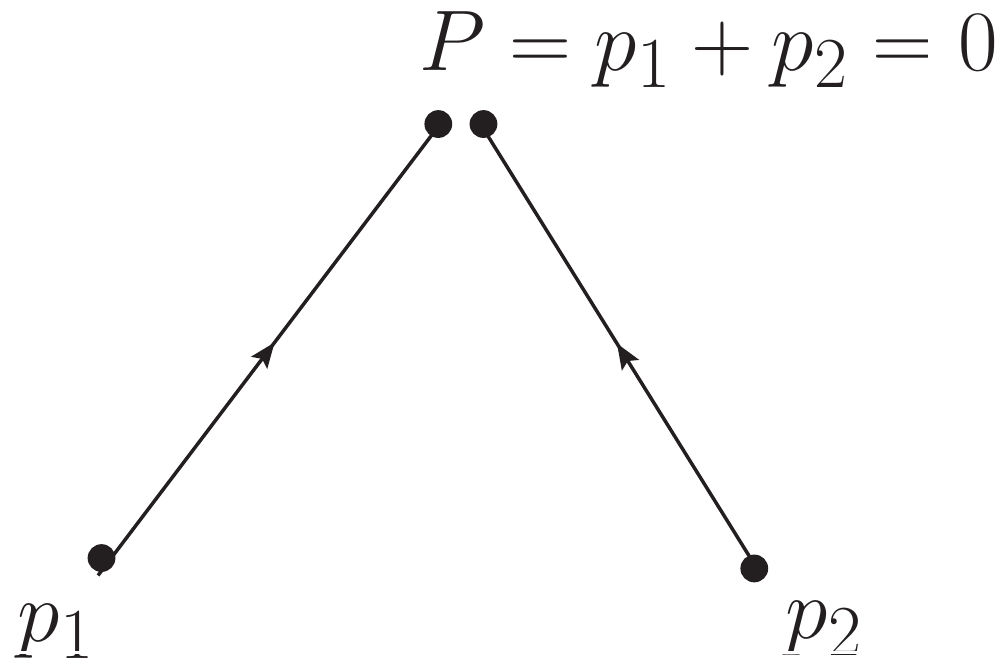}
\end{minipage}
\end{align}
\caption{Taking the limit $P=0$ in momentum space corresponds to integrating the correlator over $x_3$.}
\label{fig:limitPequal0}
\end{figure}

\notocsubsection{Collinear Limit of the Three-Point Correlator}\label{sec:limit-P-to-0}
Another possibility is to take the momentum $P$ going in at $\hat{\op}_j$ to zero, i.e. $P=p_1+p_2 \to 0$. 
This limit can be implemented in position space by integrating the three-point function over $x_3$, which can be seen as follows. \\
The Fourier transformed correlator is
\begin{align}
 \langle \op_1(x_1) \op_2(x_2) \op_3 (x_3) \rangle &= \int \frac{d^dp_1}{(2\pi)^d} \frac{d^dp_2}{(2\pi)^d} \frac{d^dP}{(2\pi)^d} ~e^{i(p_1 x_1+p_2 x_2+ P x_3)}  \\ \nonumber
&\phantom{=}\qquad \langle \op_1(p_1) \op_2(p_2) \op_3 (P)\rangle (2\pi)^d \delta(p_1+p_2+P)\,,
\end{align}
where the momentum conserving delta function is due to translation invariance of the correlator. Integrating this expression over $x_3$ leads to
\begin{align}\label{eqn:implementation-of-limit-P=0}\nn
 &\int d^dx_3 \langle \op_1(x_1) \op_2(x_2) \op_3 (x_3) \rangle  \\ 
 &= \int \frac{d^dp_1}{(2\pi)^d}\frac{d^dp_2}{(2\pi)^d} ~e^{i(p_1 x_{1}+p_2 x_{2})}(2\pi)^d\delta(p_1+p_2)  
 \langle \op_1(p_1) \op_2(p_2) \op_3 (-p_1-p_2)\rangle \\ 
&= \int \frac{d^dp_1}{(2\pi)^d} e^{ip_1 x_{12}}  \langle \op_1(p_1) \op_2(-p_1) \op_3 (0)\rangle\,,
\end{align}
where we used $\int d^dx e^{i (p_1+p_2)x_3} = (2\pi)^d \delta(p_1+p_2)$. Thus, integration of the correlator over $x_3$ is equivalent to setting $P=0$ in momentum space, see figure \ref{fig:limitPequal0}. Taking this limit largely simplifies the perturbative calculations that we will perform in this chapter.

\section{Definition of the Operators}\label{sec:definition-operators}
We use the mostly minus metric and the conventions as given in Appendix \ref{app:Conventions} and adapt the notation from  \cite{Belitsky:2007jp}. The action of $\sym$ is given in \eqref{eqn:N=4SYM-Lagrangian}.
We choose the following BPS-operators
\begin{align}\label{eqn:BPS-operators}
\op(x) = \tr \left( \bar{\phi}_{12}(x)  \bar{\phi}_{13}(x) \right),\qquad
 \tilde{\op}(x) = \tr \left( \bar{\phi}_{12} (x) \phi^{13}(x) \right)\,.
\end{align}
where $\bar{\phi}_{AB}=\bar{\phi}_{AB}^{a}T^a$ and we normalise the generators $T^a$ in the fundamental representation according to $\tr(T^a T^b) = \delta^{ab}/2$.
The twist-two operator $\hat{\op}_j$ projected to the light-cone, see also section \ref{sec:light-cone-projection}, is given by
\begin{equation}\label{eqn:def-twist-two}
\hat{\op}_j (x) = \sum_{k=0}^j a_{jk}^{1/2}\,\tr\left( \hat{D}^k \phi^{12}(x) \hat{D}^{j-k} \phi^{12}(x)\right)\,,
\end{equation}
where $\hat{D}=D^\mu z_\mu$ is the light-cone projected covariant derivative  $(z^2=0)$ and $a_{jk}^{1/2}$ are numerical coefficients, that are related to the \emph{Gegenbauer polynomials} as explained in the following section. Due to the property $a_{j,j-k}=(-1)^j a_{j,k}$ we consider $j$ even, since the operator vanishes otherwise.
As we  will explicitly see  in section \ref{sec:two-point-function-twist-op},  these operators have conformal two-point functions \eqref{eqn:two-point-function-spin-j-lighcone-projection}
\begin{equation}\label{eqn:conformal two-point functions-tree}
 \langle \hat{\op}_j(x_1) \hat{\op}_k(x_2) \rangle = \delta_{jk} C_j 2^{2j} \frac{(  \hat{x}_{12} )^{2j}}{(-x_{12}^2)^{2j+\theta}}\qquad (j~ \text{even})\,,
\end{equation}
where  $\theta = \Delta_j - j$ is the twist\index{twist}
 (dimension minus spin) of the operators. For the twist-two operators considered here $\theta=2$.
As reviewed in section \ref{sec:light-cone-projection}, conformal symmetry constrains this type of three-point functions to the form
\begin{equation}\label{eqn:conformal three-point functions-tree}
 \langle \op(x_1)  \tilde{\op}(x_2) \hat{\op}_j(x_3) \rangle = C_{\op \tilde{\op} j}(g^2) \frac{(  \hat{Y}_{12,3}) ^j}{|x_{12}|^{\Delta_1 + \Delta_2 - \theta} |x_{13}|^{\Delta_1 +  \theta- \Delta_2} |x_{23}|^{ \Delta_2 + \theta - \Delta_1 }}\,,
\end{equation}
where $|x_{ij}|=(-x_{ij}^2)^{1/2}$ and 
\begin{equation}
\hat Y_{12;3} = Y^\mu(x_{13},x_{23}) z_\mu, \qquad Y^\mu_{12,3}= \left(\frac{x_{13}^\mu}{x_{13}^2} - \frac{x_{23}^\mu}{x_{23}^2} \right), \qquad z^2=0\,.
\end{equation}
We will calculate the one-loop correction to the structure constants $C_{\op \tilde{\op} j}(g^2)$ and normalise them in a way suitable for the operator product expansion with the result \eqref{eqn:result-normalization-invariant-structure-constants}.

\subsection{Operator in Terms of Gegenbauer Polynomials}
At tree-level, the gauge fields do not enter the calculations, and therefore the covariant derivatives in \eqref{eqn:def-twist-two} can be reduced to ordinary derivatives
\begin{equation}\label{eqn:ajk-representation-twist-operator}
\hat{\op}_j^{\text{tree}} = \sum_{k=0}^j a_{jk}^{1/2}\,\tr( \hat{\partial}^k \phi^{12} \hat{\partial}^{j-k} \phi^{12})\,.
\end{equation}
The numerical coefficients $a_{jk}^\nu$ are related to the so-called \emph{Gegenbauer polynomials} $C_j^\nu(x)$ such that
\begin{equation}
\sum_k a_{jk}^\nu\, x^k y^{j-k} = (x+y)^j C_j^{\nu} \left(\frac{x-y}{x+y}\right)
\end{equation}
and $\nu= h-3/2$, where $h=d/2$. We list some properties on Gegenbauer polynomials in Appendix \ref{sec:properties-of-gegenbauer-polynomials}.
Therefore, we can rewrite the operators in the bi-local form \cite{Ohrndorf:1981qv,Makeenko:1980bh}
\begin{align}\label{eqn:definition-with-gegenbauer-polynomials}
\hat{\op}_j^{\text{tree}} &= \underbrace{\left(\hat{\partial}_a+\hat{\partial}_b \right)^j C_j^{1/2} \left( \frac{\hat{\partial}_a-\hat{\partial}_b}{\hat{\partial}_a+\hat{\partial}_b} \right)}_{=: D^j_{ab} } \tr \left( \phi^{12}(x_a)  \phi^{12}(x_b) \right) \Big|_{x_a=x_b}\,.
%\\ \nn &=: D^i_{ab} \tr \left( \phi^{12}(x_a)  \phi^{12}(x_b) \right)\Big|_{x_a=x_b}
\end{align}
As we will see in the remainder of this section, the formulation in terms of Gegenbauer polynomials is very useful in practical calculations, since this allows to use properties such as the orthogonality of Gegenbauer polynomials. In section \ref{sec:two-point-function-twist-op} we show how the tree-level two-point function  can been calculated for general index $\nu$ of the Gegenbauer polynomials. %(comment on $\phi^3$ and mixing there:) 

\section{Structure of the Calculation and Results}\label{structure-results}
In the conformal scheme of section \ref{conformal scheme}, the form \eqref{eqn:conformal two-point functions-tree},\eqref{eqn:conformal three-point functions-tree} of the two- and three-point functions of the renormalised operators is  preserved at loop-level under the replacement of the classical dimensions of the operators  \eqref{eqn:BPS-operators}, \eqref{eqn:def-twist-two} with their anomalous dimensions, i.e.
\begin{align}\label{eqn:twist-and-anomalous-dimension}
\theta(g^2) = \Delta_j - j= \Delta_j^{(0)} +  \gamma_j(g^2) - j= d-2+ g^2 \gamma_j^{(1)} + \op(g^4)
\end{align}
and the dimension of the protected operators  $\Delta_\op =\Delta_{\bar{\op}}=d-2$ remains fixed\footnote{The (anomalous) dimension of the operators is defined at $\epsilon=0$. We keep writing $d$ here and in the following however, because the arguments are also valid for scalar $(\phi^3)_6$ theory in $d=6$; integer $d$ is however required.}. The renormalised twist-two operators mix with descendants of lower spin operators as
\begin{equation}
 \hat{\mathbb{O}}_j = \sum_k \mathbb{Z}_{jk} \hat{\partial}^{j-k} \hat{\op}_k\,.
\end{equation}
As we will see in section \ref{eqn:two-point-functions} at one loop $\mathbb{Z}_{jk}$ is given through 
\begin{equation}\label{eqn:expansion-mixing-matrix-one-loop}
\mathbb{Z}_{jk} =\delta_{jk} + g^2\left(-B_{jk}^{(1)}+\frac{1}{\epsilon}\delta_{jk} Z_{j}^{(1)}\right) + \op(g^4)\,.
\end{equation}
where $B_{jk}^{(1)}$ is the finite one-loop mixing matrix specified in \eqref{eqn:N=4-mixing-matrix} and it has only entries for $k<j$ and $Z_j^{(1)}$ is the renormalisation constant given in \eqref{eqn:renormalization-constant}. 
Therefore, the renormalised three-point function of two BPS and one twist-operator reads
\begin{equation}\label{eqn:renormalized-three-point-function-structure}
 \langle \op(x_1) \tilde{\op}(x_2) \hat{\mathbb{O}}_j (x_3) \rangle  = C_{\op \tilde{\op} j} (g^2) \frac{(\hat{Y}_{12,3})^j}{((-x_{13}^2)(-x_{23}^2))^{\theta/2}( -x_{12}^2)^{\Delta_\op-\theta/2} } \,.
\end{equation}
Applying the limit described in section \ref{sec:limit-P-to-0} and therefore integrating over $x_3$, see Appendix \ref{app:integral-over-conformal-structure}, yields
\begin{equation*}
 \int d^dx_3 \frac{(\hat{Y}_{12,3})^j}{((-x_{13}^2)(-x_{23}^2))^{\theta/2}} 
= N(g^2,d) \frac{(\hat{x}_{12})^j}{(-x_{12}^2)^{\theta-d/2+j}}\,,
\end{equation*}
where the normalisation factor is
\begin{equation}\label{eqn:N(g^2)-for-d-dimensions}
N(g^2,d)= -i\frac{\Gamma(\theta-d/2+j) \Gamma((d-\theta)/2)^2 \Gamma \left( j+ (\theta -1)/2\right)}{\Gamma(d-\theta) \Gamma \left(j+\frac{\theta }{2}\right) \Gamma (j+\theta -1)} \frac{2^{\theta +2 j-2} }{\pi^{\frac{1}{2}-\frac{d}{2}} }\,.
 \end{equation}
As can be seen from \eqref{eqn:twist-and-anomalous-dimension}, this function has an expansion\footnote{It does not have an expansion in $\epsilon$, since the renormalised three-point function is defined at $\epsilon=0$.} in $g$
and for $d=4$ we explicitly get
\begin{align}\label{eqn:N(g^2)-at-epsilon=0}
N(g^2) &= N^{(0)} + g^2 N^{(1)} \\ \nn
%&=\frac{\Gamma (2 j)2\pi^{2}}{\Gamma (j+1)^2 }\left( 1+  \frac{\gamma_j(g^2)}{2} \left(-\frac{3}{j}- H_{j-1}+H_{j-\frac{1}{2}}+2+\log (4)\right)\right)  + \op(g^4) \\ \nn
  &= -i\frac{\Gamma (2 j)2\pi^{2}}{\Gamma (j+1)^2 }\left( 1+  \frac{\gamma_j(g^2)}{2} \left(-\frac{1}{j}- 2H_{j}+2H_{2j-1}+2
  \right)\right)  + \op(g^4)  \,,
\end{align}where the anomalous dimension of the twist-two operators will be given in section \ref{sec:anomalous-dimension-twist-two}.
By calculating the left-hand side of the integrated three-point function
\begin{equation}\label{eqn:renormalized-three-point-function}
 \int d^dx_3 \langle \op \tilde{\op} \hat{\mathbb{O}}_j\rangle = N(g^2) \left(C_{\op\tilde{\op}j}^{(0)}+g^2 C_{\op\tilde{\op}j}^{(1)} \right)\frac{(\hat{x}_{12})^j}{(-x_{12}^2)^{j+d-3+\gamma_j(g^2)/2}} 
\end{equation}
in terms of Feynman diagrams, we can thus easily read off the structure constants. In the following section we explain how to read off the one-loop structure constant from the perturbative calculations.

\subsection{The One-Loop Structure Constant}\label{sec:renormalized-three-point-function}
We calculate the correlator in the limit described in section \ref{sec:limit-P-to-0} which is equivalent to the integration of the correlator over $x_3$ and thus all the descendants drop out, since we can write
\begin{equation}
\int d^dx_3 \langle \op \tilde\op \hat{\mathbb{O}}_j \rangle= \sum_k \mathbb{Z}_{jk} \int d^dx_3 \hat{\partial}_3^{j-k} \langle \op \tilde\op \hat{\op}_k \rangle= \sum_k \mathbb{Z}_{jk} \delta_{jk} \int d^dx_3\langle \op \tilde\op \hat{\op}_k \rangle\,.
\end{equation}
Thus from \eqref{eqn:expansion-mixing-matrix-one-loop} one can see that the contributions of order $g^2$ to the renormalised three-point function in terms of bare Feynman diagrams are
\begin{align}\label{eqn:three-point-order-g2-terms}
\langle \op \tilde{\op} \hat{\mathbb{O}}_j \rangle \Big |_{g^2} 
&= \lim_{\epsilon\to 0} \left(\langle \op \tilde{\op} \hat{\op}_j \rangle^{(1)} + g^2  \langle \op \tilde{\op} \hat{\op}_j \rangle^{(0,\epsilon)} \frac{Z_j^{(1)}}{\epsilon}\right) \,,
\end{align}
where $ \langle \op \tilde{\op} \hat{\op}_j \rangle^{(0,\epsilon)}$ denotes the tree-level correlator calculated in $d=4-2\eps$ dimensions, including its $\op(\epsilon)$ expansion. Therefore, our calculation does not rely on the knowledge of the mixing matrix $B$. 

In order to read off the finite one-loop structure constant, we can combine the known structure in the conformal scheme \eqref{eqn:renormalized-three-point-function} and the result of the explicit one-loop calculation \eqref{eqn:three-point-order-g2-terms} and solve for $C_{\op\tilde{\op}j}^{(1)}$
\begin{align}\nn \label{eqn:bare-three-point-structure-constant}
C_{\op\tilde{\op}j}^{(1)} &=   \lim_{\epsilon \to 0} \left[\int \langle \op \tilde{\op} \hat{\op}_j \rangle^{(1)} + \int \langle \op \tilde{\op} \hat{\op}_j \rangle^{(0,\epsilon)} \frac{Z_j^{(1)}}{\epsilon} g^2 \right] /\left( N^{(0)} \frac{(\hat{x})^j}{(-x^2)^{3+j}} \right) \\ 
&- g^2 \frac{N^{(1)}}{N^{(0)}}   C_{\op\tilde{\op}j}^{(0)} \,,
\end{align}
where the division through the space-time part is to be understood symbolically.
In the following, we calculate all these contributions and \eqref{eqn:bare-three-point-structure-constant} becomes a simple plug-in formula.
The first term is given\footnote{We have calculated it in momentum space and thus an additional Fourier transformation using \eqref{eqn:FT-to-position-space} for $d=4$ is necessary.} through \eqref{eqn:bare-three-point-function}, the second term is given by \eqref{eqn:tree-level-three-point-function-in-x-space-calculated}  and \eqref{eqn:renormalization-constant} and here we have to take into account the expansion of all these expressions to finite order. The constants $N^{(0)}, N^{(1)}$ were given in \eqref{eqn:N(g^2)-at-epsilon=0}. The tree-level structure constant is given in \eqref{eqn:tree-level-strcuture-constant-N=4}. 
%From $\langle \op \tilde{\op} \hat{\op}_j\rangle^{(\epsilon)}$ we do not take into account the expansion in $\log(\mu^2 x^2)$, this term yields the exponent of the distances in the denominator.

\subsection{Normalisation Invariant Structure Constants}
In section \ref{sec:OPE-in-CFT} we have shown the equivalence of the coefficients in the operator product expansion and the structure constants of three-point functions. As mentioned there, we assumed that the operators are normalised such that the two-point functions have $C_j=1$. The operators \eqref{eqn:definition-with-gegenbauer-polynomials} in terms of Gegenbauer polynomials are not normalised to one, but as we will see in section \ref{sec:renormalized-two-point-function} are normalised as \eqref{eqn:conformal two-point functions}
\begin{equation}
 \langle \hat{\mathbb{O}}_j(x_1) \hat{\bar{\mathbb{O}}}_j(x_2) \rangle = \delta_{jk}  \left( C_j^{(0)} + g^2 C_j^{(1)} \right)2^{2j} \frac{  (\hat{x}_{12} )^{2j}}{(-x_{12}^2)^{2j+\theta}} \,,
\end{equation}
where $C_j^{(0)}, C_j^{(1)}$ are explicitly calculated in section \ref{eqn:two-point-functions}.

~\\
As can be seen from \eqref{eqn:renormalized-three-point-function-structure}, \eqref{eqn:conformal two-point functions-tree}, under rescalings of the operators 
\begin{equation}\label{eqn:rescalings}
 \hat{\mathbb{O}}_j \to \lambda_j \hat{\mathbb{O}}_j ,\qquad \op \to \lambda \op , \qquad \tilde{\op} \to \tilde{\lambda} \tilde{\op}\,,
\end{equation}
the structure constant and the normalisation of the two-point function changes as
\begin{equation*}
C_{12j}(g^2) \to \lambda \tilde{\lambda} \lambda_j C_{12j}(g^2), \qquad C_j\to \lambda_j^2 C_j, \qquad C\to \lambda^2 C , \qquad \tilde C \to {\tilde\lambda}^2 \tilde C\,.
\end{equation*}
Clearly, the ratio
\begin{align}\label{eqn:normalization-invariant-expression}
C_{\op\tilde{\op}j}^\prime = \frac{C_{\op\tilde{\op}j}}{\sqrt{C \tilde{C} C_j}}
\end{align}
is invariant under the rescalings \eqref{eqn:rescalings}. 
For operators with two-point functions that are normalised to one, the factor in the denominator of \eqref{eqn:normalization-invariant-expression} is equal to one and it is thus this structure constant which is relevant in the operator product expansion. 
 In other words, $C_{\op\tilde{\op}j}^\prime$ is the structure constant for operators which have two-point functions that are normalised to one. In our case, the BPS operators \eqref{eqn:BPS-operators} are normalised to $C=\tilde C=2^{-6}/\pi^4$ and do not get quantum corrections and thus 
$C_{\op\tilde{\op}j}^\prime$ is related to  $C_{\op\tilde{\op}j}$ through 
\begin{align}\label{eqn:normalization invariant structure constant}
 C_{\op\tilde{\op}j}^\prime &= \frac{ C_{\op\tilde{\op}j}(g^2)}{\sqrt{C_j(g^2)}C}= \frac{C_{\op\tilde{\op}j}^{(0)}}{\sqrt{C_j^{(0)}}C} \left(1+ g^2\left( \frac{C_{\op\tilde{\op}j}^{(1)}}{C_{\op\tilde{\op}j}^{(0)}}-\frac{1}{2}\frac{C^{(1)}_j}{C^{(0)}_j} \right) \right) = C_{\op\tilde{\op}j}^{\prime (0)} + g^2 C_{\op\tilde{\op}j}^{\prime (1)}\,.
\end{align}
The two-point structure constants are calculated in section \ref{sec:renormalized-two-point-function}.

\subsubsection{Result for the structure constants}
\begin{figure}[t]
\centering
\includegraphics[width=.8 \textwidth]{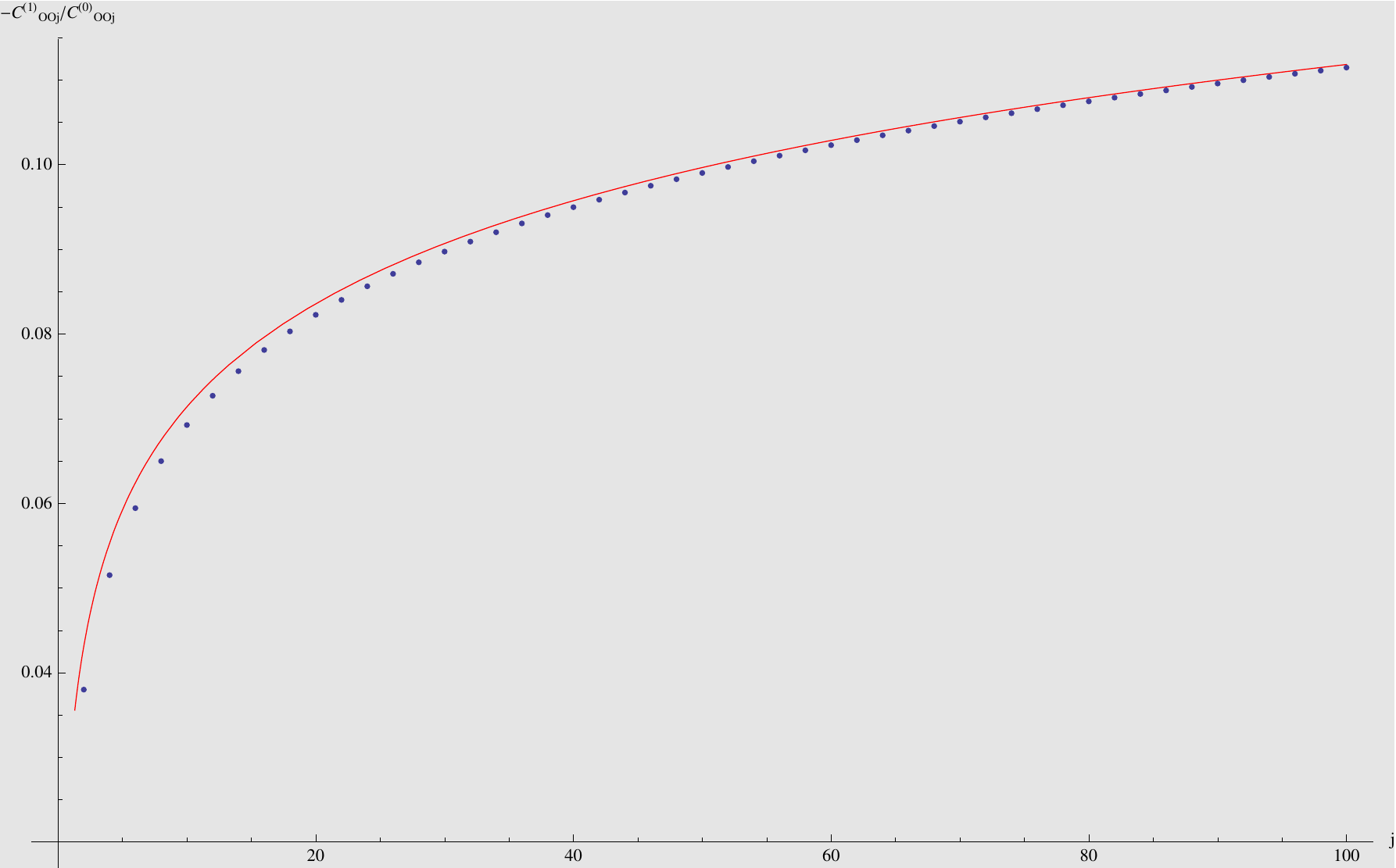}
\caption{Here we show the dependence of the ratio of the one-loop structure constant and its tree-level value on the spin $j$. The points are given by \eqref{eqn:result} and the curve represents the large $j$ limit \eqref{eqn:large-j-limit}. Note, that for graphical reasons we plot the negative value of the ratio here.}
\label{fig:ratio-structure-constant}
\end{figure}

Putting together \eqref{eqn:normalization invariant structure constant} using \eqref{eqn:normalization-tree-level-two-point}, \eqref{eqn:normalization-one-loop-two-point} for the normalisation of the two-point functions as well as \eqref{eqn:tree-level-strcuture-constant-N=4}, \eqref{eqn:bare-three-point-structure-constant} for the structure constants of the three-point functions, we find that the normalisation invariant structure constant is\footnote{We have put all contributions into a Mathematica file and verified the result \eqref{eqn:result}. One could proceed and solve some of the finite sums that remain in the solutions of the integrals or prove them via induction from the known result.}
\begin{align}\label{eqn:result}
C_{\op\tilde{\op}j}^{\prime (1)}/C_{\op\tilde{\op}j}^{\prime (0)} = \frac{g^2 N}{8\pi^2} \left(2H(j)^2 -2 H(j) H(2j)- \sum_{r=1}^j \frac{1}{r^2} \right)\,.
\end{align}
In the large $j$ limit we find
\begin{align}\label{eqn:large-j-limit}
\lim_{j\to \infty }C_{\op\tilde{\op}j}^{\prime (1)}/C_{\op\tilde{\op}j}^{\prime (0)} = \frac{g^2 N}{8\pi^2}\left(  -2 \ln(2)\left(\ln(j)+\gamma_E \right) - \frac{\pi^2}{6}\right)\,.
\end{align}
The one-loop correction thus scales with $\ln(j)$ for large $j$, as shown in figure \ref{fig:ratio-structure-constant}. This result consists of the contributions from the two-point functions
\begin{align}\label{eqn:result}
\frac{1}{2} C_j^{(1)}/C_j^{(0)} = \frac{g^2 N}{8\pi^2} \left(3H(j)^2 -2 H(j) H(2j) \right)
\end{align}
and the contribution from the three-point functions
\begin{align}\label{eqn:result}
C_{\op\tilde{\op}j}^{(1)}/C_{\op\tilde{\op}j}^{(0)} = \frac{g^2 N}{8\pi^2} \left(5 H(j)^2 - 4 H(j) H(2j)- \sum_{r=1}^j \frac{1}{r^2} \right)\,,
\end{align}
which both scale as $\ln(j)^2+...$ for $j\to \infty$ where the ellipses stand for subleading terms.

\newpage

\section{Two-Point Functions of Twist-Two Operators}\label{eqn:two-point-functions}
Here, we briefly list the results for the two-point function calculations of the operators  \eqref{eqn:definition-with-gegenbauer-polynomials}. The calculations are rather tedious and are given in detail in the appendix. We will give more details on the three-point calculation in the following section, since the focus of this chapter is on three-point functions.

\subsection{Tree-Level Two-Point Functions of Twist-Two Operators}\label{sec:two-point-function-twist-op}
In Appendix \ref{sec:details-two-point-functions-tree-level} we calculate the tree-level two-point function of the operators \eqref{eqn:definition-with-gegenbauer-polynomials} for scalar fields in general\footnote{We do however require $d$ to be integer. Choosing $d=4-2\epsilon$ leads to $\op(\epsilon)$ corrections and the unrenormalised correlator also gets non-diagonal terms.} dimension $d$   using the Schwinger parametrisation of the scalar propagators and some properties of the Gegenbauer polynomials. The result for $d=4$ is given in \eqref{eqn:two-point-tree-level-N-equal4SYM} and reads
\begin{align}\label{eqn:two-point-tree-level-N-equal4SYM-in-main-text}
  \langle \hat{\op}_j \hat{\bar{\op}}_k \rangle^{d=4} = \delta_{jk} g^4 \delta^{aa} \frac{2^{2j-1}}{(4\pi^2)^2} \Gamma(2j + 1) \frac{(\hat{x}_{12})^{2j}}{(-x_{12}^2)^{2j + 2}}\,,
\end{align}
where $\delta^{aa}=N^2-1$ is the dimension of the gauge group $SU(N)$.

%\subsubsection{Non-Diagonal Tree-Level Correlator at $\op(\epsilon)$ in $d=6$} \label{sec:non-diagnoal-tree-level-two-point-correlator}
%For the  normalization of the 3-point structure constants and the mixing matrix we also need the $\op(\epsilon)$ contributions to the tree-level correlator. Repeating the calculation of sec. \ref{sec:two-point-function-twist-op} in $d=6-2\epsilon$ but with the index of the Gegenbauer polynomials fixed at $\nu=3/2$ we find, see sections \eqref{sec:one-loop-contibutions-using-gegenbauer-polynomials} and \eqref{eqn:tree-level-expression-epsilon}, {\bf valid for $\nu=3/2$}
%\begin{align}
% \langle \op_i \op_j  \rangle &= \frac{(-1)^i}{(4\pi^h)^2} 2^{i+j+3-2h} \int dz z^{i+j-3+2h}\exp (z x^2) \\ \nn & \qquad \qquad \qquad \qquad \int_{-1}^{1} (1-y^2)^{3/2-\epsilon-1/2} C^{3/2}_i(y) C^{3/2}_j(y) \\ \nn
%&= \frac{(-1)^j}{(4\pi^h)^2} 2^{i+j+3-2h} \Gamma(i+j+2h -2) \frac{\hat{x}^{i+j}}{(-x^2)^{i+j+2h-2}} \left( \delta_{ij}N(j) - \epsilon I_{ij} \right)
%\end{align}
%where $N(j)$ is the normalisation of the Gegenbauer polynomials for $\nu=3/2$, i.e. $N(j)=2 (j+1)(j+2)/(2j+3)$ and the integral 
%\begin{equation}
% I_{ij}= \int_{-1}^1 dy (1-y^2) \log(1-y^2) C_i^{3/2}(y) C_j^{3/2}(y)
%\end{equation}
%contains non-diagonal $(i \neq j)$ contributions and is solved in appendix \ref{sec:integral-non-diagonal-tree-level}.

\subsection{Bare One-Loop Two-Point Function}
The one-loop corrections to the two-point function are shown in figure \ref{fig:two-point-function}. The self-energy diagram  \ref{fig:twopointfunctionselfenergy} can be calculated in a similar way using the orthogonality properties of the Gegenbauer polynomials. For the other diagrams it is however more convenient to use the representations \eqref{eqn:ajk-representation-twist-operator} of the operators.
For the calculation of the \emph{kite diagram} in figure \ref{fig:twopointcourt} we use the integration by parts (IBP) technique and the four-scalar diagram in figure \ref{fig:twopointfourscalar} can be shown to cancel exactly against a contribution from the kite diagram. This cancellation can be understood, by investigating the decomposition of the kite diagram into scalar basis integrals as shown in figure \ref{fig:gluon-kite-diagram-decompostion}.
For the diagrams in figures \ref{fig:twopointonecovderivative}, \ref{fig:twopointtwocovderivative} we need the representation with one gauge field in the operator \eqref{eqn:twist-operator-one-gauge-field-intro}. The diagram in figure \ref{fig:twopointtwocovderivative} vanishes in the light-cone projection, since it contains a gluon propagator contracted with the light-like vectors $z_\mu$ and thus  $\langle A_\mu A_\nu \rangle z^\mu z^\nu=0$  in our gauge.

Adding up all non-vanishing contributions we thus find
\begin{align}\label{eqn:sum-of-one-loop-two-point-contributions}
\langle \hat{\op}_j \hat{\bar{\op}}_j \rangle^{(1)} 
&= \langle \op_j \hat{\bar\op}_j \rangle^{(\ref{fig:twopointfunctionselfenergy})} +\langle \op_j \hat{\bar\op}_j \rangle^{(\ref{fig:twopointcourt}+ \ref{fig:twopointfourscalar} )} +\langle \op_j \hat{\bar\op}_j \rangle^{(\ref{fig:twopointonecovderivative})} \\ \nn
&= \eqref{eqn:diagram-a-final-version-to-expand} + \eqref{eqn:kite+scalar-version-to-expand}+ \eqref{eqn:FT-of-momentum-space-calculaiton-of-diagram-c}
\end{align}
and  the explicitly solved integrals can be found in the appendix in the equations specified in the last line of \eqref{eqn:sum-of-one-loop-two-point-contributions}

\subsection{Anomalous Dimension of Twist-Two Operators}\label{sec:anomalous-dimension-twist-two}
\begin{figure}[t]
\centering
\subfloat[~]{
\includegraphics[width=.25 \textwidth]{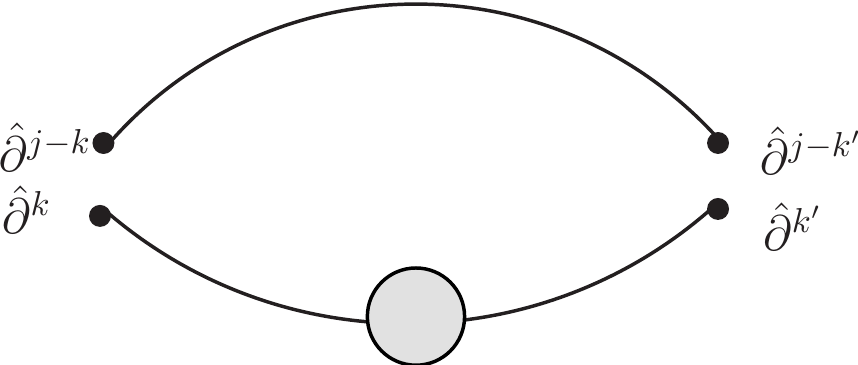}
\label{fig:twopointfunctionselfenergy}} 
~~~~~~~~~~~
\subfloat[~]{
\includegraphics[width=.25 \textwidth]{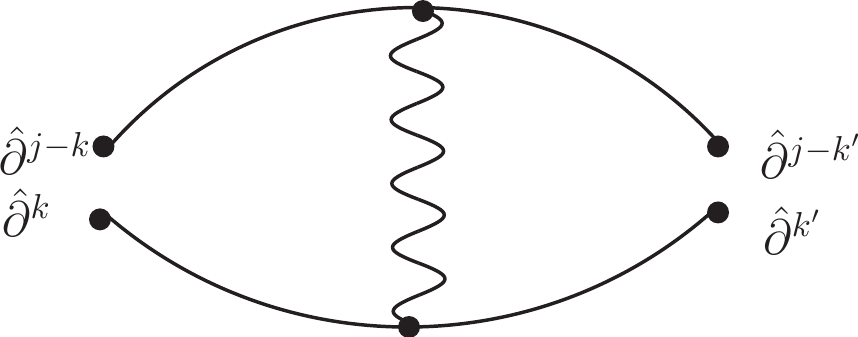}
\label{fig:twopointcourt}} \\
\subfloat[~]{
\includegraphics[width=.30 \textwidth]{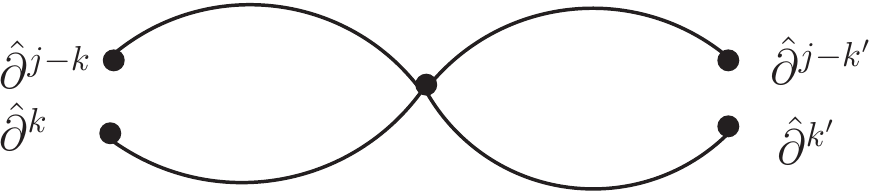}
\label{fig:twopointfourscalar}}
~~~~~~~~~~~
\subfloat[~]{
\includegraphics[width=.25 \textwidth]{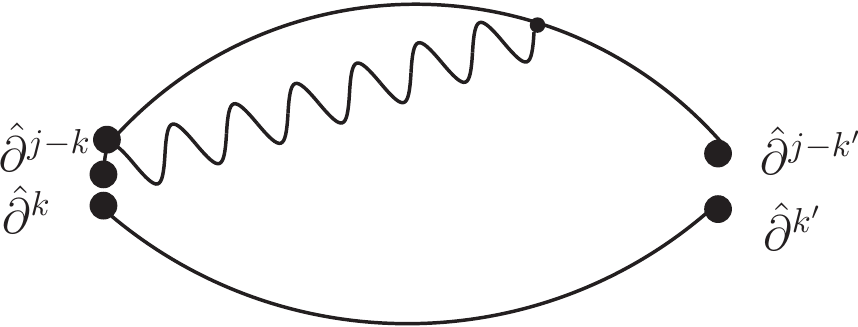}
\label{fig:twopointonecovderivative}} \\
\subfloat[~]{
\includegraphics[width=.25 \textwidth]{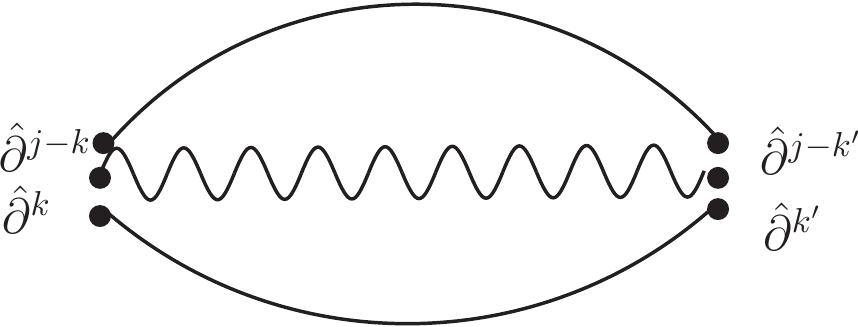}
\label{fig:twopointtwocovderivative}} \caption{One-loop corrections to the two-point function of $\langle \hat{\op}_j \hat{\op}_j \rangle$.}
\label{fig:two-point-function}
\end{figure}

\index{anomalous dimension!of twist-two operators}
We can extract the divergent contributions of \eqref{eqn:sum-of-one-loop-two-point-contributions} and find the following contributions \eqref{eqn:anomalous-dim-self-energy}, \eqref{eqn:anomalous-dim-kite}, \eqref{eqn:anomalous-dim-one-gluon-vertex}  to the anomalous dimension
\begin{equation}
\gamma_j^{\eqref{fig:twopointfunctionselfenergy}}= \frac{g^2N}{4 \pi^2},\qquad \gamma_j^{\eqref{fig:twopointcourt}}=-\frac{g^2N}{4 \pi^2} \frac{1}{(j+1)},\qquad \gamma_j^{\eqref{fig:twopointonecovderivative}} = \frac{g^2N}{4 \pi^2}  \left(2 H_j + \frac{1}{(j+1)} -1 \right)\,.
\end{equation}
Adding up the contributions we get the well-known result \cite{Kotikov:2000pm,Kotikov:2001sc} for the leading order contribution to the anomalous dimension
\begin{equation}\label{eqn:anomalous-dimension-twist-two}
\gamma_j = \left(\frac{g^2N}{8 \pi^2}\right) 4 H_j +\op(g^4)\,, \qquad\qquad H_j = \sum_{k=1}^j \frac{1}{k}\,.
\end{equation}
In order to avoid terms of $\log(\pi)$ and $\gamma_E$ in the final results, we define the renormalisation constant in the $\overline{\text{MS}}$-scheme \eqref{eqn:renormalization-with-mixing}, \eqref{eqn:Z-matrix-expansion}
\begin{equation}\label{eqn:renormalization-constant} 
Z_j^{(1)}=  \frac{H_j}{4\pi^h}  e^{\epsilon \gamma_E}\,, \qquad\qquad H_j = \sum_{k=1}^j \frac{1}{k}\,.
\end{equation}

\subsection{Renormalised Two-Point Function}\label{sec:renormalized-two-point-function}
The renormalised two-point function has the form
\begin{equation}\label{eqn:conformal two-point functions}
 \langle \hat{\mathbb{O}}_j(x_1) \hat{\mathbb{O}}_k(x_2) \rangle = \delta_{jk} C_j (g^2) \frac{(  \hat{I}_{12} )^j}{(-x_{12}^2)^{j+\theta}} = \delta_{jk}  \left( C_j^{(0)} + g^2 C_j^{(1)} \right)2^{2j} \frac{  (\hat{x}_{12} )^{2j}}{(-x_{12}^2)^{j+\theta}} \,.
\end{equation}
The tree-level normalisation constant $C_j^{(0)}$ can be read off from \eqref{eqn:two-point-tree-level-N-equal4SYM-in-main-text} and we have
\begin{equation}\label{eqn:normalization-tree-level-two-point}
C_j^{(0)} = g^4 \delta^{aa} \frac{\Gamma(2j+1)}{2^{5} \pi^4 }\,.
\end{equation}
Reading off the finite one-loop normalisation constant $C_j^{(1)}$  is a little more complicated. The renormalised two-point function for $j=k$ reads
\begin{align}\label{eqn:MS-renormalized-two-point-function}
 \langle \hat{O}^R_j(x_1) \hat{O}^R_j(x_2) \rangle&=\langle \hat{\op}_{j} \hat{\op}_{j} \rangle^{(0,\epsilon)}\left(1+ \frac{(-\mu^2 x^2)^{\epsilon}}{\epsilon} g^2 2 Z_j^{(1)} \right) + \langle \hat{\op}_{j} \hat{\op}_{j} \rangle^{(1)} \,.
\end{align}
The $\op(\epsilon^0)$ expansion of the tree-level correlator multiplied with the renormalisation constant exactly cancels the pole term from the one loop diagrams, whereas its $\op(\epsilon)$ expansion multiplied with the pole term yields another finite contribution to the normalisation constant. Thus, the one-loop normalisation is
\begin{align}\label{eqn:normalization-one-loop-two-point}
C_j^{(1)} = \lim_{\epsilon\to0} \left( g^2 \langle \hat{\op}_j \hat{\op}_j \rangle^{(0,\epsilon)} \frac{2 Z_j^{(1)}}{\epsilon} + \langle \hat{\op}_j \hat{\op}_j \rangle^{(1)} \right)/ \left(2^{2j}  \frac{\hat{x}^{2j}}{(-x^2)^{j+\theta}}\right)\,,
\end{align}
where the division by the space-time structure is to be understood symbolically. The diagonal tree-level correlator including its $\op(\epsilon)$ expansion is given in \ref{eqn:tree-level-expression-epsilon-diagonal} and the one-loop contribution is given in \eqref{eqn:sum-of-one-loop-two-point-contributions}.

\subsection{Mixing Matrix}
The renormalised two-point function \eqref{eqn:MS-renormalized-two-point-function} in the $\overline{\text{MS}}$-scheme yields finite but non-diagonal correlators.
\begin{equation}
 \langle \hat{O}^R_j(x_1) \hat{O}^R_k(x_2) \rangle  \neq 0 \qquad \text{for $j \neq k$}\,.
\end{equation}
As explained in section \ref{conformal scheme}, in order to have a diagonal basis of renormalised twist-two operators
\begin{equation}
 \langle \hat{\mathbb{O}}_j(x_1) \hat{\mathbb{O}}_k(x_2) \rangle = \delta_{jk}  \langle \hat{\mathbb{O}}_j(x_1) \hat{\mathbb{O}}_j(x_2) \rangle\,,
\end{equation}
it is necessary to subtract the finite non-diagonal terms by introducing another finite renormalisation matrix $B$. 
\begin{equation}
\mathbb{Z}_{jk} =\delta_{jk} + g^2\left(-B_{jk}^{(1)}+\frac{1}{\epsilon}\delta_{jk} Z_{j}^{(1)}\right) + \op(g^4)\,.
\end{equation}
The result is well-known %(::CITATIONS) 
and was redetermined in a very simple way in \cite{Belitsky:2007jp} in the same conventions that we use and reads
\begin{equation}\label{eqn:N=4-mixing-matrix}
B_{jk}^{(1)} = 2 d_{jk}^{1/2}\left( H_{j} + H_k + H_{\frac{j-k}{2}} - H_{\frac{j+k}{2}} - 2H_{j-k} \right)\,,
\end{equation}
where $d_{jk}^{1/2}$ is related to the expansion of the Gegenbauer polynomials $C_j^\nu$ with respect to the index $\nu$ and is given in \eqref{eqn:gegenbauer-expansion-coefficients}. 
One can determine the mixing matrix by requiring the above orthogonality, i.e. by calculating the finite non-diagonal ($j\neq k$) terms of the one-loop correlator. We do not go into the details, since the mixing matrix drops out in our limit. 

\subsection{Protected Operator for $j=0$}\label{sec:protected-operator}
\begin{figure}[b]
\begin{minipage}{17.5cm}
\includegraphics[width= .08 \textwidth]{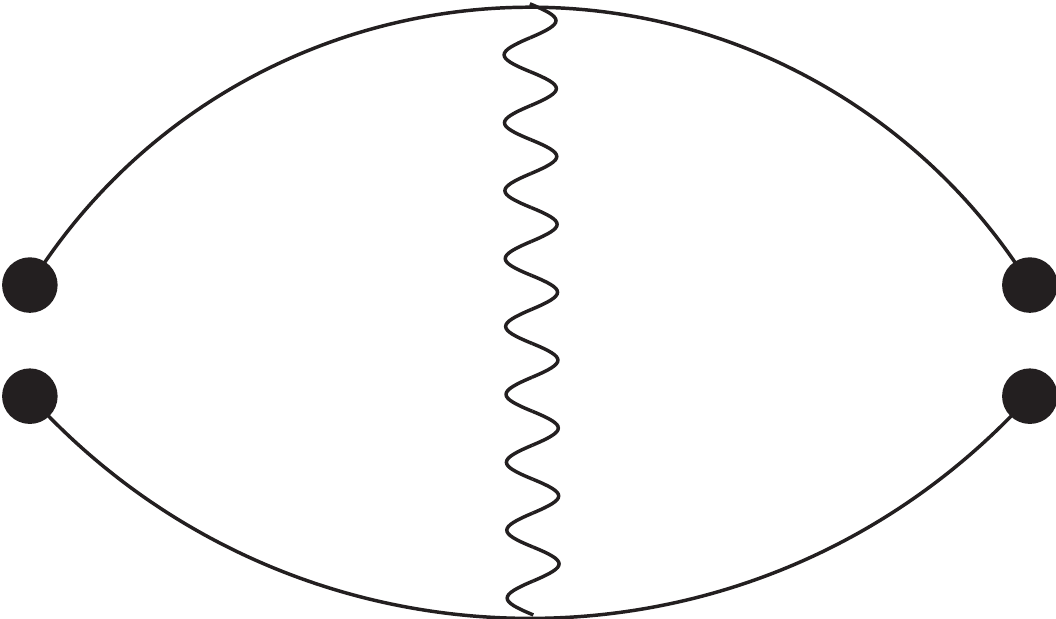} \raisebox{.32cm}{~$=~ - 2p^2$} 
\includegraphics[width= .08 \textwidth]{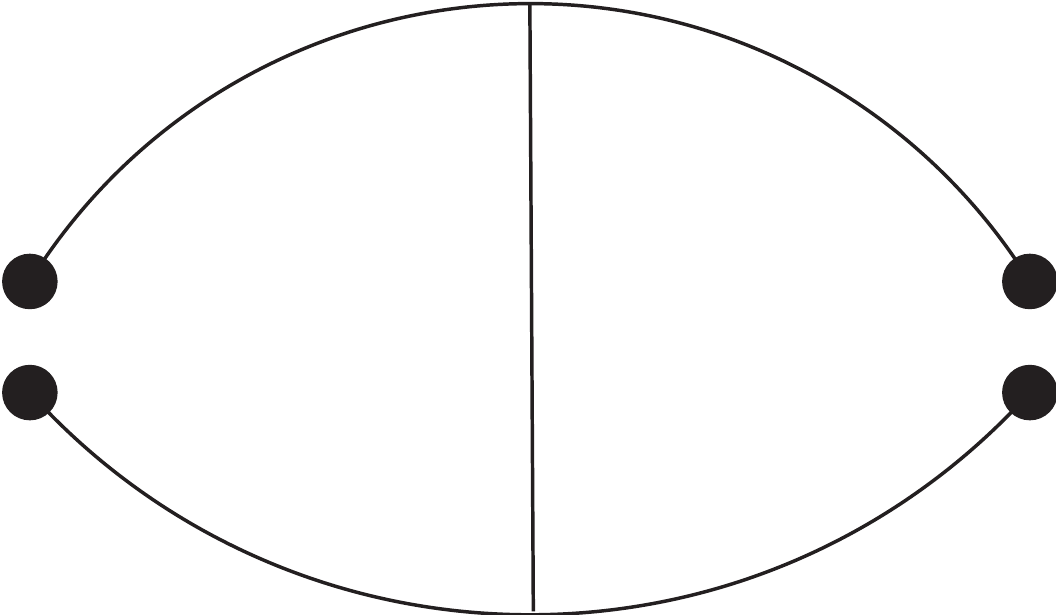} \raisebox{.32cm}{+}
\raisebox{.1cm}{\includegraphics[width= .1 \textwidth]{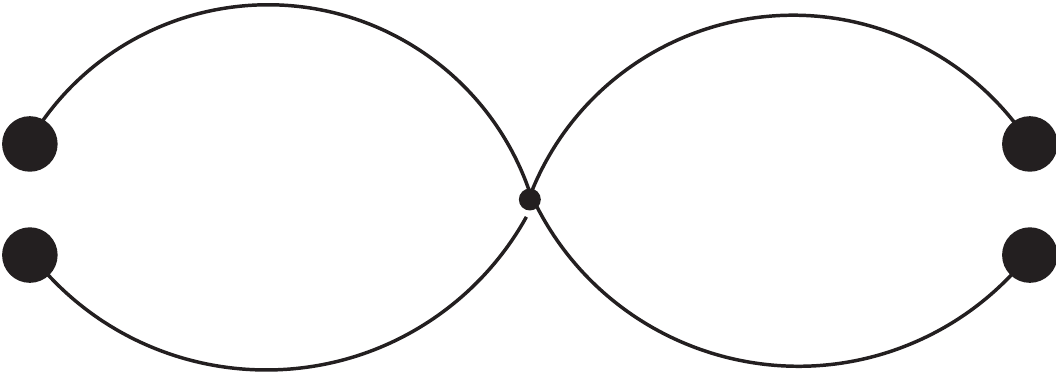}} \raisebox{.32cm}{+}
\includegraphics[width= .08 \textwidth]{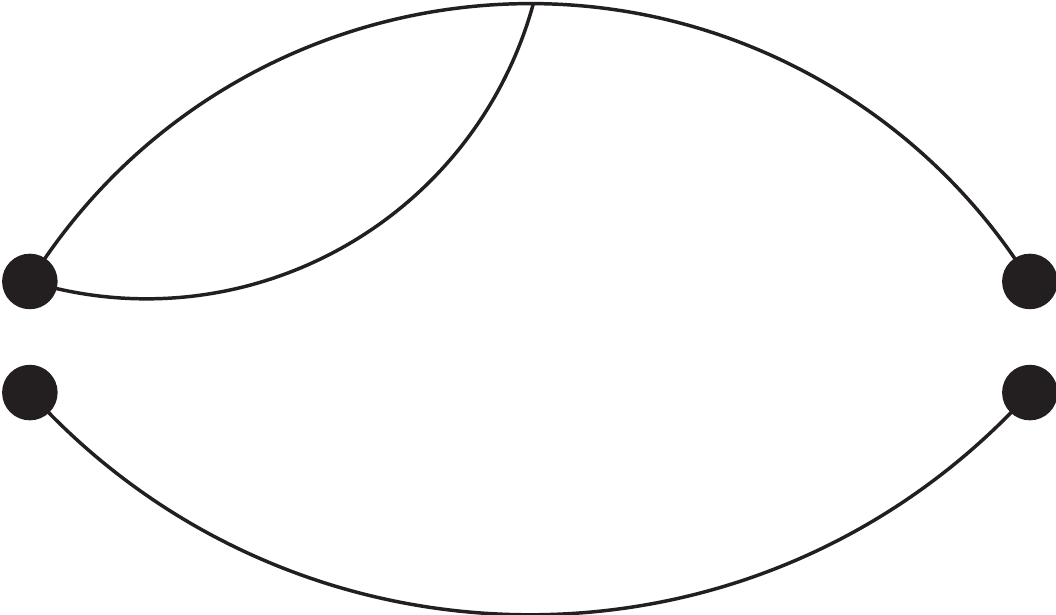} \raisebox{.32cm}{+} 
\includegraphics[width= .08 \textwidth]{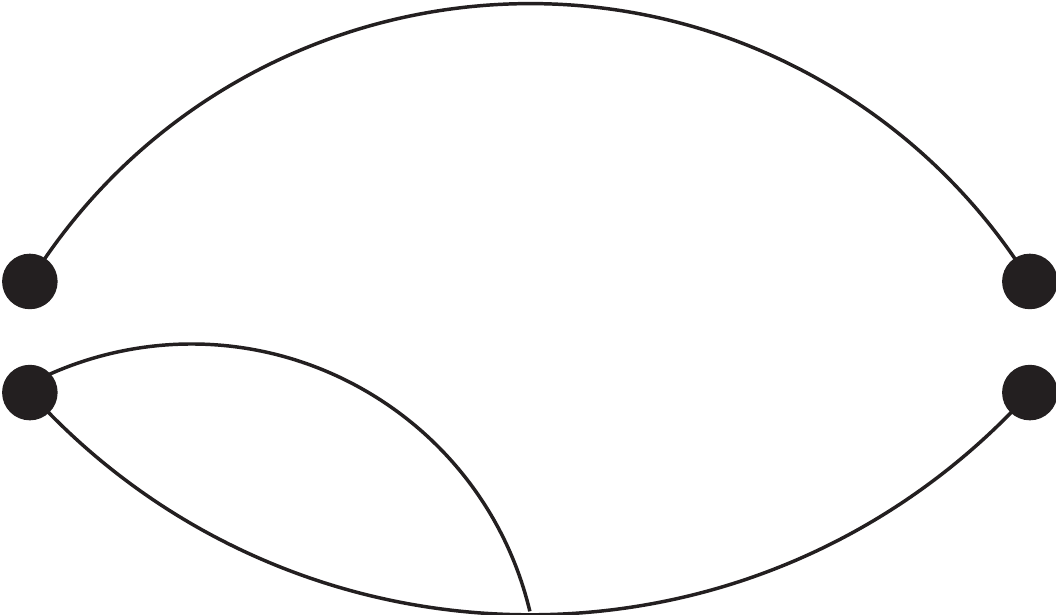} \raisebox{.32cm}{+} 
\includegraphics[width= .08 \textwidth]{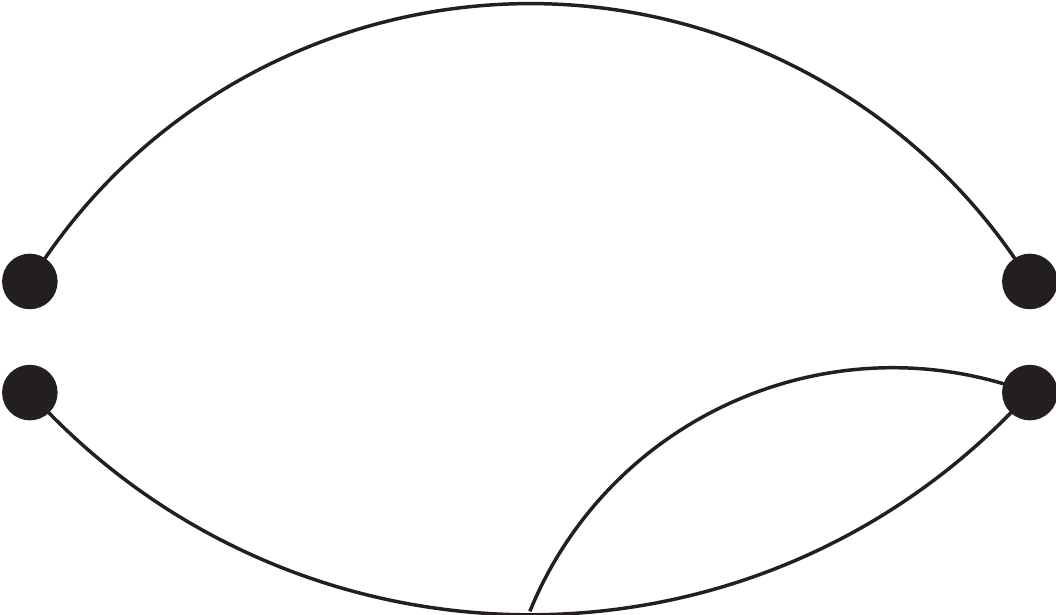} \raisebox{.32cm}{+}
\includegraphics[width= .08 \textwidth]{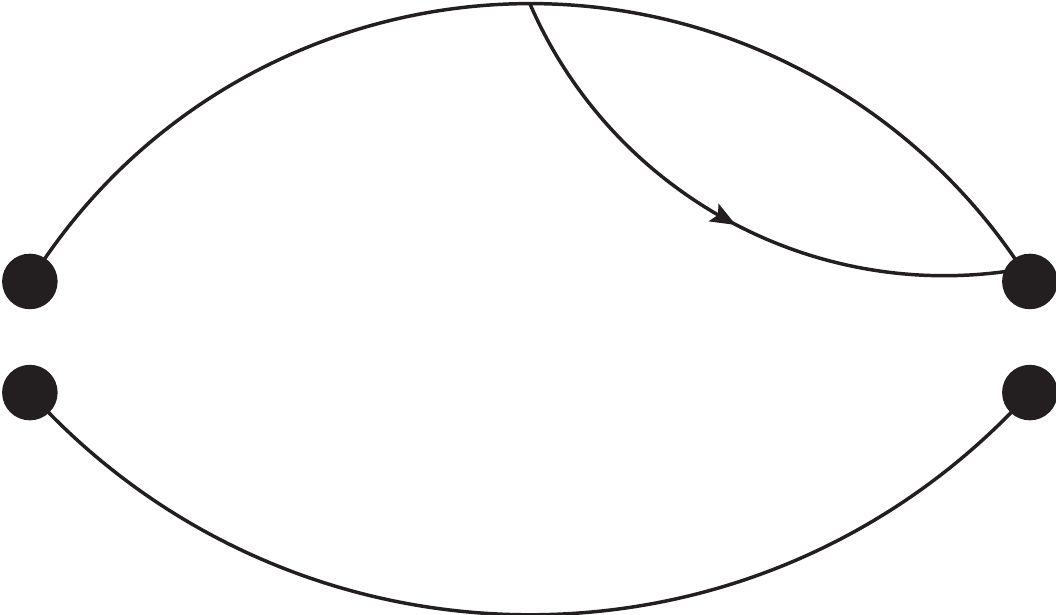} 
\end{minipage}
\caption{Due to the momenta in the numerator, that originate from the gluon-vertices, the kite diagram is decomposed into these scalar integrals.}
\label{fig:gluon-kite-diagram-decompostion}
\end{figure}

For $j=0$ the one-loop corrections cancel and thus the operator is protected and has anomalous dimension zero, this can be seen as follows. For $j=0$ there are no covariant derivatives and the operator \eqref{eqn:def-twist-two} is
\begin{equation}\label{eqn:protected operator}
\hat{\op}_{j=0} (x) = \tr \left( \phi^{12}(x)  \phi^{12}(x)\right)\,.
\end{equation}
Therefore, the diagrams shown in figures \ref{fig:twopointonecovderivative}, \ref{fig:twopointtwocovderivative} are absent.
To understand the cancellation among the remaining diagrams, it is instructive to consider  the decomposition of the kite diagram into scalar basis integrals, see \ref{sec:kite-diagram}, which is diagrammatically represented in figure \ref{fig:gluon-kite-diagram-decompostion}. The last four integrals in figure \ref{fig:gluon-kite-diagram-decompostion} are all identical and cancel  with the self-energy diagrams. The second diagram cancels with the four-scalar diagram (this is also true for $j\neq0$) and the only remaining contribution is the first diagram shown in figure \ref{fig:gluon-kite-diagram-decompostion} which  is finite in momentum space and 
yields the result
\begin{align}\label{eqn:result-protected-ops}
\langle \op_{j=0}(p) \bar{\op}_{j=0}(-p) \rangle^{(1)} 
%&= g^6 \frac{i\delta^{aa} b_0(1,1)}{4 (-p^2)^{4-d}} \Big[\frac{4}{(d-4)}(b_0(3-h,2)-b_0(1,2)) \Big]\\ \nn
&= - \frac{1}{(4\pi)^4}\frac{i g^6 N \delta^{aa}}{(-p^2)^{4-d}} 6\zeta(3) + \op(\epsilon)\,.
\end{align}
The Fourier transformation to configuration space yields a factor of $1/\Gamma(4-d) \propto \epsilon$ and thus up to order epsilon terms the correlator does not get quantum corrections
\begin{equation}
\langle \op_{j=0}(x_1) \op_{j=0}(x_2) \rangle^{(1)} = \op(\epsilon)\,.
\end{equation}
Thus, the anomalous dimension of the operator \eqref{eqn:protected operator} is zero: $\gamma_{j=0}=0$. For an operator with two different flavours like the ones we chose in \eqref{eqn:BPS-operators}, e.g.
\begin{equation}
\op(x)= \tr(\bar{\phi}_{12}(x)\bar{\phi}_{13}(x))\,,
\end{equation}
we have the same type of diagrams as for \eqref{eqn:protected operator}, but slightly different overall combinatorics. Due to the fact that the flavours are distinct, we get a factor of $1/2$ as compared to  \eqref{eqn:result-protected-ops}.

\section{Tree-Level Three-Point Functions}\label{eqn:tree-level-structure-constant}
As explained in section \ref{sec:limits-of-three-point-correlators} we can calculate the three-point function in different limits in order to simplify the calculations.  Even though the the tree-level calculation is slightly simpler in the limit given in \ref{sec:Limit-x2-to-infty}, see the calculation in Appendix \ref{sec:three-point-limit-x2-to-infinity}, we calculate the tree-level three-point function in the limit $p_1+p_2\to0$ given in section \ref{sec:limit-P-to-0}. This is, because the limit is better suited for the calculation at one-loop level and we would like to illustrate the method at tree-level.
\begin{figure}[h]
\centering
 \includegraphics[width=.3\textwidth]{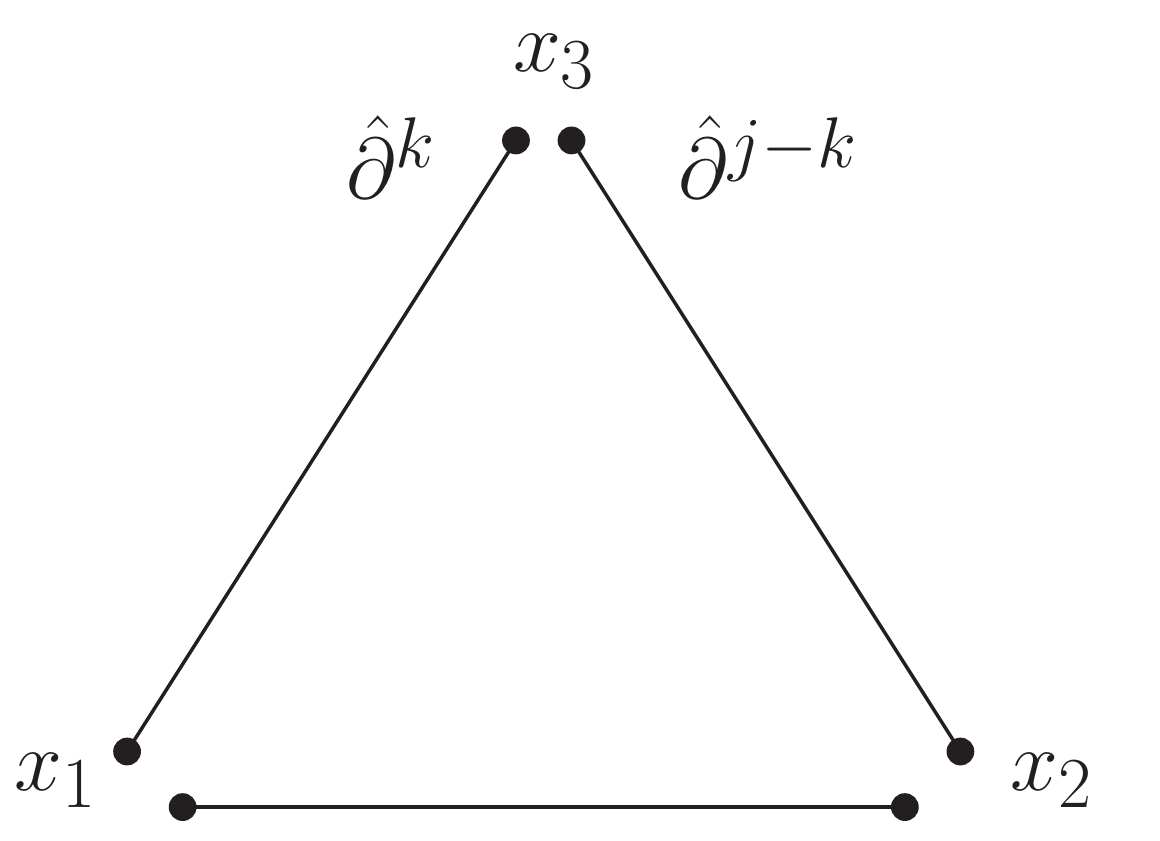}
\caption{3-point function at tree-level}
\label{fig:three-point function at tree-main-text}
\end{figure}
~\\
In momentum space, the three-point function with the twist operator in the representation \eqref{eqn:definition-with-gegenbauer-polynomials} reads
\begin{align}
\langle \op(p_1) \tilde{\op}(p_2) \hat{\op}_j \rangle &= \frac{1}{8} 2\, i^{3+j}g^6 \delta^{aa} \int \frac{d^dk}{(2\pi)^d} \frac{(\hat{p}_1+\hat{p}_2)^j C_j^{1/2}\left( \frac{2\hat{k}-\hat{p}_1-\hat{p}_2}{\hat{p}_1+\hat{p}_2} \right)}{k^2 (p_1-k)^2(p_1+p_2-k)^2}\,,
\end{align}
where the factor of $\frac{1}{8}$ is due to the traces over the gauge group and the factor of 2 is due to the two different possible contractions with $\hat{\op}_j$.

Now we take the limit $p_1+p_2 \to 0$ in momentum space. 
Then, due to the factor $(p_1+p_2)^j$ only the term with the highest power, i.e. $j$, in the Gegenbauer polynomial can survive. The corresponding coefficient is given in \eqref{eqn:gegenbauer-coefficient-higherst-power} and reads
\begin{equation}\label{eqn:highest-power-gegenbauer}
c_{jj}^{1/2}  = 2^{1-j} \frac{\Gamma(2j)}{\Gamma(j)\Gamma(j+1)} \quad \text{where} \quad C_j^{1/2} (x) = \sum_{k=0}^j c^{1/2}_{jk} x^k \,.
\end{equation}
Thus, in this limit we get
\begin{align}\label{eqn:tree-level-three-point-function-in-momentum-space}
\langle \op(p) \tilde{\op}(-p) \hat{\op}_j(0) \rangle &= g^6 \delta^{aa} 2^{j-2} i^{3+j} c_{jj}^{1/2} \int \frac{d^dk}{(2\pi)^d} \frac{(\hat{k})^j }{k^4 (p+k)^2} \\ \nn
&= -2^{j-2}g^6 \delta^{aa} i^{3+j} c_{jj}^{1/2} b_j(2,1)  \frac{\hat{p}^j}{(-p^2)^{3-h}}\,,
\end{align}
where the bubble integrals $b_j(\alpha_1,\alpha_2)$ are defined in \eqref{sec:bubble-integrals}.
Fourier transformation to x-space using \eqref{eqn:result-fourier-trafo-mink} and insertion of the coefficients yields
\begin{align}\label{eqn:tree-level-three-point-function-in-x-space-calculated}
\int d^dx_3 \langle \op \tilde\op \hat{\op}_j\rangle &= \int \frac{d^dp}{(2\pi)^d} \langle \op(p) \tilde{\op}(-p) \hat{\op}_j(0) \rangle e^{- i p x_{12}} \\ \nn
&= -ig^6 \delta^{aa} \frac{2^{d-7+j}}{\pi^{\frac{d}{2}}}\frac{\Gamma(2j)\Gamma(h-1)\Gamma(j-2+h)}{\Gamma(j)\Gamma(j+1)} \frac{(\hat{x}_{12})^j}{(-x_{12}^2)^{2h-3+j}}\,.
\end{align}
Comparing with \eqref{eqn:N(g^2)-at-epsilon=0}, \eqref{eqn:renormalized-three-point-function} we can easily read off the tree-level structure constant for $d=4$
\begin{align}\label{eqn:tree-level-strcuture-constant-N=4}
C_{\op\tilde{\op}j}^{(0)} = g^6 \delta^{aa} \frac{2^{j-8} \Gamma (j+1)}{\pi ^6}\,.
\end{align}
Note that the three-point function for a scalar theory in general dimension $d$ has the same expression \eqref{eqn:tree-level-three-point-function-in-momentum-space} with $c_{jj}^{1/2}$ replaced with $c_{jj}^\nu$ and $\nu=d/2-3/2$. The corresponding result has also been calculated in the other limit described in section \ref{sec:Limit-x2-to-infty} in Appendix \ref{sec:three-point-limit-x2-to-infinity}.
%\subsection{Gegenbauer Polynomials from three-point function}
%As in Yu. M. Makeenko 'On conformal operators in QCD', show how demanding conformal three point functions for $\langle \phi \bar{\phi} \hat{\op}_j \rangle$ yields direcly something proportional to Gegenbauer Polynomials! Very simple, just assume conformal structure and then get coefficients by comparing expressions. Very nice!

\section{One-Loop Three-Point Functions}
\begin{figure}[h]
\centering
\subfloat[]{\begin{minipage}[c]{3.3cm}
		\includegraphics[width=1 \textwidth]{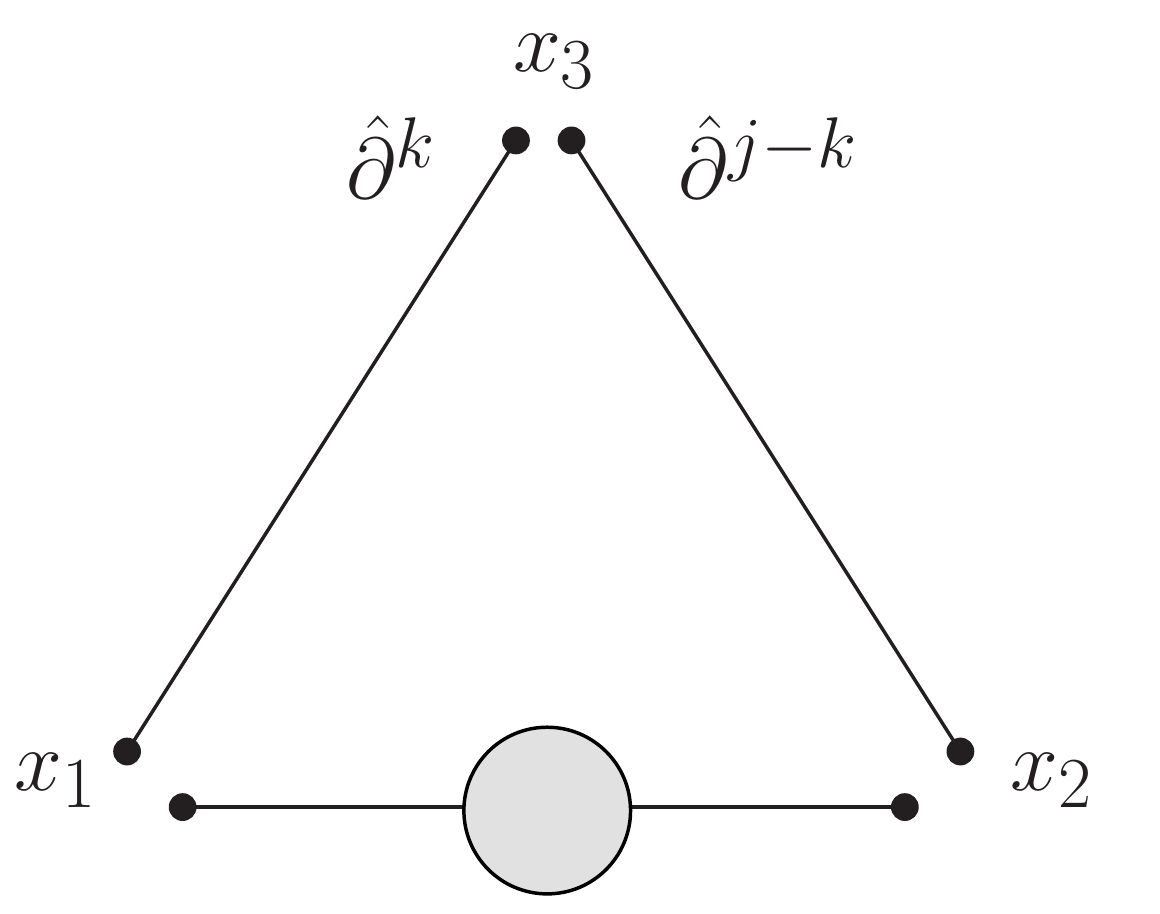}\label{fig:3point1loopb}
		\end{minipage}
		
}
\subfloat[]{\begin{minipage}[c]{3cm}
		\includegraphics[width=1 \textwidth]{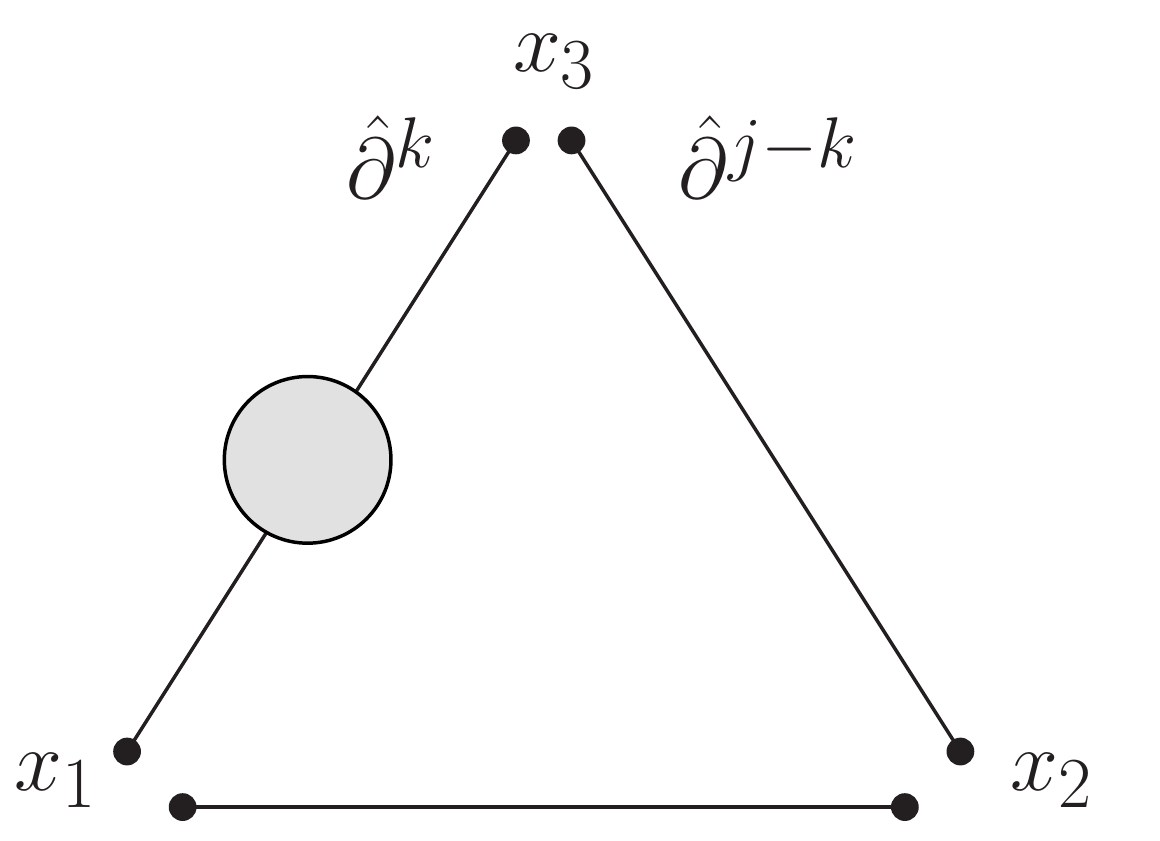}\label{fig:3point1loopa}
		\end{minipage}
		
}
\subfloat[]{\begin{minipage}[c]{3cm}
		\includegraphics[width=1 \textwidth]{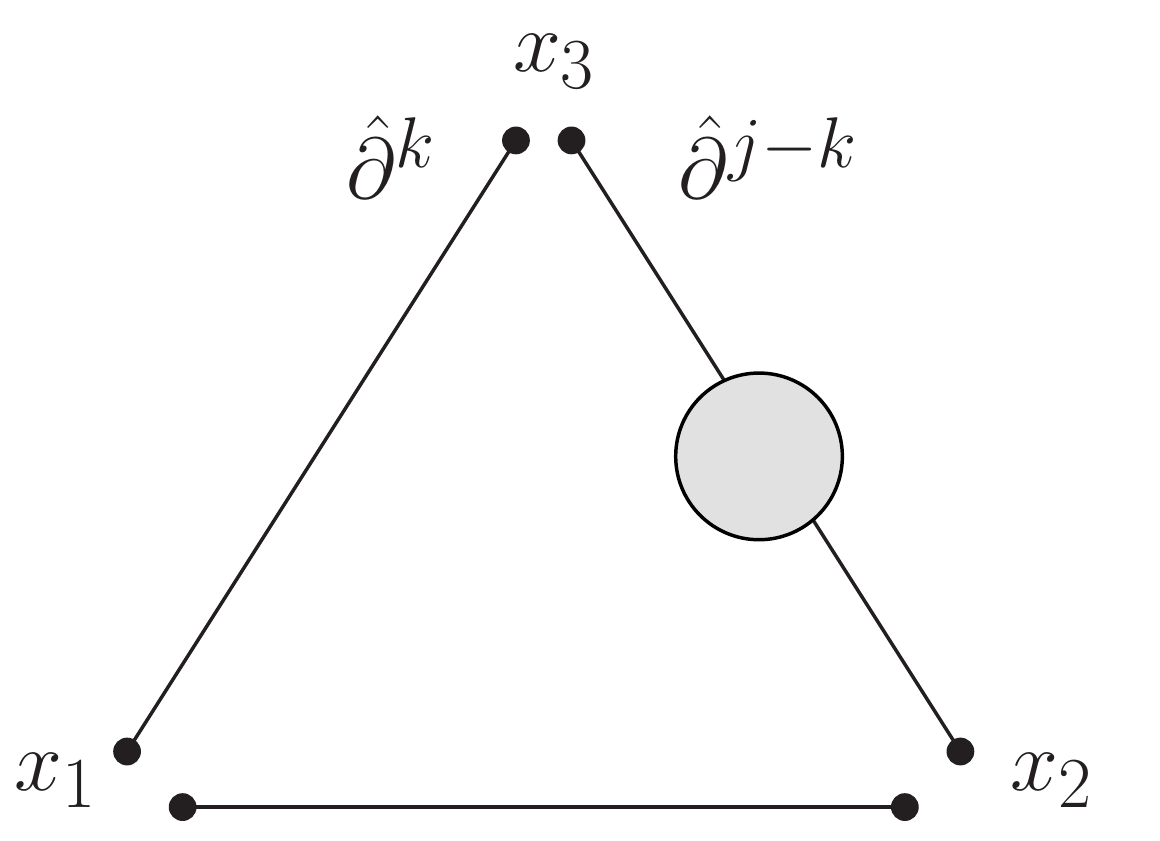}\label{fig:3point1loopc}
		\end{minipage}
		
}
\\
\subfloat[]{\begin{minipage}[c]{3cm}
		\includegraphics[width=1 \textwidth]{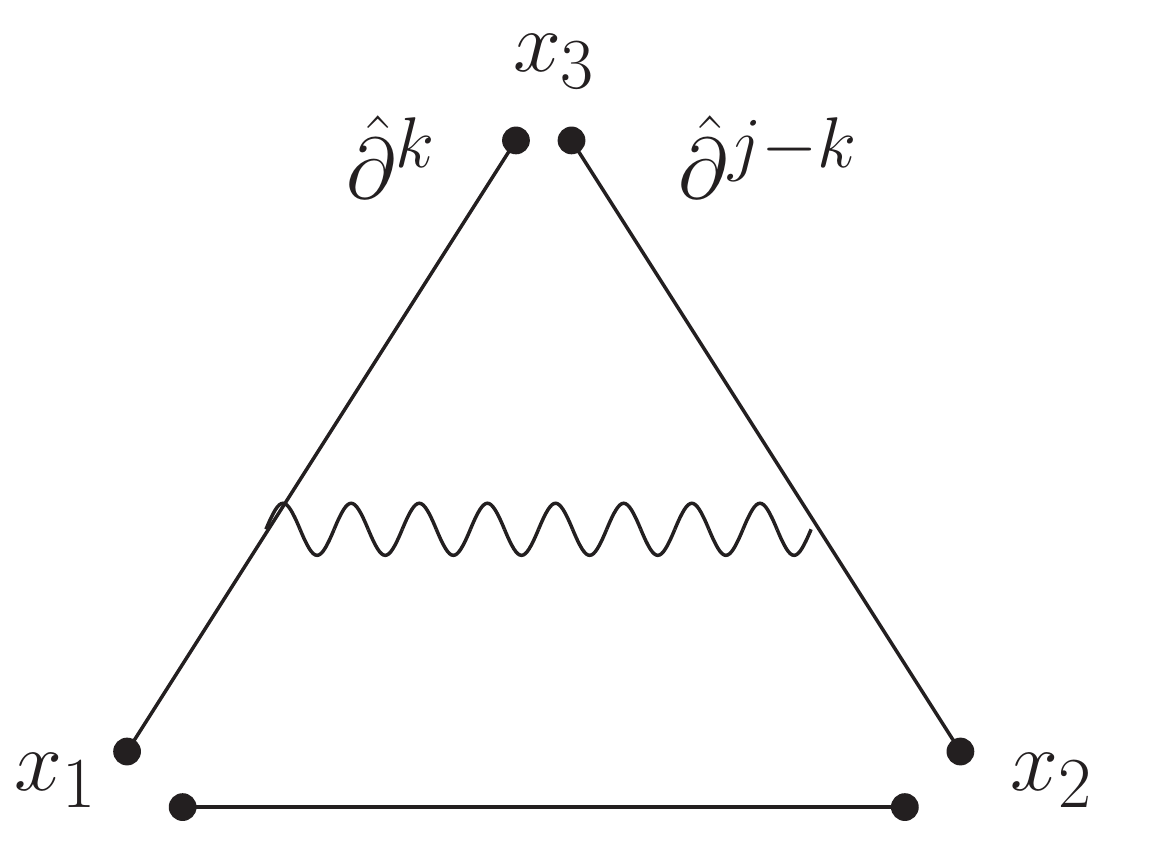}\label{fig:3point1loopg}
		\end{minipage}
		
}
\subfloat[]{\begin{minipage}[c]{3cm}
		\includegraphics[width=1 \textwidth]{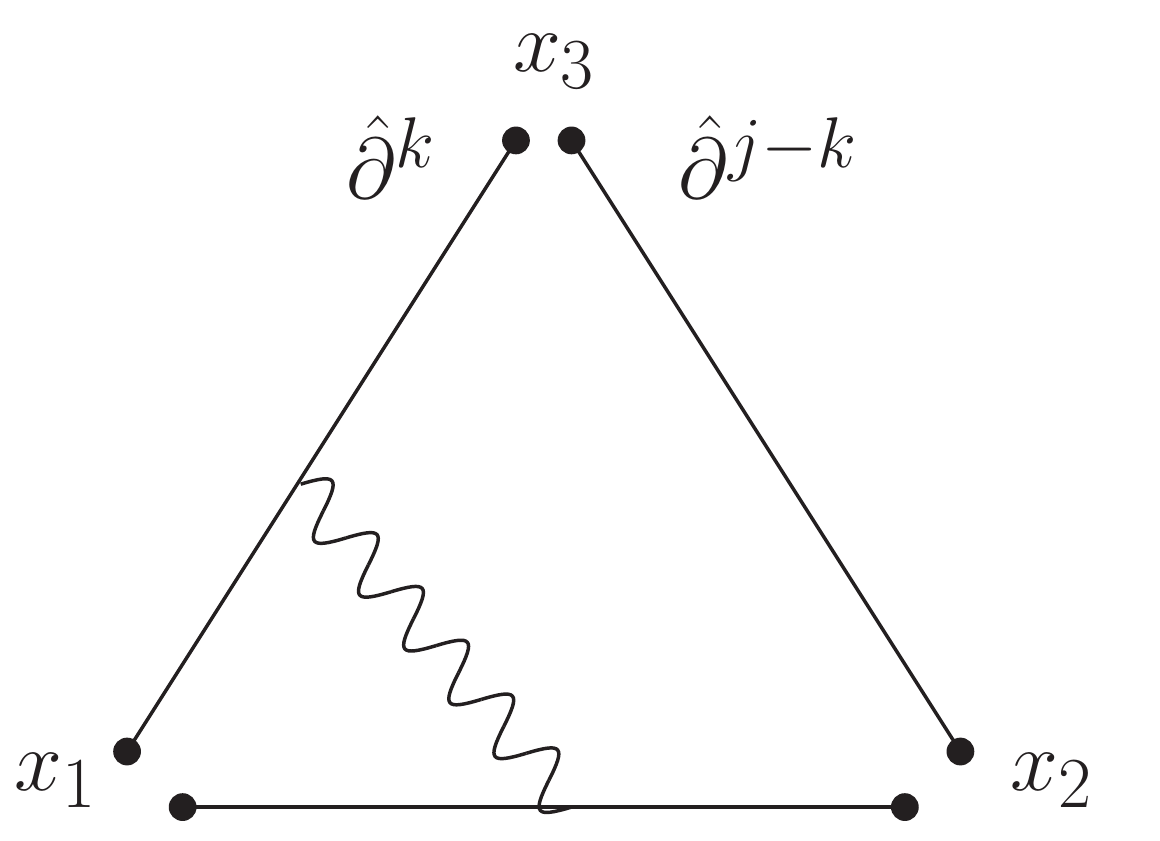}\label{fig:3point1looph}
		\end{minipage}
		
}
\subfloat[]{\begin{minipage}[c]{3cm}
		\includegraphics[width=1 \textwidth]{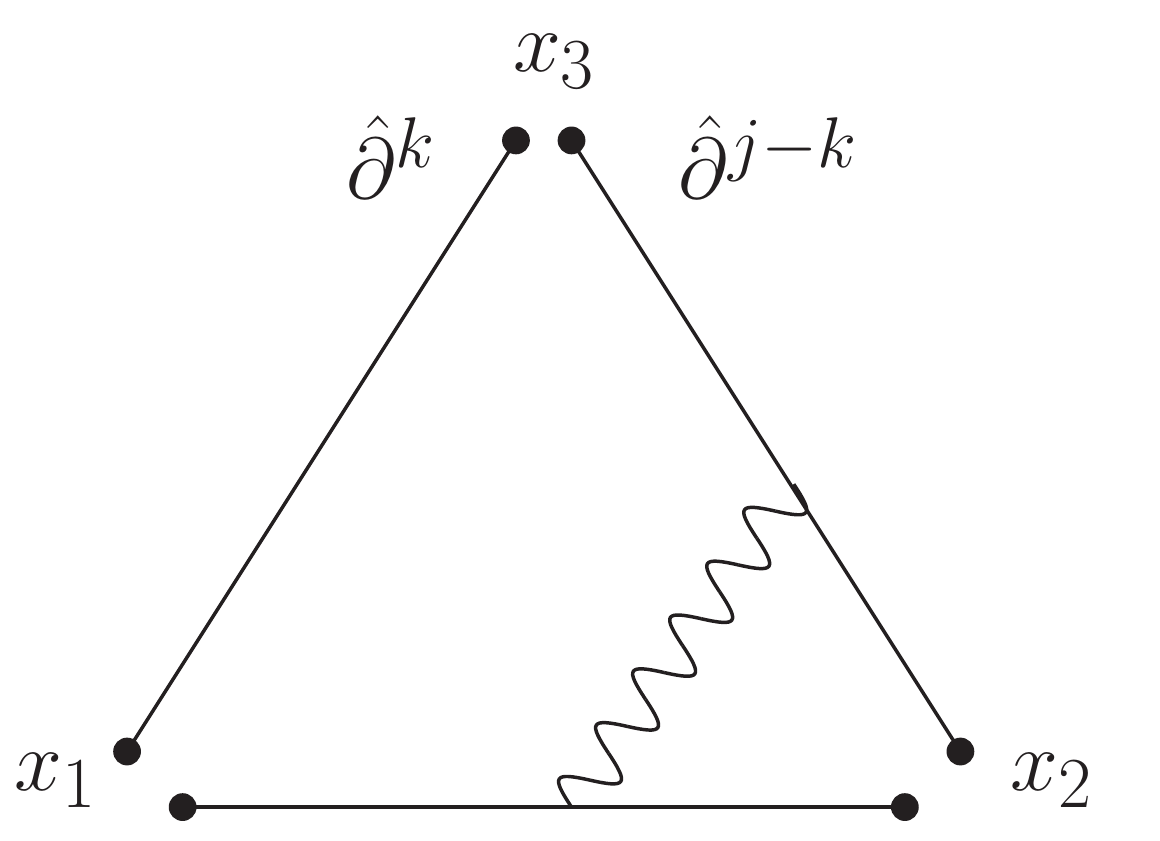}\label{fig:3point1loopi}
		\end{minipage}
		
}
\\
\subfloat[]{\begin{minipage}[c]{3cm}
		\includegraphics[width=1 \textwidth]{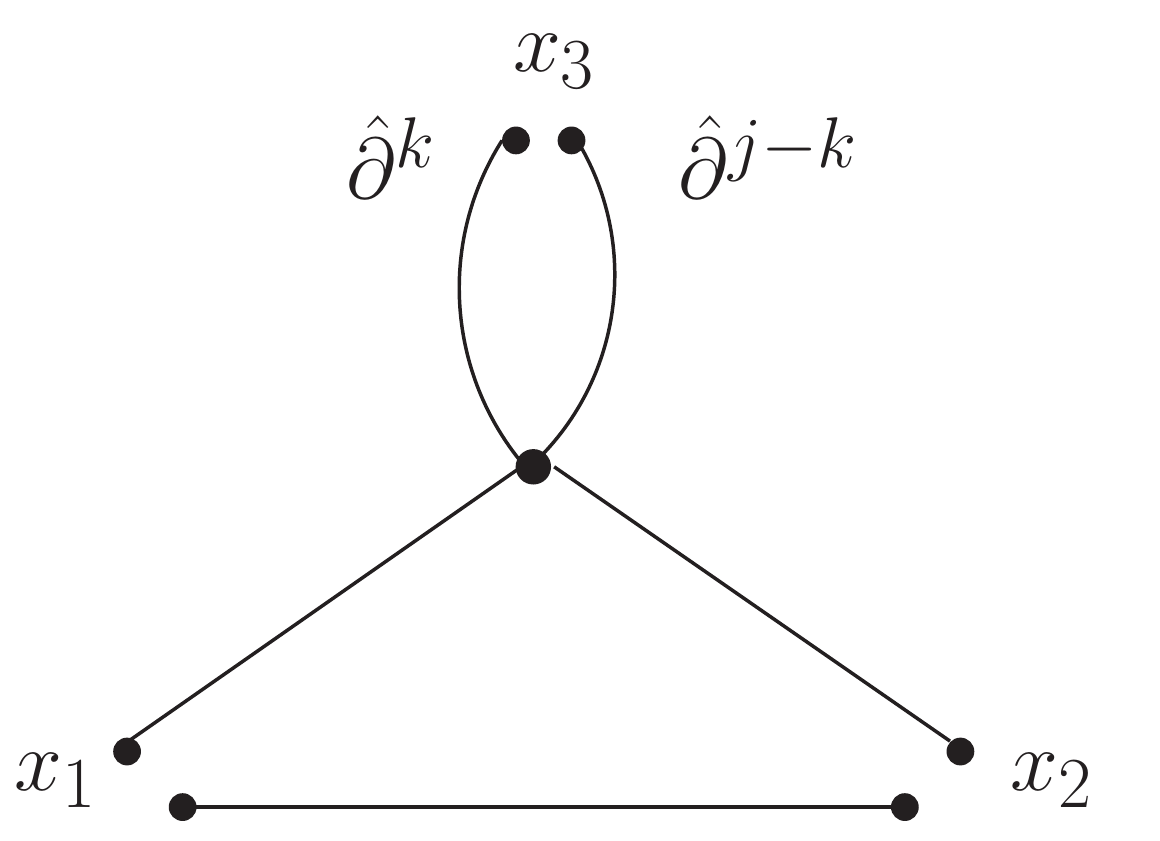}\label{fig:3point1loopl}
		\end{minipage}
		
}
\subfloat[]{\begin{minipage}[c]{3cm}
		\includegraphics[width=1 \textwidth]{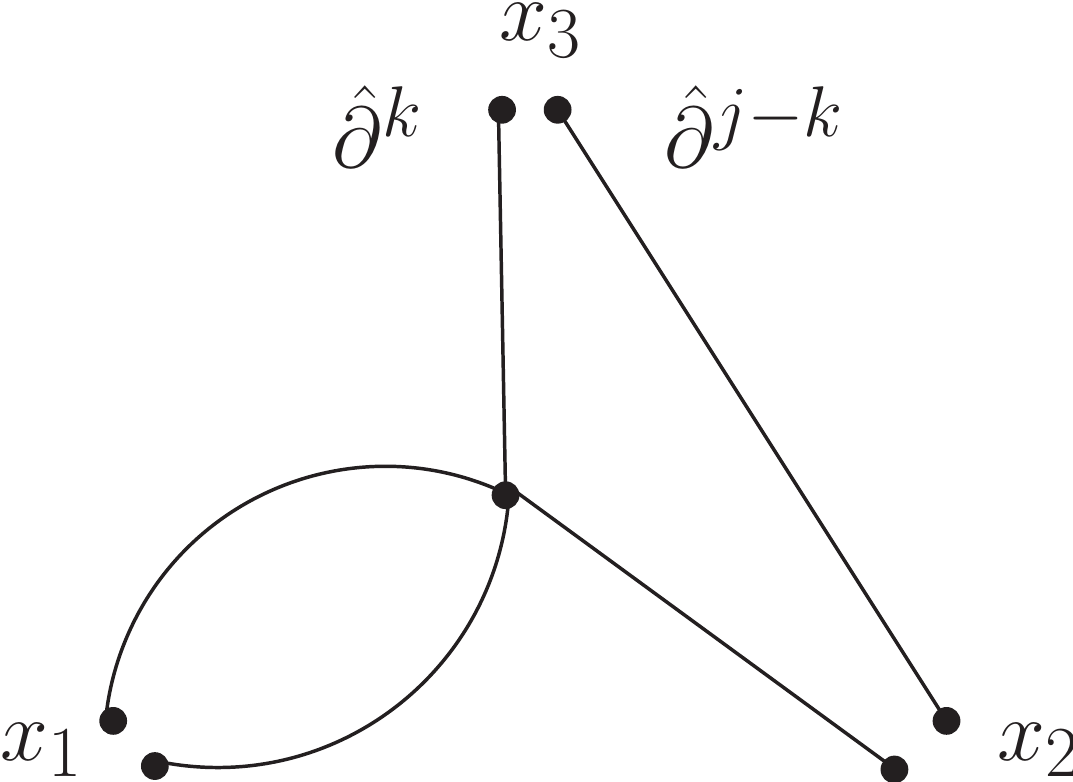}\label{fig:3point1loopj}
		\end{minipage}
		
}
\subfloat[]{\begin{minipage}[c]{3cm}
		\includegraphics[width=1 \textwidth]{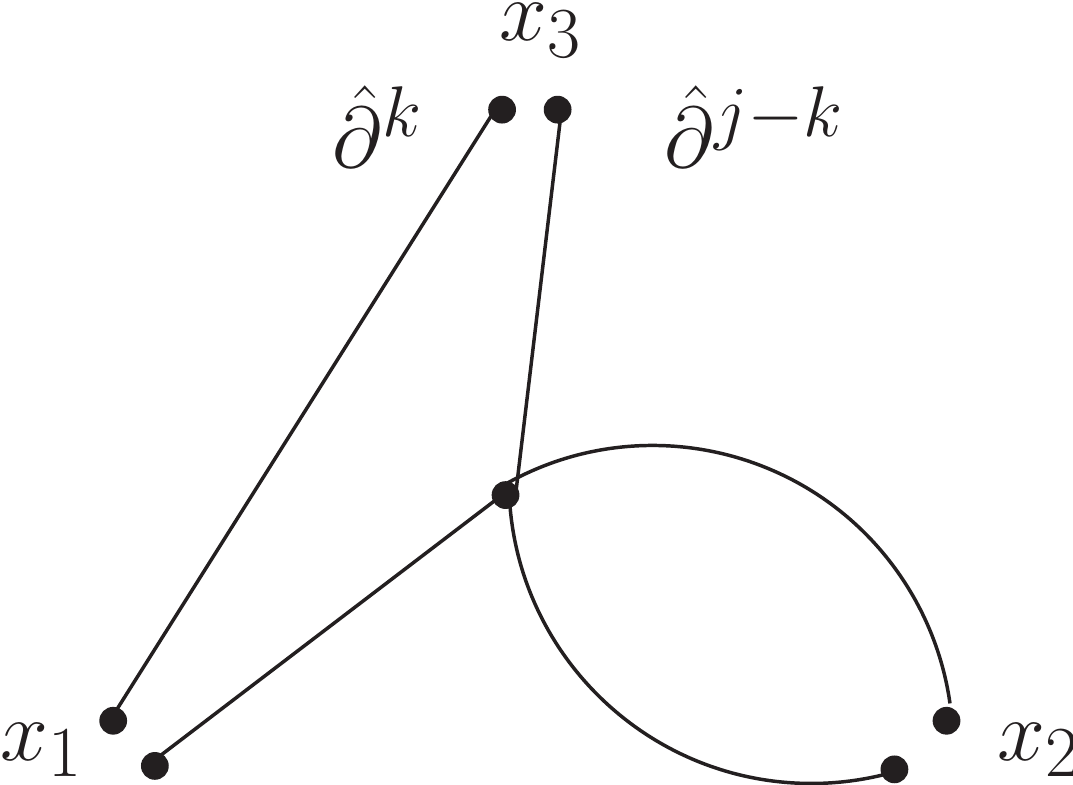}\label{fig:3point1loopk}
		\end{minipage}
		
}
\\
\subfloat[]{\begin{minipage}[c]{3cm}
		\includegraphics[width=1 \textwidth]{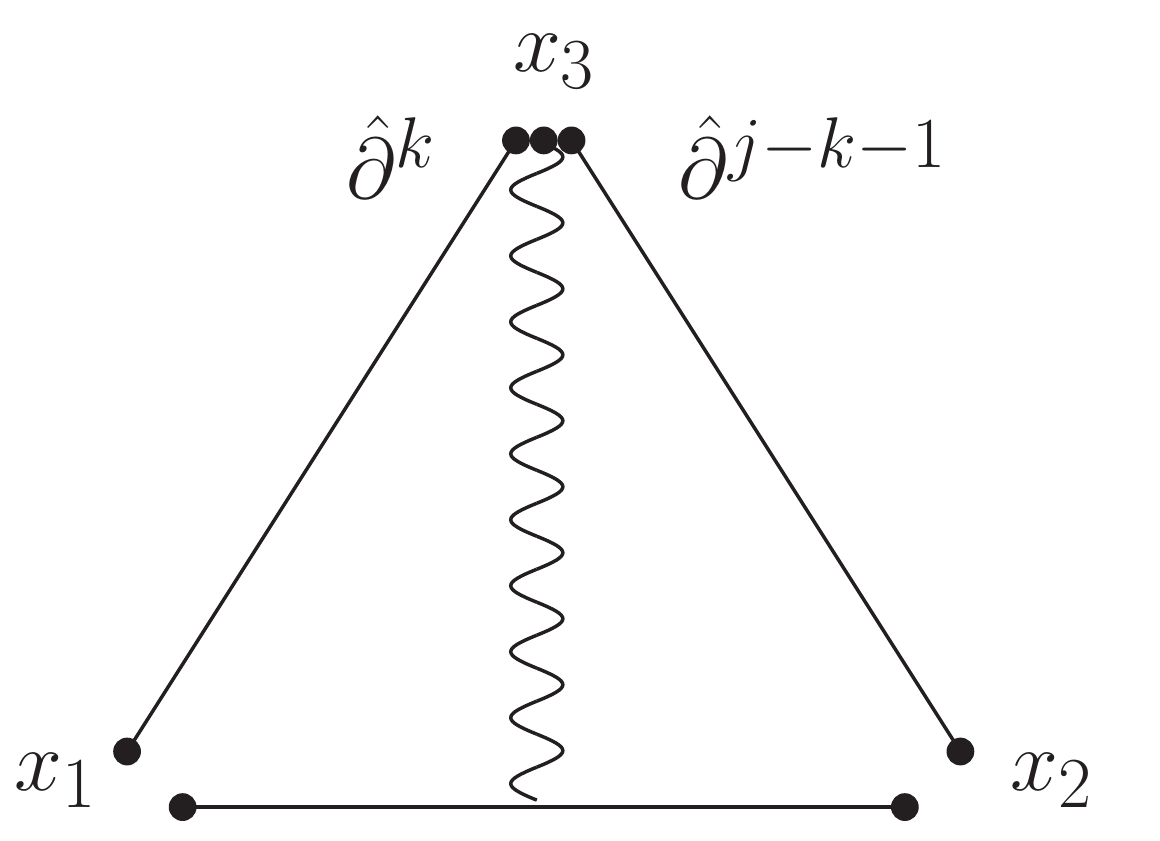}\label{fig:3point1loopf}
		\end{minipage}
		
}
\subfloat[]{\begin{minipage}[c]{3cm}
		\includegraphics[width=1 \textwidth]{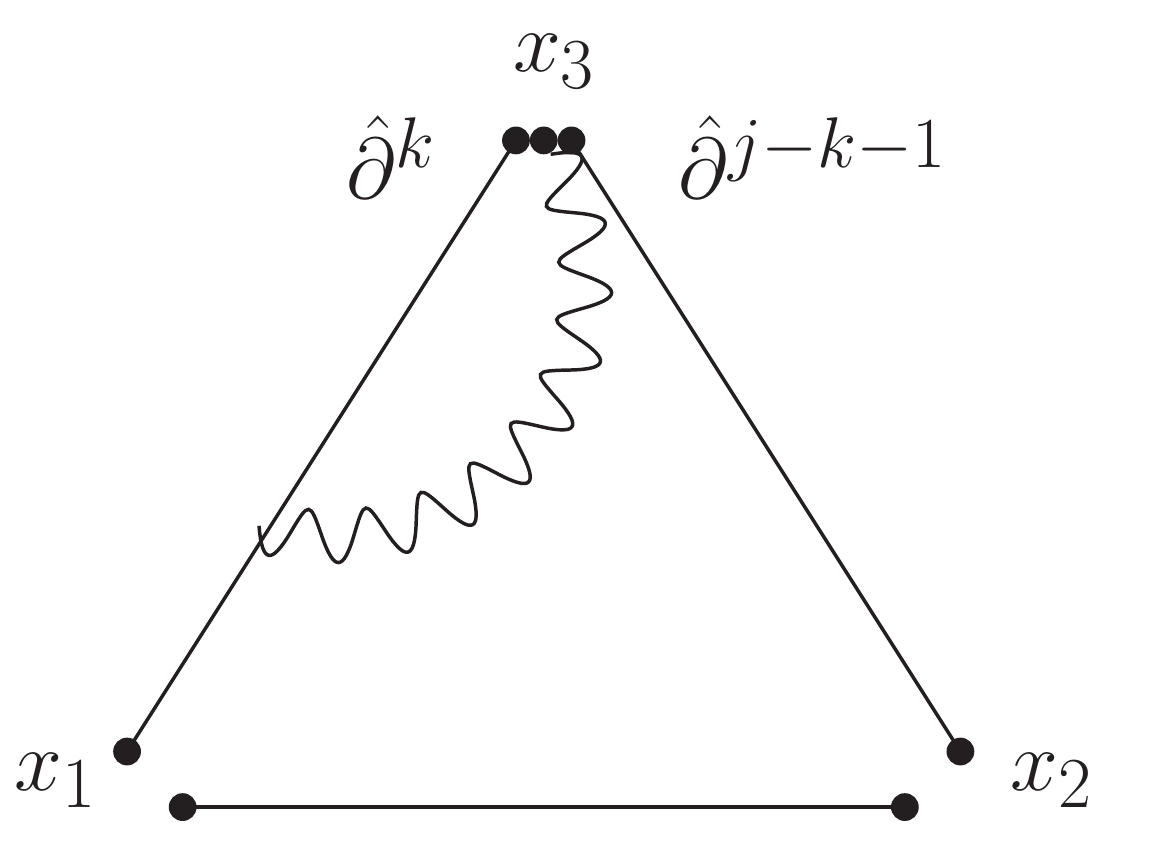}\label{fig:3point1loope}
		\end{minipage}
}		
\subfloat[]{\begin{minipage}[c]{3cm}
		\includegraphics[width=1 \textwidth]{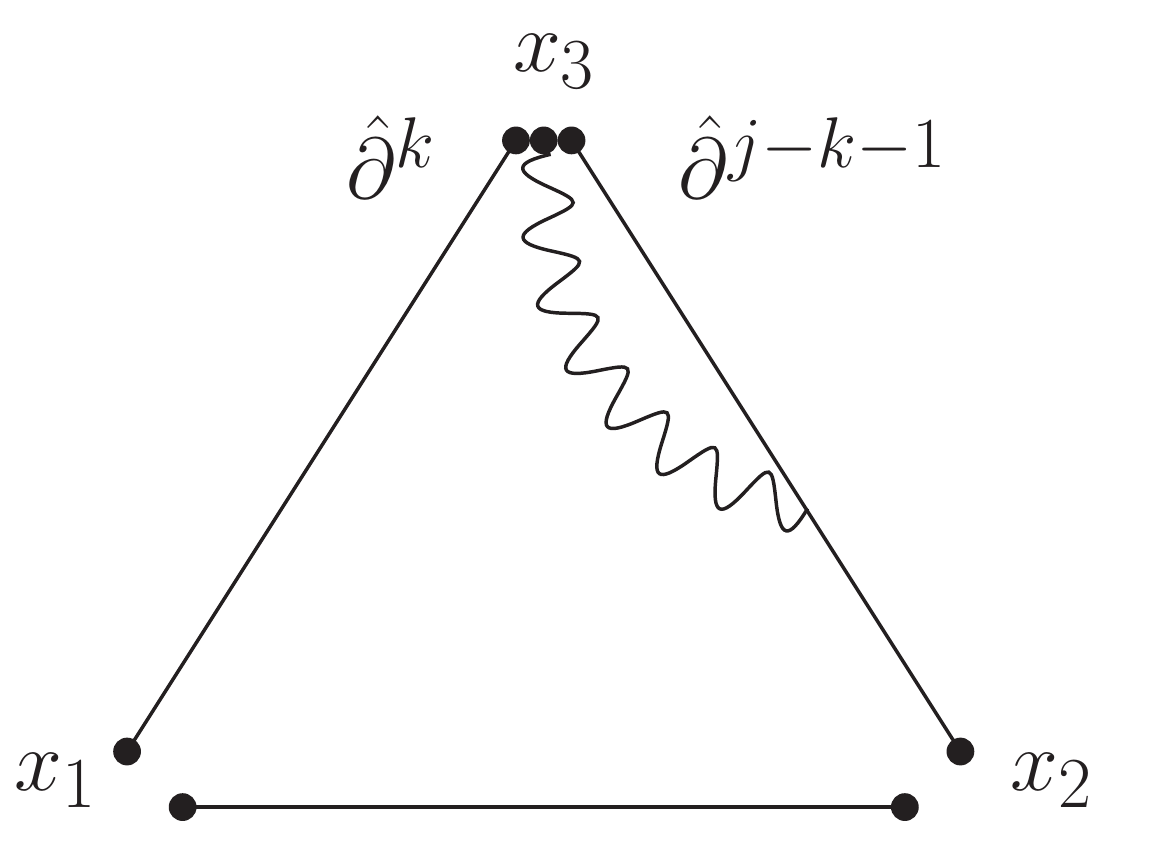}\label{fig:3point1loopd}
		\end{minipage}
		
}
\caption{Feynman diagrams contributing to the three-point function at one loop. For $j$ even the diagrams in the second and third column are identical.}
\label{fig:three-point function at 1-loop}
\end{figure}

It is advantageous to write down the full diagrams, group the contributions to see cancellations that are independent of the limit and only then take the limit $p_1+p_2\to 0$. In this way we avoid possible ambiguities in taking the limit for diagram \ref{fig:3point1loopl}, because it cancels with a contribution of diagram \ref{fig:3point1loopg} as we will see in the next section.
We will make use of the fact that we consider $j$ even such that factors like $(-1)^j$ are not displayed.

\notocsubsection{Cancellations between Diagrams \ref{fig:3point1loopg} and \ref{fig:3point1loopl} }\label{sec:cancellations-three-point function-simple}
\begin{figure}[t]
\center
 \includegraphics[width=.28\textwidth]{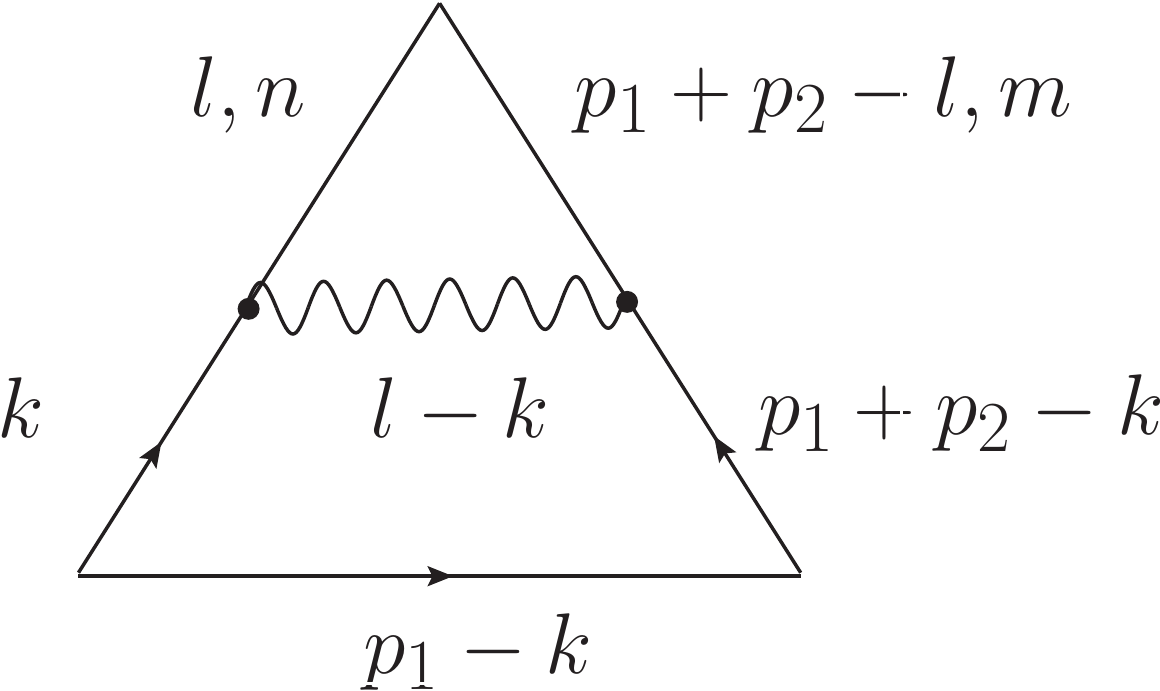} 
\caption{Momentum routing for diagram \ref{fig:3point1loopg} to obtain \eqref{eqn:diagrams-g}, the labels $m$ and $n$ designate the power of the momenta in the numerator of the expression, i.e. $(\hat{l})^n$ and $({\hat{p}_1+\hat{p}_2-\hat{l}})^m$.}
 \label{fig:momentum-routing-g}
\end{figure}

Choosing the momentum routing as shown in figure \ref{fig:momentum-routing-g}, diagram \ref{fig:3point1loopg} reads
\begin{align}\label{eqn:diagrams-g}
& \langle \op(p_1) \tilde{\op}(p_2) \hat{\op}_j \rangle^{\ref{fig:3point1loopg}}= \\ \nn
 &= \frac{1}{8}
4 \frac{i^2}{2!} i^{6+j} \left(i^2 f^{abc} f^{abc} \right) \int d^d \tilde{k},\tilde{l} ~ (\hat{p}_1+\hat{p}_2)^j C_j^{1/2} \left( \frac{2\hat{l}-(\hat{p}_1+\hat{p}_2)}{\hat{p}_2+\hat{p}_2} \right) \\ \nn
& \qquad\qquad\qquad\qquad\qquad \frac{(k+l)^\mu(-\eta_{\mu\nu})(2P-k-l)^\nu}{k^2 (P-k)^2 (l-k)^2 l^2 (P-l)^2 (p_1-k)^2}
\\ \nn
&= \frac{1}{4} g^8 N \delta^{aa} i^{j+2} \int d^d\tilde{k},\tilde{l}~ \frac{\hat{C}_j(p_1+p_2,l)}{(p_1+k)^2} \Big(-2P^2i(1,1,1,1,1,P) - i(1,1,0,1,1,P)\\ \nn
& \hspace{2.4cm}+i(0,1,1,1,1,P)  + i(1,0,1,1,1,P) + i(1,1,1,0,1,P) + i(1,1,1,1,0,P) \Big)\,,
\end{align}
where we have used $(l+k)\cdot(l+k+2P)=-2P^2 - (l-k)^2 +(P+l)^2+(P+k)^2+k^2 +l^2$, shifted $k\to-k$, $l\to-l$ and  introduced the notation 
\begin{align}\label{eqn:abbreviations-three-point-calculation}
i(a_1,a_2,a_3,a_4,a_5,P) &:= \frac{1}{k^{2a_1} (P+k)^{2a_2} (k-l)^{2a_3} l^{2a_4} (P+l)^{2a_5}}\,, \\ \nn
\hat{C}_j(p_1+p_2,l) &:=(\hat{p}_1+\hat{p}_2)^j C_j^{1/2}\left( \frac{2\hat{l} - (\hat{p}_1+\hat{p}_2)}{\hat{p}_1+\hat{p}_2} \right)\,,
\end{align}
in order to get a more compact notation.
The four-scalar diagram \ref{fig:3point1loopl} reads
\begin{align}\label{eqn:diagrams-l}
\langle \op(p_1) \tilde{\op}(p_2) \hat{\op}_j \rangle^{\ref{fig:3point1loopl}} 
&= \frac{1}{4}  g^8 N \delta^{aa}   i^{j+2} \int d^d\tilde{k},\tilde{l}~ \frac{\hat{C}_j(p_1+p_2,l)}{(p_1+k)^2} i(1,1,0,1,1,P)
\end{align}
and thus exactly cancels the second term in \eqref{eqn:diagrams-g}. 

\notocsubsection{Diagrams \ref{fig:3point1loopa}, \ref{fig:3point1loopc}}
Using the 1-loop corrected propagator \eqref{eqn:one-loop-correction-scalar-rpopagator-in-momentum-space}  the self-energy diagram \ref{fig:3point1loopa} multiplied with $1/2$ reads
\begin{equation}\label{eqn:three-point-self-energy-a-half}
\frac{1}{2}\langle  \op \tilde{\op} \hat{\op}_j\rangle^{\ref{fig:3point1loopa}} = \frac{g^8}{4}  N \delta^{aa}  i^{2+j} b_0(1,1) \int d^d \tilde{l} \frac{\hat{C}_j(p_1+p_2,l)}{(-l^2)^{3-d/2} (p_1+l)^2 (p_1+p_2+l)^2}\,.
\end{equation}
We can obtain the self-energy diagram \ref{fig:3point1loopc} by exchanging $p_1 \leftrightarrow p_2$
\begin{equation}\label{eqn:three-point-self-energy-c-half}
\frac{1}{2}\langle  \op \tilde{\op} \hat{\op}_j\rangle^{\ref{fig:3point1loopc}} = \frac{g^8}{4}   N \delta^{aa} i^{2+j} b_0(1,1) \int d^d\tilde{l} \frac{\hat{C}_j(p_1+p_2,l)}{(-l^2)^{3-d/2} (p_2+l)^2 (p_1+p_2+l)^2}\,.
\end{equation}
We will proceed with the analysis of these diagrams in section \ref{sec:j-not-0}.

\notocsubsection{Cancellations between diagrams \ref{fig:3point1loopa}, \ref{fig:3point1loopb}, \ref{fig:3point1loopj} and \ref{fig:3point1looph}}\label{sec:cancellation-of-BPS-diagrams}\label{sec:cancellations-three-point function}
The reduction of diagram \ref{fig:3point1loopj} cancels the contributions from half of the self-energy diagrams and the four-scalar interaction term such that only finite integrals remain. Since we consider $j$ even, the analysis is the same for \ref{fig:3point1loopi} and the corresponding self-energy and four-scalar diagrams.

\begin{figure}[t]
\center
 \includegraphics[width=.25\textwidth]{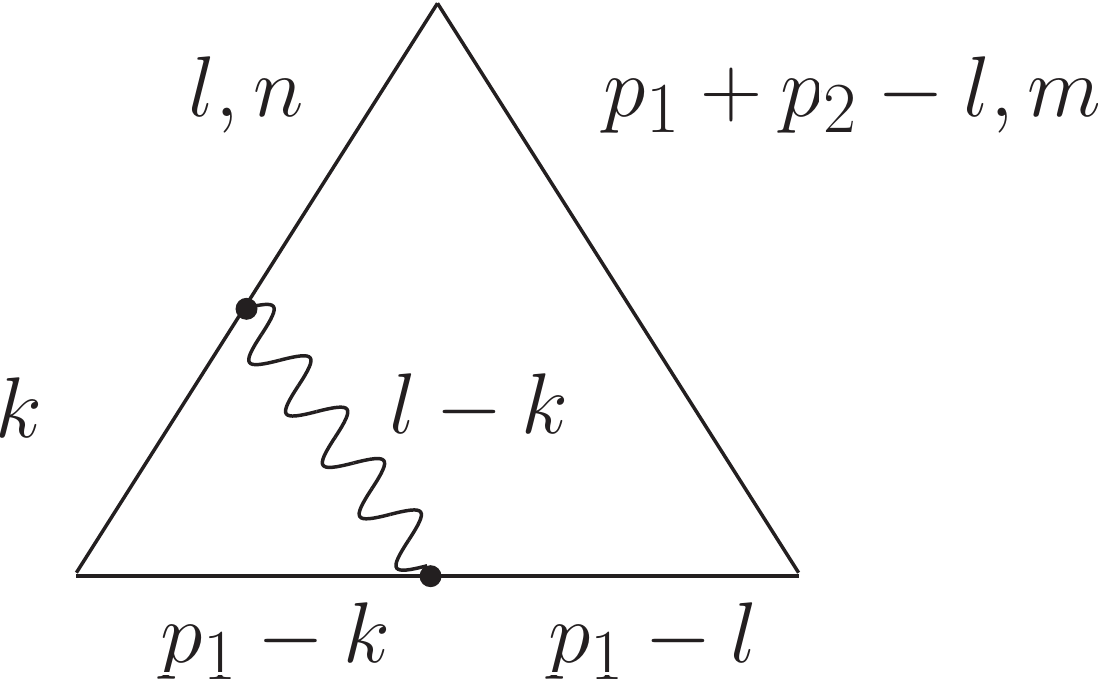} ~~~~~~~~~~~~~~~~~~~
 \includegraphics[width=.24\textwidth]{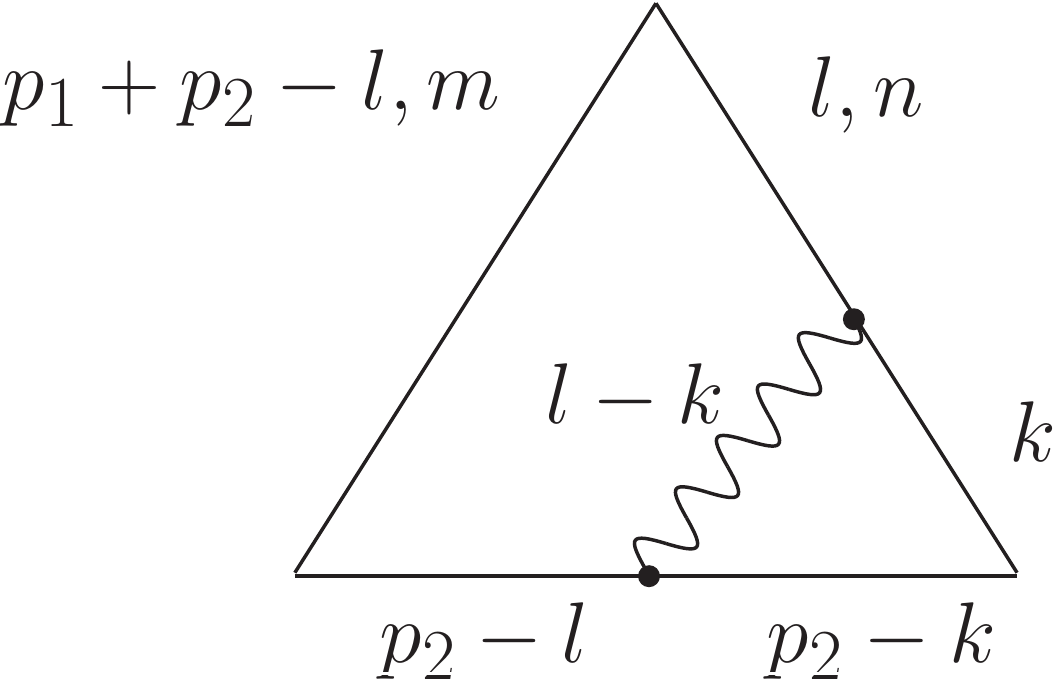} 
\caption{Momentum routing for diagram \ref{fig:3point1looph} to obtain \eqref{eqn:diagrams-h}, the labels $m$ and $n$ designate the power of the momenta in the numerator of the expression, i.e. $(\hat{l})^n$ and $({\hat{p}_1+\hat{p}_2-\hat{l}})^m$.}
 \label{fig:momentum-routing-h}
\end{figure}

\begin{figure}[b]
\center
 \includegraphics[width=.10\textwidth]{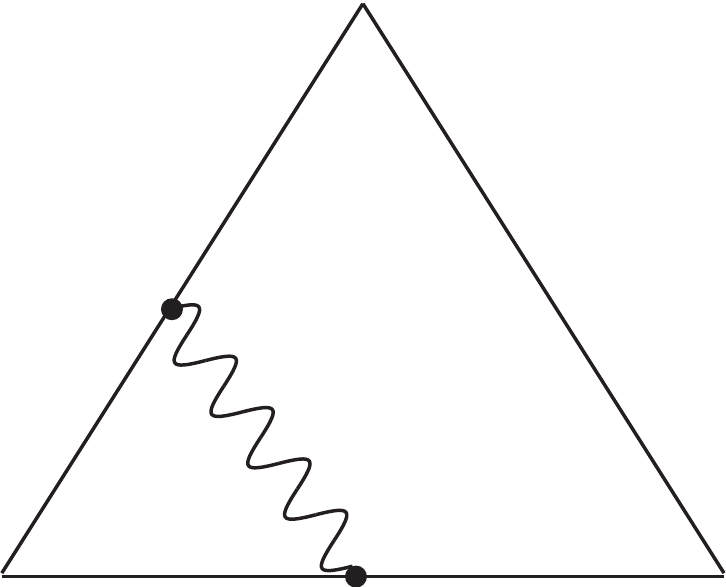} 
\raisebox{.45cm}{$= -2p_1^2$}
 \includegraphics[width=.10\textwidth]{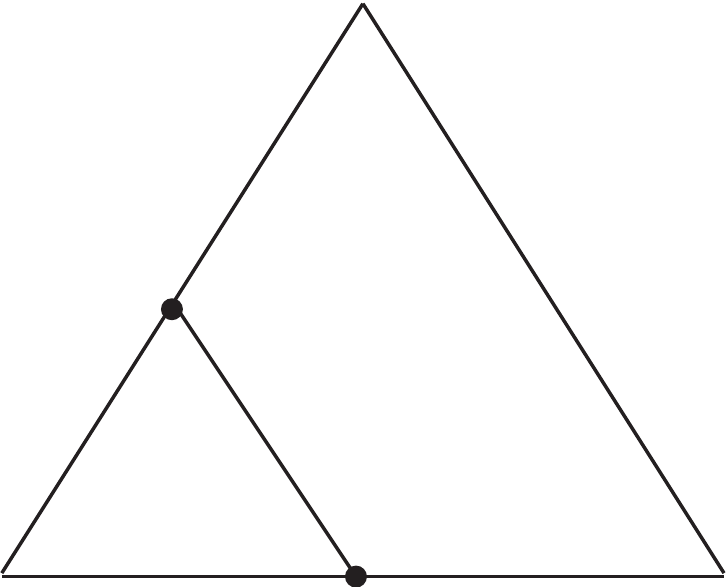} 
   \raisebox{.45cm}{$ -$}
 \includegraphics[width=.10\textwidth]{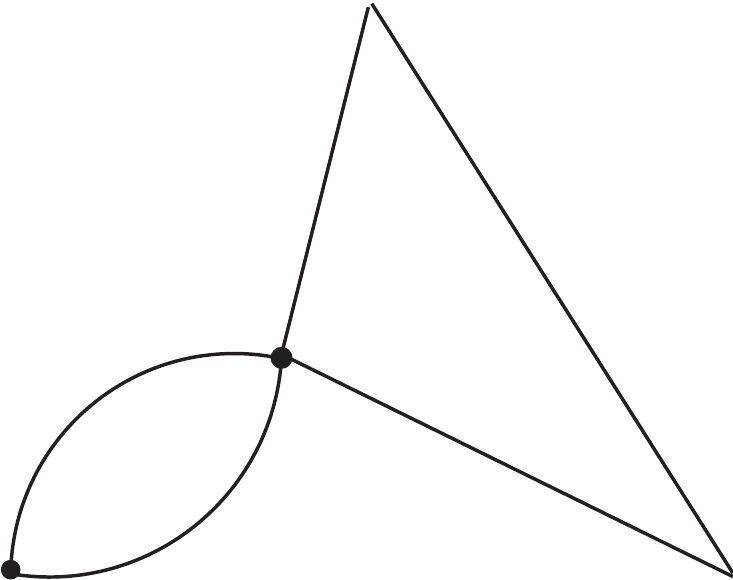} 
   \raisebox{.45cm}{$ +$}
 \includegraphics[width=.10\textwidth]{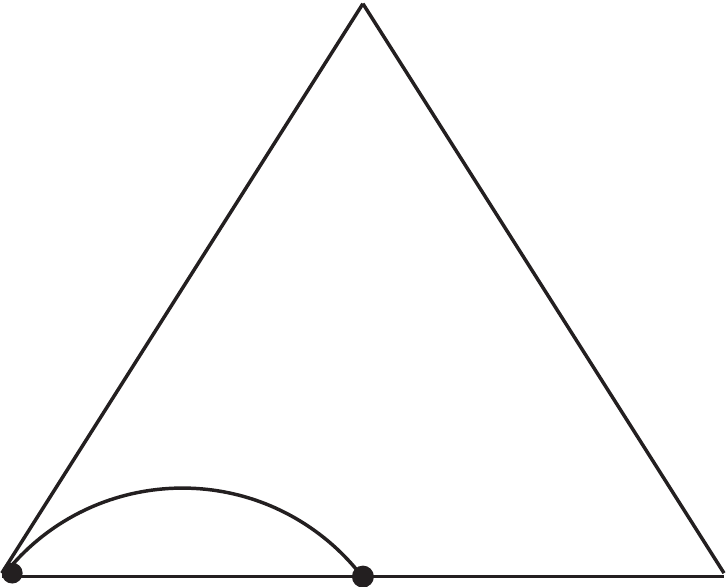} 
  \raisebox{.45cm}{$ +$}
 \includegraphics[width=.10\textwidth]{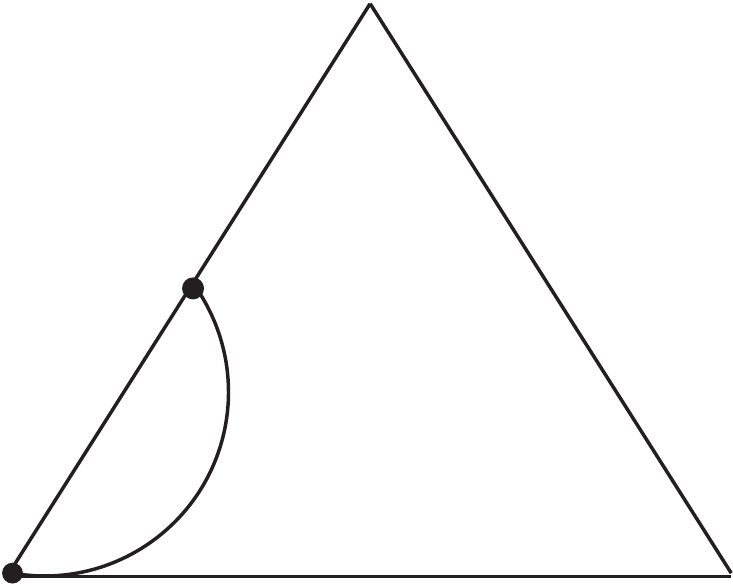} 
 \raisebox{.45cm}{$ +$}
 \includegraphics[width=.10\textwidth]{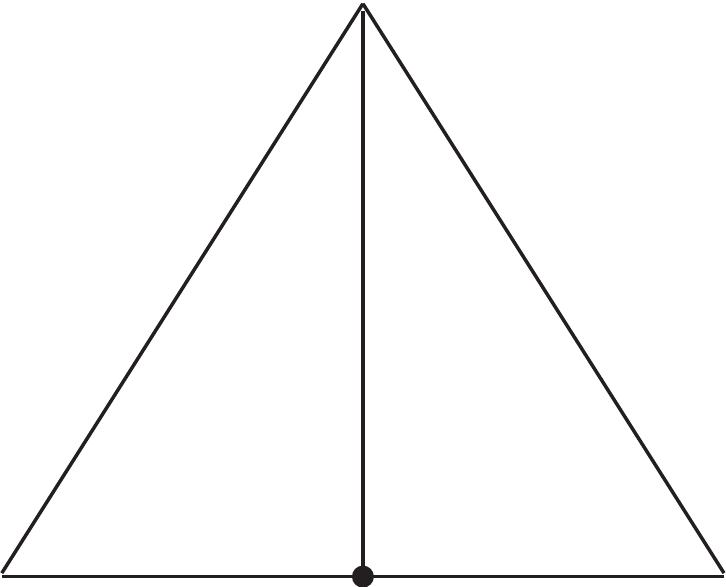} 
  \raisebox{.45cm}{$ +$}
 \includegraphics[width=.10\textwidth]{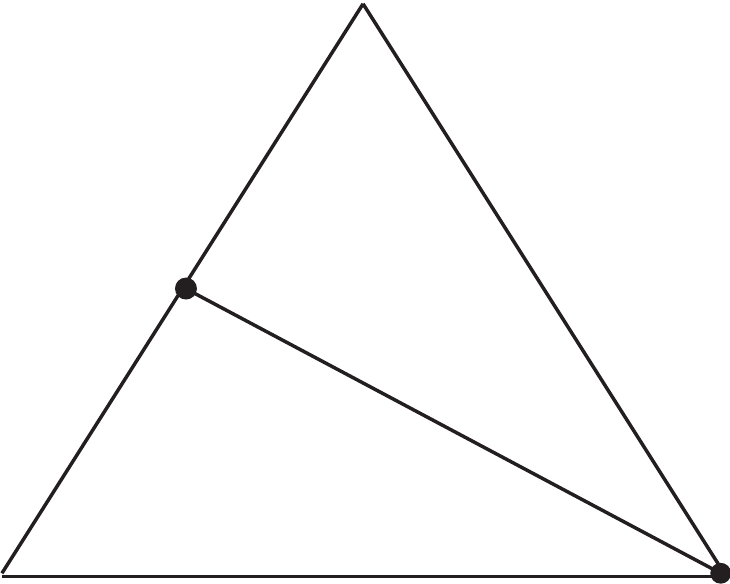} 
\caption{Due to the momenta from the gluon vertices, the integrand decomposes into simpler integrals, which cancel with the self-energy and four-scalar interaction terms. Note, that this has no effect on the other numerator momenta.}
\label{fig:decomposition-gluon-diagrams}
\end{figure}

We use the momentum routing shown in figure \ref{fig:momentum-routing-h} to obtain
\begin{align}\label{eqn:diagrams-h}
\langle \op(&p_1) \tilde{\op}(p_2) \hat{\op}_j\rangle^{\ref{fig:3point1looph}}=\\ \nn
 &= \frac{1}{8}4 \frac{(i)^2}{2!} i^j (i^2 f^{abd} f^{abd}) i^6 \int d^d \tilde{k},\tilde{l}\, (\hat{p}_1+\hat{p}_2)^j C_j^{1/2}\left( \frac{2\hat{l} - (\hat{p}_1+\hat{p}_2)}{\hat{p}_1+\hat{p}_2} \right) \\ \nn
 & \qquad\qquad\qquad\qquad\qquad\qquad \frac{(k+l)^\mu (-\eta_{\mu\nu}) (2p_1-k-l)^\nu}{k^2 (p_1-k)^2 (k-l)^2 l^2 (p_1-l)^2(p_1+p_2-l)^2} \\ \nn
&= \frac{1}{4} i^{2+j}  g^8 N \delta^{aa}  \int d^d \tilde{k},\tilde{l}\,  \frac{\hat{C}_j(p_1+p_2,l)}{(p_1+p_2+l)^2} \,i(1,1,1,1,1,p_1) (l+k)\cdot(l+k+2p_1) \\\nn
&= \frac{1}{4} i^{2+j}  g^8 N \delta^{aa}  \int d^d \tilde{k},\tilde{l}\,  \frac{\hat{C}_j(p_1+p_2,l)}{(p_1+p_2+l)^2}\Big(-2p_1^2 i(1,1,1,1,1,p_1) - i(1,1,0,1,1,p_1)\\ \nn
& \hspace{1cm}+i(0,1,1,1,1,p_1)  + i(1,0,1,1,1,p_1) + i(1,1,1,0,1,p_1) + i(1,1,1,1,0,p_1) \Big)\,,
\end{align}
where we used $(l+k)\cdot(l+k+2p_1)=-2p_1^2 - (l-k)^2 +k^2+(p_1+k)^2 +l^2+(p_1+l)^2$ in the last line as well as the abbreviations \eqref{eqn:abbreviations-three-point-calculation}. The origin of this decomposition is illustrated in figure \ref{fig:decomposition-gluon-diagrams}. As we will see, the second, third and fourth term in the brackets cancels against contributions from the self-energy and four-scalar interaction diagrams.
~\\
The four-scalar interaction diagram \ref{fig:3point1loopj} is
\begin{align}\label{eqn:diagrams-j}\nn
\langle  \op(p_1) & \tilde{\op}(p_2) \hat{\op}_j \rangle^{\ref{fig:3point1loopj}}=\\ \nn &= \frac{1}{8} 2 \frac{i}{1!} \left(- \frac{1}{g^2} (f^{abc}f^{acb}+0)\right) i^5 i^j \int d^d \tilde{k},\tilde{l}~ \hat{C}_j(p_1+p_2,l) \frac{i(1,1,0,1,1,p_1)}{(p_1+p_2+l)^2} \\ \nn
&= \frac{1}{4}i^{2+j}  g^8 N \delta^{aa}  \int d^d \tilde{k},\tilde{l}~ \hat{C}_j(p_1+p_2,l)  \frac{i(1,1,0,1,1,p_1)}{(p_1+p_2+l)^2}
\end{align}
and cancels exactly against the second contribution in the brackets in the last line of \eqref{eqn:diagrams-h}.

We have calculated the self-energy diagram \ref{fig:3point1loopa}  in \eqref{eqn:three-point-self-energy-a-half}
and now see that half of this term cancels with the fourth contribution in the brackets in the last line of \eqref{eqn:diagrams-h} by either solving the integral in \eqref{eqn:diagrams-h} or \emph{undoing} the integral from the one-loop corrected propagator. Choosing the latter way and using therefore
\begin{equation}
\int d^d\tilde{k} \frac{1}{(-k^2)(-(l-k)^2)(-l^2)} = \frac{b_0(1,1)}{(-l^2)^{3-d/2}}\,,
\end{equation}
we can thus rewrite \eqref{eqn:three-point-self-energy-a-half} as
\begin{equation}\label{eqn:three-point-self-energy-a-half-undone}
\frac{1}{2}\langle \op(p_1) \tilde{\op}(p_2) \hat{\op}_j \rangle^{\ref{fig:3point1loopa}} = -\frac{1}{4}i^{2+j}  g^8 N \delta^{aa}  \int d^d \tilde{k},\tilde{l}~ \hat{C}_j(p_1+p_2,l) \frac{i(1,0,1,1,1,p_1)}{ (p_1+p_2+l)^2}
\end{equation}
and see that this term indeed cancels with the fourth contribution in the brackets in the last line of \eqref{eqn:diagrams-h}. We repeat the same steps for the self-energy diagram \ref{fig:3point1loopb} multiplied with $1/2$ and find
\begin{equation}\label{eqn:three-point-self-energy-b-half}
\frac{1}{2}\langle  \op(p_1) \tilde{\op}(p_2) \hat{\op}_j \rangle^{\ref{fig:3point1loopb}} = -\frac{1}{4} i^{2+j}   g^8 N \delta^{aa}   \int d^d \tilde{k},\tilde{l}~ \hat{C}_j(p_1+p_2,l) \frac{i(0,1,1,1,1,p_1)}{ (p_1+p_2+l)^2}\,,
\end{equation}
which thus cancels exactly with the third contribution in the brackets in the last line of \eqref{eqn:diagrams-h}. Thus the only terms that remain in \eqref{eqn:diagrams-h} are the first and the last two terms, i.e. we can write
\begin{align}\label{eqn:diagramsh-k-a-b-remainder}
& \langle \op \tilde{\op} \hat{\op}_j \rangle^{\ref{fig:3point1looph}} + \langle \op \tilde{\op} \hat{\op}_j \rangle^{\ref{fig:3point1loopj}} + \frac{1}{2}\langle \op \tilde{\op} \hat{\op}_j \rangle^{\ref{fig:3point1loopa}} + \frac{1}{2}\langle \op \tilde{\op} \hat{\op}_j \rangle^{\ref{fig:3point1loopb}} \\ \nn
&= \frac{1}{4} i^{2+j}  g^8 N \delta^{aa}  \int d^d \tilde{k},\tilde{l}\,  \frac{\hat{C}_j(p_1+p_2,l)}{(p_1+p_2+l)^2}\\ \nn
& \qquad\qquad \qquad\qquad \qquad \Big(-2p_1^2 i(1,1,1,1,1,p_1)  + i(1,1,1,0,1,p_1) + i(1,1,1,1,0,p_1) \Big)\,.
\end{align}

\notocsubsection{Cancellations between Diagrams \ref{fig:3point1loopb}, \ref{fig:3point1loopc}, \ref{fig:3point1loopk} and \ref{fig:3point1loopi}}\label{sec:cancellations-three-point function mirrored}
The same analysis as in \ref{sec:cancellation-of-BPS-diagrams} can be repeated for diagrams \ref{fig:3point1loopb}, \ref{fig:3point1loopc}, \ref{fig:3point1loopk} and \ref{fig:3point1loopi} which have the same expressions under the exchange $p_1 \leftrightarrow p_2$ and thus we find 
\begin{align}\label{eqn:remainder-l-j-b-c}
& \langle \op \tilde{\op} \hat{\op}_j \rangle^{\ref{fig:3point1loopk}} + \langle \op \tilde{\op} \hat{\op}_j \rangle^{\ref{fig:3point1loopi}} + \frac{1}{2}\langle \op \tilde{\op} \hat{\op}_j \rangle^{\ref{fig:3point1loopb}} + \frac{1}{2}\langle \op \tilde{\op} \hat{\op}_j \rangle^{\ref{fig:3point1loopc}} \\ \nn
&= \frac{1}{4} i^{2+j}  g^8 N \delta^{aa}  \int d^d \tilde{k},\tilde{l}\,  \frac{\hat{C}_j(p_1+p_2,l)}{(p_1+p_2+l)^2}\\ \nn 
& \qquad\qquad\qquad\qquad \Big(-2p_2^2 i(1,1,1,1,1,p_2)  + i(1,1,1,0,1,p_2) + i(1,1,1,1,0,p_2) \Big)\,.
\end{align}

\notocsubsection{Protected Three-Point Function for $j=0$}
For the special case $j=0$, all contributions cancel, since then the operator at $x_3$ is also a protected operator, see section \ref{sec:protected-operator}. 
In this case diagrams \ref{fig:3point1loope}, \ref{fig:3point1loopd}, \ref{fig:3point1loopf} do not appear, since there is no covariant derivative from the operator at $x_3$ and furthermore $\hat{C}_{j=0} = 1$. Furthermore, for $j=0$, the third and fifth term in \eqref{eqn:diagrams-g} exactly cancel with the self-energy diagrams. If we rewrite \ref{fig:3point1loopa} in terms of \eqref{eqn:three-point-self-energy-a-half-undone} and correspondingly  for \ref{fig:3point1loopc} the cancellation for $j=0$  is trivial. Then, the remaining contributions are
\begin{align}\nn
&\langle  \op \tilde{\op} \hat{\op}_{j=0} \rangle^{(1)} \\ \nn
&= \frac{1}{4}(i)^2  g^8 N \delta^{aa} \int d^d\tilde{k},\tilde{l} ~\left(\frac{-2P^2i(1,1,1,1,1,P)}{(p_1+k)^2}   + \frac{i(0,1,1,1,1,P)}{(p_1+k)^2}  + \frac{i(1,0,1,1,1,P)}{(p_1+k)^2} \right) \\ \nn
&\qquad\qquad +\left(\frac{-2p_1^2 i(1,1,1,1,1,p_1)}{(p_1+p_2+l)^2}+\frac{i(1,1,1,0,1,p_1) }{(p_1+p_2+l)^2}+\frac{ i(1,1,1,1,0,p_1)}{(p_1+p_2+l)^2} + (p_1 \leftrightarrow p_2) \right)\,.
\end{align}
From the definition \eqref{eqn:abbreviations-three-point-calculation} one easily verifies
\begin{align}\label{eqn:id-one}
\int d^d\tilde{k},\tilde{l}~ \frac{i(1,0,1,1,1,P)}{(p_1+k)^2} = \int d^d\tilde{k},\tilde{l}~ \frac{i(1,1,1,1,0,p_1)}{(p_1+p_2+l)^2}\,, \\ 
\label{eqn:id-two}
\int d^d\tilde{k},\tilde{l}~ \frac{i(0,1,1,1,1,P)}{(p_1+k)^2} = \int d^d\tilde{k},\tilde{l}~ \frac{i(1,1,1,1,0,p_2)}{(p_1+p_2+l)^2}\,, \\ 
\label{eqn:id-three}
\int d^d\tilde{k},\tilde{l}~ \frac{i(1,1,1,0,1,p_1)}{(p_1+p_2+l)^2} = \int d^d\tilde{k},\tilde{l}~ \frac{i(1,1,1,0,1,p_2)}{(p_1+p_2+l)^2} \,.
\end{align}
The second equality can be obtained by changing variables $k \to -k-p_1-p_2 $, $l \to -l-p_1-p_2 $ and the last equality by the same change of variables and an additional renaming $k \leftrightarrow l$.

\begin{figure}[h]
\center
 \includegraphics[width=.45\textwidth]{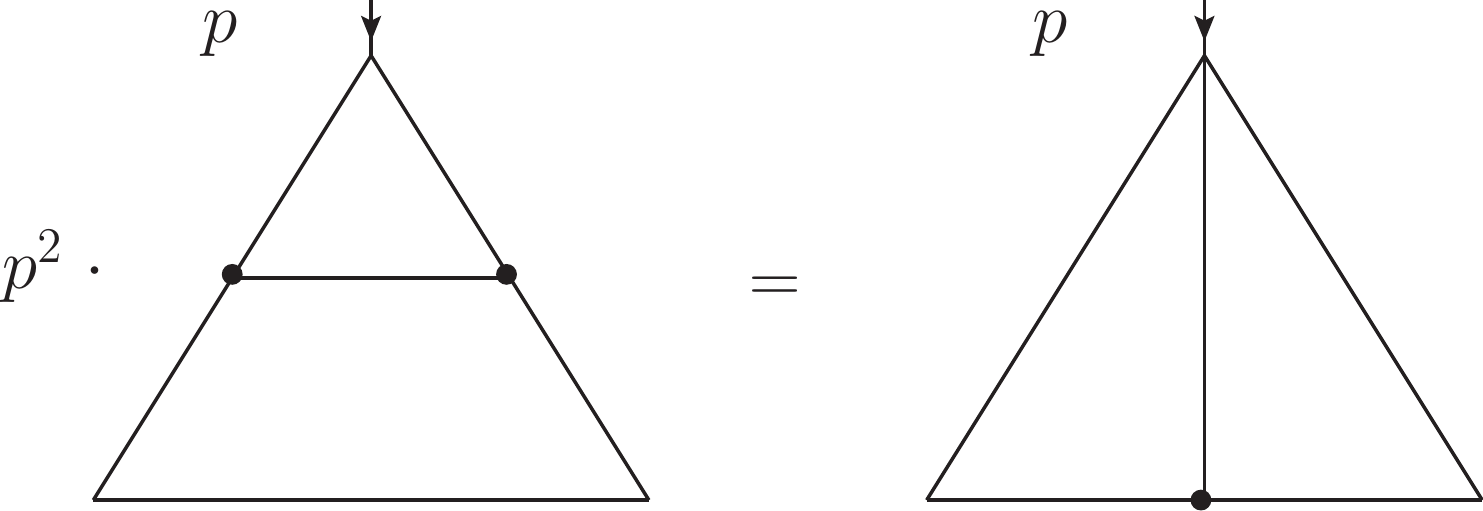} 
\caption{Relation for scalar three-point integrals derived in \cite{Usyukina:1994iw}. It would be desirable to derive a modified identity for diagrams with numerator momenta.}
\label{fig:relation_three-point_integrals}
\end{figure}

Now we can use the relation shown in figure \ref{fig:relation_three-point_integrals}
that was derived in \cite{Usyukina:1994iw} and all diagrams cancel, e.g.
\begin{equation}
\int d^d\tilde{k},\tilde{l}~ \left(\frac{-2P^2i(1,1,1,1,1,P)}{(p_1+k)^2} +  2\frac{i(1,1,1,0,1,p_2)}{(p_1+p_2+l)^2}  \right) =0
\end{equation}
and thus all one-loop contributions cancel
\begin{equation}
\langle  \op \tilde{\op} \hat{\op}_{j=0} \rangle^{(1)} = 0\,.
\end{equation}

\notocsubsection{Implementation of the Limit $p_1+p_2\to0$}\label{sec:j-not-0}
We can now safely take the limit $p_1+p_2\to 0$, since the only diagram that could lead to ambiguities in this limit is the four-scalar diagram \ref{fig:3point1loopl} which has been cancelled by a contribution from \ref{fig:3point1loopg}. For the same reason as in \eqref{eqn:tree-level-structure-constant} we only need one term from the Gegenbauer polynomials, since
\begin{equation}
\lim_{p_1+p_2\to0}\hat{C}_j(p_1+p_2,l) = c_{jj}^{1/2} (2\hat{l})^j\,,
\end{equation}
where $c_{jj}^{1/2}$ is the coefficient with the highest power of the Gegenbauer polynomial that was given in \eqref{eqn:highest-power-gegenbauer}. Furthermore, we have
\begin{equation}
\frac{i(a_1,a_2,a_3,a_4,a_5,P\to0)}{(k+p_1)^2} = i(a_1+a_2,1,a_3,a_4+a_5,0,p_1)
\end{equation}
and the integrals turn into simple two-point bubble integrals that are easily solved, see Appendix \ref{sec:bubble-integrals}. Denoting from now on $p_1=p$ we find
\begin{align}\label{eqn:diagram-g-plus-j}
\langle \op\tilde{\op} \hat{\op}_j \rangle^{\ref{fig:3point1loopg}} +\langle & \op\tilde{\op} \hat{\op}_j \rangle^{\ref{fig:3point1loopl}} = \\ \nn 
&\qquad= - c_{jj}^{1/2} 2^{j-1} i^{2+j}  g^8 N \delta^{aa}  b_j(4-\frac{d}{2},1)\left(b_j(2,1)+b_j(1,1) \right) \frac{\hat{p}^j}{(-p^2)^{5-d}}\,.
\end{align}
The self-energy diagram multiplied with one half \eqref{eqn:three-point-self-energy-a-half} becomes
\begin{equation}\label{eqn:diagram-a-in-limit}
\frac{1}{2} \langle \op\tilde{\op} \hat{\op}_j \rangle^{\ref{fig:3point1loopa}} = c_{jj}^{1/2} i^{2+j}  g^8 N \delta^{aa}  2^{j-2} b_0(1,1) b_j(4-\frac{d}{2},1) \frac{\hat{p}^j}{(-p^2)^{5-d}}\,.
\end{equation}
Since $p_1=-p_2=p$ and $j$ even we find the same result for the other remaining half self-energy diagram 
\begin{equation}
\frac{1}{2} \langle \op\tilde{\op} \hat{\op}_j \rangle^{\ref{fig:3point1loopc}} = c_{jj}^{1/2} i^{2+j}  g^8 N \delta^{aa}  2^{j-2} b_0(1,1) b_j(4-\frac{d}{2},1) \frac{\hat{p}^j}{(-p^2)^{5-d}}\,.
\end{equation}
Applying the limit to the diagrams in \eqref{eqn:diagramsh-k-a-b-remainder} we find
\begin{align}\label{eqn:diagram-l-j-b-c}
& \langle \op \tilde{\op} \hat{\op}_j \rangle^{\ref{fig:3point1looph}} + \langle \op \tilde{\op} \hat{\op}_j \rangle^{\ref{fig:3point1loopj}} + \frac{1}{2}\langle \op \tilde{\op} \hat{\op}_j \rangle^{\ref{fig:3point1loopa}} + \frac{1}{2}\langle \op \tilde{\op} \hat{\op}_j \rangle^{\ref{fig:3point1loopb}} \\ \nn
& \quad =  i^{2+j}  g^8 N \delta^{aa}  2^{j-2} c_{jj}^{1/2} \left( 2c_{0j}(1,1,1,2,1) + c_{0j}(1,1,1,1,1)+ c_{0j}(1,1,1,2,0) \right) \frac{\hat{p}^j}{(-p^2)^{5-d}}\,.
\end{align}
The integrals $c_{nm}(a_1,a_2,a_3,a_4,a_5)$ are defined in \eqref{eqn:definition-cmn} and solved in Appendix \ref{sec:appendix-integrals}  using the IBP technique.
All these integrals are finite.  Since $j$ is even we get the same contribution for $p_2=-p$ when taking the limit in \eqref{eqn:remainder-l-j-b-c}.

\notocsubsection{Diagrams with One Covariant Derivative}\label{sec:additional-diagrams}
For $j\neq 0$ we have additional diagrams with exactly one gauge field from the covariant derivative in $\hat\op_j$. We derive the form of the operator with exactly one gauge field in Appendix \ref{sec:derivation-operator-with-exactly-one-gaueg-field} and find that it can be written as
\begin{align}\label{eqn:twist-operator-one-gauge-field-intro}
\hat{\op}_j^{\text{A}} = \frac{1}{2} f^{abc} \left(1+(-1)^j\right) 
\left(  \sum_{k=1}^{j}a_{jk}^{1/2}\sum_{m=1}^k{k \choose m} \hat{\partial}^{m-1} \hat{A}^a \,\hat{\partial}^{k-m} \phi^{12,b} \, \hat{\partial}^{j-k} \phi^{12,c} 
 \right)\,.
\end{align}
The diagrams are shown in figure \ref{fig:3point1loopf},\ref{fig:3point1loope}, \ref{fig:3point1loopd}. We will directly calculate these diagrams in the limit $p_1+p_2\to0$. \\

\notocsubsection{Diagrams \ref{fig:3point1loope} and \ref{fig:3point1loopd}}
Diagram \ref{fig:3point1loope} and \ref{fig:3point1loopd} have the same expression under the exchange $p_1 \leftrightarrow p_2$ and in the limit $p_1 = -p_2$ they yield identical contributions, since we consider $j$ even.
We start with diagram \ref{fig:3point1loope}, which arises from two different contractions with the operator with one gauge field \eqref{eqn:twist-operator-one-gauge-field} for $j$ even.
\begin{equation}
\hat{\op}_j^{\text{A}} = f^{abc} 
\left(  \sum_{k=1}^{j}a_{jk}^{1/2}\sum_{m=1}^k{k \choose m} \hat{\partial}^{m-1} \hat{A}^a \hat{\partial}^{k-m} \phi^{12,b}  \hat{\partial}^{j-k} \phi^{12,c} 
 \right)\,.
\end{equation}
We thus get
\begin{align}\nn
\langle \op(p) & \tilde{\op}(-p) \hat{\op}_j(0) \rangle^{\ref{fig:3point1loope}} =\\ \nn
&= i^{6+j} \frac{ g^8 N \delta^{aa} }{4} \sum_{k=1}^j a_{jk} \sum_{m=1}^k {k \choose m} \int d^d\tilde{k},\tilde{l}~ \frac{(\hat{k}+\hat{l})\left((\hat{l}-\hat{k})^{m-1} (-\hat{l})^{k-m}\hat{k}^{j-k} +(k \leftrightarrow l)\right)}{k^4 (p+k)^2 (l-k)^2 l^2}\\ \nn
&= i^{6+j} \frac{ g^8 N \delta^{aa} }{4} \sum_{k=1}^j a_{jk} \sum_{m=1,n=0}^{k,m-1} {k \choose m} {m-1 \choose n} (-1)^{n+k-m} \\ \nn
& \hspace{5cm} \int d^d\tilde{k},\tilde{l}~ \frac{\left(\hat{l}^{k-n}\hat{k}^{j-k+n} +\hat{l}^{k-n-1}\hat{k}^{j-k+n+1} +(k \leftrightarrow l)\right)}{-(-k^2)^2 (-(p+k)^2) (-(l-k))^2 (-l^2)}\,.
\end{align}
The integrals are simple two-point integrals
\begin{equation}\label{eqn:diagram-7-d-in-limit}
\langle \op(p) \tilde{\op}(-p) \hat{\op}_j(0) \rangle^{\ref{fig:3point1loope}} = - i^{6+j} \frac{ g^8 N \delta^{aa} }{4} b_j(4-d/2,1) \frac{\hat{p}^j}{(-p^2)^{5-d}} ~ S^{\ref{fig:3point1loope}}[j]\,,
\end{equation}
where
\begin{align}
S^{\ref{fig:3point1loope}}[j]= \sum_{k=1}^j a_{jk} \sum_{m=1}^{k} & \sum_{n=0}^{m-1}{k \choose m} {m-1 \choose n}(-1)^m \\ \nn 
&\qquad \left[ b_{k-n}(1,1)-b_{k-n-1}(1,1) +b_{j-k+n}(1,1)-b_{j-k+n+1}(1,1)  \right]\,.
\end{align}
We can solve at least the sum over $n$ using the relation \eqref{eqn:identity-beta-function}. 

%Now we extract the leading pole. We find that the sum is consistent with
%\begin{equation}
%S^{\ref{fig:3point1loope}}[j] = - i \frac{1}{\epsilon} \frac{2 \Gamma(2j)}{\Gamma(j)\Gamma(j+1)} \left(H_j +H_{j+1} -1 \right) + \op(\epsilon^0)\,.
%\end{equation}
%Using $b_j(4-d/2,1)= i (-1)^j \Gamma(j)/\Gamma(j+1) + \op(\epsilon) $ we can thus write
%\begin{equation}\label{eqn:pole-term-diagram-d}
%\langle \op(p) \tilde{\op}(-p) \hat{\op}_j(0) \rangle^{\ref{fig:3point1loope}} =  \frac{i^{j}}{\epsilon} \frac{ g^8 N \delta^{aa} }{2} \frac{\Gamma(2j)}{\Gamma(j+1)\Gamma(j+1)} \left(H_j +H_{j+1} -1 \right) \frac{1}{(4\pi)^{d}}  \frac{\hat{p}^j}{(-p^2)^{5-d}} + \op(\epsilon^0) 
%\end{equation}

\notocsubsection{Diagram  \ref{fig:3point1loopf}}
Diagram  \ref{fig:3point1loopf} is calculated in a very similar way and the two possible contractions yield
\begin{align}
\langle \op(p) \tilde{\op}(-p) \hat{\op}_j(0) \rangle^{\ref{fig:3point1loopf}}  &= i^{6+j} \frac{ g^8 N \delta^{aa} }{4} \sum_{k=1}^j a_{jk} \sum_{m=1}^k {k \choose m}(-1)^{k-m} \\ \nn
& \qquad \int d^d\tilde{k},\tilde{l}~ \frac{(2\hat{p}+\hat{k}+\hat{l})\left((\hat{k}-\hat{l})^{m-1} \hat{k}^{k-m}\hat{l}^{j-k} +(k \leftrightarrow l)\right)}{k^2 (p+k)^2 (l-k)^2 l^2(p+l)^2}\,.
\end{align}
Introducing another binomial sum for rewriting $(k-l)^{m-1}$ and noting that due to $f_{nm}(1,1,1,1,1)$ the term with $(k \leftrightarrow l)$ yields the same contribution we have
\begin{align}\label{eqn:diagram-7-f-in-limit} \nn
\langle \op(p) \tilde{\op}(-p) \hat{\op}_j(0) \rangle^{\ref{fig:3point1loopf}}  &= i^{6+j} \frac{ g^8 N \delta^{aa} }{2} \sum_{k=1}^j a_{jk} \sum_{m=1}^k \sum_{n=0}^{m-1} {k \choose m} {m-1 \choose n}(-1)^{k-m+n} \\
& \qquad \Big( 2 \hat{p} f_{k-1-n,j-k+n}+f_{k-n,j-k+n}+f_{k-1-n,j-k+n+1}\Big)
\end{align}
and all integrals $f_{nm}(1,1,1,1,1)$ are finite and were solved in \ref{sec:Integrals for diagrams b}.

\subsection{Full Bare Three-Point Function}
Taking into account the exact cancellations between diagrams, as well as the fact that diagrams with $p_1 \leftrightarrow p_2$ are identical in the limit $p_1\to -p_2$ for $j$ even, the remaining bare one-loop contribution to the three-point function is given by
\begin{align}\label{eqn:bare-three-point-function}
\langle \op \tilde{\op} \hat{\op}_j \rangle^{(1)} &= \sum_{\alpha= a..l} \langle \op \tilde{\op} \hat{\op}_j \rangle^{\ref{fig:three-point function at 1-loop}\alpha} \\ \nn
&= 2 \times \eqref{eqn:diagram-a-in-limit} + 2   \times \eqref{eqn:diagram-7-d-in-limit} +   \eqref{eqn:diagram-7-f-in-limit} + \eqref{eqn:diagram-g-plus-j}+ 2  \times  \eqref{eqn:diagram-l-j-b-c}\,.
\end{align}
We have calculated all diagrams in momentum space. In order to read off the structure constant we Fourier transform the expression to position space using
\begin{equation}\label{eqn:FT-to-position-space}
\int \frac{d^dp}{(2\pi)^d} \frac{\hat{p}^j}{(-p^2)^{5-d}} e^{-i p x} = i \frac{\Gamma(3d/2-5+j)}{\Gamma(5-d) } \frac{(2i)^j }{4^{5-d}\pi^{d/2}} \frac{\hat{x}^j}{(-x^2)^{3h-5+j}}
\end{equation}
We simply plug in all these equations into a Mathematica file and find that the result for the normalisation invariant structure constant is given by \eqref{eqn:result}.

\subsection{Anomalous Dimension from Three-Point Function}
As a check of the relative factors and signs between the diagrams we can read off the anomalous dimension from the three-point function. 
It is interesting to note that even though all diagrams in \ref{fig:three-point function at 1-loop} except the one in \ref{fig:3point1loopf} are divergent, the divergences \emph{localised} at the BPS operators exactly cancel due to the cancellations described in sections \ref{sec:cancellations-three-point function}, \ref{sec:cancellations-three-point function mirrored}. This is what one would expect, since the only non-BPS operator is $\hat{\op}_j$ and the divergences should be \emph{localised} at this operator. 
It turns out that we get the following contributions to the anomalous dimension
\begin{equation}\nn
\gamma_j^{(\ref{fig:3point1loopa}+\ref{fig:3point1loopc})/2}  = \frac{g^2N}{4\pi^2},~~   \gamma_j^{\ref{fig:3point1loopg}+\ref{fig:3point1loopl}}  = \frac{g^2N}{4\pi^2}\left(-\frac{1}{j+1}\right),~~  \gamma_j^{\ref{fig:3point1loope}+\ref{fig:3point1loopd}}  = \frac{g^2N}{4\pi^2}\left(2 H_j + \frac{1}{j+1}-1 \right)\,,
\end{equation} 
very similar to the two-point case in section \ref{sec:anomalous-dimension-twist-two}. The sum of all contributions correctly yields the first order of the anomalous dimension
\begin{align}\label{eqn:anomalous-dimension-Z-factor}
\gamma_j =  \left(\frac{g^2N}{4\pi^2}\right) 2H_j + \op(g^4)\,.
\end{align}
Some more details can be found in Appendix \ref{sec:details-anomalous-dimension-from-three-point-function}.

\chapter{Conclusions and Outlook}\label{sec:conclusions}
In the first part of this thesis, we have analysed light-like polygonal Wilson loops in Chern-Simons and ABJM theory and found remarkable similarities to the Wilson loop in $\syml$. We found that the two-loop expectation value of the $n$-sided light-like Wilson loop in ABJM theory \eqref{eqn:scs-two-loop-wl} has the same functional form as the $n$-sided light-like polygonal Wilson loop in $\syml$ theory at one-loop order \eqref{eqn:n-4-one-loop-wl}. Furthermore, for $n=4$, the expectation value precisely agrees with the form of the expectation value of the two-loop calculation of four-point scattering amplitudes in ABJM theory.
We have shown, that the form of the Wilson loop is fixed by anomalous conformal Ward identities. Starting from $n=6$, an arbitrary \emph{remainder function} of the conformally invariant cross ratios could potentially contribute. Using numerical analysis we have found however, that this function is a trivial constant. We have also checked that this statement is true for $n=8,10,12,14$ and expect it to hold for all $n$. This fact is decisive for the equality to Wilson loops in $\syml$ theory, where the remainder function at one loop is also trivial. Interestingly, this follows a similar pattern in the spectrum of planar anomalous dimensions persisting to all orders in the coupling.

Furthermore, recent one-loop six-point amplitude calculations in ABJM theory \cite{Bargheer:2012cp,Bianchi:2012cq}  have revealed non-vanishing results in contrast to the vanishing of the Wilson loop in ABJM theory. Together with the fact, that no T-self-duality of the type IIA string on $AdS_4 \times \mathbb{CP}^3$ could be found so far, this might indicate that there is no amplitude / Wilson loop duality in ABJM theory.  

On the other hand, in the $\sym$ duality the bosonic Wilson loop matches the tree-level stripped MHV amplitudes and for the matching of non-MHV amplitudes the introduction of a supersymmetric Wilson loop is necessary.
Due to the lack of helicity in the three-dimensional ABJM theory, one may thus suspect, that a suitable supersymmetric Wilson loop is necessary in order to find a relation between the Wilson loop and higher-point amplitudes. It would be interesting, whether one can find a Wilson loop that matches the six-point amplitudes at one loop.

In this context one should also keep in mind, that non-trivial evidence for a duality between Wilson loops and correlators in ABJM theory was found  at one-loop level in \cite{Bianchi:2011rn}. 

Furthermore, it would be desirable to extend the knowledge on amplitudes in ABJM theory to higher loops and legs. It would also be interesting to perform a four-point amplitude or Wilson loop computation at four loops, to check whether the conjectured BDS-like ansatz \cite{Bianchi:2011dg} for the four-point amplitude in ABJM theory indeed holds. Additionally, it would be very interesting to investigate the symmetry of the integrands found in \cite{Eden:2011we} for $\mathcal{N}=4$ SYM also in ABJM theory and possibly make use of this in order to determine higher-loop correlation functions.
The possible duality between correlators and Wilson loops in ABJM theory has only been tested at one-loop order and it would certainly be interesting to compare a two-loop calculation of correlators to the Wilson loop result.

Furthermore, it is fair to say, that it has not yet been fully understood, why the duality between amplitudes and Wilson loops in $\sym$ works at all. It would be interesting to see, what makes the integrands identical and whether there are additional hidden symmetries that fully fix the perturbative calculations.\\

In the second part of this thesis we have calculated three-point functions of two half-BPS operators  and one spin $j$ operator. It turns out that the one-loop corrections \eqref{eqn:result-normalization-invariant-structure-constants} to the structure constants have a simple result  in terms of generalised harmonic sums. 

For the calculation we have used a special infrared limit, where the momentum $P$ going in at the twist-two operator $\hat{\op}_j$ is zero. One simplification in this limit is, that the three-point calculation is effectively reduced to the calculation of two-point functions. Furthermore, a large simplification occurs due to the fact, that the one-loop mixing matrix drops out completely in this limit. we have shown that the result matches a previous result in the literature, which is obtained from an application of the operator product expansion on four-point correlators and thus independently confirm this result by our direct calculation. Furthermore, our procedure can easily be applied to the calculation of other structure constants, e.g. the calculation of three-point functions involving two operators with spin. It would also be interesting to apply integrability methods to the calculation of three-point functions of operators with spin.

\chapter*{Acknowledgements}\addcontentsline{toc}{chapter}{Acknowledgements}
First of all, I would like to thank my supervisor Jan Plefka for the possibility to write this thesis on my subject of choice, for supporting this work and always being available for helpful discussions. Furthermore, I am grateful for the many opportunities to take part in international summer schools and conferences that were given to me and my fellow PhD students. The atmosphere was very collaborative and I will remember the time in this research group as a very pleasant one.

Furthermore, I am highly indebted to Johannes Henn who let me profit from his knowledge, answered many questions on his previous work and always had helpful comments for performing calculations more efficiently. Furthermore, I especially appreciate discussions on the calculation of three-point functions in $\phi^3$ theory, similar to the ones that are performed in this thesis for $\syml$ theory. I am also grateful to Gregory Korchemsky, who suggested the interesting project on three-point functions of twist-two operators and for welcoming me for a discussion about this project in Paris.
I enjoyed the lectures of Harald Dorn on conformal quantum field theory very much and appreciate that he always had a  prompt answer to many of my questions in the past years.
I am very thankful to Dietmar Ebert for initiating a journal club, in which we studied different publications and books on quantum gravity over the last years, which has certainly significantly contributed to my general scientific education.

I have profited from discussions and correspondence with Mohammad Assadsolimani, Niklas Beisert, Marco S. Bianchi, Burkhardt Eden, Simon Caron-Huot, Wei-Ming Chen, Harald Dorn, Nadav Drukker, Dietmar Ebert, Livia Ferro, Valentina Forini, Rouven Frassek, Nikolay Gromov, Andr\'e Gros\-sardt, Martin Heinze, Johannes Henn, Yu-tin Huang, George Jorjadze,  Chrysostomos Kalousi\-os, Gregory Korchemsky, Josua Groeger, Anatoly~V.~Kotikov, Olaf Lechtenfeld, Matias Leoni, Andrea Mauri, Carlo Mene\-ghelli, V.~P.~Nair, Filippo Passerini, Silvia Penati, Jan Plefka, Andreas Rodigast, Radu Roiban, Alberto Santambrogio, Ralf Sattler, Theodor Schuster, Chris\-toph Sieg, Dominik Stoeckinger, Per Sundin, V\'aclav Tlap\'ak, Pedro Viera, Valentin Verschinin, Sebastian Wuttke and Donovan Young. 

In particular, I would like to thank Nikolay Gromov and Pedro Viera for their excellent lectures and tutorials on programming with Mathematica, which I extensively used in this thesis. Additionally, I enjoyed the joint seminars and discussions with the group of Matthias Staudacher.

Furthermore, I am also thankful to Sylvia Richter for competent and friendly administrative support.
My work was supported by the Volkswagenstiftung and the SFB 647 "Raum - Zeit - Materie. Analytische und Geometrische Strukturen". Furthermore, I gratefully acknowledge financial support for the participation in summer-schools and conferences from the research training group "Mass, Spectrum, Symmetry (GK 1504)" and the Helmholtz-Stiftung. 

I would like to thank my colleagues Benedikt Biedermann, Livia Ferro, Martin Heinze, Jonas Pollok  and Sebastian Wuttke for helpful comments on the manuscript, which certainly improved the presentation of the material.

Furthermore, I would like to thank my family and friends for constant support over the last years. I am particularly thankful to Nina Richter for supporting me enormously in the final phase of this work.

\appendix

\addcontentsline{toc}{chapter}{Appendix}

\chapter{Notation and Conventions}
\label{app:Conventions}

\subsubsection{Transposition and conjugation}
\begin{enumerate}
	\item $(A)^*$ denotes the complex conjugate of $A$
	\item $(A)^T$ denotes the transposition of $A$, i.e. $(A_{ij})^T=A_{ji}$
	\item $(A)^\dagger$ is the hermitian conjugate of $A$, i.e. $(A)^\dagger = ((A)^*)^T$
\end{enumerate}

\subsubsection{Metric-Convention}
For the calculations\footnote{For simplicity of notation we will however just write $1/x^2$ in the introductory chapters.} throughout this thesis, we will use the mostly minus metric of $d$-dimensional Minkowski space 
\begin{equation}
\eta_{\mu\nu}= \text{diag } (1,-1,-1,...,-1) 
\end{equation}
and write Lorentz invariants $k^2, x^2$ in integrals in such a way that the integrals are real in the \emph{euclidean (space-like) regime}, i.e. they are real for $-k^2>0, -x^2>0$. Therefore, the x-space propagators in explicit calculations are $1/(-x^2+ i \epsilon)^{a}$, but we will suppress the $i \epsilon$ prescription.  Furthermore, we will frequently rewrite $h=d/2$ where $d$ is the dimension of space-time.

\subsubsection{Derivatives}
For derivatives with respect to $x^\mu$, respectively several points $x_i^\mu$ we will use the notation
\begin{equation}
\partial_\mu = \frac{\partial}{\partial x^\mu} \qquad \text{and} \qquad \partial_{i,\mu} = \frac{\partial}{\partial x_i^\mu}\,.
\end{equation}

\subsubsection{Expectation values}
We denote the expectation value of operators by
\begin{equation}
\langle \op_1...\op_n \rangle := \langle 0| T(\op_1...\op_n)|0 \rangle = Z^{-1} \int \mathcal{D} \Phi ~e^{iS[\Phi]} \op_1...\op_n \,,
\end{equation}
where $Z= \int \mathcal{D} \Phi ~e^{iS[\Phi]}$.

\subsubsection{Normalisation of generators}
We normalise the generators $T^a$ $(a=1\,...\, N^2-1)$ of the $SU(N)$ algebra
\begin{equation}\label{eqn:algebra}
[T^a,T^b] = i f^{abc}T^c
\end{equation}
in the fundamental representation as
\begin{equation}\label{eqn:normalization-generators}
\tr(T^a T^b) =\frac{1}{2} \delta^{ab}
\end{equation} 
and we will abbreviate the number of generators as $\delta^{aa}=N^2-1$ in the calculations. The generators satisfy the relation
\begin{equation}\label{eqn:relation-structure-constants}
\sum_{c,d} f^{acd} f^{bcd} = N \delta^{ab}\,.
\end{equation}
The latter relation is independent of the representation and can can easily be derived in the adjoint representation in the standard normalisation
\begin{equation}
(T^a_A )^{bc} = i f^{abc}\,,
\end{equation}
where we have
\begin{equation}\label{eqn:normalization-adjoint}
\tr ( T_A^a T_A^b ) = N \delta^{ab}\,.
\end{equation}
One can thus easily derive the relation \eqref{eqn:relation-structure-constants}
\begin{equation}
\sum_{c,d}  f^{acd} f^{bcd}= - \sum_{c,d}  f^{acd} f^{bdc}= \sum_{c,d}  (T^a_A)^{cd} (T^b_A)^{dc} =  \tr( T_A^a T_A^b) = N \delta^{ab}\,.
\end{equation}

\subsubsection{Definition of Polygons}
An $n$-sided polygon can be defined by $n$ points $x_i$ ($i=1,..., n$), with the edge $i$ being the line connecting $x_i$ and $x_{i+1}$. Defining
\begin{equation}
p_i^{\mu}=x^{\mu}_{i+1}-x^{\mu}_i
\end{equation}
and parametrising the position $z^{\mu}_i$ on edge $i$ with the parameter $s_i \in [0,1]$ we have
\begin{equation} \label{eqn:z-parametrization}
z^{\mu}_i(s_i)= x^{\mu}_i + p^{\mu}_i s_i \,.
\end{equation}
Furthermore, we use the notation
\begin{equation}
\epsilon(p,q,r) = \epsilon_{\mu\nu\rho} p^{\mu} q^{\nu} r^{\rho} \qquad 
 \text{and} \qquad \bar{s}_i= 1- s_i.
\end{equation}
One can easily check that with the definition $x^{\mu}_{ij} = x^{\mu}_i-x^{\mu}_j $
\begin{equation}\label{eqn:scalprodx_ijx_mn}
2\, x_{ij}\cdot x_{mn}= x^2_{in}+x^2_{jm} - x^2_{im} - x^2_{jn}\,.
\end{equation}
Using \eqref{eqn:scalprodx_ijx_mn} we can rewrite the scalar products
\begin{equation} \label{eqn:products:p_ip_m}
2 p_i \cdot p_j  = x^2_{i,j+1} +  x^2_{i+1,j} -  x^2_{i,j} -  x^2_{i+1,j+1} \,.
\end{equation}
Furthermore, using the definition \eqref{eqn:z-parametrization}, one can easily show that
\begin{equation}\label{eqn:z_i-z_j^2}
(z_i-z_j)^2=x^2_{ij}\bar{s}_i\bar{s}_j + x^2_{i+1,j} s_i \bar{s}_j + x^2_{i,j+1}\bar{s}_i s_j+x^2_{i+1,j+1}s_i s_j\, .
\end{equation}

\section{Fourier Transformation}
In Minkowski-space with signature $(+ - - - ...)$ we have 
\begin{equation}\label{eqn:result-fourier-trafo-mink}
 \int \frac{d^dp}{(2\pi)^d} \frac{e^{-ip \cdot x}}{(-p^2-i\epsilon)^k} =  \frac{\Gamma(\frac{d}{2}-k)}{\Gamma(k)} \frac{1}{4^k \pi ^\frac{d}{2}} \frac{i}{(-x^2+i\epsilon)^{\frac{d}{2}-k}} \,.
\end{equation}
Note, that this can be derived from the Euclidean Fourier-transform by Wick-rotating. The sign of the Wick-rotation depends on the location of the poles. Since the sign of $i\epsilon$ changes after the Fourier-transform, for the transformation back from x-space to p-space, we have to Wick-rotate into the other direction, getting an overall sign.
\begin{equation}\label{eqn:result-fourier-trafo-mink-inverse}
 \int d^dx \frac{e^{+ip \cdot x}}{(-x^2+i\epsilon)^k} =  \frac{\Gamma(\frac{d}{2}-k)}{\Gamma(k)} \frac{1}{4^k \pi ^\frac{d}{2}} \frac{(-i)}{(-p^2-i\epsilon)^{\frac{d}{2}-k}} \,.
\end{equation}
This yields the identity when transforming and transforming back. 

\section{Bubble Integrals}\label{sec:bubble-integrals}
We define the scalar bubble integral as
\begin{align}\label{eqn:bubble-integral}
B_0(\alpha_1,\alpha_2) = \int \frac{d^dk}{(2\pi)^d} \frac{1}{(-k^2)^{\alpha_1} (-(p+k)^2)^{\alpha_2}} = b_0(\alpha_1,\alpha_2) \frac{1}{(-p^2)^{\alpha_1+\alpha_2-h}}\,,
\end{align}
where $h=d/2$ and
\begin{equation}
 b_0(\alpha_1,\alpha_2) = \frac{i}{(4\pi)^{d/2}} \frac{\Gamma(\frac{d}{2}-\alpha_1)\Gamma(\frac{d}{2}-\alpha_2)}{\Gamma(d-\alpha_1-\alpha_2)} \frac{\Gamma(\alpha_1+\alpha_2-\frac{d}{2})}{\Gamma(\alpha_1)\Gamma(\alpha_2)}\,.
\end{equation}
%This is the same as in \cite{Smirnov:2006ry} in (A.7). 
Furthermore, the bubble integral with momenta in the numerator is
\begin{align}\label{eqn:bubble-integral}
B_n(\alpha_1,\alpha_2)&= \int \frac{d^dk}{(2\pi)^d} \frac{\hat{k}^n}{(-k^2-i\epsilon)^{\alpha_1}(-(p+k)^2-i\epsilon)^{\alpha_2}} \\\nn
&= b_n(\alpha_1, \alpha_2) \frac{\hat{p}^n}{(-p^2-i\epsilon)^{\alpha_1+\alpha_2-\frac{d}{2}}}\,,
\end{align}
where
\begin{equation}
 b_n(\alpha_1,\alpha_2) = i\frac{(-1)^n}{(4\pi)^{d/2}} \frac{\Gamma(\frac{d}{2}+n-\alpha_1)\Gamma(\frac{d}{2}-\alpha_2)}{\Gamma(d+n-\alpha_1-\alpha_2)} \frac{\Gamma(\alpha_1+\alpha_2-\frac{d}{2})}{\Gamma(\alpha_1)\Gamma(\alpha_2)}\,.
\end{equation}
For the analogous integrals in x-space, one has to Wick-rotate into the other direction, since the poles are located differently in the complex plane, thus giving the integral
\begin{equation}\label{eqn:bubble-integral-x-space}
\int \frac{d^dx}{(2\pi)^d} \frac{\hat{x}^n}{(-x^2+i\epsilon)^{\alpha_1}(-(x+y)^2+i\epsilon)^{\alpha_2}} = - b_n(\alpha_1, \alpha_2) \frac{\hat{y}^n}{(-y^2+i\epsilon)^{\alpha_1+\alpha_2-\frac{d}{2}}}\,.
\end{equation}

\chapter{Technical Details on Conformal Symmetry}\label{sec:details-conformal-symmetry}

\section{Derivation of Conformal Killing Vectors}\label{sec:derivation-conformal-transformations}
Here, we present some details on the derivation of infinitesimal conformal transformations, roughly following chapter four of the book \cite{DiFrancesco:1997nk}. Under conformal transformations,  by definition, the metric transforms as
\begin{equation}\label{eqn:conformal-trafo-of-metric-detail}
g^\prime_{\mu\nu}(x^\prime) = \frac{\partial {x^\prime}^\rho}{\partial x^\mu} \frac{\partial {x^\prime}^\sigma}{\partial x^\nu} g_{\rho \sigma}(x^\prime) = \rho(x) g_{\mu\nu}(x)\,.
\end{equation}
Angles between two tangent vectors $u^\mu(x), v^\mu(x)$ defined by
\begin{equation}\label{eqn:angles}
\cos \theta = \frac{u \cdot v}{|u||v|} = \frac{g_{\alpha\beta}(x) u^\alpha v^\beta}{\sqrt{g_{\alpha\beta}(x) u^\alpha u^\beta}\sqrt{g_{\alpha\beta}(x) v^\alpha v^\beta}}
\end{equation}
are thus preserved under conformal transformations.
Inserting \eqref{eqn:trafo-killing-vector} into \eqref{eqn:conformal-trafo-of-metric-detail} and specialising to flat Minkowski space with metric $\eta_{\mu\nu}$, the terms linear in $k^\mu$ read
\begin{equation}\label{eqn:way-to-killing}
 \partial_\mu k_\nu + \partial_\nu k_\mu = \omega(x) \eta_{\mu\nu} \,,
\end{equation}
where we have written $\rho(x)=1+ \omega(x)$. Taking the trace of \eqref{eqn:way-to-killing} solving for $\omega(x)=2/d\,\partial^\alpha k_\alpha$ and reinserting it into  \eqref{eqn:way-to-killing}  leads to the \emph{conformal killing equation} in flat space
\begin{equation}\label{eqn:conformal-killing-equation}
 \partial_\mu k_\nu + \partial_\nu k_\mu =\frac{2}{d} \eta_{\mu\nu} \partial^\alpha k_\alpha \,.
\end{equation}
For $d=1$ this equation does not lead to any restrictions on $k_\mu(x)$ and thus every smooth transformation is conformal. The case $d=2$ is special, since then  \eqref{eqn:conformal-killing-equation} has an infinite number of solutions and the conformal group is infinite-dimensional. In this thesis we will only explicitly make use of conformal transformations for $d\geq3$ and thus specialise to this case from now on.
The derivative of \eqref{eqn:conformal-killing-equation} in the following linear combination with permuted indices leads to the second order differential equation
\begin{equation}\label{eqn:second-order-diff-equ}
\left( \eta_{\mu\rho} \partial_\nu + \eta_{\nu\rho} \partial_\mu-\eta_{\mu\nu} \partial_\rho \right)(\partial \cdot k) = d\, \partial_\mu \partial_\nu k_\rho \,.
\end{equation}
where we have written $(\partial^\alpha k_\alpha)=(\partial \cdot k)$.
Contraction of \eqref{eqn:second-order-diff-equ} with $\partial^{\rho}$ yields
\begin{equation}\label{eqn:k-constraint-1}
(2-d) \partial_\mu \partial_\nu (\partial \cdot k) = \eta_{\mu\nu} \partial^2(\partial \cdot k)\,.
\end{equation}
Contracting this equation with $\eta^{\mu\nu}$ yields
\begin{equation}\label{eqn:k-constraint-2}
2(1-d) \partial^2 (\partial \cdot k) =0\,.
\end{equation}
Combining \eqref{eqn:k-constraint-1} and \eqref{eqn:k-constraint-2} we find $\partial_\mu \partial_\nu (\partial \cdot k) =0$ which is thus at most linear in $x^\mu$. Combining this fact with \eqref{eqn:second-order-diff-equ} it is clear that $\partial_\mu \partial_\nu k_\rho(x)= \text{constant}$ and thus $k_\rho$ is at most quadratic in $x^\mu$, i.e. we can write
\begin{equation}
k_\mu(x)= a_\mu + b_{\mu\nu} x^\nu + c_{\mu\nu\rho} x^\nu x^\rho\,.
\end{equation}
Inserting this ansatz into the first \eqref{eqn:way-to-killing} respectively second \eqref{eqn:second-order-diff-equ} order differential equations for $k^\mu$ yields constraints for these coefficients. Each order in $x^\mu$ can be investigated separately. There are no constraints on $a^\mu$ which has thus $d$ independent components. Inserting the first order term in $x^\mu$ into \eqref{eqn:conformal-killing-equation} we get
\begin{equation}
b_{\nu\mu}+b_{\mu\nu} = \frac{2}{d} \eta_{\mu \nu} b^{\rho}_{~\rho}
\end{equation}
and thus a combination of a trace term proportional to the metric and an antisymmetric part $\omega_{\nu\mu}=-\omega_{\mu\nu}$ satisfies this equation, i.e.
\begin{equation}
b_{\mu\nu} = \lambda \eta_{\mu\nu} + \omega_{\mu\nu} \,.
\end{equation}
The first term describes an infinitesimal dilatation 
\begin{equation}
{x^\prime}^\mu = x^\mu + \lambda\, x^\mu
\end{equation}
and has one free parameter $\lambda$ and the second term generates infinitesimal rotations (Lorentz transformations)
\begin{equation}
{x^\prime}^\mu = x^\mu + \omega_{\mu\nu} \,x^\nu
\end{equation}
and has $d(d-1)/2$ free parameters. Inserting the second order term of $k^\mu$ into \eqref{eqn:second-order-diff-equ} we find
\begin{equation}
c_{\mu\nu\rho} = \eta_{\nu\mu} c_\rho + \eta_{\rho\mu}c_\nu -\eta_{\nu\rho} c_\mu \qquad \text{with}\qquad c_\mu = \frac{1}{d}c^{\lambda}_{~\lambda\mu}\,,
\end{equation}
which generates infinitesimal special conformal transformations
\begin{equation}
{x^\prime}^\mu = x^\mu +2(x \cdot c) x^\mu - c^\mu x^2
\end{equation}
and $c^\mu$ has $d$ independent components that parametrise the transformation.
All these transformations form the most general vector $k^\mu(x)$ given in \eqref{eqn:Killing-vector}.

\section{Correlators in Conformal Field Theory}\label{sec:correlators-in-conformal-field-theory}
The form of the correlation functions can be found by analysing the behaviour of the fields in the correlation functions under conformal transformations. To achieve this, it is useful to represent the change of a conformal primary operator $\op(x)$ by the action of unitary operators $U$ in the state space. Assuming the invariance of the vacuum under these transformations $U | 0 \rangle = | 0 \rangle$, we can deduce the following relation between the Green's functions of the transformed and the original operators
\begin{align}\label{eqn:unitary-trafo-of-correlator}
\langle 0 | \op_1(x_1)... \op_n(x_n) |0\rangle &= \langle 0 | (U^{-1} U) \op_1(x_1) (U^{-1} U) ... (U^{-1} U) \op_n(x_n) (U^{-1} U)  |0\rangle \\ \nn
&= \langle 0 | \op^\prime_1(x_1)... \op_n^\prime(x_n) |0\rangle \,.
\end{align}

\notocsubsection{Transformation of Scalar Operators}\label{sec:transformations-of-scalar-conformal-primaries-inversions}
We use the following transformation properties of scalar conformal primary operators
\begin{align}\label{eqn:dilatation-as-unitary-operator}
\op^\prime(x) &= U_D \op(x) U_D^{-1} = \lambda^\Delta \op(\lambda x)\,, \\ \label{eqn:inversion-as-unitary-operator}
\op^\prime(x) &= U_I \op(x) U_I^{-1} = \frac{1}{(x^2)^{\Delta}} \op(I( x))\,,
\end{align}
under dilatations respectively inversions.  The right-hand side of \eqref{eqn:dilatation-as-unitary-operator} is obtained from the generalization of \eqref{eqn:trafo-primary} to scalar conformal primary operators $\op(x)$ and the transformation under inversions \eqref{eqn:inversion-as-unitary-operator} can be deduced from \eqref{eqn:trafo-primary} and the fact that special conformal transformations can be represented as inversions followed by a translation and another inversion \eqref{eqn:I-P-I gives K}. 

\notocsubsection{Two-Point Functions of Scalar Operators}\label{sec:derivation-two-point-function}
Dilatations \eqref{eqn:dilatation-as-unitary-operator} for the two-point function in \eqref{eqn:unitary-trafo-of-correlator} yield
\begin{equation}\label{eqn:unitary-trafo-of-correlator}
\langle 0 | \op_1(x_1)\op_2(x_2) |0\rangle = \lambda^{\Delta_1+\Delta_2}\langle 0 | \op_1(\lambda x_1)\op_2( \lambda x_2) |0\rangle\,,
\end{equation}
which requires
\begin{equation}\label{eqn:two-point-function-constraint-dilatation}
\langle 0 | \op_1(x_1)\op_2(x_2) |0\rangle = \frac{c}{(x_{12}^2)^{(\Delta_1+\Delta_2)/2}}\,,
\end{equation}
where $c$ is a constant.  Inversions \eqref{eqn:inversion-as-unitary-operator} lead to
\begin{equation}
\langle 0 | \op_1(x_1)\op_2(x_2) |0\rangle = \frac{1}{(x_1^2)^{\Delta_1}} \frac{1}{(x_2^2)^{\Delta_2}} \langle 0 | \op_1(I(x_1))\op_2( I(x_2)) |0\rangle\,.
\end{equation}
Using \eqref{eqn:two-point-function-constraint-dilatation} and the inversion of distances \eqref{eqn:inversions-xij2} we thus find
\begin{equation}
\langle 0 | \op_1(x_1)\op_2(x_2) |0\rangle = \frac{1}{(x_1^2)^{\Delta_1}} \frac{1}{(x_2^2)^{\Delta_2}} \frac{c (x_{1}^2 x_{2}^2)^{(\Delta_1+\Delta_2)/2}}{(x_{12}^2)^{(\Delta_1+\Delta_2)/2}}\,,
\end{equation}
which can only be satisfied for $\Delta_1=\Delta_2$, otherwise the correlation function has to vanish. Thus we find
\begin{equation}
\langle 0 | \op_1(x_1)\op_2(x_2) |0\rangle =  \frac{c \delta_{\Delta_1\Delta_2}}{(x_{12}^2)^{\Delta_1}}\,.
\end{equation}
It is easy to see, that this two-point function indeed satisfies the Ward-identities  \eqref{eqn:dilatation-Ward-identity-general}, \eqref{eqn:special-conformal-Ward-identity-general}. 

\notocsubsection{Three-Point Functions}\label{sec:derivation-three-point-function}
Dilatations \eqref{eqn:dilatation-as-unitary-operator} for the three-point function in \eqref{eqn:unitary-trafo-of-correlator} yield
\begin{equation}\label{eqn:unitary-trafo-of-correlator-three-point}
\langle 0 | \op_1(x_1)\op_2(x_2) \op_3(x_3) |0\rangle = \lambda^{\Delta_1+\Delta_2+\Delta_3}\langle 0 | \op_1(\lambda x_1)\op_2( \lambda x_2) \op_3( \lambda x_3) |0\rangle\,,
\end{equation}
which requires
\begin{equation}\label{eqn:three-point-function-constraint-dilatation}
\langle 0 | \op_1(x_1)\op_2(x_2) \op_3(x_3) |0\rangle = \frac{c}{(x_{12}^2)^{a}(x_{13}^2)^{b}(x_{23}^2)^{c}}\,,
\end{equation}
where $c$ is a constant and $a+b+c = (\Delta_1+\Delta_2+\Delta_3)/2$.  Inversions \eqref{eqn:inversion-as-unitary-operator} lead to
\begin{equation}\nn
\langle 0 | \op_1(x_1)\op_2(x_2)\op_3(x_3) |0\rangle = \frac{1}{(x_1^2)^{\Delta_1}} \frac{1}{(x_2^2)^{\Delta_2}}   \frac{1}{(x_3^2)^{\Delta_3}} \langle 0 | \op_1(I(x_1))\op_2( I(x_2))  \op_3( I(x_3)) |0\rangle
\end{equation}
Using \eqref{eqn:three-point-function-constraint-dilatation} and the inversion of distances \eqref{eqn:inversions-xij2} we thus find
\begin{equation}
\langle 0 | \op_1(x_1)\op_2(x_2)\op_3(x_3) |0\rangle = \frac{1}{(x_1^2)^{\Delta_1}} \frac{1}{(x_2^2)^{\Delta_2}} \frac{1}{(x_3^2)^{\Delta_3}} \frac{c (x_1^2)^{a+b} (x_2^2)^{a+c} (x_3^2)^{b+c}}{(x_{12}^2)^{a}(x_{13}^2)^{b}(x_{23}^2)^{c}}\,,
\end{equation}
which yields $2a=\Delta_1+\Delta_2-\Delta_3$, $2b=\Delta_1+\Delta_3-\Delta_2$, $2a=\Delta_2+\Delta_3-\Delta_1$
and thus we find
\begin{equation}
\langle 0 | \op_1(x_1)\op_2(x_2) \op_3(x_3) |0\rangle = \frac{c}{|x_{12}|^{\Delta_1+\Delta_2-\Delta_3} |x_{13}|^{\Delta_1+\Delta_3-\Delta_2} |x_{23}|^{\Delta_2+\Delta_3-\Delta_1}}\,.
\end{equation}
It is easy to see, that this two-point function indeed satisfies the Ward-identities  \eqref{eqn:dilatation-Ward-identity-general}, \eqref{eqn:special-conformal-Ward-identity-general}. 

\notocsubsection{Transformation of Operators with Spin}\label{sec:transformations-of-scalar-conformal-primaries-inversions}
In the same way as for scalar operators \eqref{eqn:dilatation-as-unitary-operator}, one can derive the transformation properties for primary conformal operators with spin 
\begin{align}\label{eqn:dilatation-inversion-spin-operator}
\op^\prime_{\mu_1...\mu_j}(x) &= U_D \op_{\mu_1...\mu_j}(x) U_D^{-1} = \lambda^{\Delta_j }\op_{\mu_1...\mu_j}(\lambda x)\,, \\ \nn
\op_{\mu_1...\mu_j}^\prime(x) &= U_I \op_{\mu_1...\mu_j}(x) U_I^{-1} = \frac{1}{(x^2)^{\Delta_j } } I_{\mu_1\nu_1}(x)... I_{\mu_j\nu_j}(x) \op_{\nu_1...\nu_j}(I( x))\,,
\end{align}
under dilatations respectively inversions. Here,  $I_{\mu\nu}(x)$ is the inversion tensor
\begin{equation}\label{eqn:Imunu}
I_{\mu\nu}(x)=\eta_{\mu\nu} - 2\frac{x_\mu x_\nu}{x^2}\,,
\end{equation}
which satisfies the property
\begin{equation}
I_{\mu\sigma}(x_1) I^{\sigma\tau}(I(x_{12})) I_{\tau\nu}(x_2)= I_{\mu\nu}(x_{12})
\end{equation}
and can be used for explicitly working out the transformation properties below.

\notocsubsection{Two-Point Functions of Operators with Spin}
The transformations \eqref{eqn:dilatation-inversion-spin-operator} lead to the conformal symmetry conditions
\begin{align}
\langle \op_{\mu_1...\mu_j}(x_1) \op_{\nu_1...\nu_j}(x_2)  \rangle &=  \lambda^{2\Delta_j}\langle \op_{\mu_1...\mu_j}(\lambda x_1) \op_{\nu_1...\nu_j}(\lambda x_2)  \rangle\,, \\ \nn
\langle \op_{\mu_1...\mu_j}(x_1) \op_{\nu_1...\nu_j}(x_2)  \rangle &=  \frac{1}{(x_{1}^2)^{\Delta_j}} \frac{1}{(x_{2}^2)^{\Delta_j}} I_{\mu_1\sigma_1}(x_1)... I_{\mu_j\sigma_j}(x_1) \times \\ \nn 
&  \qquad \qquad \times I_{\nu_1\tau_1}(x_2)... I_{\nu_j\tau_j}(x_2) \langle    \op_{\sigma_1...\sigma_j}(I(x_1)) \op_{\tau_1...\tau_j}(I(x_2))  \rangle \,.
\end{align}
The solution for a traceless, totally symmetric operator as introduced in \eqref{eqn:correlation-functions-spin} is
\begin{equation}
\langle \op_{\mu_1...\mu_j}(x_1) \op_{\nu_1...\nu_j}(x_2)  \rangle = \frac{C_j}{(x_{12}^2)^{\Delta_j}} \left( I_{\{\mu_1\nu_1}...I_{\mu_j\}\nu_j}(x_{12}) - \text{traces} \right)\,,
\end{equation}
where $C_j$ is a constant, $\{..\}$ indicates that the right-hand side is totally symmetrised in all indices $\mu_k, \nu_l$ and we subtract all traces in $\mu_i$ resp. $\nu_i$ in accord with the symmetries of the operators on the left-hand side. 

\notocsubsection{Three-Point Functions of Operators with Spin}\label{sec:three-point-function-with-spin}
The transformations \eqref{eqn:dilatation-inversion-spin-operator} lead to the conformal symmetry conditions
\begin{align}
\langle \op_A( x_1)   \op_B( x_2)  \op_{\mu_1...\mu_j}(x_3)  \rangle &=  \lambda^{\Delta_A+\Delta_B+\Delta_j}\langle  \op_A(\lambda x_1)   \op_B(\lambda x_2)   \op_{\mu_1...\mu_j}(\lambda x_3)\rangle\,, \\ \nn
\langle \op_A( x_1)   \op_B( x_2)  \op_{\mu_1...\mu_j}(x_3)   \rangle &=  \frac{I_{\mu_1\nu_1}(x_3)... I_{\mu_j\nu_j}(x_3)}{(x_{1}^2)^{\Delta_A}(x_{2}^2)^{\Delta_B}(x_{3}^2)^{\Delta_j}}  \langle  \op_A(I(x_1))  \op_B(I(x_2)) \op_{\nu_1...\nu_j}(I(x_3))  \rangle \,.
\end{align}
The solution of these constraints is
\begin{align}
\langle \op_A(x_1) \op_B(x_2)  \op_{\mu_1...\mu_j}(x_3) \rangle = \frac{ C_{ABj} \left(Y_{\mu_1}...Y_{\mu_j}(x_{13},x_{23}) -\text{traces}\right)}{|x_{12}|^{\Delta_A+\Delta_B-\theta}|x_{13}|^{\Delta_A+\theta-\Delta_B}|x_{23}|^{\Delta_B+\theta-\Delta_A}}\,,
\end{align}
where $C_{ABj}$ is a constant, $\theta=\Delta_j-j$ is the \emph{twist}\index{twist} of the operator $\op_{\mu_1...\mu_j}$ and
\begin{equation}
Y^\mu (x_{13},x_{23}) = \frac{x_{13}^\mu}{x_{13}^2} - \frac{x_{23}^\mu}{x_{23}^2} = \frac{1}{2}\partial_{x_3}^\mu \ln \left(\frac{x_{13}^2}{x_{23}^2}  \right)\,.
\end{equation}
For three-point functions with more operators with spin we can have several separately invariant space-time structures and the general solution is a linear combination of these with a priori arbitrary coefficients for each term.

\chapter{$\mathcal{N} = 4$ SYM from Dimensional Reduction}\label{sec:N-4-SYM-from-reduction}

In this section we describe how to obtain $\mathcal{N}=4$ Super Yang-Mills theory in four dimensions from dimensional reduction of $\mathcal{N}=1$ SYM in ten dimensions. We will closely follow the presentation and notation of \cite{Belitsky:2003sh}, but give a few more details.

\section{$\mathcal{N}=1$ Super Yang-Mills Theory in Ten Dimensions}
We start from the  $\mathcal{N}=1$ Super Yang-Mills action in $D=10$ with the {\itshape mostly minus} metric $g_{MN} = \text{diag}(+, -,..., -) $
\begin{align}
S = \int d^D x\, \tr \left( -\frac{1}{2} F_{MN} F^{MN} + i \bar{\psi} \Gamma^M  D_M \psi \right)\,,
\end{align}
where all fields are matrix valued in SU(N), i.e. $\psi = \psi^a T^a, A_M = A_M^a T^a$ and
\begin{align}
F_{MN} &= \partial_M A_N-\partial_N A_M - i g  [ A_M , A_N]\,, \\ \nn
D_M (\cdot) &= \partial_M (\cdot) - i g [ A_M , (\cdot)]\,.
\end{align}
For this action to be supersymmetric, it is necessary that $\psi$ be a Majorana-Weyl spinor, i.e.  it satisfies\footnote{These two conditions reduce the number of fermionic degrees of freedom from 32  to 8, which equals the degrees of freedom of the massless gauge field in ten dimensions. For a linear realization of supersymmetry without auxiliary fields it is required that the bosonic gauge field carries the same number of degrees of freedom as the fermionic field.} the Weyl condition
\begin{equation}
\Gamma \psi =  \psi , \qquad \Gamma= i \,\Gamma^0 ... \,\Gamma^{D-1}
\end{equation}
and the Majorana condition
\begin{equation}
\psi^T C_{10}= \psi^\dagger \Gamma^0 = \bar{\psi}\,,
\end{equation}
where the ten-dimensional Dirac matrices $\Gamma^M$ and the charge conjugation matrix $C_{10}$ are constructed from a four-dimensional and a six-dimensional represenation of the Clifford algebra, the explicit representation is given in section \ref{sec:Dirac-matrices in D=10}.
The Majorana-Weyl conditions are satisfied if the spinor has the form
\begin{equation}\label{eqn:10d-Majorana-Weyl-spinor}
\psi = \underbrace{\begin{pmatrix} 1 \\ 0 \end{pmatrix} \otimes  \begin{pmatrix} \lambda_\alpha^A \\ 0 \end{pmatrix}}_{\text{4 left-handed Weyl spinors}} +
\underbrace{\begin{pmatrix} 0 \\ 1 \end{pmatrix} \otimes  \begin{pmatrix} 0 \\ \bar{\lambda}_{\dot{\alpha},A}  \end{pmatrix}}_{\text{4 right-handed Weyl spinors}} \,,
\end{equation}
where $A=1...4$ denotes four left-handed respectively right-handed Weyl spinors and where $(\lambda_\alpha^A)^* = \bar{\lambda}_{\dot{\alpha}}^A$ and $(\lambda^{\alpha A})^* = \bar{\lambda}^{\dot{\alpha} A}$. Weyl-spinor indices are raised and lowered using the antisymmetric tensor $\epsilon^{\alpha\beta} = - \epsilon_{\dot{\alpha}\dot{\beta}}$ with components $\epsilon^{12}=\epsilon_{12}=-\epsilon_{\dot{1}\dot{2}}=-\epsilon^{\dot{1}\dot{2}}=1$
\begin{equation}
\lambda^\alpha = \epsilon^{\alpha\beta} \lambda_\beta, \qquad
\bar{\lambda}_{\dot{\alpha}} = \epsilon_{\dot{\alpha}\dot{\beta}} \lambda^{\dot{\beta}}, \qquad
\lambda_\alpha = \lambda^\beta \epsilon_{\beta\alpha} , \qquad
\bar{\lambda}^{\dot{\alpha}} = \bar{\lambda}_{\dot{\beta}} \epsilon^{\dot{\beta}\dot{\alpha}} \,.
\end{equation}
Note that $(\epsilon_{\dot{\alpha}\dot{\beta}})^* = -\epsilon_{\alpha\beta}$.

One can show  that the action is invariant under the supersymmetry transformations
\begin{equation}\label{eqn:susy-trafo-in-d=10}
\delta A_M = -i \bar{\xi} \Gamma_M \psi, \qquad \delta \psi = \frac{i}{2} \Gamma_{MN} F^{MN} \xi\,,
\end{equation}
where $\Gamma_{MN} = \frac{i}{2} ( \Gamma_{M}\Gamma_{N} -  \Gamma_{N}\Gamma_{M}) $.

%\notocsubsection{Supersymmetry of the ten dimensional action}\label{sec:susy-proof-ten-dimensions}
%use fierz transformations as given by scherk, gliozzi, olive

\section{Reduction to $\mathcal{N}=4$ SYM in Four Dimensions}
Here, we follow the procedure of dimensional reduction given in \cite{Belitsky:2003sh}.  We split up the ten-dimensional coordinates labelled by capital latin letters $M=0,...,9$ into greek indices $\mu=0,1,2,3$ and small latin indices $m=4,...,9$. Now, we let the fields only depend on the coordinates $x_\mu$, i.e
\begin{align}\label{eqn:reduction-coordinate-dependence}
 \partial_m A_M(x_\mu) = 0,   
 \qquad \partial_m \psi(x_\mu) = 0, \qquad m=4,...,9\,.
 \end{align}
The remaining non-vanishing components are then
\begin{align}
- \frac{1}{2} F_{MN}F^{MN} &= -\frac{1}{2} F_{\mu\nu}F^{\mu\nu} - D_\mu A_a D^\mu A^a+ \frac{1}{2} g^2 [A_a,A_b] [A^a,A^b]  \\
i \bar{\psi} \Gamma_M D^M \psi& = i \bar{\psi} \Gamma_\mu D^\mu \psi  + g \bar{\psi} \Gamma_a [A^a,\psi]\,.
\end{align}
We can now rewrite these terms using the form of the Majorana-Weyl spinor \eqref{eqn:10d-Majorana-Weyl-spinor} and the representation of the $\Gamma$ matrices in terms of four- and six-dimensional Dirac matrices as given in section \ref{sec:Dirac-matrices in D=10}.

We define a complex scalar field, which is related to the six real components $A^a$ of the gauge field via
\begin{equation}\label{eqn:definition-complex-scalar-fields}
\phi^{AB} = \frac{1}{\sqrt{2}} \Sigma^{a,AB} A^a,\qquad \bar{\phi}_{AB} = (\phi^{AB} )^* = \frac{1}{2}\epsilon_{ABCD} \phi^{CD} = \frac{1}{\sqrt{2}} \bar{\Sigma}^{a}_{AB} A^a\,,
\end{equation}
where $\Sigma^a = (\bar{\Sigma}^a)^*$ is defined in \ref{sec:Dirac-matrices in D=10} and satisfies $\Sigma^a \bar{\Sigma}^b+\Sigma^b \bar{\Sigma}^a= -2 \delta^{ab}$. The above defined scalar field is antisymmetric in $A \leftrightarrow B$. 
Thus we have\footnote{Note that $a,b,c,d$ are the euclidean space-time indices, not the $SU(N)$ group indices. Furthermore, we use cyclic invariance which is due to the trace over the group indices, which is not shown explicitly here.}
\begin{align}
[\phi^{AB},\phi^{CD}][\bar{\phi}_{AB},\bar{\phi}_{CD}] &= \frac{1}{4} [A^a,A^b] [A^c, A^d] \Sigma^{a,AB}\Sigma^{b,CD}\bar{\Sigma}^{c}_{AB}\bar{\Sigma}^{d}_{CD} \\ \nn
&= \frac{1}{16} [A^a,A^b] [A^c, A^d] \tr( \Sigma^a \bar{\Sigma}^c+\Sigma^c \bar{\Sigma}^a ) \tr( \Sigma^b \bar{\Sigma}^d+\Sigma^d \bar{\Sigma}^b )  \\ \nn
&= \frac{1}{4} [A^a,A^b] [A^c, A^d] \delta^{ac}\delta^{bd}\tr( I_4) \tr(I_4 )\\ \nn
&= 4 [A_a,A_b] [A^a, A^b]\,.
\end{align}
Note, that due to the metric for the components $a=4,....,6$ we have $A_a=-A^a$.
An analogous calculation leads to
\begin{align}
D_\mu \phi^{AB} D^\mu \bar{\phi}_{AB} &= \frac{1}{2} D_\mu A_a D^\mu A_b \Sigma^{a,AB}\bar{\Sigma}^{b}_{AB} = - \frac{1}{4}  D_\mu A_a D^\mu A_b (-2\delta^{ab}) \tr(I_4) \\ \nn
&= -2 D_\mu A_a D^\mu A^a\,.
\end{align}
In order to simplify the fermionic terms we note that $\Gamma^a=(\gamma^5 \otimes \hat{\gamma}^a)$ and $\Gamma^0 =  (\gamma^0 \otimes I_8 )$ and thus
\begin{equation}
\Gamma^0 \Gamma^a = \begin{pmatrix} 0 & -1 \\ 1 & 0 \end{pmatrix} \otimes
\begin{pmatrix} 0 & \Sigma^{a,AB} \\ \bar{\Sigma}^{a}_{AB} & 0 \end{pmatrix}\,.
\end{equation}
Then 
\begin{align}
\bar{\psi} \Gamma^a [ A_a, \psi] &= -\psi^\dagger \Gamma^0 \Gamma^a [A^a,\psi]  \\ \nn
&= -
 \begin{pmatrix}  ( \lambda_\alpha^A )^* \\ (\bar{\lambda}^{\dot{\alpha}}_{A})^* \end{pmatrix}^T   
  \begin{pmatrix} 0 & \Sigma^{a,AB} \\ \bar{\Sigma}^{a}_{AB} & 0 \end{pmatrix}
 \begin{pmatrix}  [A^a , \lambda_\alpha^B] \\  
 -[A^a,\bar{\lambda}^{\dot{\alpha}}_{B}]  \end{pmatrix}   \\ \nn
&= +\sqrt{2} \bar{\lambda}_{\dot{\alpha},A} [\phi^{AB},\bar{\lambda}^{\dot{\alpha}}_{B}] 
- \sqrt{2} \lambda^{\alpha,A} [\bar{\phi}_{AB},\lambda_{\alpha}^{B}] \,.
 \end{align}
 The last term is
 \begin{align}
 i \bar{\psi} \Gamma^\mu D_\mu \psi &=  i \begin{pmatrix} \bar{\lambda}^{\dot{\alpha}}_{A}, \lambda^{A,\alpha}  \end{pmatrix} \begin{pmatrix} (\sigma^\mu)^{\dot{\alpha}\beta} & 0 \\ 0 & (\bar{\sigma}^\mu)_{\alpha\dot{\beta}}\end{pmatrix} 
 \begin{pmatrix}
 D_\mu \lambda^A_\beta \\ D_\mu \bar{\lambda}^{\dot{\beta}}_A
 \end{pmatrix} \\ \nn
 &= i \bar{\lambda}^{\dot{\alpha},A} (\sigma^\mu)^{\dot{\alpha}\beta}D_\mu \lambda^A_\beta + i \lambda^{A,\alpha} (\bar{\sigma}^\mu)_{\alpha\dot{\beta}} D_\mu \bar{\lambda}^{\dot{\beta}}_A \\ \nn
  &= 2 i \bar{\lambda}_{\dot{\alpha},A} (\sigma^\mu)^{\dot{\alpha}\beta}D_\mu \lambda^A_\beta\,,
 \end{align}
 where we have used the identity $(\bar{\sigma}^\mu)_{\alpha\dot{\beta}} = \epsilon_{\dot{\beta}\dot{\gamma}} (\sigma^\mu)^{\dot{\gamma}\delta} \epsilon_{\delta\alpha} $ and the rules for raising and lowering Weyl spinor indices in the last line. 

Putting everything together we find the maximally supersymmetric gauge theory $\mathcal{N}=4 ~\text{SYM}$
\begin{align} \label{eqn:N=4SYM-group-matrices} \nn
S = \int d^dx \,  & \tr \Big(- \frac{1}{2} F_{\mu\nu}F^{\mu\nu} + \frac{1}{2} D_\mu \phi^{AB} D^\mu \bar{\phi}_{AB} + \frac{1}{8} g^2 [\phi^{AB},\phi^{CD}][\bar{\phi}_{AB},\bar{\phi}_{CD}] \\ 
&\phantom{\tr} +  2 i \bar{\lambda}_{\dot{\alpha}A} (\sigma_\mu)^{\dot{\alpha}\beta} D^\mu \lambda_\beta^A -\sqrt{2} g \lambda^{\alpha A} [\bar{\phi}_{AB},\lambda_\alpha^B] + \sqrt{2} g \bar{\lambda}_{\dot{\alpha}A}[\phi^{AB},\bar{\lambda}^{\dot{\alpha}}_B] \Big)\,,
\end{align}
where $d=4$ and the remaining volume integrals $V=\int dx_4...dx_9$ have been absorbed into a redefinition of all fields $\Phi\rightarrow V^{1/2}\Phi$ and the coupling constant $g \rightarrow V^{-1/2}g $. Then, the fields have classical dimension $[A_\mu]=[\phi^{AB}]= d/2-1$ and $[\lambda]=(d-1)/2$ and the coupling constant is dimensionless $[g]=0$. 

The Lagrangian is invariant under the supersymmetry transformations that follow from the reduction of the supersymmetry transformations \eqref{eqn:susy-trafo-in-d=10} in $D=10$ to $d=4$
\begin{eqnarray}
\label{N=4SUSYrules}
&&\delta A^\mu
=
- i \xi^{\alpha \; A} \bar\sigma^\mu{}_{\alpha\dot\beta}
\bar\lambda^{\dot\beta}_A
- i \bar\xi_{\dot\alpha \; A} \sigma^{\mu\; \dot\alpha\beta}
\lambda_{\beta}^A
\, , \nonumber\\
&&\delta \phi^{AB}
=
- i \sqrt{2}
\left\{
\xi^{\alpha \; A} \lambda_\alpha^B - \xi^{\alpha \; B} \lambda_\alpha^A
-
\varepsilon^{ABCD} \bar\xi_{\dot\alpha \; C} \bar\lambda^{\dot\alpha}_D
\right\}
\, , \nonumber\\
&&\delta \lambda^A_\alpha
=
\ft{i}2 F_{\mu\nu} \sigma^{\mu\nu}{}_\alpha{}^\beta \xi_\beta^A
-
\sqrt{2} \left( {\cal D}_\mu \phi^{AB} \right)
\bar\sigma^\mu{}_{\alpha\dot\beta} \bar\xi^{\dot\beta}_B
+
i g [\phi^{AB} , \bar\phi_{BC}] \xi_\alpha^C
\, , \nonumber\\
&&\delta \bar\lambda_A^{\dot\alpha}
=
\ft{i}2 F_{\mu\nu} \bar\sigma^{\mu\nu}{}^{\dot\alpha}{}_{\dot\beta}
\bar\xi^{\dot\beta}_A
+
\sqrt{2} \left( {\cal D}_\mu \bar\phi_{AB} \right)
\sigma^{\mu \, \dot\alpha\beta} \xi_\beta^B
+
i g[\bar\phi_{AB} , \phi^{BC}] \bar\xi^{\dot\alpha}_C
\, .
\end{eqnarray}
The theory is said to be the maximally supersymmetric {\itshape gauge theory}, since any theory with $\mathcal{N}>4$ would necessarily contain fields of spin $j>1$.

\notocsubsection{Action in $SU(N)$ Components}
Using $[T^a,T^b]=i f^{abc}T^c$ and taking the generators to be normalised as $\tr T^a T^b= \frac{1}{2}\delta^{ab} $ we can take the trace and rewrite the action as
\begin{equation}\label{eqn:action-N=1 SYM}
S= \int d^d x \left( -\frac{1}{4} F_{MN}^a F^{MN, a} + \frac{i}{2} \bar{\psi}^a \Gamma^M  D_M \psi^a \right)
\end{equation}
with
\begin{align}
F_{MN}^a &= \partial_M A_N^a -\partial_N A_M^a + g f^{abc} A_M^b A_N^c \,,\\ \nn
D_M \psi^a &= \partial_M \psi^a + g f^{abc} A_M^b \psi^c \,.
\end{align}
Taking the trace and rescaling the fields $\Phi \to \frac{1}{g} \Phi$ we get from \eqref{eqn:N=4SYM-group-matrices}  
\begin{align} \label{eqn:N=4SYM-Lagrangian} 
S = \frac{1}{g^2}\int d^dx \,   \Big( &- \frac{1}{4} F_{\mu\nu}^aF^{a,\mu\nu} + \frac{1}{4} D_\mu \phi^{a,AB} D^\mu \bar{\phi}^{a}_{AB} \\ \nn
& 
- \frac{1}{16} f^{abc}f^{ade} \phi^{b,AB} \phi^{c,CD} \bar{\phi}^{d}_{AB} \bar{\phi}^{e}_{CD}  +   i \bar{\lambda}^a_{\dot{\alpha}A} (\sigma_\mu)^{\dot{\alpha}\beta} D^\mu \lambda_\beta^{a,A} 
\\ \nn
& - \frac{i}{\sqrt{2}} f^{abc} \lambda^{a,\alpha A} \bar{\phi}^{b}_{AB} \lambda_\alpha^{c,B} + \frac{i}{\sqrt{2}} f^{abc} \bar{\lambda}^{a}_{A,\dot{\alpha}} \phi^{b,AB} \bar{\lambda}^{c,\dot{\alpha}}_{B} \Big)\,.
\end{align}
Taking into account that $\phi^{12}=-\phi^{21}$ and $\bar{\phi}_{34}=\phi^{12}$, the kinetic term for $\phi^{12}$ reads $\partial_\mu \phi^{12} \partial^\mu \bar{\phi}_{12}$
such that we have the standard propagator $i/p^2$ for a complex scalar field. Summation over all repeated indices is implied and we do not display ghost-fields and gauge terms.  The field strength and covariant derivative read
\begin{align}
F_{\mu\nu}^a &= \partial_\mu A_\nu^a -\partial_\nu A_\mu^a +  f^{abc} A_\mu^b A_\nu^c\,, \\ \nn
D_\mu \psi^a &= \partial_\mu \psi^a +  f^{abc} A_\mu^b \psi^c \,.
\end{align}

%%%%%%%%%%%%%%%%%%%%%%%%%%%%%%%%%%%%%

\notocsubsection{Dirac Matrices in $D=10$}
\label{sec:Dirac-matrices in D=10}
We construct the ten-dimensional $32\times 32$ Dirac matrices in terms of a representation of four- and six-dimensional Clifford algebra which are explicitly given in the following sections. From these lower dimensional representations we build up the ten-dimensional Dirac matrices in the following way
\begin{equation}
\Gamma^M = ( \gamma^\mu \otimes I_8, \gamma^5 \otimes \hat{\gamma}^a), \qquad \mu=0,1,2,3 \qquad  a=4,...,9\,.
\end{equation}
They satisfy the Clifford algebra $\{ \Gamma^M, \Gamma^N\} = 2 \eta^{MN}$ with metric $\eta^{MN}= \text{diag}(+-...-)$. The chiral and charge conjugation matrix are
\begin{equation}
\Gamma = \gamma^5  \otimes \hat{\gamma}^7, \qquad C_{10} = C_4 \otimes C_6\,,
\end{equation}
where the four- and six-dimensional chiral and charge conjugation matrices are defined in the following.
 
\subsubsection{Four-Dimensional Clifford Algebra}
We take the four-dimensional matrices in the Weyl (chiral) representation
\begin{equation}
\gamma^\mu = \begin{pmatrix} 0 & \bar{\sigma}^{\mu,\alpha\dot{\beta}} \\ \sigma^\mu_{\dot{\alpha}\beta} & 0 \end{pmatrix}, \qquad \{\gamma^\mu , \gamma^\nu \} = 2 \eta^{\mu\nu}, \qquad \eta^{\mu\nu} = \text{diag}(+---)\,,
\end{equation}
where $\sigma^\mu = (1,\vec{\sigma})$, $\bar{\sigma}^\mu = (1,-\vec{\sigma})$ and $\vec{\sigma}$ comprise the three standard Pauli matrices. In this representation we have
\begin{equation}
\gamma^5 = i \gamma^0 \gamma^1 \gamma^2 \gamma^3 = \begin{pmatrix} 1 & 0 \\ 0 & -1 \end{pmatrix} ,\qquad C_4 = i \gamma^2 \gamma^0 = \begin{pmatrix} -\epsilon^{\alpha\beta} & 0 \\ 0 & -\epsilon_{\dot{\alpha}\dot{\beta}} \end{pmatrix}\,,
\end{equation}
where $\epsilon^{\alpha\beta} = - \epsilon_{\dot{\alpha}\dot{\beta}}$ and $\epsilon^{12}=\epsilon_{12}=-\epsilon_{\dot{1}\dot{2}}=-\epsilon^{\dot{1}\dot{2}}=1$. 

%%%%%%%%%%%%%%%%
\subsubsection{Six-Dimensional Clifford Algebra}
Furthermore, we take the Dirac matrices in six-dimensional Euclidean space to be
\begin{equation}
\hat{\gamma}^a = \begin{pmatrix} 0 & \Sigma^{a, AB} \\ \bar{\Sigma}^{a}_{AB} & 0\end{pmatrix}, \qquad \{ \hat{\gamma}^a , \hat{\gamma}^b  \} = - 2\delta^{ab}, \qquad \delta^{ab} = \text{diag}(+...+)\,,
\end{equation}
where the $4 \times 4$ matrices $\Sigma^a$ are
\begin{align}
\Sigma^{a} &= (\eta^{1},\eta^{2},\eta^{3},i \bar{\eta}^{1},i \bar{\eta}^{2},i \bar{\eta}^{3}), \\
 \bar{\Sigma}^{a} &= (\Sigma^{a})^*, \\
( \eta^{i})_{AB} &= \epsilon_{iAB} + \delta_{iA}\delta_{4B}- \delta_{iB}\delta_{4A}\,, \\
 (\bar{\eta}^{i})_{AB} &= \epsilon_{iAB} - \delta_{iA}\delta_{4B}+ \delta_{iB}\delta_{4A} \,,
\end{align}
where $\epsilon^{iAB}$ is defined to be zero if $A,B>3$.  Furthermore, the chiral and charge conjugation matrices are defined as
\begin{equation}
\hat{\gamma}^7=i \hat{\gamma}^1\hat{\gamma}^2\hat{\gamma}^3\hat{\gamma}^4\hat{\gamma}^5\hat{\gamma}^6 = \begin{pmatrix} 1 & 0 \\ 0 & -1 \end{pmatrix},\qquad  C_6 = \hat{\gamma}^1 \hat{\gamma}^2 \hat{\gamma}^3  = \begin{pmatrix} 0 & \delta_A^{~B} \\ \delta^A_{~B} & 0 \end{pmatrix}\,.
\end{equation}

%%%%%%%%%%%%%%%%%%%%%%%%%%%%%%%%%%%
\section{Feynman Rules of $\mathcal{N}=4$ SYM}
The gauge field propagator in Feynman gauge, the complex scalar field and the fermion propagators following from \eqref{eqn:N=4SYM-Lagrangian} are
\begin{align}\label{eqn:N-4-SYM-propagators}
\begin{minipage}[t]{0.3\textwidth}
\includegraphics[width=1.0 \textwidth]{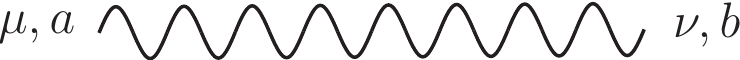}
\end{minipage}  
&=
\langle A_\mu^a(p) A_\nu^b(-p) \rangle  = - \frac{i \delta^{ab} g_{\mu\nu}}{p^2} g^2\,, \\ \nn
\begin{minipage}[t]{0.3\textwidth}
\includegraphics[width=1.0 \textwidth]{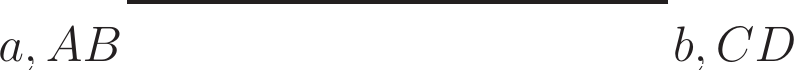}
\end{minipage}  
 &= \langle \phi^{a,AB}(p) \bar{\phi}_{b,CD} (-p)\rangle = \frac{i \delta^{ab}}{p^2}g^2 (\delta^{A}_{C}\delta^{B}_{D}-\delta^{A}_{D}\delta^{B}_{C})\,, \\ \nn
\begin{minipage}[t]{0.3\textwidth}
\includegraphics[width=1.0 \textwidth]{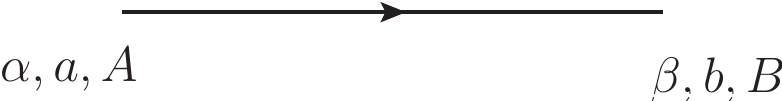}
\end{minipage}  
&=\langle \lambda^{aA}_\alpha(p) \bar{\lambda}_{\dot{\beta},B}^{b}(-p) \rangle = \frac{i\delta^{ab} p_\mu (\bar{\sigma}^\mu)_{\alpha\dot{\beta}} }{p^2}g^2  \delta^{A}_{B}\,.
\end{align}
These propagators can easily be Fourier transformed to position space by using \eqref{eqn:result-fourier-trafo-mink}, e.g. the scalar propagator is
\begin{equation}\label{eqn:scalar-propagator}
 \langle \phi^{a,12} (x_1) \bar{\phi}^{b}_{12}(x_2) \rangle = g^2 \delta^{ab} \int \frac{d^dp}{(2\pi)^d} \frac{i}{p^2} e^{- i px_{12}} = g^2 \frac{\Gamma(h-1)}{4 \pi^h} \frac{\delta^{ab} }{(-x_{12}^2)^{h-1}}\,,
\end{equation}
where $h=d/2$.

Furthermore, we list the Feynman rules for some interaction terms that will be of use in the calculations. From the scalar-scalar-gluon interaction term $\frac{1}{4} D_\mu \phi^{a,AB} D^\mu \bar{\phi}_{a,AB} $ we get
\begin{align}\label{eqn:feynman-rule-scalar-gluon-vertex}
\begin{minipage}[h]{7cm}
\includegraphics[ width=1\textwidth ]{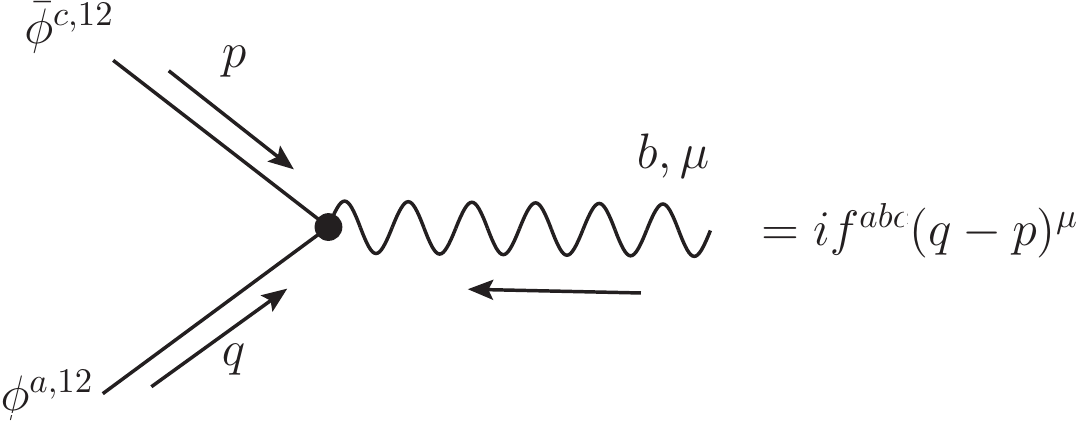}
\end{minipage}
\end{align}
in momentum space.
The interaction-term of four scalars is
\begin{align}
- \frac{1}{16 g^2} f^{abc}f^{ade} \phi^{b,AB} \phi^{c,CD} \bar{\phi}_{d,AB} \bar{\phi}_{e,CD}\,.
\end{align}
Taking into account that $\bar{\phi}_{12}=\phi^{34}$ we find the interaction term with four scalars of the same flavour
\begin{align}
 - \frac{1}{2g^2}  f^{abc}f^{ade} \phi^{b,12}\phi^{e,12}\bar{\phi}_{c,12}\bar{\phi}_{d,12}\,.
\end{align}
However, taking two different flavours, e.g. $\phi^{12}=\bar{\phi}_{34},\phi^{13}=-\bar{\phi}_{24}$ we get
\begin{align}
 - \frac{1}{g^2}  \left( f^{abc}f^{ade}+f^{adc}f^{abe} \right) \phi^{b,12}\phi^{c,13}\bar{\phi}_{d,12}\bar{\phi}_{e,13} \,.
\end{align}
There are two terms since the exchange of $ \{A,B\} \leftrightarrow \{C,D\}$ does not yield identical terms if we choose different flavours.

\begin{align}\label{eqn:feynman-rule-scalar-four-vertex}
\begin{minipage}[h]{3cm}
\includegraphics[ width=1\textwidth ]{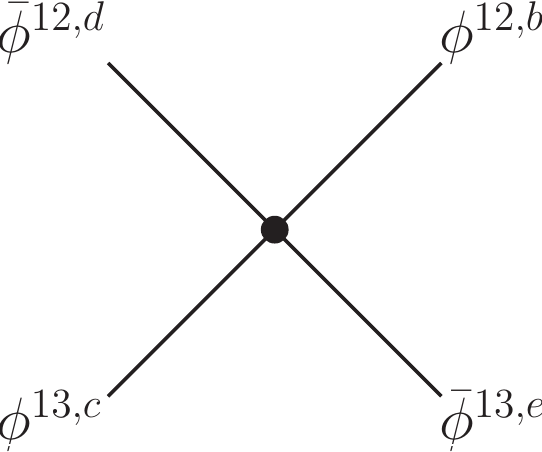} 
\end{minipage}
=  - \frac{1}{g^2}  \left( f^{abc}f^{ade}+f^{adc}f^{abe} \right)\,.
\end{align}

\section{One-Loop Corrections to the Scalar Propagator}
Here, we review the calculation of one-loop corrections to the scalar propagator
\begin{equation}
\langle \phi^{a,12}(p) \bar{\phi}^{b}_{12}(-p) \rangle =  \frac{i\delta^{ab}}{p^2} \,.
\end{equation}
For the correction from the gluon loop we get a symmetry factor of two, since there are two vertices 
\begin{align}
\langle \phi^{a,12}  \bar{\phi}^b_{12} \rangle^{(1)}_{\text{gluons}} &= 2 \frac{(i)^2}{2!} \frac{(i)^4}{p^4} f^{cda} f^{bdc} \left(i\right)^2 \int \frac{d^dl}{(2\pi)^d}   \frac{(l-2p)_\mu(-\eta^{\mu\nu})(l-2p)_\nu}{l^2 (l-p)^2}  \\ \nn
&= f^{acd}f^{bcd} \frac{1}{p^4} \int \frac{d^dl}{(2\pi)^d} \frac{(l-2p)^2}{l^2 (l-p)^2} \\ \nn
&= 2N\delta^{ab} \frac{1}{p^2} \int \frac{d^dl}{(2\pi)^d} \frac{1}{l^2 (l-p)^2} \\ \nn
&=  -\frac{2N\delta^{ab}}{(-p^2)^{3-\frac{d}{2}}} b_0(1,1)
\end{align}
and where we used $f^{acd}f^{bcd}=N\delta^{ab}$. Furthermore, we have to take into account the corrections from the fermion loop. The relevant part of the interaction Lagrangian is
\begin{equation}\label{eqn:interaction-term-fermion-scalar}
\mathcal{L}_{int} = \frac{1}{g^2} \frac{i}{\sqrt{2}} f^{abc}\left(\bar{\lambda}_{\dot{\alpha}}^{a,A}\phi^{b,AB}\bar{\lambda}^{c,\dot{\alpha}B}  -  \lambda^{a,\alpha A} \bar{\phi}^{b,AB} \lambda_\alpha^{c,B} \right)\,.
\end{equation}
There are four terms in each vertex, two with $\phi^{12}$ and two with $\bar{\phi}_{12}$. We get a  symmetry factor of 2 for the vertices and always need both terms in \eqref{eqn:interaction-term-fermion-scalar}. Another factor of 2 is due to $\bar{\phi}_{34}=\phi^{12}$
\begin{align}
\langle \phi^{a,12}  \bar{\phi}^b_{12} \rangle^{(1)}_{\text{fermions}} &= - 4 \frac{i^2}{2!} \left(\frac{i}{\sqrt{2}} \right)^2 f^{def}  f^{ghi} \delta^{ae} \delta^{hb} \frac{(i)^2}{p^4} \int \frac{d^dl}{(2\pi)^d} \\ \nn
&\phantom{=} \epsilon^{\alpha\beta}\epsilon^{\dot{\beta}\dot{\alpha}} \Big(\langle \lambda^{d,1}_\beta \bar{\lambda}^{g,1}_{\dot{\alpha}} \rangle \langle \lambda^{f,2}_\alpha \bar{\lambda}^{i,2}_{\dot{\beta}} \rangle + \langle \lambda^{d,1}_\beta \bar{\lambda}^{i,1}_{\dot{\beta}} \rangle \langle \lambda^{f,2}_\alpha \bar{\lambda}^{g,2}_{\dot{\alpha}} \rangle  \\ \nn
 &\phantom{= } \qquad +  \langle \lambda^{d,2}_\beta \bar{\lambda}^{i,2}_{\dot{\beta}} \rangle \langle \lambda^{f,1}_\alpha \bar{\lambda}^{g,1}_{\dot{\alpha}} \rangle  +\langle \lambda^{d,2}_\beta \bar{\lambda}^{g,2}_{\dot{\alpha}} \rangle \langle \lambda^{f,1}_\alpha \bar{\lambda}^{i,1}_{\dot{\beta}} \rangle \Big) \\ \nn
 &= -2g^4 f^{daf} f^{gbi} \frac{1}{p^4} \int \frac{d^dl}{(2\pi)^d} \frac{p_\mu  (p-l)_\nu}{l^2 (p-l)^2} \\ \nn
 &\phantom{=} \epsilon^{\alpha\beta}\epsilon^{\dot{\beta}\dot{\alpha}} \left( \delta^{dg}\delta^{fi} \bar{\sigma}^\mu_{\beta\dot{\alpha}}\bar{\sigma}^\nu_{\alpha\dot{\beta}}+\delta^{di}\delta^{fg} \bar{\sigma}^\mu_{\beta\dot{\beta}}\bar{\sigma}^\nu_{\alpha\dot{\alpha}}   \right) \\ \nn
 &= -2 g^4 f^{afd} f^{bfd} 2 \,\tr(\sigma^\mu \bar{\sigma}^\nu) \frac{1}{p^4} \int \frac{d^dl}{(2\pi)^d} \frac{p_\mu  (p-l)_\nu}{l^2 (p-l)^2} \\ \nn
 &= -4 g^4 N \delta^{ab} 2 \frac{1}{p^4} \int \frac{d^dl}{(2\pi)^d} \frac{p \cdot  (p-l)}{l^2 (p-l)^2} =  4  g^4 N \frac{\delta^{ab}}{(-p^2)^{3-\frac{d}{2} } } b_0(1,1)\,.
\end{align}
Thus, the one-loop correction to the scalar propagator reads
\begin{align}\label{eqn:one-loop-correction-scalar-rpopagator-in-momentum-space}
\langle \phi^{a,12}(p) \bar{\phi}^{b}_{12}(-p) \rangle^{(1)} = \frac{2 N \delta^{ab}}{(-p^2)^{3-\frac{d}{2}}} b_0(1,1)\,.
\end{align}
Fourier transformation  yields
\begin{equation}
\label{eqn:scalar-propagator-one-loop}
\langle \phi^{a,12}(x_1) \bar{\phi}^{b}_{12}(x_2) \rangle^{(1)} = - g^4 \frac{N\delta^{ab}}{(-x_{12}^2)^{2h-3}} \frac{1}{2} \frac{1}{(4\pi^h)^2} \frac{\Gamma(2h-3)\Gamma(h-1)^2 \Gamma(2-h)}{\Gamma(2h-2)\Gamma(3-h)}\,,
\end{equation}
which has also been calculated in \cite{Erickson:2000af} with a different metric.

\chapter{Introduction to the Mellin-Barnes Technique}\label{sec:introd-mellin-barnes}
In the calculation of Feynman diagrams one often has to deal with integrals of the type
\begin{align}
\int_0^1 ds_{1}...ds_n \frac{1}{\left(X_1(s_1,...s_n)+...+X_m(s_1,...s_n)\right)^\lambda}\,,
\end{align}
where $X_m$  are products of the integration variables $s_i$ and $\bar{s}_i =1-s_i$. There is a possibility to simplify the integrations over the parameters $s_i$ for the price of introducing integrations in the complex plane. Depending on how many terms the sum in the denominator has, this technique may be very useful.

In the following we shall prove the integral representation
\begin{align}\label{eqn:Mellin-rep}
\frac{1}{(X+Y)^\lambda} = \frac{1}{2 \pi i}\frac{1}{\Gamma(\lambda)} \int_{\beta-i\infty}^{\beta+i\infty} \left( \frac{Y^z}{X^{\lambda+z}}\Gamma(\lambda + z) \Gamma(-z) \right)dz\,,
\end{align}
where $-\text{Re}(\lambda)< \beta <0$. Note that the poles of $\Gamma(-z)$ are located at $z_0=0,1,2,...n$ and $\Gamma(\lambda + z)$ has poles at $z_0= -(\lambda),-(\lambda+1),...-(\lambda+n)$ where $n \in \mathbb{N}$.
\begin{itemize}
	\item We first consider the case $Y<X$:\\
By expanding the left-hand side of \eqref{eqn:Mellin-rep} as a Taylor series, one can rewrite
\begin{align}\label{eqn:taylor-series}
(X+Y)^{-\lambda}&=X^{-\lambda}\left(1+\frac{Y}{X}\right)^{-\lambda} \\ \nonumber
&= X^{-\lambda}\sum_{n=0}^{\infty} \frac{(-1)^n}{\Gamma(n+1)} \lambda (\lambda+1)...(\lambda + n -1 ) \left(\frac{Y}{X}\right)^n \\ \nonumber
&= X^{-\lambda}\sum_{n=0}^{\infty} (-1)^n \frac{\Gamma(\lambda+n)}{\Gamma(n+1)\Gamma(\lambda)}\left(\frac{Y}{X}\right)^n\,.
\end{align}
One can evaluate the right-hand side of \eqref{eqn:Mellin-rep} by using of the residue theorem. Since $Y<X$ we can close the contour to the right at $z\rightarrow+\infty$ without adding any contributions to the integral. Note that we get an extra minus sign by reversing the orientation of the integration to the mathematically positive direction. Since $\beta<0$ we enclose the simple poles of $\Gamma(-z)$ at $z_0=0,1,2,...n$. 

By repeated use of 
$$\Gamma(-z)= \frac{1}{-z}\Gamma(-z+1)\,,$$ it is easy to see that the residue of $\Gamma(-z)$ at $z_0=n \in N$ is
\begin{equation}
\text{Res}(z_0=n,\Gamma(-z))= - \frac{(-1)^{n}}{n!}=- \frac{(-1)^n}{\Gamma(n+1)}\,.
\end{equation}
Then we have
\begin{align}
&\phantom{=}\oint_{Re(z)\geq\beta} dz \underbrace{\frac{1}{2 \pi i}\frac{1}{\Gamma(\lambda)}\left( \frac{Y^z}{X^{\lambda+z}}\Gamma(\lambda + z) \Gamma(-z) \right)}_{f(z)} \\ \nonumber
&= -2 \pi i \sum_{n=0}^{\infty}\text{Res}(z_o=n,f(z))\\ \nonumber
&= \sum_{n=0}^{\infty} (-1)^n \frac{\Gamma(\lambda + n)}{\Gamma(\lambda)\Gamma(n+1)} \frac{Y^n}{X^{\lambda+n}}\,.
\end{align}
Note, that the extra minus sign in the second line is due to reversing the orientation of the integration contour to the mathematically \emph{positive} sense.
Comparing this to \eqref{eqn:taylor-series} we find that \eqref{eqn:Mellin-rep} is true for $Y<X$.

\item Now we consider the case $Y>X$:\\
In this case we can close the contour to the left and the orientation of the integration path has mathematically positive sense automatically.
Then we have poles at $z_0=-(\lambda+n)$ for $n \in N$, which yield the residue
\begin{equation}
	\text{Res}\left(z_0=-(\lambda+n),\Gamma(\lambda + z)\right)=Res(z_0=-n,\Gamma(z))=\frac{(-1)^n}{n!}\,.
\end{equation}
Taylor expansion in $X/Y$ therefore yields same result as the integral.
\end{itemize}
\subsubsection{Remarks}
Sometimes it is necessary to pick the real part of the integration contour differently, therefore we pick up different poles. Taking $0<\beta<1$ for example yields
\begin{align}\label{eqn:Mellin-rep-shifted}
\frac{1}{(X+Y)^\lambda} - \frac{1}{(X)^\lambda} = \frac{1}{2 \pi i}\frac{1}{\Gamma(\lambda)} \int_{\beta-i\infty}^{\beta+i\infty} \left( \frac{Y^z}{X^{\lambda+z}}\Gamma(\lambda + z) \Gamma(-z) \right)dz\,.
\end{align}

\chapter{The Wilson Loop Operator}\label{sec:wilson-loop-op-and-gauge-invariance}

The Wilson loop operator defined by
\begin{align}
W(\mathcal{C}) = \frac{1}{N}\tr \mathcal{P} \exp \left( i \oint_{\mathcal{C}} A_\mu dz^\mu \right)
\end{align}
is invariant under $SU(N)$ gauge transformations
\begin{align}\label{eqn:gauge-trafo-A}
A_\mu \rightarrow A_\mu^\prime= g(x)\left(A_\mu + i \partial_\mu \right)g^{-1}(x)\,, \qquad g(x) \in SU(N)\,,
\end{align}
if the path ordering\footnote{There are different conventions on the path ordering. They are equivalent, it is however important to choose the sign in the Wilson loop such that it is gauge invariant. Reversing the sign in the exponent of the Wilson loop does not just reverse the integration contour, due to the path ordering.} is defined as
\begin{align}\label{eqn:definition-path-ordering}
 \mathcal{P} \left(A_\mu(z(t))A_\nu(z(t^\prime)) \right) &= A_\mu(z(t))A_\nu(z(t^\prime)) \qquad \text{for} \qquad t>t^\prime \,,\\ \nonumber
 \mathcal{P} \left(A_\mu(z(t))A_\nu(z(t^\prime)) \right) &= A_\nu(z(t^\prime))A_\mu(z(t)) \qquad \text{for} \qquad t^\prime>t
\end{align}
and $s \in (0,1)$ parametrises the path along the curve $\mathcal{C}$. The path ordered exponential in the Wilson loop operator then has the expansion
\begin{align}\label{eqn:Wilson-Loop-after-path-ordering}
\mathcal{P} \exp{\left(i \oint_{\mathcal{C}} A_{\mu} dz^{\mu}\right)}&=  {\bf1} + i \oint_{\mathcal{C}}dz^{\mu}_i A_{\mu}  + (i)^2 \oint_{\mathcal{C}}dz^{\mu}_i \int^{z_i} dz_j^{\nu} A_{\mu} A_{\nu} \\ \nonumber
&\phantom{=}+ (i)^3 \oint_{\mathcal{C}}dz^{\mu}_i \int^{z_i} dz_j^{\nu}\int^{z_j} dz_k^{\rho} A_{\mu} A_{\nu} A_{\rho} \\ \nonumber
&\phantom{=}+ (i)^4\oint_{\mathcal{C}}dz^{\mu}_1 \int^{z_i} dz_j^{\nu}\int^{z_j} dz_k^{\rho}\int^{z_k} dz_l^{\sigma} A_{\mu} A_{\nu} A_{\rho}A_{\sigma} +~...\,.
\end{align} 
The sign in the exponent of the Wilson line 
\begin{align}
 W(x,y) &= \mathcal{P} \exp \left(i \int_y^x A_\mu dz^\mu \right)
\end{align}
is fixed by requiring the transformation property $W(x,y) \rightarrow g(x) W(x,y) g^{-1}(y)$ under local gauge transformations \eqref{eqn:gauge-trafo-A} of $A_\mu$. \\

\chapter{$\Scs$ (ABJM)  theory}\label{sec:ABJM}

\section{Cherns-Simons Theory}
The Chern-Simons action
\begin{equation}\label{eqn:pure-CS-Lagrangian}
S_{\text{CS}}=  \frac{k}{4 \pi} \int d^dx\, \epsilon^{\mu\nu\rho}\, \tr \left( A_\mu \partial_\nu A_\rho - \frac{2}{3} i A_\mu A_\nu A_\rho \right) 
\end{equation}
is invariant  under $SU(N)$ gauge transformations
\begin{align}
A_\mu \rightarrow A_\mu^\prime= g(x)\left(A_\mu + i \partial_\mu \right)g^{-1}(x)\,, \qquad g(x) \in SU(N)\,.
\end{align}
More precisely, the action is invariant under infinitesimal transformations $g(x)=1 + i \alpha(x)$ and transforms like 
\begin{equation} 
S \rightarrow S^\prime + (2\pi k ) \delta S \,,
\end{equation} 
where
\begin{equation} 
\delta S=- \frac{1}{24\pi^2}\int d^dx\, \epsilon^{\mu\nu\rho}\tr \left( (\partial_\mu g^{-1})g(\partial_\nu g^{-1})g(\partial_\rho g^{-1})g \right)
\end{equation}
under finite transformations. Since $\delta S$ takes integer values, $\exp ({i S})$ is invariant under large gauge transformations for $k \in \mathbb{N}$.

Quantising the theory with the standard Fadeev-Popov procedure yields the gauge fixing and ghost action
\begin{equation}
\mathcal{S}_{\text{g.f.}}=  \frac{k}{4\pi}\int d^dx\,   \tr \left(\frac{1}{\xi} \left( \partial^\mu A_\mu \right)^2 +  \bar{c} \left( \partial^\mu D_\mu \right)c \right)\,,
\end{equation}
where $ D_\mu c= \partial_\mu c +i [A_\mu, c\, ]$. In Landau gauge $(\xi=0)$ the gauge field
propagator reads
\begin{align}%\label{eqn:gluon-prop}
\langle \left(A_\mu\right)_{ij}(x) \left(A_\nu\right)_{kl}(y) \rangle &= \delta_{il} \delta_{jk} \frac{1}{k}\left( \frac{\Gamma\left(\frac{d}{2}\right)}{\pi^{\frac{d-2}{2}}}\right) \epsilon_{\mu\nu\rho}\frac{(x-y)^\rho}{\left(-(x-y)^2\right)^{\frac{d}{2}}}\,,
\end{align}
where we have rescaled the coupling constant $k \rightarrow \mu^{-2\epsilon}k$ and restore the dependence on the regularisation scale $\mu$ only in the final results.
The ghost propagator is
\begin{equation}
\langle c(x) \bar{c}(y) \rangle =  \delta_{il} \delta_{jk} \frac{1}{k}\left( \frac{\Gamma\left(\frac{d}{2}-1\right)}{\pi^{\frac{d-2}{2}}}\right)\frac{1}{\left(-(x-y)^2\right)^{\frac{d}{2}-1}}\,.
\end{equation}
Note that the gauge field propagator is related to the ghost propagator by
\begin{equation}\label{eqn:ghost-gluon-propagator-relation}
\langle \left(A_\mu\right)_{ij}(x) \left(A_\nu\right)_{kl}(y) \rangle = \frac{1}{2}\epsilon_{\mu\nu\rho} \partial^\rho \langle c(x) \bar{c}(y) \rangle \,,
\end{equation}
which can be used to see the cancellation of gauge field 
and ghost loop contributions to the one-loop gauge field propagator in a simple way.
\section{Other Conventions}
The different conventions found in the literature on Chern-Simons theory deserve a short comment. One can show, that 
\begin{align}
\mathcal{S}&= \frac{k}{4 \pi} \int d^dx\, \tr \left( A_\mu \partial_\nu A_\rho - \frac{2}{3} s A_\mu A_\nu A_\rho \right) \epsilon^{\mu\nu\rho}\,,\\
W(\mathcal{C}) &= \frac{1}{N}\tr \mathcal{P} \exp \left( s \oint_{\mathcal{C}} A_\mu dz^\mu \right)
\end{align}
are invariant  ({in the sense mentioned above}) under gauge transformations 
\begin{equation}
A_\mu \rightarrow A_\mu^\prime = g(x)\left(A_\mu - \frac{1}{s} \partial_\mu \right)g^{-1}(x)\,,
\end{equation}
where $s$ is a real or imaginary parameter. The sign in the Wilson loop and the Lagrangian are thus correlated through gauge invariance, see \ref{sec:gauge-invariance-Wilson-loop}. Taking a hermitian gauge field $(A_\mu)^\dagger=A_\mu$ we can choose $s=i$, which corresponds to the choice used throughout this document.\\

All other conventions found in the literature can be obtained by rescaling the gauge field $A_\mu \rightarrow -A_\mu$, $A_\mu \rightarrow i A_\mu$ etc. Note however, that this changes factors in the Lagrangian, the gauge transformation, the covariant derivative and the Wilson loop. A sign difference in the Wilson loop only may also be due to a different definition of the path ordering \eqref{eqn:definition-path-ordering}.

\section{Lagrangian of ABJM Theory}\label{sec:Lagrangian of ABJM theory}
The action of ABJM theory is
\begin{equation}\label{eqn:ABJM-Lagrangian}
\mathcal{S}_{\text{ABJM}}=  \mathcal{S}_{\text{CS}}+\mathcal{S}_{\text{g.f.}}+\mathcal{\hat{S}}_{\text{CS}}+\mathcal{\hat{S}}_{\text{g.f.}}+\mathcal{S}_{\text{matter}}\,,
\end{equation}
where $\mathcal{\hat{S}}$ is obtained from $\mathcal{S}$ by replacing $A_\mu$ with the gauge field in the anti-fundamental representation $\hat{A}_\mu$ and letting $k\rightarrow -k$. Explicitly, we have
\begin{align} \nn
S_{\text{CS}} + \hat{S}_{\text{CS}} & = \frac{k}{4\pi}\,\int d^3x\, \varepsilon^{\mu\nu\rho} \,\Bigl [\,
\Tr (A_\mu\partial_\nu A_\rho-\frac{2}{3}i\,A_\mu A_\nu A_\rho)- 
\Tr (\hat A_\mu\partial_\nu \hat A_\rho-\frac{2}{3}i\, \hat A_\mu \hat A_\nu \hat A_\rho)\, \Bigr ]\,,  \\ \nn
S_\text{gf}+\hat S_\text{gf} & = \frac{k}{4\pi}\, \int d^3 x\, \Bigl [\,\frac{1}{\xi}\, \Tr(\partial_\mu A^\mu)^2
-\Tr(\partial_\mu \bar c\, D_\mu c) - \frac{1}{\xi}\, \Tr(\partial_\mu \hat A^\mu)^2
+\Tr(\partial_\mu \bar{ \hat c}\, D_\mu \hat c) \, \Bigr ]\,, \\ \label{eqn:ABJM-part-Lagrangian}
S_\text{matter} & = \int d^3 x\, \Bigl [\, \Tr(D_\mu\, C_I\, D^\mu {\bar C}^{I}) + i\, \Tr(\bar\psi^I
\, \slsh{D}\, \psi_I)\, \Bigr ] + S_\text{int}\,.
\end{align}
The field content consists of two $U(N)$ gauge fields $(A_\mu)_{ij}$ and
$(\hat A_\mu)_{\hi\hj}$, the complex fields $(C_I)_{i\hi}$ and $({\bar C}^{I})_{\hi i}$ as
well as the fermions $(\psi_I)_{\hi i}$ and $({\bar\psi}^{I})_{i \hi }$ in 
the $({\bf N},{\bf \bar N})$ and $({\bf \bar N},{\bf N})$ of $U(N)$ respectively,
$I=1,2,3,4$ is the $SU(4)_R$ index. We employ the covariant gauge fixing function 
$\partial_\mu A^\mu$ for both gauge fields and have two sets of ghosts $(\bar c,c)$ and
$(\bar{\hat c},\hat c)$. $S_\text{{int}}$ are the sextic scalar potential and $\Psi^2C^2$ Yukawa type potentials spelled out in \cite{Aharony:2008ug} or in components in \cite{Benna:2008zy}. The covariant derivative is given by $D_\mu c = \partial_\mu c + i [A_\mu, .]$. It's action on $\Psi^I, C^I$ is given in \cite{Drukker:2008zx} and not needed here. The Lagrangian has been  been recast into a manifestly $SU(4)$ R-Symmetry invariant form and can be found in \cite{Bandres:2008ry,Lipstein:2011ej}.

\section{One-loop Gauge Field Propagator in ABJM Theory}\label{sec:one-loop-correction-propagator}

Here we calculate the one-loop correction to the gauge field propagator in $d$ dimensions, see also \cite{Drukker:2008zx}. We have fermionic and bosonic contributions in the loop and thus
\begin{equation}
G_{\mu\nu}^{(1)}(p) =G_{\mu\nu}^{(F,1)}(p) +G_{\mu\nu}^{(B,1)}(p) \,,
\end{equation}
where 
\begin{equation}\label{eqn:Gmunu}
G_{\mu\nu}^{(1)}(p) = \left(\frac{2\pi}{k}\right)^2 \frac{\epsilon_{\mu\rho\kappa}p^\kappa}{p^2}\left( \Pi_{\rho\lambda}^{(B)}(p) + \Pi_{\rho\lambda}^{(F)}(p) \right) \frac{\epsilon_{\lambda\nu\delta}p^\delta}{p^2}
\end{equation}
and
\begin{align}
 \Pi_{\mu\nu}^{(B)}(p) &=+N \delta^I_I \mu^{2\epsilon}\int \frac{ d^dk}{(2\pi)^d} \frac{(2k+p)_\mu(2k+p)_\nu}{k^2(p+k)^2}\,, \\ \nonumber
 \Pi_{\mu\nu}^{(F)}(p) &= -N \delta^I_I \mu^{2\epsilon}\int \frac{ d^dk}{(2\pi)^d} \frac{\tr\left(\gamma_\mu(\pslash +\kslash) \gamma_\nu \kslash \right)}{k^2(p+k)^2}\,.
\end{align}
We use the DRED scheme for Dirac matrix operations as well as for Levi-Civita tensor contractions, i.e. we work in strictly $d=3$ to obtain scalar integrands and only then continue the loop momenta to d-dimensional space to perform the integrals in d dimensions. This scheme has been shown to respect the Slavnov-Taylor identities up to two loop order in \cite{Chen:1992ee}.
 Then we have
\begin{equation}
\tr\left(\gamma_\mu(\pslash +\kslash) \gamma_\nu \kslash \right) = 2\left(-\eta_{\mu\nu} (p+k) \cdot k+ 2k_\mu k_\nu+ p_\mu k_\nu + p_\mu k_\nu \right)\,.
\end{equation}
The last two terms can be dropped, since they vanish when contracted with \eqref{eqn:Gmunu}. The same is true for terms proportional to $p_\mu, p_\nu$ in the bosonic term.
Summing up all remaining terms we get
\begin{equation}
+N \delta^I_I \mu^{2\epsilon} 2 \eta_{\mu\nu}\int \frac{ d^dk}{(2\pi)^d} \frac{ k \cdot (p+k)}{k^2(p+k)^2}\,.
\end{equation}
Introducing Feynman parameters, we have
\begin{align}
+N \delta^I_I \mu^{2\epsilon}2 \eta_{\mu\nu} \int_0^1 d\alpha \int \frac{ d^dk}{(2\pi)^d}  \frac{ k \cdot (p+k)}{[(k+\bar{\alpha}p)^2 - \Delta]^2}\,,
\end{align}
where $\Delta= - \alpha \bar{\alpha}p^2$. Then, we shift $k=l-\bar{\alpha} p$ and drop terms linear in $l_\mu$
\begin{align}
+N \delta^I_I \mu^{2\epsilon}2 \eta_{\mu\nu} \int_0^1 d\alpha \int \frac{ d^dl}{(2\pi)^d}  \frac{ l^2-\alpha \bar{\alpha} p^2}{[l^2 - \Delta]^2}\,.
\end{align}
Using the standard integrals
\begin{equation}
\int \frac{ d^dl}{(2\pi)^d} \frac{l^2}{[l^2-\Delta]^2}= -\frac{i}{(4 \pi)^{\frac{d}{2}}} \frac{d}{2} \frac{\Gamma(1-\frac{d}{2})}{\left( \Delta \right)^{1-\frac{d}{2}}}\,, \qquad \int \frac{ d^dl}{(2\pi)^d} \frac{1}{[l^2-\Delta]^2}= \frac{i}{(4 \pi)^{\frac{d}{2}}}  \frac{\Gamma(2-\frac{d}{2})}{\left( \Delta \right)^{2-\frac{d}{2}}}\,
\end{equation}
and $\epsilon_{\lambda\kappa\mu}\epsilon_{\lambda\nu\delta} = \eta_{\kappa\nu}\eta_{\mu\delta}-\eta_{\kappa\delta}\eta_{\mu\nu}$ we get
\begin{align}\label{eqn:Gmunuinserted}
G_{\mu\nu}^{(1)}(p) 
&= \left( N \delta^I_I \mu^{2\epsilon}2  \frac{(-i)}{(4\pi)^{\frac{d}{2}}}  \frac{\Gamma(1-\frac{d}{2}) \Gamma(\frac{d}{2})^2}{\Gamma(d-1)} \right) \left(\frac{2\pi}{k}\right)^2  \frac{1}{{(-p^2)^{3-\frac{d}{2}}}}\left(p_\mu p_\nu - \eta_{\mu\nu}p^2 \right)\,.
\end{align}
Fourier transformation \eqref{eqn:result-fourier-trafo-mink} of  \eqref{eqn:Gmunuinserted}
leads to 
\begin{align}\label{eqn:one-loop-correctd-gluon-prop.}
G_{\mu\nu}^{(1)}&(x) = \mu^{2\epsilon} \int \frac{d^dp}{(2\pi)^d}  G_{\mu\nu}^{(1)}(p)  e^{-ipx} \\ \nonumber
&= \left(\frac{2\pi}{k}\right)^2  \frac{N \delta^I_I}{8} \frac{\Gamma(1-\frac{d}{2}) \Gamma(\frac{d}{2})^2}{\Gamma(d-1)}   \frac{(\mu^{2\epsilon})^2}{\pi^d} \left(  \frac{\Gamma(d-2)}{\Gamma(2-\frac{d}{2})} \frac{\eta_{\mu\nu}}{(-x^2)^{d-2}} -\partial_\mu \partial_\nu\left( \frac{\Gamma(d-3)}{\Gamma(3-\frac{d}{2})} \frac{1}{4}  \frac{1}{(-x^2)^{d-3}}  \right) \right)\,.
\end{align}

\chapter{Details of the Wilson Loop Calculations}
\label{app:Details}

\section{Vertex Diagram for the Tetragon}\label{app:Vertex}
Using $\epsilon(p_1,p_2,z_{12})=\epsilon(p_3,p_2,z_{32})=0$ the first line in \eqref{eqn:vertex-Iijk} can be rewritten as
\begin{equation}
I_{321}=\int d^3s_{1,2,3} \epsilon(p_1,p_2,\partial_{z_1})\epsilon(p_3,p_2,\partial_{z_3}) \int d^d w \frac{(d-2)^{-2}}{|w|^d|w-z_{12}|^{d-2}|w-z_{32}|^{d-2}}\,.
\end{equation}
We begin by introducing Feynman parameters,
\begin{equation}
\int d^d w \frac{1}{|w|^d|w-z_{12}|^{d-2}|w-z_{32}|^{d-2}} = \int [d\beta]_3 \int d^d w \frac{1}{\left(-(w-\beta_1 z_{12}-\beta_3 z_{32})^2 + \Delta\right)^{(3d-4)/2}}\,,
\end{equation}
where
\begin{equation}
\int [d\beta]_3 = \int_0^1 d\beta_1d\beta_2d\beta_3 (\beta_1\beta_2\beta_3)^{(d-2)/2-1}\beta_2 \delta(\sum_i \beta_i-1) \frac{\Gamma(\frac{3d}{2}-2)}{\Gamma(\frac{d}{2})\Gamma(\frac{d}{2}-1)^2}
\end{equation}
and
\begin{equation}\label{eqn:Delta-denominator-vertex}
\Delta= 2 \beta_1\beta_3 (z_{12}\cdot z_{32})- z_{12}^2 \beta_1 \bar{\beta}_1- z_{32}^2 \beta_3 \bar{\beta}_3\,.
\end{equation}
Shifting the integration contour $w\rightarrow l=w - \beta_1 z_{12}-\beta_3 z_{32}$ we have a standard integral
\begin{align}
\int d^d l  \frac{1}{[l^2 - \Delta]^{n}}=(-1)^n  i \pi^{d/2}\frac{\Gamma(n-\frac{d}{2})}{\Gamma(n)}\left(\frac{1}{\Delta}\right)^{n-\frac{d}{2}}\,.
\end{align}
Thus we get
\begin{equation}
I_{321}= \frac{c_1 }{(d-2)^2} \int [d\beta]_3d^3s_{1,2,3} \epsilon(p_1,p_2,\partial_{z_1})\epsilon(p_3,p_2,\partial_{z_3})\frac{1}{\Delta^{d-2}} \,,
\end{equation}
where
%\begin{align}
$c_1= i \pi^{\frac{d}{2}}{\Gamma\left(d-2\right)}/{\Gamma\left({3d}/{2}-2\right)} $.
%\end{align}
Evaluating the action of the derivatives  and abbreviating $x_{13}^2=s, x_{24}^2=t$ we obtain
\begin{align}\label{eqn:I123}
I_{321}&= c_2 s t  \int_0^1 d^3s_{1,2,3} d^3\beta_{1,2,3}(\beta_1\beta_2\beta_3)^{(d-2)/2} \delta(\sum_i \beta_i-1) \\ \nn 
& \qquad \qquad \qquad \qquad \left( \frac{ 1}{\Delta^{d-1}}- 2\frac{(d-1)}{\Delta^{d}}\beta_1\beta_3 \bar{s}_1 s_3 (s+t) ) \right)\,,
\end{align}
where 
%\begin{equation}\label{eqn:c2}
$c_2={i \pi^{{d}/{2}}}{\Gamma(d-1)}/( 8 {\Gamma^{3}({d}/{2})})
%=i + \mathcal{O}(\epsilon)
$
%\end{equation}
and both terms are separately symmetric under $s \leftrightarrow t$. 
Performing the change of variables 
\begin{equation}\label{eqn:variable-change-four-point}
\beta_1= x y\,,\qquad \beta_2= \bar{x} y\,,\qquad\beta_3= \bar{y}\,,\qquad \sum_i \beta_i=1\,\qquad x,y \in [0,1]
\end{equation}
with Jacobian $y$
we can rewrite \eqref{eqn:I123} in a form where all integrations range from $0$ to $1$,
\begin{align}\label{eqn:IAIB}
I_A&= c_2s t  \int_0^1 d^3s dx dy  \frac{(x \bar{x} \bar{y})^{\frac{d-2}{2}}}{\Delta_y^{d-1}} \,, \\ \nonumber
I_B&= - 2 (d-1) c_2 s t (s+t) \int_0^1 d^3s dx dy  \frac{(x \bar{y})^{\frac{d}{2}} \bar{x}^{\frac{d-2}{2}} \bar{s}_1 s_3}{\Delta_y^{d}} \,,
\end{align}
where
\begin{align}
\Delta_y &= - \left( s x \bar{s}_1 (\bar{s}_3 \bar{y} + s_2 \bar{x}y) + t \bar{y}s_3(xs_1+\bar{s}_2 \bar{x} ) \right)
\end{align}
and $I_{321}=I_A + I_B$. 
The integral $I_A$ is divergent as $\bar{s}_1, s_3\rightarrow 0$, see also fig. \ref{fig:divergence-vertex}

\notocsubsection{Numerical Evaluation using the Mellin-Barnes Method}\label{sec:tetragon-MB}
In this section, we switch from the Feynman parametrisation in equation \eqref{eqn:IAIB} 
to a Mellin-Barnes representation, as the latter is very convenient to perform a systematic
expansion in $\epsilon$.
An introduction to the Mellin Barnes technique 
can be found in section \ref{sec:introd-mellin-barnes}, see also \cite{smirnov2006feynman} for a more complete discussion.

In the first step the sum in the denominator is transformed into an integral over a product of terms.  
Since the denominator in \eqref{eqn:I123} consists of a sum of four terms, we will introduce 3 Mellin parameters $z_1,z_2,z_3$.
By repeated use of the Mellin-Barnes representation
\begin{equation}
\frac{1}{(X+Y)^\lambda}= \frac{1}{\Gamma(\lambda)}\frac{1}{2 \pi i} \int_{\beta - i \infty}^{\beta + i \infty}\frac{Y^z}{X^{\lambda +z}}  \Gamma(z+\lambda) \Gamma(-z)dz\,,
\end{equation}
where $- \text{Re} (\lambda) < \beta < 0$, one obtains
\begin{equation}\label{3-fold-MB}
\frac{(2 \pi i)^3 \Gamma(\lambda)}{(a+b+c+d)^\lambda}=\int dz_{1,2,3} a^{z_1}b^{z_2}c^{z_3}d^{-\lambda-z_1-z_2-z_3}\Gamma(-z_1)\Gamma(-z_2)\Gamma(-z_3)\Gamma(\lambda +z_1+z_2+z_3)\,,
\end{equation}
where the real parts $\beta_i$ of the integration contour have to be chosen such that  the arguments in all $\Gamma$ functions have positive real part.
Applying \eqref{3-fold-MB} to the denominator of $I_A$ \eqref{eqn:IAIB}, we can rewrite $I_A$ as
{\small
\begin{align}\nn
 I_A&= \frac{c_2}{\Gamma(d-1)}\int d\tilde{z}_{1,2,3} \Gamma(-z_1)\Gamma(-z_2)\Gamma(-z_3)\Gamma(d-1 +z_1+z_2+z_3) (-s)^{z_1+z_2+1}(-t)^{-d - z_1-z_2+2} \\
&\phantom{=} \int d^3s dx dy  s_1^{z_3}\bar{s}_1^{z_1+z_2} s_2^{z_2} \bar{s}_2^{-z_1-z_2-z_3-d+1}s_3^{-z_1-z_2-d+1}\bar{s}_3^{z_1} x^{z_1+z_2+z_3+d/2-1} \bar{x}^{-z_1-z_3-d/2} y^{z_2}\bar{y}^{-z_2-d/2} \,,
\end{align}
}
where $d\tilde{z} = (2 \pi i)^{-1}dz$. 
The integrals over $s_1,s_2,s_3,x,y$  can be carried out using 
\begin{equation}
\int_0^1 s_i^{a-1}(1-s_i)^{b-1}dt %= B(a,b) 
={\Gamma(a)\Gamma(b)}/{\Gamma(a+b)}\,,
\end{equation}
and we arrive at 
% an expression that contains only $\Gamma$ functions and is very convenient for the numerical extraction of divergent terms:
{\small
\begin{align}\label{eqn:I_Aind3-2eps}
 I_A&= \frac{c_2}{\Gamma(d-1)} \int d\tilde{z}_{1,2,3}{(-s)}^{z_1+z_2+1}{(-t)}^{-d - z_1-z_2+2} \Gamma(-z_1)\Gamma(-z_2)\Gamma(-z_3) \\ \nonumber 
&\times   \Gamma (z_1+1) \Gamma (z_2+1) \Gamma (z_3+1)  \Gamma(d-1 +z_1+z_2+z_3)\Gamma \left(-{d}/{2}-z_2+1\right) \Gamma (z_1+z_2+1) \\ \nonumber
&\times \Gamma(-d-z_1-z_2+2) \Gamma \left(-{d}/{2}-z_1-z_3+1\right) \Gamma (-d-z_1-z_2-z_3+2) \Gamma
  \left({d}/{2}+z_1+z_2+z_3\right)\\ \nonumber
&\times \left[\Gamma \left(2-{d}/{2}\right) \Gamma (-d-z_2+3) \Gamma (-d-z_1-z_3+3)   \Gamma (z_1+z_2+z_3+2)\right]^{-1}\,.
\end{align}}
Recall that we investigate the kinematical region where $s,t<0$.
One can see that this integral is divergent as $\epsilon \rightarrow 0$ by noticing that 
for $\epsilon =0$ it is impossible
to choose the integration contours such that all poles of $\Gamma(...+z_1)$ are to the left of the integration contour 
and all poles of $\Gamma(...-z_1)$ are to the right of the contour. \footnote{This is necessary in order for the previous steps to be well-defined.} 
The reason is that the poles of $\Gamma(z_1 + z_2 +1)$ and $\Gamma (-d-z_1-z_2-z_3+2) =\Gamma(-z_1 - z_2 -1+2 \epsilon)$ ``glue together'' at $z_1=-z_2-1$ for $\epsilon=0$. 
However, one can find allowed contours for $\epsilon \neq 0$.

By shifting the contour left to the pole at $z_1=-z_2-1$ we pick up a residue. 
The factor of $\Gamma(-z_1 - z_2 -1+2 \epsilon)$ evaluated at the residue, results in a divergent 
factor of $\Gamma(2 \epsilon)$. The remaining integral over the shifted contour yields a finite contribution. 

The steps of shifting contours and taking residues have been automatised in \cite{Czakon:2005rk} 
and we used this package to systematically extract the pole terms.
%Feeding the package with
Applying this procedure to \eqref{eqn:I_Aind3-2eps} and expanding in $\epsilon$ yields 3 integrals:
\begin{align}\label{eqn:intsA1}
I_A^{(1)} &= c_2 \int \frac{dz_1}{2 \pi i}\left(\frac{1}{\epsilon}+2 \log(-s) - g_1(z_1)\right) f_1(z_1) + \mathcal{O}(\epsilon)\,,\\ \label{eqn:intsA2}
 I_A^{(2)} &= c_2   \int \frac{dz_1 dz_3}{(2 \pi i)^2}\left(\frac{1}{\epsilon}+2 \log(-s) - g_2(z_1,z_3)\right) f_2(z_1,z_3) + \mathcal{O}(\epsilon)\,,\\ \label{eqn:intsA3}
 I_A^{(3)} &= c_2  \int \frac{dz_1dz_2 dz_3}{(2 \pi i)^3} \left(\frac{s}{t}\right)^{1+z_1+z_2}f_3(z_1,z_2,z_3)\,.
\end{align}
We do not specify the values of the real parts $\beta_i$ as well as the functions $f_i,g_i$ here, which are lengthy expressions of products of $\Gamma$ functions and can be obtained automatically by expanding   \eqref{eqn:I_Aind3-2eps} with the help of \cite{Czakon:2005rk}.
Adding up the divergent part of \eqref{eqn:intsA1} and \eqref{eqn:intsA2} we get
\begin{equation}
I^{\text{vertex}}_{\text{div}}= \frac{a_1}{\epsilon} c_2 \,,
\end{equation}
where by numerical integration one finds
\begin{align}
a_1 = \int \frac{dz_1}{2 \pi i}f_1(z_1) +\int \frac{dz_1dz_3}{(2 \pi i)^2}f_2(z_1,z_3)
= 8.710344 \pm 10^{-6}
\approx 4 \pi \ln(2)=8.710344361\ldots
\end{align}
accurately approximated by our analytic guess.
Further numerical evaluation of the finite part of $I_A^{(1)}$, $I_A^{(2)}$, $I_A^{(3)}$ yields
\begin{align}
I_{A,\text{finite}} = c_2 \left( a_1 \ln (-s) + a_1 \ln (-t) +a_2 \ln^2 \left(\frac{s}{t}\right) + a_4 \right)\,.
\end{align}
where $a_1$ has the same value as above and 
\begin{align}
a_2 = - 0.84 \pm 0.01\, \qquad  a_4 = -14.375216465 \pm 10^{-9}\,.
\end{align}
Numerical analysis for $I_B$ suggests
\begin{align}
I_B=c_2 \left(a_3\ln^2\left(\frac{s}{t}\right) + a_5 \right)\,,
\end{align}
where
\begin{align}
a_3=3.97\pm0.01\,,\qquad a_5 = 40.620843911 \pm 10^{-9}\,.
\end{align}
Adding up $a_2$, $a_3$ we obtain
\begin{equation}
a_2+a_3=3.136\pm 0.02 \approx \pi\,.
\end{equation}
Thus suggesting
\begin{align}\nn
I_A+I_B&=c_2  \left(\frac{a_1}{\epsilon}+a_1\left(\ln(-s)+\ln(-t) \right) + (a_2+a_3) \ln \left(\frac{s}{t}\right) + (a_4+a_5) + \mathcal{O}(\epsilon) \right)\\ \nonumber
&\approx c_2 \pi \left(\frac{4 \ln(2)}{\epsilon}+4 \ln(2) \left(\ln(-s)+\ln(-t) \right) + \ln^2 \left(\frac{s}{t}\right) + a_6 + \mathcal{O}(\epsilon) \right)\,,
\end{align}
where
\begin{equation}\label{eqn:a6}
a_6=(a_4+a_5)/\pi= 8.354242685 \pm 2\cdot 10^{-9}\,.
\end{equation}
The constant fits the value $a_6 \approx -\frac{2}{3}\pi^2 + 16 \ln(2) + 8 \ln^2 (2) =8.35424273...$ . The result can be rewritten in the form
\begin{align}
I_A+I_B
&\approx c_2 \pi \left(2 \ln(2)\frac{ ((-s) ^{2\epsilon}+(-t)^{2\epsilon})}{\epsilon}  + \ln^2 \left(\frac{s}{t}\right) + a_6 + \mathcal{O}(\epsilon) \right)\,.
\end{align}

\section{Gauge Field and Ghost Loops}\label{app:gluon-and-ghost-loops}
It is well known \cite{Chen:1992ee}, that the contributions of ghost and gauge field loops to the gauge field self energy cancel. We briefly review the cancellation of the gauge field and ghost loop corrections in the Wilson loop, since from this it is easy to see, how the cancellation for the insertions in the conformal Ward identities takes place.  
%The relevant part of the action is
%\begin{align}
% S=  \frac{k}{4\pi}\int d^dx\tr\left(A_\mu \partial_\nu A_\rho - \frac{2}{3} i A_\mu A_\nu A_%rho\right)- \frac{k}{4\pi}\int d^dx\tr\left( \partial^\mu \bar{c}\, D_\mu c \right)+...
%\end{align}
%where
%\begin{equation}
%D_\mu c=\partial_\mu c + i [A^\mu,c] \,.
%\end{equation}
\notocsubsection{Gauge Field Loop}
The gauge field loop-diagram arises at second order in perturbation theory
\begin{align}
&\langle W_n \rangle^\text{gluon-loop} \\ \nn
&= \frac{1}{N}\langle \tr(-\oint dz_i^\mu dz_j^\nu A_\mu A_\nu ) \left(-\frac{1}{2!}\right)\left( \frac{k}{4\pi}  \int d^dx\epsilon^{\alpha\beta\gamma}  \tr(\frac{2}{3}iA_\alpha A_\beta A_\gamma)\right)^2 \rangle \\ \nonumber
&=c_8 \left(\frac{2}{3}\right)^2\oint dz_i^\mu dz_j^\nu \int d^dx d^dy  \epsilon^{\alpha\beta\gamma} \epsilon^{\delta\sigma\tau} \langle \tr(A_\mu A_\nu) \tr (A_\alpha A_\beta A_\gamma)(x)\tr (A_\delta A_\sigma A_\tau)(y) \rangle \,,
\end{align}
where 
\begin{equation}
c_8 = -\frac{1}{N}\frac{1}{2} \left(\frac{k}{4\pi} \right)^2\,.
\end{equation}
Taking into account that $A_\mu$, $A_\nu$ give 3 identical contractions with one of the vertex terms and that we can contract them either with the $x$- or $y$-dependent vertex, we get a symmetry factor of $3\cdot 3\cdot 2$. The remaining contractions of the gauge fields are dictated by taking into account only planar diagrams. Thus we get
\begin{align}\label{eqn:gluon-loop}
\langle W_n \rangle^\text{gluon-loop}&= 8 c_8  \oint dz_i^\mu dz_j^\nu \int d^dx d^dy \epsilon^{\alpha\beta\gamma} \epsilon^{\delta\sigma\tau} \langle A_\mu A_\alpha \rangle \langle A_\nu A_\delta \rangle \langle A_\beta A_\tau \rangle \langle A_\gamma A_\sigma \rangle \,.
\end{align}
To proceed, we recall the relation \eqref{eqn:ghost-gluon-propagator-relation} between 
gauge field and ghost propagator and write 
\begin{align}
\epsilon^{\alpha\beta\gamma} \epsilon^{\delta\sigma\tau}  \langle A_\beta A_\tau \rangle \langle A_\gamma A_\sigma \rangle 
&= \frac{1}{4}\, {\epsilon^{\alpha\beta\gamma} \epsilon^{\delta\sigma\tau} \epsilon_{\beta\tau\kappa} \epsilon_{\gamma\sigma\rho}}\partial^\rho_{x} \langle c(x) \bar{c}(y) \rangle \partial^\kappa_{x} \langle c(x) \bar{c}(y) \rangle \nn\\
&= \frac{1}{2} \partial^\rho_{x} \langle c(x) \bar{c}(y) \rangle \partial^\kappa_{y} \langle c(x) \bar{c}(y) \rangle\,,
\end{align}
where we used $
\epsilon^{\alpha\beta\gamma} \epsilon^{\delta\sigma\tau} \epsilon_{\beta\tau\kappa}\epsilon_{\gamma\sigma\rho} ={- \left(\eta^\alpha_\kappa \eta^\delta_\rho+\eta^\alpha_\rho \eta^\delta_\kappa \right)}$ and
$\partial_{x} F(x-y)= - \partial_{y} F(x-y)$ in the last step.

\notocsubsection{Ghost Loop}
The ghost loop diagram arises from contraction of the second order perturbation theory expansion of the gauge-field-ghost vertex term
\begin{align}
\langle W_n \rangle^\text{ghost loop}&= 
\frac{1}{N}\langle \tr(-\oint dz_i^\mu dz_j^\nu A_\mu A_\nu ) \left(-\frac{1}{2!}\right)\left( \frac{k}{4\pi}  \int d^dx  \tr(\partial^\mu \bar{c}\,i[A_\mu,c] ) \right)^2 \rangle \\ \nonumber
&= 2c_8 \oint dz_i^\mu dz_j^\nu \int d^dx d^dy  \langle \tr(A_\mu A_\nu) \tr(\partial^\rho_{x} \bar{c} A_\rho c) \tr(\partial^\sigma_{y} \bar{c} A_\sigma c)  \rangle \,,
\end{align}
where $c_8$ ist the same as defined above and the factor of $2$ is due to 
the fact that the evaluation of the first line yields two identical planar diagrams 
that are kept and two identical non-planar diagrams that we drop. 
Contracting $A_\mu$ either with the $x$- or $y$-dependent vertex, we get a symmetry factor of 2. 
There is only one way for the remaining contractions and thus we get
\begin{align}\label{eqn:ghost-loop}
\langle W_n \rangle^\text{ghost loop}&=-4 c_8\,  \oint dz_i^\mu dz_j^\nu \int d^dx d^dy \langle A_\mu A_\sigma \rangle  \langle A_\nu A_\rho \rangle \partial_x^\rho \langle c(y) \bar{c}(x) \rangle \partial_y^\sigma \langle c(x) \bar{c}(y) \rangle\,,
\end{align}
where a factor of $-1$ due to the anti-commuting ghost fields in the loop was taken into account.
Summing up \eqref{eqn:gluon-loop} and \eqref{eqn:ghost-loop} we get
\begin{align} \langle W_n \rangle^\text{gauge field loop}+\langle W_n \rangle^\text{ghost loop}=0\,.
\end{align}
The same relation \eqref{eqn:ghost-gluon-propagator-relation} can be used to show the vanishing for the dilatation and special conformal Ward identities.

\section{Conformal Ward identity}\label{app:conf-wi-two-loop}

\notocsubsection{Insertion of the Interaction Term} \label{app:conf-wi-insertion-interaction-term}
We can rewrite \eqref{eqn:I321prime} as
\begin{align}
I_{321}^\prime&=\frac{1}{(d-2)^2}\int d^3s_{1,2,3} \epsilon(p_1,p_2,\partial_{z_1})\epsilon(p_3,p_2,\partial_{z_2}) \int d^dx  \frac{(x+z_2)^\nu}{|x|^{d}|x-z_{12}|^{d-2}|x-z_{32}|^{d-2}}\,.
\end{align}
Introducing Feynman parameters, changing the integration variable to $l=x-\beta_1 z_{12}-\beta_3 z_{32}$, using the same notation as in app. \ref{app:Vertex}, integrating over $l$ 
%\begin{align}
%I_{321}^\prime&=\frac{(-1)^{\frac{3d}{2}-2}}{(d-2)^2}\int d^3s_{1,2,3} %\epsilon(p_1,p_2,\partial_{z_1})\epsilon(p_3,p_2,\partial_{z_2}) \int d[\beta]_3\int %d^dl  \frac{(l + \beta_1 z_1+ \beta_2 z_2+ \beta_3 %z_3)^\nu}{(l^2-\Delta)^{\frac{3d}{2}-2}}\\ \nonumber
%&= \frac{1}{(d-2)^2}\int d^3s_{1,2,3} %\epsilon(p_1,p_2,\partial_{z_1})\epsilon(p_3,p_2,\partial_{z_2}) \int d[\beta]_3 %(\beta_1 z_1+ \beta_2 z_2+ \beta_3 z_3)^\nu \frac{c_1}{\Delta^{d-2}}
%\end{align}
and evaluating the action of the derivatives yields
\begin{align}\nonumber 
I_{321}^\prime &=\frac{c_1}{(d-2)^2}\int d^3s_{1,2,3} \int d[\beta]_3\Big( (\beta_1 z_1+ \beta_2 z_2+ \beta_3 z_3)^\nu  \epsilon(p_1,p_2,\partial_{z_1})\epsilon(p_3,p_2,\partial_{z_2}) \frac{1}{\Delta^{d-2}}
\\ 
&\phantom{=}\quad  \qquad\qquad\qquad +\epsilon(p_1,p_2,p_3)2 \beta_1 \beta_3 \epsilon_{\alpha\beta}^{~~~\nu}p_2^\beta \left(\beta_1p_1^\alpha \bar{s}_1+\beta_3 p_3^\alpha s_3 \right)\frac{(2-d)}{\Delta^{d-1}} \Big)\,,
\end{align}
The last term can be shown to be finite and the first term is very similar to the vertex diagram. Evaluation of the derivatives as in \ref{app:Vertex} yields
\begin{align}\label{eqn:I123}
I_{321}^\prime &= c_2  s t  \int_0^1 d^3s_{1,2,3} d^3\beta_{1,2,3}(\beta_1\beta_2\beta_3)^{(d-2)/2} \delta(\sum_i \beta_i-1)\\ \nonumber
&\phantom{=}\qquad \qquad (\beta_1 z_1+ \beta_2 z_2+ \beta_3 z_3)^\nu \left( \frac{ 1}{\Delta^{d-1}}- 2\frac{(d-1)}{\Delta^{d}}\beta_1\beta_3 \bar{s}_1 s_3 (s+t)  \right) + \text{finite}\,.
\end{align}
It can be shown, e.g. using the Mellin Barnes technique as in \ref{app:Vertex}, that all divergent contributions are due to the first term. We have the following divergent contributions:
\begin{align}
st \int d^3s d[\beta]_3 \beta_1\beta_2\beta_3 \left(\frac{\beta_1 z_1}{\Delta^{d-1}}\right)&= \frac{1}{\epsilon} \, {a\, x_2^\nu } + \mathcal{O}(\epsilon^0)\,,\\ \nonumber
st \int d^3s d[\beta]_3 \beta_1\beta_2\beta_3 \left(\frac{\beta_2 z_2}{\Delta^{d-1}}\right)&= \frac{1}{\epsilon}\, {b \, (x_2^\nu+x_3^\nu) } + \mathcal{O}(\epsilon^0)\,, \\ \nonumber
st\int d^3s d[\beta]_3 \beta_1\beta_2\beta_3 \left(\frac{\beta_3 z_3}{\Delta^{d-1}}\right)&= \frac{1}{\epsilon}\left(a\, x_3^\nu  \right) + \mathcal{O}(\epsilon^0)\,.
\end{align}
Numerical evaluation yields
\begin{align}
a= 1.8562\pm 0.0001 \,,\quad
b= 2.4989 \pm 0.0001\,.
\end{align}
To good accuracy we find
\begin{equation}
a+b=4.35517 \pm 0.0002 \approx 2 \pi \ln(2) =4.35517...\,.
\end{equation}
Summarising $I_{321}^\prime$ then reads
\begin{equation}
I_{321}^\prime =  \frac{c_2}{\epsilon} (a+b)(x_2+x_3)^\nu + \mathcal{O}(\epsilon^0)
\approx  \frac{2 \pi i\ln(2)}{\epsilon}(x_2+x_3)^\nu+ \mathcal{O}(\epsilon^0)\,.
\end{equation}

\notocsubsection{Insertion of the Kinetic Term into the Vertex Diagram}\label{app:conf-wi-insertion-kinetic-term-in-vertex}
For the kinetic insertion into the vertex diagram we have \eqref{eqn:kinetic-insertion-in-vertex}
\begin{align}\langle \mathcal{L}(x) W_4 \rangle^{(2)}_{\text{(c)}} \nonumber
&=  \underbrace{\left( \frac{N}{k}\right)^2 \frac{i}{8 \pi^2} \left( \frac{\Gamma\left(\frac{d}{2}\right)}{\pi^{\frac{d-2}{2}}}\right)^4}_{=: c_3} \int d^dw  \oint dz_{i,j,k}^{\mu\nu\rho} \epsilon^{\delta\sigma\tau} I_{\nu\sigma}\, G_{\mu\tau}(z_i-w) G_{\rho\delta}(z_k-w) \\ 
& \qquad \qquad \qquad  \qquad \qquad \qquad + \text{cyclic}(\mu,\nu,\rho;z_i,z_j,z_k)\,,
\end{align}
where $G_{\mu\nu}(x-y)=\epsilon_{\mu\nu\rho} \frac{(x-y)^\rho}{\left(-(x-y)^2\right)^{\frac{d}{2}}}$ and
\begin{align}\label{eqn:integration-yields-prop}
I_{\nu\sigma}(x- z_j,x-w) &=\epsilon^{\alpha\beta\gamma}\left(G_{\alpha\nu}(x-z_j) \partial_\beta^{(x)} G_{\gamma\sigma}(x-w) + G_{\alpha\sigma}(x-w) \partial_\beta^{(x)} G_{\gamma\nu}(x-z_j) \rangle \right) \,.
\end{align} 
and the two other contractions are contained in cyclic$(\mu,\nu,\rho;z_i,z_j,z_k)$.
For the dilatation Ward identity the integration over $x$ can be performed by introducing two Feynman parameters. The result simply yields a propagator
\begin{align}
\int d^dx\, I_{\nu\sigma}
&= - \frac{4  \pi^{\frac{d}{2}}}{\Gamma\left(\frac{d}{2}\right)} \epsilon_{\nu\sigma\varphi}  \frac{(z_j-w)^\varphi}{(-(z_j-w)^2)^{\frac{d}{2}}} = - \frac{4   \pi^{\frac{d}{2}}}{\Gamma\left(\frac{d}{2}\right)} G_{\nu\sigma}(z_j-w)\,,
\end{align}
In the case of the special conformal Ward identity the integration is a little more involved. The integral over $d^dx$ can be solved by introducing Feynman parameters and after some algebra one finds
\begin{align}
\int d^d x \,x^\lambda  \langle \mathcal{L}(x) W_4\rangle^{(2)}_{(c)} &= 2 c_3 c_4 \oint dz_{i,j,k}^{\mu,\nu,\rho}\int d^dw  \epsilon^{\delta\sigma\tau}\epsilon^{\alpha\beta\gamma}\epsilon_{\alpha\nu\xi} \epsilon_{\gamma\sigma\varphi} \\ \nonumber
&\phantom{=} \partial^\xi \left((2 \eta^{\varphi\lambda}\partial_\beta + \eta_\beta^{\lambda}\partial^\varphi) \left(\frac{1}{((z_j-w)^2)^{\frac{d}{2}-2}}\right)- \partial^\varphi \partial_\beta\left(\frac{(z_j+w)^{\lambda}}{((z_j-w)^2)^{\frac{d}{2}-2}}\right)\right) \\ \nonumber
&\phantom{=} G_{\rho\delta} G_{\mu\tau} + \text{cyclic} (\mu,\nu,\rho; z_i,z_j,z_k)\,,
\end{align}
where all derivatives are taken with respect to $z_j$ and $c_4$ is a constant obtained through integration over $d^dx$. Inserting the propagators we can write this in a form convenient to solve the integral over $d^dw$
\begin{align}
&2 c_3 c_4 \oint dz_{i,j,k}^{\mu,\nu,\rho} \epsilon^{\delta\sigma\tau}\epsilon^{\alpha\beta\gamma}\epsilon_{\alpha\nu\xi} \epsilon_{\gamma\sigma\varphi}\epsilon_{\rho \delta \chi} \epsilon_{\mu \tau \theta} \partial^\theta_i \partial^\xi_j \partial^\chi_k \left((2 \eta^{\varphi\lambda}\partial_{j,\beta} + \eta_\beta^{\lambda}\partial^\varphi_j) J_j- \partial^\varphi_j \partial_{\beta,j} J_j^\lambda \right) 
+ \text{cyclic}\,,
\end{align}
where the integrals read
\begin{align}
J_j &= \frac{1}{ (2-d)^2} \int d^dw \frac{1}{((z_j-w)^2)^{\frac{d}{2}-2} ((z_i-w)^2)^{\frac{d}{2}-1} ((z_k-w)^2)^{\frac{d}{2}-1}} \,, \\ \nonumber
J_j^{\lambda}&= \frac{1}{ (2-d)^2} \int d^dw \frac{(z_j+w)^{\lambda}}{((z_j-w)^2)^{\frac{d}{2}-2} ((z_i-w)^2)^{\frac{d}{2}-1} ((z_k-w)^2)^{\frac{d}{2}-1}}\,.
\end{align}
The integrations over $w$ can be performed by introducing three Feynman parameters $\beta_i$ and we get
\begin{align}
J_j &=  c_5\int d[\beta]_{3,j} \left(\frac{1}{\Delta}\right)^{d-4}\,, \qquad
J_j^\lambda = c_5\int d[\beta]_{3,j}  (z_j + \sum_i \beta_i z_i )^\lambda \left(\frac{1}{\Delta}\right)^{d-4}\,,
\end{align}
where
\begin{align}\label{eqn:feynman-measure}
\int d[\beta]_{3,j} &= \int_0^1 d\beta_i d\beta_j d\beta_k \delta\left(\sum_i \beta_i  -1\right) \left(\beta_i \beta_j \beta_k \right)^{\frac{d}{2}-2} \beta_j^{-1}
\end{align}
and $c_5$ is a constant obtained by integrating over $w$, the product $c_3 c_4 c_5 $ is explicitly given below.
The expression for $\Delta$ is the same as in \eqref{eqn:Delta-denominator-vertex}
For the cyclic permutations we can use the same expression, replacing the measure with $d[\beta]_{3,i}$ respectively $d[\beta]_{3,k}$, i.e. exchanging $\beta_j^{-1}$ with $\beta_i^{-1}$ respectively $ \beta_k^{-1}$ in \eqref{eqn:feynman-measure}. 
~\\
All three contributions can then be written as
\begin{align}
2 c_3 c_4 c_5 \oint dz_{i,j,k}^{\mu,\nu,\rho} &\epsilon^{\delta\sigma\tau}\epsilon^{\alpha\beta\gamma}\epsilon_{\gamma\sigma\varphi} 
 \\ \nn 
 &\Big( \epsilon_{\alpha\mu\xi} \epsilon_{\nu \delta \chi} \epsilon_{\rho \tau \theta} \partial^\theta_k \partial^\xi_i \partial^\chi_j \left((2 \eta^{\varphi\lambda}\partial_{i,\beta} + \eta_\beta^{\lambda}\partial^\varphi_i) J_i- \partial^\varphi_i \partial_{\beta,i} J_i^\lambda \right) \\ \nonumber
& \phantom{\Big(}\epsilon_{\alpha\nu\xi} \epsilon_{\rho \delta \chi} \epsilon_{\mu \tau \theta} \partial^\theta_i \partial^\xi_j \partial^\chi_k \left((2 \eta^{\varphi\lambda}\partial_{j,\beta} + \eta_\beta^{\lambda}\partial^\varphi_j) J_j- \partial^\varphi_j \partial_{\beta,j} J_j^\lambda \right) \\ \nn
& \phantom{\Big(} \epsilon_{\alpha\rho\xi} \epsilon_{\mu \delta \chi} \epsilon_{\nu \tau \theta} \partial^\theta_j \partial^\xi_k \partial^\chi_i \left((2 \eta^{\varphi\lambda}\partial_{k,\beta} + \eta_\beta^{\lambda}\partial^\varphi_k) J_k- \partial^\varphi_k \partial_{\beta,k} J_k^\lambda \right) 
\Big)\,,
\end{align}
where 
\begin{align}
2 c_3 c_4 c_5 = i\frac{ \pi^{2-d}}{128} \left(\frac{N}{k}\right)^2 \Gamma(d-4)\,.
\end{align}
We can evaluate the derivatives and contractions with the computer and find that the non-vanishing contributions have the structure
\begin{align}
&\phantom{=} \int d^dx\,x^\lambda \langle \mathcal{L}(x) W_4 \rangle^{\text{vertex-insertion}} \\ \nn
&=\left(\frac{N}{k}\right)^2 \sum_{i>j>k}\int_0^1 ds_{i,j,k}\int d[\beta]_3\left(I_{ijk,-d-1}^\lambda+I_{ijk,-d}^\lambda+I_{ijk,-d+1}^\lambda \right)
\end{align}
where $I_{ijk,p}$ are lengthy terms proportional to $1/\Delta^p$.

For the conformal Ward identity we are only interested in the divergent part of the above quantities, which can be automatically extracted with the Mellin-Barnes technique. We find that all terms vanish except for $i\neq j \neq k$. Specialising to the case $i=3, j=2, k=1$ we find
\begin{align}
\int I_{312,-d-1}^\lambda&= \mathcal{O}(\epsilon^0)\,,\\ \nonumber
\int I_{312,-d}^\lambda&= \frac{i}{\epsilon} ( a_1 x_2^\lambda+a_2 x_3^\lambda)+ \mathcal{O}(\epsilon^0)\,, \\ \nonumber
\int I_{321,-d+1}^\lambda&= \frac{i}{\epsilon} ( b_1 x_2^\lambda+b_2 x_3^\lambda)+  \mathcal{O}(\epsilon^0)\,.
\end{align}
Numerical evaluation of the integrals yields
\begin{align}
a_1 &= 0.3465735 \pm 10^{-6}  \approx \frac{1}{2}\ln(2) = 0.3465735...\,, \\ \nonumber
a_2 &= 0.3465735 \pm 8\cdot 10^{-7}  \approx \frac{1}{2}\ln(2) = 0.3465735...\,, \\ \nonumber
b_1 &=-0.8664339 \pm 14\cdot 10^{-7}\approx -\frac{5}{4}\ln(2) = -0.86643397...\,, \\ \nonumber
b_2 &=-0.8664339 \pm 10^{-6} \approx -\frac{5}{4}\ln(2) = -0.86643397...\,. 
\end{align}
Adding up the results, summing over all four diagrams and taking into account the corresponding prefactors we get
\begin{align}
&\phantom{=} \int d^dx\,x^\lambda \langle \mathcal{L}(x) W_4 \rangle^{(2)}_{(c)} \approx -i\frac{3}{4} \frac{\ln(2)}{\epsilon}\left(\sum_i x_i^\lambda\right)+  \mathcal{O}(\epsilon^0)\,.
\end{align}

\section{Vertex Diagram for Higher Polygons}\label{app:Vertex-n-gon}
We start from \eqref{eqn:vertex-diagram}
\begin{equation}\label{eqn:vertex-diagram}
\langle W_n \rangle^{(2)}_{\text{vertex}}
=   \left(\frac{N}{k}\right)^2 \frac{i}{2 \pi}  \left( \frac{\Gamma\left(\frac{d}{2}\right)}{\pi^{\frac{d-2}{2}}}\right)^3 \sum_{i>j>k} I_{ijk} \,,
\end{equation}
where
\begin{equation}
I_{ijk}= \frac{1}{(d-2)^3}\int dz_{i,j,k}^{\mu,\nu\rho}  \epsilon^{\alpha\beta\gamma}\epsilon_{\mu\alpha\sigma}\epsilon_{\nu\beta\lambda}\epsilon_{\rho\gamma\tau} \partial_i^\sigma \partial_j^\lambda \partial_k^\tau \left( J \right)
\,,
\end{equation}
and
\begin{equation}
J= \int d^dw \frac{1}{\left((-w^2)(-(w-z_{ij})^2)(-(w-z_{kj})^2)\right)^{d/2-1}}\,.
\end{equation}
We begin by introducing Feynman parameters,
\begin{equation}
J = \int d^d w \frac{1}{\left(|w| |w-z_{ij}| |w-z_{kj}|\right)^{d-2}} = \int [d\beta]_3 \int d^d w \frac{1}{\left(-(w-\beta_1 z_{ij}-\beta_3 z_{kj})^2 + \Delta\right)^{3d/2-3}}\,,
\end{equation}
where $|w|=(-w^2)^{1/2}$ and
\begin{equation}
\int [d\beta]_3 = \int_0^1 d\beta_id\beta_j d\beta_k (\beta_i\beta_j\beta_k)^{(d-2)/2-1}\delta(\sum_i \beta_i-1) \frac{\Gamma(\frac{3d}{2}-3)}{\Gamma(\frac{d}{2}-1)^3}
\end{equation}
and
\begin{equation}
\Delta= 2 \beta_i\beta_k (z_{ij}\cdot z_{kj})- z_{ij}^2 \beta_i \bar{\beta}_i- z_{kj}^2 \beta_k \bar{\beta}_k
\end{equation}
or when taking into account the delta function
\begin{equation}\label{eqn:Delta-denominator-vertex}
\Delta= - z_{ij}^2 \beta_i \beta_j- z_{ik}^2 \beta_i {\beta}_k- z_{kj}^2 \beta_k {\beta}_j\,.
\end{equation}
Shifting the integration contour $w\rightarrow l=w - \beta_1 z_{ij}-\beta_3 z_{kj}$ we have a standard integral
\begin{align}
\int d^d l  \frac{1}{[l^2 - \Delta]^{n}}=(-1)^n  i \pi^{d/2}\frac{\Gamma(n-\frac{d}{2})}{\Gamma(n)}\left(\frac{1}{\Delta}\right)^{n-\frac{d}{2}}
\end{align}
and thus
\begin{equation}
J = \int [d\beta]_3 i \pi^{d/2} \frac{\Gamma(d-3)}{\Gamma(3d/2-3)} \frac{1}{\Delta^{d-3}}\,.
\end{equation}
Since only second and first derivatives  $ \partial_i^\sigma  \partial_k^\tau \Delta = 2 \eta^{\tau\sigma} \beta_i \beta_k$ and $\partial_i^\sigma \Delta = -2 (z_{ij}^\sigma  \beta_j + z_{ik}^\sigma \beta_k) \beta_i$ are non-vanishing we have
\begin{align}
\partial_i^\sigma \partial_j^\lambda \partial_k^\tau \Delta^{3-d} &= \left(3-d\right)\left(2-d\right) \Delta^{1-d}  \left[  \partial_j^\lambda  \Delta  \partial_i^\sigma  \partial_k^\tau \Delta  + 2\, terms   \right] \\ \nn
 &+ \left(3-d\right)\left(2-d\right) \left(1-d\right)\Delta^{-d} \partial_i^\sigma \Delta \partial_j^\lambda \Delta \partial_k^\tau  \Delta \,.
\end{align}
We thus split up the integral as
\begin{align}
I_{ijk} = c \left( I^{(a)}_{ijk} + I^{(b)}_{ijk}\right)\,,
\end{align}
where 
\begin{align}\label{eqn:original-expression-for-numerics}
I^{(a)}_{ijk} &= \int d[s,\beta]_6  \epsilon^{\alpha\beta\gamma}\epsilon_{\mu\alpha\sigma}\epsilon_{\nu\beta\lambda}\epsilon_{\rho\gamma\tau}  p_i^\mu p_j^\nu p_k^\rho  \left[  \partial_j^\lambda  \Delta  \partial_i^\sigma  \partial_k^\tau \Delta  + 2\, \text{terms}   \right] \Delta^{1-d}\,, \\ \nn
I^{(b)}_{ijk} &= \int d[s,\beta]_6 \epsilon^{\alpha\beta\gamma}\epsilon_{\mu\alpha\sigma}\epsilon_{\nu\beta\lambda}\epsilon_{\rho\gamma\tau}   p_i^\mu p_j^\nu p_k^\rho   \left[\partial_i^\sigma \Delta \partial_j^\lambda \Delta \partial_k^\tau  \Delta \right] \Delta^{-d} (1-d)
\end{align}
and 
\begin{equation}
 \int d[s,\beta]_6 = \int_0^1 ds_{i,j,k}\int_0^1 d\beta_{i,j,k}\delta(\sum_m \beta_m-1)(\beta_i \beta_j \beta_k)^{d/2-2}\,,~~   c=   \frac{i \pi^{d/2}}{8} \frac{\Gamma(d-1)}{\Gamma(\frac{d}{2})^3 }\,.
\end{equation}
Inserting this into \eqref{eqn:vertex-diagram} yields
\begin{equation}
\langle W_n \rangle^{(2)}_{\text{vertex}}
=  -\frac{1}{4} \left(\frac{N}{k}\right)^2  \pi^{3-d} \Gamma(d-1) \frac{1}{4 \pi}  \sum_{i>j>k} \left( I^{(a)}_{ijk} + I^{(b)}_{ijk}\right) \,.
\end{equation}
For $n=4$ the expressions \eqref{eqn:original-expression-for-numerics} are identical to \eqref{eqn:IAIB}. For $n>4$ the structure of the denominator $\Delta$ involves more terms and thus for an analogous evaluation of the finite parts of the divergent diagrams using Mellin-Barnes representations as in \ref{sec:tetragon-MB} we would need more Mellin-Barnes parameters. Therefore we will evaluate the divergent integrals in a different manner in the following section, which is better suited for a numerical evaluation of the finite part and as a side effect allows for the analytical evaluation of the pole term.

\section{Simplifications of the Divergent Vertex Diagram}\label{app:divergent-vertex-diagram}
As in the four-point case, diagrams with $i=k+2,j=k+1,k$ are divergent. It is sufficient to evaluate one of these diagrams, since the others have the same structure and can be obtained by shifting the index $k$. We choose  $i=3, j=2, k=1$ in the following. After performing the variable change $\beta_i=x y, \beta_j= \bar{x}y, \beta_k=\bar{y}$ (where $\bar{x}=1-x, \bar{y}=1-y$) we get
\begin{align}
 I_{321}^{(a)} &= x^2_{13} x^2_{24} \int_0^1 ds_{i,j,k} dx dy \\ \nn
&
  \frac{(\bar{x} x \bar{y})^{d/2-1} }{ (- s_j \bar{s}_k \bar{x} \bar{y}  x^2_{13}-x  (\bar{y} (\bar{s}_i \bar{s}_k x^2_{13}+s_i (\bar{s}_k
   x^2_{14}+s_k x^2_{24}))+s_i \bar{s}_j \bar{x} y x^2_{24}))^{d-1}}\,.
\end{align}
The diagram is divergent as $s_i \to 0, s_k \to 1$ and it is convenient to introduce new variables $s_i = r \sin \theta , \bar{s}_k = r \cos \theta $ such that
\begin{align}\label{eqn:divergent-diagram-polar}
 I_{321}^{(a)} &=  \int_0^1 ds_{j} dx dy (\bar{x} x \bar{y})^{d/2-1} \left( \int_{0}^{\pi/4} d\theta \int_0^{1/\cos \theta} r dr + \int_{\pi/4}^{\pi/2} d\theta \int_0^{1/\sin \theta} r dr\right) \\ \nn
&
  \frac{r^{1-d} x^2_{13} x^2_{24}}{(-x_{13}^2 \cos \theta \bar{y} (x+s_j \bar{x})-x_{24}^2 \sin \theta x (\bar{s}_j \bar{x} y + \bar{y})  +r x \bar{y} \cos \theta \sin \theta (x_{13}^2+x_{24}^2-x_{14}^2) )^{d-1}}\,.
\end{align}
The integrand has the structure
\begin{equation}\label{eqn:expansion-inetgrand}
 f(r)= \frac{r^{2-d}}{[a+b r]^{d-1}} = \frac{r^{2-d}}{a^{d-1}}+ \sum_{n=1}^\infty \frac{f^{(n)}}{n!}\Big|_{r=0} r^{n+2-d}\,,
\end{equation}
where  $a+br >0$ everywhere except at the boundary. For the following analysis we use the short-hand notations
\begin{equation}\label{eqn:shothand-notation-a-b}
a= e \cos \theta + f \sin \theta, \qquad b = \cos \theta \sin \theta g\,,
\end{equation}
where
\begin{equation}
e  = -x_{13}^2 \bar{y}(x+s_j \bar{x}) , \quad
f   = -x_{24}^2 x(\bar{y} +y \bar{s}_j \bar{x}) , \quad
g  =- x \bar{y} (-x_{13}^2-x_{24}^2+x_{14}^2)\,.
\end{equation}

\notocsubsection{Analytical Result for the Pole Term}\label{sec:analytical-evaluation-of-pole-term}

For extracting the leading pole we can discard terms of order $r$ in the denominator, they contribute to the finite part of the integral. Thus we integrate only the first term in \eqref{eqn:expansion-inetgrand} and use
\begin{align}\label{eqn:only-leading-pole}
\int r^{2-d}= \frac{1}{3-d} r^{3-d}\Big|_0^{1/\cos{\theta}} = \frac{1}{2 \epsilon} + \mathcal{O}(\epsilon^0)\,.
\end{align}
Setting $d=3$ in the remaining finite integrals we thus have
\begin{align}\label{eqn:solvable-integral}
  I_{321}^{(a,div)} &= \frac{1}{2 \epsilon} \int_0^1 ds_{j} dx dy (\bar{x} x \bar{y})^{1/2} \\ \nn
& \qquad \int_{0}^{\pi/2} d\theta \frac{x_{13}^2 x_{24}^2}{\left(\sin \theta (-x_{13}^2 x (\bar{y} + s_j \bar{x} y))+ \cos \theta (-x_{24}^2 \bar{y}(x+ \bar{s}_j \bar{x}))\right)^2} + \mathcal{O}(\epsilon^0)\,.
\end{align}
The integral over $\theta$ can be solved
\begin{equation}
 \int_0^{\pi/2} d\theta \frac{1}{(e \sin \theta + f  \cos \theta )^2} = \frac{1}{e f}\,,
\end{equation}
such that we get
\begin{align}
  I_{321}^{(a,div)} &= \frac{1}{2 \epsilon} \int_0^1 ds_{j} dx dy \frac{(\bar{x} x \bar{y})^{1/2}}{ x(\bar{y} + s_2 \bar{x} y)\bar{y}(x+ \bar{s}_2 \bar{x})} + \mathcal{O}(\epsilon^0)\,.
\end{align}
Integration over $x$ and $y$ yields
\begin{align}\label{eqn:result-divergent-part}
   I_{321}^{(a,div)} &= \frac{\pi}{\epsilon}\int_0^1 ds_j \frac{1}{s_j} \left(\frac{1}{\sqrt{1-s_j}
   }-\frac{1}{\sqrt{s_j}+1} \right) =  \frac{4 \pi \ln 2}{\epsilon} + \mathcal{O}(\epsilon^0)\,.
\end{align}

\notocsubsection{Finite Terms of $I_A$}
There are two types of finite contributions that we have to take into account: one comes from taking into account the second term in \eqref{eqn:expansion-inetgrand}  
and the other one from considering the $\mathcal{O}(\epsilon)$ expansions in \eqref{eqn:divergent-diagram-polar} multiplied with the divergence \eqref{eqn:only-leading-pole} as well as the $\mathcal{O}(\epsilon^0)$ expansion of \eqref{eqn:only-leading-pole}.

\subsubsection{Order $\epsilon$ expansions}

Considering the $\epsilon$ expansion of the terms that are multiplied with the pole term,  instead of \eqref{eqn:only-leading-pole} we thus get
\begin{align}\label{eqn:Ia1,321}
 I_{321}^{(a,1)} &= x_{13}^2 x_{24}^2 \int (x \bar{x} \bar{y})^{1/2-\epsilon} \frac{1}{2\epsilon} \left( \int_0^{\pi/4} \left(\frac{1}{\cos \theta}\right)^{2\epsilon} +\int_{\pi/4}^{\pi/2} \left(\frac{1}{\sin \theta}\right)^{2\epsilon} \right) \frac{1}{a^{2-2\epsilon}} \\ \nn
&= I_{321}^{(a,div)}+ \text{ (I) + (II) + (III)}\,,
\end{align}
where $ I_{321}^{(a,div)}$ is given in \eqref{eqn:result-divergent-part} and
\begin{align}
\text{(I)} &= -\frac{1}{2}x_{13}^2 x_{24}^2 \int ds_j dx dy \log (x \bar{x} \bar{y}) (x \bar{x} \bar{y})^{1/2}\int_0^{\pi/2} d\theta \frac{1}{a^2}\,,  \\ \nn
 \text{(II)} &= x_{13}^2 x_{24}^2 \int ds_j dx dy  (x \bar{x} \bar{y})^{1/2}\int_0^{\pi/2} d\theta \frac{\log{a}}{a^2}\,,  \\ \nn
 \text{(III)} &= -x_{13}^2 x_{24}^2 \int ds_j dx dy  (x \bar{x} \bar{y})^{1/2} \left( \int_0^{\pi/4} d\theta \frac{\log{\cos \theta}}{a^2}+ \int_{\pi/4}^{\pi/2} d\theta \frac{\log{\cos \theta}}{a^2} \right)
\end{align}
and where $a$ is given in \eqref{eqn:shothand-notation-a-b}. Integration over $\theta$ yields
\begin{align}
 \text{(I)}  &= -\frac{1}{2}x_{13}^2 x_{24}^2 \int ds_j dx dy \log (x \bar{x} \bar{y}) (x \bar{x} \bar{y})^{1/2}  \frac{1}{e f} %\\ \nn
%&= -\frac{1}{2} \int ds_j dx dy \log (x \bar{x} \bar{y}) (x \bar{x} \bar{y})^{1/2}  \frac{1}{e^\prime f^\prime}
\end{align}
\begin{equation}\label{eqn:Iadiv(II)(III)}
 \text{(II)} + \text{(III)} =  x_{13}^2 x_{24}^2  \int ds_j dx dy (x \bar{x} \bar{y})^{1/2} \left( \frac{1}{e f}+ \frac{\log\left( \frac{e f}{e+f} \right)}{ e f} \right)\,.
\end{equation}

%Solving further integrations and performing the last integration numerically yields
%\begin{equation}\label{eqn:Iadiv(I)}
% \text{(I)} =  56.204191749674.. = \pi  \left(\frac{2 \pi ^2}{3}+12 \log ^2(2)+8 \log (2)\right) = 56.204191749674..
%\end{equation}
%For the remaining terms the integrations over $\theta$ can be performed as well and we find

\subsubsection{Finite term}

The finite contribution from the second term in \eqref{eqn:expansion-inetgrand} can be integrated over $r$ and resummed. We have
\begin{align}
 \int_0^{1/\cos \theta} dr \sum_{n=1}^\infty \frac{f^{(n)}}{n!}\Big|_{r=0} r^{n-1} &=  \int_0^{1/\cos \theta} dr \sum_{n=1}^\infty (-1)^n (n+1) r^{n-1} \\ \nn 
&= \sum_{n=1}^\infty (-1)^n \left(1+\frac{1}{n} \right) r^n\Big|_0^{1/\cos \theta} \left(\frac{b}{a}\right)^n \frac{1}{a^2} \\ \nn
&= \frac{1}{a^2} \sum_{n=1}^\infty (-1)^n \left(1+\frac{1}{n} \right) \left(\frac{b/\cos \theta}{a}\right)^n \\
&= -\left( \frac{1}{a^2} \right) \left(\frac{b/\cos \theta}{b/\cos \theta+a} + \log \left(1+\frac{b/\cos \theta}{a} \right)\right)\,.
\end{align}
We used that $-1<(b/ \cos \theta)/a<0$.
An analagous term results from integration with upper boundary $1/\sin \theta$. Both terms can then be integrated over $\theta$ and we find
\begin{align}\label{eqn:vertex-diag-start2}
 I_{321}^{(a,2)} =\int ds_j dx dy (x \bar{x} \bar{y})^{1/2} \frac{\log\left(\frac{e+f}{e+f+g} \right)}{e f}\,, %= -\int ds_j dx dy (x \bar{x} \bar{y})^{1/2}\frac{\log\left(1+\frac{g}{e+f} \right)}{e f}
\end{align}
where $b= \cos \theta \sin \theta g$ and $g=- x \bar{y} (-x_{13}^2-x_{24}^2 +x_{14}^2)$. We always have $g<0$, but $e+f+g >0$, thus $|e+f|>|g|$.

\subsubsection{Sum of all terms}
Adding up \eqref{eqn:Ia1,321} and \eqref{eqn:vertex-diag-start2} we get
\begin{align} \nn
&\phantom{=} I_{321}^{(a,1)}+I_{321}^{(a,2)} \\ \nn
&= \frac{4 \pi \ln(2)}{\epsilon} + \text{(I)} +     \int ds_j dx dy \frac{(x \bar{x} \bar{y})^{1/2}}{ e^\prime f^\prime} \left( 1+ \log\left( \frac{e f}{e+f} \right)+ \log\left( \frac{e + f}{e+f+g} \right) \right) \\ \nn
%&=  \frac{4 \pi \ln(2)}{\epsilon} + \text{(I)} + 8 \pi \ln(2)+  \int ds_j dx dy \frac{(x \bar{x} \bar{y})^{1/2}}{ e^\prime f^\prime} \left( \log\left( \frac{e f}{e+f+g} \right) \right) \\ \nn
&= \frac{4 \pi \ln(2)}{\epsilon} + \text{(I)} %+ 8 \pi \ln(2)
+ 4 \pi \ln(2) \left( \ln(-x_{13}^2) + \ln(-x_{24}^2) \right) \\ \nn
&+  \int ds_j dx dy \frac{(x \bar{x} \bar{y})^{1/2}}{ e^\prime f^\prime} \left(1+ \ln \left( e^\prime f^\prime \right) -\frac{1}{2} \ln \left( e^\prime \frac{x_{13}^2}{x_{24}^2} + f^\prime +\frac{g}{x_{24}^2} \right)-\frac{1}{2} \ln \left( e^\prime + f^\prime \frac{x_{24}^2}{x_{13}^2}+\frac{g}{x_{13}^2} \right) \right)\,,
\end{align}
where we used \eqref{eqn:only-leading-pole} and defined  $e= -x_{13}^2 e^\prime$, $f = -x_{24}^2 f^\prime$. The first integral is a constant and the remaining two integrals depend on all three distances. \begin{align}\label{eqn:Iasplit-up}
 I_{321}^{(a)} = 2\pi \ln(2) \left( \frac{(-x_{13}^2)^{2\epsilon}}{\epsilon} + \frac{(-x_{24}^2)^{2\epsilon}}{\epsilon} \right)  + I_{321}^{(a,f)}
\end{align}
and where
\begin{align}\label{eqn:Iaf-expression-for-numerical-evaluation} \nn
  I_{321}^{(a,f)} = &-\frac{1}{2}\int_0^1 ds_j dx dy \frac{(x \bar{x} \bar{y})^{1/2}}{ e^\prime f^\prime}\left(  \ln \left( e^\prime \frac{x_{13}^2}{x_{24}^2} + f^\prime +\frac{g}{x_{24}^2} \right)+ \ln \left( e^\prime + f^\prime \frac{x_{24}^2}{x_{13}^2}+\frac{g}{x_{13}^2} \right) \right)
 \\ 
  & + \int_0^1 ds_j dx dy \frac{(x \bar{x} \bar{y})^{1/2}}{ e^\prime f^\prime} \left(1+ \ln(x\bar{x}\bar{y}) + \ln(e^\prime f^\prime)\right)\,.
\end{align}
Note, that the last term is a constant independent of $x_{ij}^2$. We can easily solve some of the integrals in the last term by using \eqref{eqn:solvable-integral}-\eqref{eqn:result-divergent-part}, it is however sufficient to use this form for the numerical integration.

\notocsubsection{Finite Integral $I_B$}
We have
\begin{align}
I^{(b)}_{321} = -4 x_{13}^2 x_{24}^2(x_{13}^2+x_{24}^2 -x_{14}^2) &\int dx dy ds_j (x \bar{y})^{3/2} \bar{x}^{1/2} \\ \nn 
& \left( \int_0^{\pi/4} d\theta dr \frac{  r^3 \cos \theta \sin \theta }{r^3(a+rb)^3 }+\int_{\pi/4}^{\pi/2}...\right)\,.
\end{align}
Expanding in a Taylor series in $r$ as for the divergent integral, the second line of the previous formula reads
\begin{align}
& \left( \int_0^{\pi/4} d\theta \int_0^{1/\cos \theta} dr \right) \sum_{n=0}^\infty (-1)^n \frac{r^n}{n!} \frac{\Gamma(d+n)}{\Gamma(d)} \left(\frac{b}{a}\right)^n \frac{1}{a^3} \cos \theta  \sin \theta + \int_{\pi/4}^{\pi/2}... \\ \nn
&= \int_0^{\pi/4}  \frac{1}{2} \frac{\sin \theta}{a^3} \sum_{n=0}^\infty (-1)^n  \left(\frac{b/\cos\theta}{a}\right)^n (n+2)  +\int_{\pi/4}^{\pi/2}.. \\ \nn
&=  \int_0^{\pi/4} \frac{1}{2} \frac{\sin \theta}{a^2} \left( \frac{a}{(a+b / \cos \theta)^2} + \frac{1}{(a+b/\cos \theta)} \right)  +\int_{\pi/4}^{\pi/2}.. \\ \nn
&= \frac{1}{2 e (e+f)(e+f+g)}+\frac{1}{2 f (e+f)(e+f+g)} = \frac{1}{2}\frac{1}{e f(e+f+g)}
\end{align}
and thus
\begin{align}\label{eqn:Ib-numerical-purpose}
I^{(b)}_{321}& = -2 x_{13}^2 x_{24}^2(x_{13}^2+x_{24}^2 -x_{14}^2) \int dx dy ds_j (x \bar{y})^{3/2} \bar{x}^{1/2} \frac{1}{ef (e+f+g)} \\ \nn
&=  2  \int_0^1 dx dy ds_j (x \bar{y})^{3/2} \bar{x}^{1/2} \frac{(x_{14}^2-x_{13}^2-x_{24}^2)}{e^\prime f^\prime (e+f+g)}\,.
\end{align}
%Note that since $g= - x \bar{y} (-x_{13}^2-x_{24}^2 +x_{14}^2))$
%\begin{equation}
%(-x_{13}^2-x_{24}^2 +x_{14}^2) \frac{d}{d x_{14}^2}   \int dx dy ds_j (x \bar{y}\bar{x})^{1/2} \frac{ \ln \left( e+f+g \right)}{ e^\prime f^\prime } = I_B
%\end{equation}
%which is an integral that also appears above.
\notocsubsection{Full Vertex Integral for Numerical Evaluation}
Using \eqref{eqn:Iasplit-up} and \eqref{eqn:Ib-numerical-purpose} the vertex integrals \eqref{eqn:vertex-diagrams} can thus be split up into a divergent and a finite piece
\begin{equation}\label{eqn:split-up-fin-div}
\langle W_n \rangle^{\text{vertex}} = \langle W_n \rangle^{\text{div}} + \langle W_n \rangle^{\text{finite}}\,,
\end{equation}
where the divergent piece is 
\begin{equation}
\langle W_n \rangle^{\text{div}} = - \left( \frac{N}{k} \right)^2 \left( \frac{\ln (2)}{2}  \sum_{i=1}^n \frac{(-x_{i,i+2}^2 \mu^2  \pi e^{\gamma_E})^{2\epsilon}}{2 \epsilon} \right)\,.
\end{equation}
The remaining finite piece is
\begin{align}\label{eqn:vertex-finite-pieces}
\langle W_n \rangle^{\text{finite}} 
=  - \frac{1}{4} \left(\frac{N}{k}\right)^2 & \left( \frac{1}{4\pi} \sum_{i>j>k} \left(I_{ijk}^{(a,f)}+I_{ijk}^{(b)} \right)- n \,2 \ln(2)\right)\,,
\end{align}
where $I^{(a,f)}_{ijk},I^{(b)}_{ijk}$ are given by \eqref{eqn:Iaf-expression-for-numerical-evaluation}, \eqref{eqn:Ib-numerical-purpose}  for the case $i=k+2,j=k+1,k$. All other cases are treated purely numerically starting from \eqref{eqn:original-expression-for-numerics}.In  particular $I_{ijk}^a$ is not divergent in this case and  $I^{(a,f)}_{ijk}=I^{(a)}_{ijk}$ in these cases. 

These are the expressions that are used for the numerical evaluation of the finite parts of the divergent diagrams.

\section{Generation of Kinematical Configurations}\label{app:generation-kinematics}
A set of $n$ light-like vectors $p_i^\mu$ satisfying momentum conservation, i.e. $p_i^2=0$ and $\sum p_i^\mu=0$, can easily be generated by choosing the $p^0_i$ components of $n-3$ vectors and the angle $\theta_i$ between $p_i^1 =p_i^0 \cos \theta_i $, $p_i^2 =p_i^0 \sin \theta_i$. The remaining components are then fixed. For $n$ even it is possible to choose configurations, where
all non-light-like distances $x_{ij}^2$ are space-like. For the numerical evaluation we make use of this type of configurations, such that all integrals are real.

The configurations used for the results shown in fig. 
\ref{fig:Wvertextwogluonplot} are obtained by continuously deforming two angles $\theta_i$, leading to conformally non-equivalent kinematical configurations. We use the angles
\begin{align}
 \theta_i(a):= \pi \left\{\frac{16}{9}a,\frac{13 }{9},\frac{5}{3}a,\frac{13  }{8},\frac{19 }{14},1\right\}
\end{align}
and choose $p_i^{\mu=0}=\{1,-3,4\}$, the remaining components are then fixed. The parameter $a$ is chosen between $a=1$ and $a=1.2$ in steps of $0.01$.

\chapter{Details of the Three-Point Function Calculation}\label{eqn:details-of-the-two-point-calculation}\chaptermark{Details of the 3-Point Function Calculation}

\section{Some Properties of Gegenbauer Polynomials}\label{sec:properties-of-gegenbauer-polynomials}
The Gegenbauer polynomials $C_j^\nu(x)$ are polynomials of degree $j$ in the variable $x$ and are solutions of the Gegenbauer differential equation, which is of second order and depends on the parameter $\nu$.
The Gegenbauer polynomials satisfy the orthogonality relation
\begin{align}\label{eqn:GP-orthogonality}
 \int_{-1}^1 (1-y^2)^{\nu - \frac{1}{2}} C_j^\nu (y) C_k^\nu (y)= \delta_{jk} 2^{1-2\nu} \pi \frac{\Gamma(j+2\nu)}{\Gamma^2(\nu)\Gamma(j+1)(j+\nu)}= N(j,\nu) \delta_{jk}\,.
\end{align}
One can directly find a representation for the coefficients related to the Gegenbauer polynomials by
\begin{align}
 (\hat{\partial}_a+\hat{\partial}_b)^j C_j^\nu \left(\frac{\hat{\partial}_a-\hat{\partial}_b}{\hat{\partial}_a+\hat{\partial}_b} \right) = \sum_{k=0}^j a_{jk}^\nu \hat{\partial}_a^k\hat{\partial}_b^{j-k} = (\hat{\partial}_a+\hat{\partial}_b)^j \sum_{k=0}^j c_{jk}^\nu\left( \frac{\hat{\partial}_a-\hat{\partial}_a}{\hat{\partial}_a+\hat{\partial}_b}\right)^k
\end{align}
and for  $\nu=1/2,3/2$ we have
\begin{align}
 a_{jk}^{1/2} &= (-1)^{k} {j \choose k} {j \choose k}\,, \\ \nonumber
 a_{jk}^{3/2} &= \frac{1}{2} (-1)^{k} (k+1){j+1 \choose k+1} {j+2 \choose k+1}\,,
\end{align}
clearly satisfying $a_{jk}=(-1)^j a_{j,j-k}$. Of course the coefficient $a_{jj}$ coincides with the one determined before. Useful identities are 
\begin{equation}\label{eqn:sum-over-coefficients-ajk32}
 \sum_{k=0}^j (-1)^{j-k} a_{jk}^{3/2} = \frac{2 (2 j+1) \Gamma (2 j)}{\Gamma (j) \Gamma (j+1)}\,,
\end{equation}
\begin{equation}\label{eqn:sum-over-coefficients-ajk12}
 \sum_{k=0}^j (-1)^{j-k} a_{jk}^{1/2} = \frac{2 \Gamma (2 j)}{\Gamma (j) \Gamma (j+1)}\,.
\end{equation}
The coefficient $c_{jj}^\lambda$ in the expansion $C_j^\nu(x)=\sum_k x^k c_{jk}^\nu$ is easy to deduce from a derivative identity of the Gegenbauer polynomials
\begin{equation}\nonumber
 \frac{d}{dx} C_n^\lambda(x) = 2 \lambda C_{n-1}^{\lambda+1}, \qquad C_0^\lambda=1\,.
\end{equation}
This leads to 
\begin{equation}\label{eqn:gegenbauer-coefficient-higherst-power}
\frac{d^j}{dx^j} C_j^\lambda(x) = 2^j \frac{\Gamma(\lambda+j)}{\Gamma(\lambda)}\qquad \Rightarrow  \qquad c_{jj}^\lambda = 2^j \frac{\Gamma(\lambda+j)}{\Gamma(\lambda)\Gamma(j+1)}\,.
\end{equation}

\notocsubsection{Expansion of Gegenbauer Polynomials}\label{sec:expansion-of-gegenbauer-polynomials}
The Gegenbauer Polynomials can be expanded with respect to their index using
\begin{align} 
\frac{\partial}{\partial \rho} C_j^{\nu + \rho}(x)|_{\rho=0} = -2 \sum_{k=0}^{j} d_{jk}^{\nu} C_k^{\nu}(x)
\end{align} 
with coefficients
\begin{align}\label{eqn:gegenbauer-expansion-coefficients}
d_{jk}^\nu &= -(1 + (-1)^{j-k}) \frac{k+\nu}{(j + k + 2\nu)(j - k)}\qquad \text{for}\qquad  j>k\,, \\ \nonumber
d_{jj}^\nu &= -1/2 \frac{\Gamma(\nu)}{\Gamma(\nu +j)}\frac{\partial}{\partial \rho} \left( \frac{\Gamma(\nu + \rho+j)}{\Gamma(\nu +\rho)}\right)_{\rho=0} = \frac{1}{2} (\psi(\lambda )-\psi(j+\lambda ))\,.
\end{align}
In $d=6,4$ we get
\begin{align}
 d_{jj}^{3/2} &=-\frac{1}{2}
   H_{j+\frac{1}{2}}-\log (2) + 1 = \frac{1}{2} \left(H_j-2 H_{2 j+1}+2\right)\,, \\ \nonumber 
 d_{jj}^{1/2} &=-\frac{1}{2} H_{j-\frac{1}{2}}-\log
   (2) = \frac{1}{2} \left(H_{j-1}-2 H_{2 j-1}\right)\,,
\end{align}
where $H_{j+1/2}$ are harmonic numbers.  One can reformulate the above equation using
\begin{equation}\label{sec:hjplus12identity}
H_{j+1/2}=2 H_{2j+1}-H_j -2\ln 2\,.
\end{equation}

%\subsection{Formulas from somewhere else, are they useful/different?}
%Try to make use of \cite{Kotikov:2000pm}, page 6:\\
%
%\begin{eqnarray}
%{\rm C}_n^0 (\cos \theta) \equiv \lim_{\lambda \to 0} \frac{1}{\lambda} {\rm
%%
%C}_n^{\lambda} (\cos \theta) = \frac{2}{n} {\rm T}_n (\cos \theta).
%\nonumber
%\end{eqnarray}
%
%To calculate the terms singular at $\varepsilon \to 0$ it is convenient to
%to use the formulae \emph{from uncomment reference in latex file:} %\cite{CheKaTka, Kotikov96} 
%for the traceless
%products 
%
%
%
%
%\begin{eqnarray}
%x^{(\mu_1,...,\mu_n)}~=~
%\sum_{p \geq 0}^{[n/2]} \frac{n!(-1)^{p} \Gamma(n-p+\lambda)}
%{2^{2p} p! (n-2p)!\Gamma(n+\lambda)}~
%g^{\mu_1\mu_2}...g^{\mu_{2p-1}\mu_{2p}}~x^{2p}~x^{\mu_{2p+1}}...x^{\mu_n}\,.
% \nonumber 
%\end{eqnarray}
%The traceless products are related to the Gegenbauer polynomials 
%${\rm C}_{n}^{\lambda}(\cos \theta )$ as follows
%
%\begin{eqnarray}
%\frac{n}{2}\frac{1}{\lambda }{\rm C}_{n}^{\lambda }(\hat{q}_{1}\hat{q}%
%_{2})=S_{n}^{(1)}(\lambda )\frac{q_{1}^{(\mu _{1},...,\mu _{n})}q_{2}^{(\mu
%_{1},...,\mu _{n})}}{{(q_{1}^{2}q_{2}^{2})}^{n/2}},
%\end{eqnarray}
%where the symbols $\hat{q}_{1}$ and $\hat{q}_{2}$ are the 
%unit vectors parallel to
%the transverse momenta $q_{1}$ and $q_{2}$,
%\begin{eqnarray}
%\hat{q}_{1}\hat{q}_{2}=\frac{(q_{1}q_{2})}{{(q_{1}^{2}q_{2}^{2})}^{1/2}}
%\end{eqnarray}
%and $S_{n}^{(1)}(\lambda )$ is the following factor
%\begin{eqnarray}
%S_{n}^{(1)}(\lambda )=\frac{2^{n-1}\Gamma (n+\lambda )}{\Gamma (n)\Gamma
%(1+\lambda )}
%\end{eqnarray}

\newpage

\section{Basis Integrals Solved by the IBP Method}\label{sec:appendix-integrals}
We define the following set of integrals\footnote{These integrals were considered before in \cite{Kazakov:1986mu}.}
\begin{equation}\label{eqn:basic-integral}
f_{mn}(a_1,a_2,a_3,a_4,a_5) = \int \frac{d^dk,l}{(2\pi)^{2d}} \frac{(\hat{k})^m (\hat{l})^n}{k^{2 a_1}(p+k)^{2a_2}(l-k)^{2a_3}l^{2 a_4}(p+l)^{2a_5}}\,,
\end{equation}
where $\hat{k}=z^\mu k_\mu$ is the contraction with a light-like vector $z^2=0$. The integral is symmetric under the simultaneous exchange  $(m,a_1,a_2) \leftrightarrow (n,a_4,a_5)$.
In particular in this thesis we need $f_{mn}(1,1,1,1,1)$ and $f_{j0}(2,1,1,1,1)$.  Since the momentum dependence is the same everywhere we define $c_{mn}$ by stripping off the equal coefficients 
\begin{equation}\label{eqn:definition-cmn}
 f_{mn}(a_1,a_2,a_3,a_4,a_5) = c_{mn}(a_1,a_2,a_3,a_4,a_5) \frac{(\hat{p})^{m+n}}{(-p^2)^{\sum_i a_i -d}} \frac{1}{(4\pi)^d}\,.
\end{equation}

\notocsubsection{Bubble Integrals}\label{sec:app-reduced-integrals}
The integrals \eqref{eqn:basic-integral} with one argument set to zero are easy to solve and are the building blocks of $f_{mn}(1,1,1,1,1)$ after using the IBP method. We have
\begin{align}
f_{mn}(a_1,a_2,0,a_4,a_5) &= (-1)^{\sum_i a_i} b_m(a_1,a_2) b_n(a_4,a_5) \frac{(\hat{p})^{m+n}}{(-p^2)^{\sum_i a_i-d}} 
\end{align}
and
\begin{equation}\nn
f_{mn}(a_1,a_2,a_3,a_4,0) = (-1)^n (-1)^{\sum_i a_i} b_n(a_4,a_3) b_{m+n}(a_1+a_3+a_4-\frac{d}{2},a_2) \frac{(\hat{p})^{m+n}}{(-p^2)^{\sum_i a_i-d}} \,.
\end{equation}
By the symmetry of the integral \eqref{eqn:basic-integral} $(m,a_1,a_2) \leftrightarrow (n,a_4,a_5)$ we also know
\begin{equation}
f_{mn}(a_1,0,a_3,a_4,a_5) = f_{nm}(a_4,a_5,a_3,a_1,0)\,.
\end{equation}
Furthermore, we have
\begin{align} \nn
 f_{mn}(a_1,a_2,a_3,0,a_5) &= (-1)^{\sum_i a_i} \frac{(\hat{p})^{n+m}}{(-p^2)^{\sum_i a_i -d}}  \quad  \sum_{j=0}^n \sum_{i=0}^{n-j} {n \choose j} {n -j \choose i} (-1)^n \\
& b_{n-j}(a_5,a_3) b_{m+n-i-j}(a_1,a_2+a_3+a_5-\frac{d}{2})
\end{align}
and $ f_{mn}(0,a_2,a_3,a_4,a_5)$ is related to this integral by the symmetry  $(m,a_1,a_2) \leftrightarrow (n,a_4,a_5)$.
We can perform the sum over $i$ by using \eqref{eqn:identity-beta-function} and find
\begin{align}\label{eqn:fmna1a2a30a5}
 f_{mn}(a_1,a_2,a_3,0,a_5) &= \frac{(\hat{p})^{n+m}}{(-p^2)^{\sum_i a_i -d}}  \frac{\Gamma(a_1-m)\Gamma(a_1+a_2+a_3+a_5-d)}{\Gamma(a_1)\Gamma(a_1+a_2+a_3+a_5-d-m)}  \\ \nn 
 & \sum_{j=0}^n  {n \choose j} (-1)^{m+n-j} b_{j}(a_5,a_3) b_{j}(a_2+a_3+a_5-\frac{d}{2},a_1-m)\,.
\end{align}
The sum can be performed using Mathematica in terms of  generalised hypergeometric functions
%
%\begin{align}\nn
%c_{mn}(a_1,1,1,0,1)&= -(-1)^{m+n} \frac{\Gamma \left(2-\frac{d}{2}\right) \Gamma \left(\frac{d}{2}-1\right) \Gamma \left(\frac{d-2}{2}\right) \Gamma (a_1-d+3) \Gamma
%   \left(-a_1+\frac{d}{2}+m\right) \, }{(d-3) \Gamma (a_1) \Gamma
%   \left(3-\frac{d}{2}\right) \Gamma \left(\frac{1}{2} (-2 a_1+3 d+2 m-6)\right)} \\ 
% & \quad _3F_2\left(\frac{d}{2}-1,d-3,-n;d-2,-\text{a1}+\frac{3 d}{2}+m-3;1\right)\,.
%\end{align}

\notocsubsection{Integrals $f_{mn}(1,1,1,1,1)$}\label{sec:Integrals for diagrams b}
First we would like to solve the integral  $f_{mn}(1,1,1,1,1)$, given through the definition
\begin{equation}
 f_{mn}(a_1,a_2,a_3,a_4,a_5) = \int  d^dk,l \underbrace{\frac{(\hat{k})^m (\hat{l})^n}{k^{2 a_1} (p+k)^{2 a_2} (k-l)^{2 a_3} l^{2 a_4} (p+l)^{2 a_5}}}_{\hat{i}_{mn}(....)}\,.
\end{equation}
The integral is symmetric under simultaneous exchange of $\{m,a_1,a_2\} \leftrightarrow \{n,a_4,a_5\}$ and can be solved by using the IBP method, i.e. using the fact that - with suitable boundary conditions - the divergence of the integral vanishes.
\begin{equation}
 \int \frac{\partial}{\partial k^\mu}(v^\mu \, \hat{i}_{mn}(1,1,1,1,1)) = 0\,.
\end{equation}
Evaluating the differentiation explicitly we find
\begin{align}\nn
 0 &= \left( \frac{\partial}{\partial k^\mu}v^\mu \right)f_{mn} + \int \hat{i}_{mn}(1,1,1,1,1) \left( -\frac{2 k\cdot v}{k^2} - \frac{2(p+k)\cdot v}{(p+k)^2} +  \frac{2(l-k)\cdot v}{(l-k)^2}  \right) \\ \nonumber
&{\phantom{=}}+ m \int \hat{v}\, \hat{i}_{m-1,n}(1,1,1,1,1)\,.
\end{align}
Choosing $v^\mu=k^\mu, z^\mu,p^\mu$ etc. one finds several relations between the integrals with different indices. These relations contain simpler integrals, that can be explicitly solved. The last term provides recursions in $m,n$.

In the following, we will use the convention, that all arguments of $f_{mn}$ are 1 if not explicitly specified, i.e. $f_{mn}:=f_{mn}(1,1,1,1,1)$.

We start by choosing $v^\mu= (l-k)^\mu$, which yields the equation
\begin{align}
 0 = &- d\, f_{mn} + \int \hat{i}_{mn} \left( -\frac{2 k \cdot (l-k)}{k^2} - \frac{2(p+k) \cdot (l-k)} {(p+k)^2}+  2  \right)\\ \nonumber  &- m \,f_{mn} + m \, f_{m-1,n+1}\,.
\end{align}
The term in the brackets can be rewritten by simply using
\begin{align}
 -2 k \cdot (l-k) & = (l-k)^2 + k^2 - l^2, \\ \nonumber
-2(p+k) \cdot (l-k) &= (l-k)^2+(k+p)^2-(l+p)^2
\end{align}
and thus we get 
we can write this as
\begin{equation}\label{eqn:IBP-relation}
(d+m- 4) f_{mn} =\text{bubbles}(m,n) + m \, f_{m-1,n+1}\,,
\end{equation}
where we abbreviate the known integrals as
\begin{align}\label{eqn:bubbles}
\text{bubbles}(m,n) &= f_{mn}(2,1,0,1,1) - f_{mn}(2,1,1,0,1)\\ \nonumber
&+  f_{mn}(1,2,0,1,1)-  f_{mn}(1,2,1,1,0)\,.
\end{align}
We can actually solve for all integrals by writing down \eqref{eqn:IBP-relation} for $m=n+1$ and $n=m-1$, i.e.
\begin{align}\label{eqn:recursion2}
(d+(n+1)- 4) f_{n+1,m-1} &= \text{bubbles}(n+1,m-1)+ (n+1) \, f_{nm}
\end{align}
and using $f_{mn}(1,1,1,1,1)=f_{nm}(1,1,1,1,1)$. Solving \eqref{eqn:recursion2} for $f_{n+1,m-1}$  and inserting it into \eqref{eqn:IBP-relation} we get $f_{nm}$ in terms of known integrals:
\begin{equation}\label{eqn:solution-fnm}
f_{mn}(1,1,1,1,1) = \frac{(n+d-3)\text{bubbles}(m,n)+m \text{bubbles}(n+1,m-1)}{(d-4)(d-3+m+n)}\,.
\end{equation}
Note, that the coefficient on the left-hand side is of order $\epsilon$ in  $d=4-2\epsilon$.
All appearing integrals were solved in section \ref{sec:app-reduced-integrals}.

\notocsubsection{$f_{00}(1,1,1,1,1)$ in $d=4$}
The recursion has been solved in  \eqref{eqn:solution-fnm} and $b(m,n)$ is given in \eqref{eqn:bubbles}. 
For $n=m=0$ we have
\begin{equation}
f_{00}(1,1,1,1,1) = \frac{\text{bubbles}(0,0)}{(d-4)}\,.
\end{equation}
The bubbles then read
\begin{align}
\text{bubbles}(0,0)= 2 b_0(1,1)\left( b_0(3-d/2,2) -b_0(2,1)\right)
\end{align}
and we would now like to investigate the case $d=4-2 \epsilon$. Furthermore, we have
\begin{align}
\left( b_0(3-d/2,2) -b_0(2,1)\right)= - \frac{i}{(4 \pi)^2}  6 \zeta(3) \epsilon^2 + \op(\epsilon^3)
\end{align}
and thus
\begin{align}
\text{bubbles}(0,0)= \frac{2}{(4 \pi)^4}  6 \zeta(3) \epsilon + \op(\epsilon^2)\,,
\end{align}
which leads to
\begin{equation}\label{eqn:f00(1,1,1,1,1)}
f_{00}(1,1,1,1,1) = - \frac{1}{(4 \pi)^4}  6 \zeta(3) + \op(\epsilon)\,.
\end{equation}

\notocsubsection{Integral $f_{nm}(2,1,1,1,1)$}\label{eqn:integral-f_{j,0}(2,1,1,1,1)}

For the three-point function calculation we need the integral
\begin{equation}
f_{j,0}(2,1,1,1,1) = \int \frac{d^dk,l}{(2\pi)^{2d}} \frac{(\hat{k})^j}{k^4 (p+k)^2(l-k)^2 l^2(p+l)^2}\,.
\end{equation}
Using the IBP method, just as in section \ref{sec:Integrals for diagrams b} for $f_{mn}(1,1,1,1,1)$, we find
\begin{equation}\label{eqn:recursion-fj0(2,1,1,1,1)}
(d-5+m) f_{m,n}(2,1,1,1,1) = m f_{m-1,n+1}(2,1,1,1,1) + \text{bubbles}_2(m,n)\,,
\end{equation}
where
\begin{align}
\text{bubbles}_2(m,n) &=  2 f_{mn} (3,1,0,1,1) - 2 f_{mn} (3,1,1,0,1) \\ \nn
&\phantom{= } +f_{mn} (2,2,0,1,1)-f_{mn} (2,2,1,1,0)
\end{align}
and all these integrals are solved in the following section. Thus we can obtain $f_{mn}$ recursively from \eqref{eqn:recursion-fj0(2,1,1,1,1)}, the recursion ends at $f_{0,m+n}$ and
\begin{equation}
f_{0,m+n}(2,1,1,1,1) =  \frac{1}{d-5}\text{bubbles}_2(0,m+n)\,.
\end{equation}
Note that the recursion for $f_{1,j}$ takes the form
\begin{equation}
 f_{1,j-1}(2,1,1,1,1) =  \left(f_{0,j}(2,1,1,1,1) + \text{bubbles}_2(1,j-1)\right)/(-2\epsilon)\,.
\end{equation}

\notocsubsection{A Useful Relation between Euler Beta-Functions}
For the Euler Beta-function defined as
\begin{align}
B(x,y)=\frac{\Gamma(x)\Gamma(y)}{\Gamma(x+y)}\,,
\end{align}
a useful relation can be derived
\begin{align}\label{eqn:beta-identity}
\sum_{n=0}^j {j \choose n} (-1)^n B(n+x,y) =  B(j+y,x) \,.
\end{align}

\subsubsection{Proof by induction $j \to j+1$}
It is easy to show that
\begin{align}\label{eqn:substep}
B(y+j+1,x) = B(y+j,x) - \sum_{n=0}^j (-1)^n { j \choose n} B(x+n+1,y)\,.
\end{align}
Using \eqref{eqn:beta-identity} for the first term on the right-hand side of \eqref{eqn:substep} we get
\begin{align}
&\phantom{=} \sum_{n=0}^j (-1)^n { j \choose n} B(x+n,y) - \sum_{n=0}^j (-1)^n { j \choose n} B(x+n+1,y) \\ \nn
&=\sum_{n=0}^{j} (-1)^n { j \choose n} B(x+n,y) - \sum_{n=1}^{j+1} (-1)^{n-1} { j \choose n-1} B(x+n,y) \\ \nn
&= \sum_{n=1}^{j} (-1)^n \left(  { j \choose n}+{ j \choose n-1} \right) B(x+n,y) \\ \nn 
&\qquad +  {j+1\choose0}B(x,y) + (-1)^{j+1}  {j+1\choose j+1}B(x+j+1,y) \\ \nn
&=\sum_{n=0}^{j+1} (-1)^n { j+1 \choose n} B(x+n,y)
\end{align}
and thus we have shown that the formula is valid for $j+1$ by using it for $j$. This completes the proof by induction\footnote{I am grateful to Martin Heinze for a discussion on this subject and suggesting the idea for working out the last and most important steps to the proof by induction.}. 

\subsubsection{Application to bubble integrals}
We can use \eqref{eqn:beta-identity} for simplifying sums as
\begin{align}
(-1)^j \sum_{n=0}^j {j \choose n} b_n(x,y) = b_j(y,x)
\end{align}
or by letting $x \to x-m$, which is equivalent to $b_n \to(-1)^m b_{n+m}$ we get
\begin{align}\label{eqn:identity-beta-function}
(-1)^{j+m} \sum_{n=0}^j {j \choose n} b_{n+m}(x,y) = b_j(y,x-m) \frac{\Gamma(x-m)\Gamma(x+y-d/2)}{\Gamma(x)\Gamma(x+y-m-d/2)}\,.
\end{align}

\section{Form of Twist-Two Operators with Exactly One Gauge Field}\label{sec:derivation-operator-with-exactly-one-gaueg-field}
Here we deduce a form of the operator \eqref{eqn:def-twist-two} with exactly one covariant derivative. The gauge field can appear in any of the derivatives in \eqref{eqn:def-twist-two} and therefore
\begin{align}
\hat{\op}_j^{\text{A}} = \sum_{k=1}^{j-1} a_{jk}^{1/2} \,\tr \Big(  
& \sum_{m=1}^{k} \hat{\partial}^{m-1} \left( -i [\hat{A}, \hat{\partial}^{k-m} \phi^{12} ] \right)\hat{\partial}^{j-k} \phi^{12} \\ \nn
+ & \sum_{m=1}^{j-k}  \hat{\partial}^{k} \phi^{12} \hat{\partial}^{m-1} \left( -i [\hat{A}, \hat{\partial}^{j-k-m} \phi^{12} ]\right)  \Big) \\ \nn
+ (a_{j0}^{1/2}+a_{jj}^{1/2}) &\sum_{m=0}^{j-1} \tr \left( \hat{\partial}^{m} \left( -i [\hat{A}, \hat{\partial}^{j-m-1} \phi^{12} ] \right)\phi^{12} \right)\,.
\end{align}
Taking the trace we can write this as
\begin{align}
\hat{\op}_j^{\text{A}} = \frac{1}{2} f^{abc} \sum_{k=1}^{j-1}a_{jk}^{1/2} \, \Big(  
& \sum_{m=1}^{k} \hat{\partial}^{m-1} \left( \hat{A}^a  \hat{\partial}^{k-m} \phi^{12,b}\right)  \hat{\partial}^{j-k} \phi^{12,c}\\ \nn
+ & \sum_{m=1}^{j-k}  \hat{\partial}^{m-1} \left( \hat{A}^a \hat{\partial}^{j-k-m} \phi^{12,b} \right)  \hat{\partial}^{k} \phi^{12,c}  \Big)
 \\ \nn
+ \frac{1}{2} f^{abc} (a_{j0}^{1/2}+a_{jj}^{1/2}) &\sum_{m=0}^{j-1}  \hat{\partial}^{m} \left( \hat{A}^a \hat{\partial}^{j-m-1} \phi^{12,b}  \right)\phi^{12,c}\,.
\end{align}
Using $a_{jk}^{1/2}=(-1)^{j} a_{j,j-k}^{1/2}$ we find
\begin{equation*} 
\hat{\op}_j^{\text{A}} = \frac{1}{2} f^{abc} \left(1+(-1)^j\right) \Big(  \sum_{k=1}^{j}a_{jk}^{1/2} \sum_{m=1}^{k} \hat{\partial}^{m-1} \left( \hat{A}^a  \hat{\partial}^{k-m} \phi^{12,b}\right)  \hat{\partial}^{j-k} \phi^{12,c} 
 \Big)
\end{equation*}
and rewriting the derivative $\hat{\partial}^{m-1}$ as a binomial sum
\begin{align}\nn
\hat{\op}_j^{\text{A}} = \frac{1}{2} f^{abc} \left(1+(-1)^j\right)
 \left(  \sum_{k=1}^{j}a_{jk}^{1/2} \sum_{m=1}^{k} \sum_{i=0}^{m-1} {m-1 \choose i} \hat{\partial}^{i} \hat{A}^a  \hat{\partial}^{k-1-i} \phi^{12,b}  \hat{\partial}^{j-k} \phi^{12,c} 
 \right)\,.
\end{align}
We can furthermore rewrite the summation
\begin{equation}\nn
\sum_{m=1}^k \sum_{i=0}^{m-1} {m-1 \choose i} \hat{\partial}^i \hat{A}^a \hat{\partial}^{k-1-i} \phi^{12,b} = \sum_{m=1}^k{k \choose m} \hat{\partial}^{m-1} \hat{A}^a \hat{\partial}^{k-m} \phi^{12,b} \,,
\end{equation}
such that we get
\begin{align}\label{eqn:twist-operator-one-gauge-field}
\hat{\op}_j^{\text{A}} = \frac{1}{2} f^{abc} \left(1+(-1)^j\right) 
\left(  \sum_{k=1}^{j}a_{jk}^{1/2}\sum_{m=1}^k{k \choose m} \hat{\partial}^{m-1} \hat{A}^a \hat{\partial}^{k-m} \phi^{12,b}  \hat{\partial}^{j-k} \phi^{12,c} 
 \right)\,.
\end{align}
This is the form of the operator, that we will insert in the perturbative calculations.

\section{Integral over Conformal Structure of the Three-Point Correlator}\label{app:integral-over-conformal-structure}
We assume a general exponent $\theta$ and $j$ even, to solve the integral
\begin{align}
 & \int d^dx_3 \frac{(\hat{Y}_{12,3})^j}{((-x_{13}^2)(- x_{23}^2))^{\theta/2}} \\ \nonumber &= \sum_{k=0}^j   {j \choose k}(-1)^{k} \int d^dx_3 \frac{(\hat{x}_{13})^k (\hat{x}_{23})^{j-k}}{(-x_{13}^2)^{\theta/2+k} (-x_{23}^2)^{\theta/2+j-k}} \\ \nonumber
&=2^{-j} \sum_k {j \choose k} (-1)^{k} \frac{\Gamma(\theta/2) \Gamma(\theta/2)}{\Gamma(\theta/2+k) \Gamma(\theta/2+j-k)} \hat{\partial}_1^k  \hat{\partial}_2^{j-k} \int d^dx_3 \frac{1}{((-x_{13}^2) (-x_{23}^2))^{\theta/2}} \\ \nonumber
&= -i 2^{-j} \sum_k {j \choose k}(-1)^{k} \frac{\Gamma((d-\theta)/2)^2 \Gamma(\theta- d/2) \pi^{d/2}}{\Gamma(\theta/2+k) \Gamma(\theta/2+j-k)\Gamma(d-\theta)} \hat{\partial}_1^k  \hat{\partial}_2^{j-k} \frac{1}{(-x_{12}^2)^{\theta-d/2}} \\ \nonumber
&= -i \frac{(\hat{x}_{12})^j\Gamma(\theta-d/2+j)}{(-x_{12}^2)^{\theta-d/2+j}} \frac{\Gamma((d-\theta)/2)^2}{\Gamma(d-\theta)\pi^{-\frac{d}{2}}}  \sum_k {j \choose k} \frac{1 }{\Gamma(\theta/2+k) \Gamma(\theta/2+j-k)} \\ \nonumber
&= -i \frac{(\hat{x}_{12})^j}{(-x_{12}^2)^{\theta-d/2+j}} \frac{\Gamma(\theta-d/2+j) \Gamma((d-\theta)/2)^2 \Gamma \left(\frac{1}{2} (2 j+\theta -1)\right)}{\Gamma(d-\theta) \Gamma \left(j+\frac{\theta }{2}\right) \Gamma (j+\theta -1)} \frac{2^{\theta +2 j-2} }{\pi^{\frac{1}{2}-\frac{d}{2}}   }\,,
\end{align}
where we used the $x$-space version \eqref{eqn:bubble-integral-x-space} of the bubble integral. 

\section{Details of the Tree-Level Two-Point Function Calculation}\label{sec:details-two-point-functions-tree-level}\sectionmark{Details of the Tree-Level 2-Point Calculation}
In the following, we suppress the gauge group indices and the trace over the gauge group, such that the calculation is also valid for other scalar theories in general dimension, e.g. for $(\phi^3)_{d=6}$ theory. Furthermore, we take general flavours $I \neq J$.
 I.e. the operator under consideration is
\begin{align}
\hat{\op}_j^{\text{tree}} &= \left(\hat{\partial}_a+\hat{\partial}_b \right)^j C_j^{\nu} \left( \frac{\hat{\partial}_a-\hat{\partial}_b}{\hat{\partial}_a+\hat{\partial}_b} \right) \left( \phi^{I}(x_a)  \bar{\phi}^{J}(x_b) \right) \Big|_{x_a=x_b}\,.
\end{align}
We will take into account the gauge group factor at then end of the calculation when specializing to the operators \eqref{eqn:definition-with-gegenbauer-polynomials} in $\sym$, where the corresponding flavours are  $I=\{12\}, J=\{34\}$.  

The two-point functions of the operators \eqref{eqn:definition-with-gegenbauer-polynomials} can straightforwardly be evaluated using the Schwinger parametrisation\footnote{Formally, this is valid in the euclidean regime $x_{12}^2<0$.} of the scalar propagators \eqref{eqn:scalar-propagator}
\begin{equation}\label{eqn:schwinger-param}
 \langle \phi_I(x_1) \bar{\phi}^J(x_2) \rangle = g^2 \delta^{IJ} \frac{\Gamma(h-1)}{4 \pi^h} \frac{1}{(-x_{12}^2)^{h-1}} = \delta^{IJ} \frac{1}{4 \pi^h} \int_{0}^\infty d\alpha \, \alpha^{h-2} \exp (-\alpha (-x_{12}^2))\,,
\end{equation}
where $I$ and $J$ denote the flavours of the fields.
\begin{align}\nn
 \langle \hat{\op}_j \hat{\bar{\op}}_k \rangle &= \left(\hat{\partial}_a + \hat{\partial}_b \right)^j C_j^\nu \left(\frac{\hat{\partial}_a - \hat{\partial}_b}{\hat{\partial}_a + \hat{\partial}_b} \right) \left(\hat{\partial}_c + \hat{\partial}_d \right)^k C_k^\nu \left(\frac{\hat{\partial}_c - \hat{\partial}_d}{\hat{\partial}_c + \hat{\partial}_d} \right)\langle (\phi^I_a \bar{\phi}^J_b) (\bar{\phi}^I_c \phi^J_d) \rangle \Big{|}_{a=b,c=d}\\ \nn
  &= \left(\hat{\partial}_a + \hat{\partial}_b \right)^j C_j^\nu \left(\frac{\hat{\partial}_a - \hat{\partial}_b}{\hat{\partial}_a + \hat{\partial}_b} \right) \left(\hat{\partial}_c + \hat{\partial}_d \right)^k C_k^\nu \left(\frac{\hat{\partial}_c - \hat{\partial}_d}{\hat{\partial}_c + \hat{\partial}_d} \right)\frac{1}{(4 \pi^{h})^2} \frac{g^4 \Gamma(h-1)^2}{((-x_{ac}^2)(-x_{bd}^2))^{h-1}}  \Big{|}_{a=b,c=d}\,,
\end{align}
where $\phi_a \equiv \phi(x_a)$. 
Due to $\hat \partial_a  \hat{x}_{ab}= z^2 =0$, we have
\begin{align}
( \hat{\partial}_a)^k \exp(\alpha x_{ac}^2) &= 2^k (\alpha)^k (\hat{x}_{ac})^k  \exp( \alpha x_{ac}^2)\,,\\ \nonumber
( \hat{\partial}_c)^k \exp(\alpha x_{ac}^2) &= 2^k (-\alpha)^k  (\hat{x}_{ac})^k \exp(\alpha x_{ac}^2)\,.
\end{align}
Thus we have
\begin{align}
 \langle \hat{\op}_j \hat{\bar{\op}}_k \rangle &= g^4(-1)^k (2)^{j+k} (\hat{x}_{12})^{j+k} \int \frac{d\alpha d\beta}{(4 \pi^{h})^2} (\alpha \beta)^{h-2} \left(\alpha +\beta \right)^{j+k} \\ \nn
 & \qquad\qquad\qquad\qquad\qquad \times C_j^\nu \left(\frac{\alpha - \beta}{\alpha +\beta} \right) C_k^\nu \left(\frac{\alpha - \beta}{\alpha +\beta} \right) \exp((\alpha+\beta)x_{12}^2)
\end{align}
Substituting $y= \frac{\alpha-\beta}{\alpha + \beta}$ and $\alpha +\beta =z$ (Jacobian $=\frac{z}{2}$), we get
\begin{align}
 \langle \hat{\op}_j \hat{\bar{\op}}_k \rangle &=g^4 (-1)^k (2)^{j+k}2^{3-2h} \frac{(\hat{x}_{12})^{j+k}}{(4 \pi^{h})^2} \\ \nonumber
& \int_0^\infty dz  z^{2h-4 +j+k+1} \exp(z x_{12}^2) \int_{-1}^1 (1-y^2)^{h-3/2-\frac{1}{2}} C_j^\nu (y) C_k^\nu (y)\,.
\end{align}
Using \eqref{eqn:schwinger-param} and \eqref{eqn:GP-orthogonality} we thus get 
\begin{align}
  \langle \hat{\op}_j \hat{\bar{\op}}_k \rangle &= g^4 \delta_{jk} 2^{2j +3-2h+1- 2\nu } \frac{1}{(4\pi^h)^2}  \frac{\pi}{\Gamma^2(\nu)} \frac{\Gamma(2j+2\nu+1) \Gamma(j + 2\nu)}{\Gamma(j+1)(j+\nu)} \frac{(\hat{x}_{12})^{2j}}{(-x_{12}^2)^{2j + 2\nu +1}} \end{align}
for integer\footnote{As we will see later in $d=4-2\epsilon$ the correlators defined in terms of Gegenbauer polynomials have non-vanishing non-diagonal contributions.} d, such that $\nu=(d-3)/2$.
For $d=2h=6$ and $I \neq J$
\begin{align}\label{eqn:two-point-function-phi-3-theory}
  \langle \hat{\op}_j \hat{\bar{\op}}_k \rangle^{d=6} = g^4 \delta_{jk} (-1)^k \frac{2^{2j-2}}{(4\pi^3)^2} (j+1)(j+2) \Gamma(2j + 3) \frac{(\hat{x}_{12})^{2j}}{(-x_{12}^2)^{2j + 4}}\,.
\end{align}
Specialising to $\sym$ in $d=2h=4$, we have to take into account a factor of $\delta^{aa}/4 $ from the traces $\tr (T^a T^b) =\frac{1}{2}\delta^{ab}$ and another factor of 2, since we have two possible identical (for $j$ even) contractions (due to $\bar{\phi}_{34}= \phi^{12}$) 
\begin{align}\label{eqn:two-point-tree-level-N-equal4SYM}
  \langle \hat{\op}_j \hat{\bar{\op}}_k \rangle^{d=4} = g^4 \delta_{jk} \delta^{aa} \frac{2^{2j-1}}{(4\pi^2)^2} \Gamma(2j + 1) \frac{(\hat{x}_{12})^{2j}}{(-x_{12}^2)^{2j + 2}}\,.
\end{align}
These results were derived in a similar way in \cite{Belitsky:2007jp}. We generalise this calculation in section \ref{sec:one-loop-contibutions-using-gegenbauer-polynomials}.

\section{Integrals with Gegenbauer Polynomials}

\notocsubsection{Two-Point Function with Two Different Propagators}\label{sec:one-loop-contibutions-using-gegenbauer-polynomials}
The following formula is useful, when evaluating the contractions of two (possibly different) propagators $G_\tau, G_\sigma$. We would like to evaluate
\begin{equation}
I_{\sigma,\tau}^{i,j}=D_{ab}^i D_{cd}^j \left(G_\tau(x_{ac}) G_\sigma(x_{bd})+ G_\sigma(x_{ac}) G_\tau(x_{bd})\right)\,,
\end{equation}
where
\begin{equation}
D_{ab}^i = (\hat{\partial}_a+\hat{\partial}_b)^i C_i^\nu \left(\frac{\hat{\partial}_a-\hat{\partial}_b}{\hat{\partial}_a+\hat{\partial}_b}\right)
\end{equation}
parametrises the operators in the Gegenbauer representation. The expression arises in the contribution to the two-point function when using general propagators in the Schwinger representation
\begin{align}
G_\tau(x_{ac}) &= g_\tau \frac{\Gamma(\tau)}{(-x_{ac}^2)^\tau} = g_\tau \int_0^\infty d\alpha \alpha^{\tau-1} \exp(\alpha x_{ac}^2)\,, \\ \nn
G_\sigma(x_{ac}) &= g_\sigma \frac{\Gamma(\sigma)}{(-x_{ac}^2)^\sigma}= g_\sigma  \int_0^\infty d\beta \beta^{\sigma-1} \exp( \beta x_{ac}^2)\,.
\end{align}
Then we can write
\begin{align}
I_{\sigma,\tau}^{i,j}= g_\tau g_\sigma 2^{i+j} (-1)^j \hat{x}_{12}^{i+j} \int d\alpha d\beta (\alpha+\beta)^{i+j} \left( \alpha^{\tau-1} \beta^{\sigma-1}+ \alpha^{\sigma-1}\beta^{\tau-1} \right) \\ \nn 
\times C_i^\nu C_j^\nu \left(\frac{\alpha-\beta}{\alpha+\beta} \right) \exp((\alpha+\beta)x_{12}^2)\,.
\end{align}
Using $C_j^\nu(-x)=(-1)^j C_j(x)$ and exchanging $\alpha \leftrightarrow \beta$ in the second expression in the brackets we find
\begin{align}
I_{\sigma,\tau}^{i,j}= g_\tau g_\sigma 2^{i+j} (-1)^j \hat{x}_{12}^{i+j} (1+(-1)^{i+j}) \int d\alpha d\beta (\alpha+\beta)^{i+j}  \alpha^{\tau-1} \beta^{\sigma-1}
\\ \nn
\qquad\qquad\qquad \times C_i^\nu C_j^\nu \left(\frac{\alpha-\beta}{\alpha+\beta} \right)\exp((\alpha+\beta)x_{12}^2)\,.
\end{align}
Changing variables to $z=\alpha+\beta$ and $y=(\alpha - \beta)/(\alpha+\beta)$ with Jacobian $z/2$ we find
\begin{align}
I_{\sigma,\tau}^{i,j}= g_\tau g_\sigma 2^{i+j+1-\tau-\sigma} (-1)^j \hat{x}_{12}^{i+j} (1+(-1)^{i+j}) & \int_{0}^\infty dz z^{i+j-1 +\tau +\sigma} \exp(z x_{12}^2) \\ \nn & \int_{-1}^1 dy (1+y)^{\tau-1} (1-y)^{\sigma-1}
C_i^\nu C_j^\nu (y)\,.
\end{align}
Using the Schwinger parametrisation backwards leads to
\begin{align}\label{eqn:Iijsigmatau}
I_{\sigma,\tau}^{i,j}= g_\tau g_\sigma 2^{i+j+1-\tau-\sigma} (-1)^j  (1+(-1)^{i+j}) \frac{\hat{x}_{12}^{i+j}\Gamma(i+j+\tau+\sigma)}{(-x_{12}^2)^{i+j+\tau+\sigma}} \\ \nn \int_{-1}^{1} dy (1+y)^{\tau-1}(1-y)^{\sigma-1} C_i^\nu(y) C_j^\nu(y)\,.
\end{align}

\notocsubsection{Two-Point Function at Tree-Level including  $\op(\epsilon)$ Terms}\label{eqn:tree-level-expression-epsilon}
We can extract the tree-level two-point function by using section \ref{sec:one-loop-contibutions-using-gegenbauer-polynomials} for two identical propagators with $\sigma=\tau=h-1=2-\epsilon$ as well as $g_\tau =g_\sigma = 1/4\pi^h$. The trace normalisation yields an additional factor of $1/4$. Then we have \eqref{eqn:Iijsigmatau}
\begin{align}
\langle \hat\op_i (x_1) \bar{\hat\op}_j(x_2)\rangle = \frac{g^4 \delta^{aa}}{4}\frac{(-1)^i}{(4\pi^h)^2} 2^{i+j+3-d}(1+(-1)^{i+j}) \frac{\hat{x}_{12}^{i+j}\Gamma(i+j+2h-2)}{(-x_{12}^2)^{i+j+2h-2}} \\ \nn\int_{-1}^1 dy (1-y^2)^{1-\epsilon}C_i^\nu(y) C_j^\nu(y)\,.
\end{align}
The integral is solved in \ref{sec:integral-non-diagonal-tree-level} and we have 
\begin{align}
 \int_{-1}^1 dy (1-y^2)^{1-\epsilon}C_i^\nu(y) C_j^\nu(y) = \delta_{ij} N(j) - \epsilon I_{ij}\,,
\end{align}
where $N(j)$ is given in \eqref{eqn:GP-orthogonality}, $I_{ij}$ in \eqref{eqn:I-ij} and the tree-level contribution thus yields \eqref{eqn:two-point-tree-level-N-equal4SYM}, since $I_{ij}=0$ for $i+j=\text{uneven}$.

\subsubsection{Diagonal contribution}\label{eqn:tree-level-expression-epsilon-diagonal}
For $i=j$ we have
\begin{align}\label{tree-level-correlator-to-order-epsilon}
\langle \hat\op_j (x_1) \bar{\hat\op}_j(x_2)\rangle = g^4 \delta^{aa} \frac{2^{2j-2} }{(2\pi)^d} \Gamma(2j+2h-2)\frac{\hat{x}_{12}^{i+j}}{(-x_{12}^2)^{i+j+2h-2}}\left(N(j,1/2)- \epsilon I_{jj}^{1/2} \right)\,,
\end{align}
where $N(j,1/2)= 2/(2j+1)$ is given in \eqref{eq:app-orthogonality} and $I_{jj}^{1/2}$ in \eqref{Ijj1/2}. Expanding the $\Gamma$ function we get
\begin{align}
\langle \hat\op_j (x_1) \bar{\hat\op}_j(x_2)\rangle &= g^4 \delta^{aa} \frac{2^{2j-1} }{(2\pi)^d} \Gamma(2j+1)\frac{(\hat{x}_{12})^{i+j}}{(-x_{12}^2)^{i+j+2h-2}} \\ \nn
& \left(1+ 2\epsilon  \left(2H_{2j-1}-H_{2j+1}-H_j-H_{j-1} + \frac{1}{2j+1} - \log(2)  + \gamma_E  \right)\right)\,.
\end{align}
%{\bf Remark:} The $\gamma_E$ coud be eliminated by using the $\overline{MS}$ scheme. The $\log(2)$ disappears if we expand $1/2^d$ as well, i.e.
%\begin{align}
%\langle \op_j (x_1) \bar{\op}_j(x_2)\rangle &= \frac{2^{2j-5} }{\pi^d} \Gamma(2j+1)\frac{\hat{x}_{12}^{i+j}}{(x_{12}^2)^{i+j+2h-2}} \\ \nn
%& \left(1+ 2\epsilon  \left(2H_{2j-1}-H_{2j+1}-H_j-H_{j-1} + \frac{1}{2j+1}  + \gamma_E  \right)\right)
%\end{align}

\notocsubsection{Integral $I_{ij}$}\label{sec:integral-non-diagonal-tree-level}

 The orthogonality relation of the Gegenbauer polynomials with arbitrary $\nu > -1/2$ is
\begin{equation}\label{eq:app-orthogonality}
\int_{-1}^1 (1-y^2)^{\nu - \frac{1}{2}} C_j^\nu (y) C_k^\nu (y)= \delta_{jk} 2^{1-2\nu} \pi \frac{\Gamma(j+2\nu)}{\Gamma^2(\nu)\Gamma(j+1)(j+\nu)}=\delta_{jk} N(j,\nu)\,.
\end{equation}
However, we calculate the correlators in $d=6- 2\epsilon$ dimensions, but with fixed index $\nu \equiv 3/2$ of the Gegenbauer polynomials. Thus in the evaluation of the tree-level correlator with Schwinger parameters we get the expression
\begin{align}
  \int_{-1}^1 (1-y^2)^{\nu^\prime-\epsilon - \frac{1}{2}} C_j^{\nu^\prime} (y) C_k^{\nu^\prime} (y) &=  \int_{-1}^1 (1-y^2)^{\nu^\prime-1/2} C_j^{\nu^\prime} (y) C_k^{\nu^\prime} (y) \\ \nn
  & -\epsilon \int_{-1}^1 (1-y^2)^{\nu^\prime-1/2}\log(1-y^2) C_j^{\nu^\prime} (y) C_k^{\nu^\prime} (y)\,.
\end{align}
which is non-diagonal at order $\op(\epsilon)$. 
This integral is easy to evaluate by using the expansion of the Gegenbauer polynomials w.r.t. their index
\begin{equation}
 C_j^{\nu-\epsilon}(x)= C_j^\nu(x) + \epsilon \left( 2 \sum_{k=0}^j d_{jk}^{\nu} C_k^\nu(x) \right)
\end{equation}
with $d_{jk}^\nu$ as given in \eqref{eqn:gegenbauer-expansion-coefficients}, also including a $d_{jj}$ term.

\subsubsection{Explicit solution for $d=4$}
Then, by using the exact expression \eqref{eq:app-orthogonality} for $\nu=1/2-\epsilon$, expanding the indices in $C_j^{1/2-\epsilon}$ w.r.t. $\epsilon$ and comparing the order $\epsilon$ terms we find
\begin{align}\label{eqn:I-ij}
I_{ij}^{1/2}=& \int_{-1}^{1}dy\, \log(1-y^2) C_i^{1/2} C_j^{1/2}  = \\  \nn
&= 2 \left(d_{ij}^{1/2}N(j,\nu=\frac{1}{2}) + d_{ji}^{1/2}N(k,\nu=\frac{1}{2})\right) - \delta_{ij} N(j,\nu=\frac{1}{2}-\epsilon)\Big|_{\epsilon}  \\ \nn
&= 4 \left(d_{ij}^{1/2}\frac{1}{(2j+1)} + d_{ji}^{3/2} \frac{1}{(2i+1)} \right) - \delta_{ij}\frac{2^{2\epsilon} \pi }{\Gamma^2(1/2-\epsilon)}\frac{\Gamma(j+1-2\epsilon)}{\Gamma(j+1)(j+1/2-\epsilon)}\Big|_{\epsilon}  \,.
\end{align}
Also note that $d_{ij}=0$ for $j>i$, such that only both terms are present when $i=j$, then we have
\begin{align}\label{Ijj1/2}
 I_{jj}^{1/2} &=-\frac{4}{(2 j+1)} \left(H_{j-\frac{1}{2}}-H_j+\frac{1}{2j+1}+\log(2) \right) \\ \nn
 %&=- 2 N(j,1/2) \left(H_{j-\frac{1}{2}}-H_j+\frac{1}{2j+1}+\log(2) \right) \\ \nn
 &= - 2 N(j,1/2) \left(-H_{j-1}-H_j+2 H_{2 j-1}+\frac{1}{2 j+1}-\log (2)\right)\,.
\end{align}
Note that here we also had to take into account the $d_{jj}^{1/2}$ terms and in the last line we used the identity \eqref{sec:hjplus12identity}.

%\subsubsection{Explicit solution for $d=6$}
%Then, by using the exact expression \eqref{eq:app-orthogonality} for $\nu=3/2-\epsilon$, expanding the indices in $C_j^{3/2-\epsilon}$ w.r.t. $\epsilon$ and comparing the order $\epsilon$ terms we find
%\begin{align}\label{eqn:I-ij=d-6}
%I_{ij}^{3/2}=& \int_{-1}^{1}dy\, (1-y^2)  \log(1-y^2) C_i^{3/2} C_j^{3/2}  = \\  \nn
%&= 4 \left(d_{ij}^{3/2}\frac{(j+1)(j+2)}{(2j+3)} + d_{ji}^{3/2}\frac{(i+1)(i+2)}{(2i+3)}\right) - \delta_{ij}\frac{2^{-2+2\epsilon} \pi }{\Gamma^2(3/2-\epsilon)}\frac{\Gamma(j+3-2\epsilon)}{\Gamma(j+1)(j+3/2-\epsilon)}\Big|_{\epsilon}  
%\end{align}
%where the index indicates that we take the $\op(\epsilon)$  part of the last term. Also note that $d_{ij}=0$ for $j>i$, such that only both terms are present when $i=j$, then we have
%\begin{align}
% I_{jj}^{3/2} =-\frac{4 (j+1) (j+2)}{(2 j+3)} \left( H_{j+\frac{1}{2}}- H_{j+2}+\frac{1}{2j +3}+\log (2)\right) \quad \text{\bf correct, checked}
%\end{align}
%Note that here we also had to take into account the $d_{jj}^{3/2}$ terms.

\section{Two-Point Correlator at One Loop}
\notocsubsection{Diagram \ref{fig:twopointfunctionselfenergy}}

For the self-energy diagrams we do not have to take into account the gauge fields from the covariant derivative and can thus start again from the tree-level operator \eqref{eqn:definition-with-gegenbauer-polynomials}. Taking into account a factor of $1/4$ from the trace over the gauge group as well as the fact that we have two identical contractions (for $i+j$ even, otherwise the correlator vanishes anyways)  we find
\begin{align}\nn
\langle \hat{\op}_i(x_1)  \hat{\bar{\op}}_j(x_2) \rangle^{(\ref{fig:twopointfunctionselfenergy})} &= 2 \frac{1}{4} D_{ab}^i D_{cd}^j  \left( G_\sigma (x_{ac}) G_\tau (x_{bd})+G_\tau (x_{ac}) G_\sigma (x_{bd}) \right) \Big|_{x_a=x_b,x_c=x_d} \\ \nn
&= \frac{1}{2} \delta^{aa} I^{i,j}_{\sigma,\tau}(x_{12})\,,
\end{align}
where $I^{i,j}_{\sigma,\tau}$ is given in \eqref{eqn:Iijsigmatau} and here we have $\sigma=2h-3, \tau=h-1$ and
\begin{equation}\nn
g_\sigma = -\frac{1}{2} \frac{g^4N}{(4 \pi^h)^2} \frac{\Gamma(h-1)^2 \Gamma(2-h)}{\Gamma(2h-2)\Gamma(3-h)}\,, \qquad g_\tau = \frac{g^2}{4\pi^h}\,.
\end{equation}
Then for $i=j$ the full correlator reads
\begin{align}\nn
\langle \hat{\op}_j(x_1)  \hat{\bar{\op}}_j(x_2) \rangle^{(\ref{fig:twopointfunctionselfenergy})} &= g^6 N \delta^{aa} g_\tau g_\sigma 2^{2j-3h+5}\Gamma(2j+3h-4) \left( N(j,1/2) - \frac{3}{2}\epsilon I_{jj}^{1/2} \right)\frac{(\hat{x}_{12})^{2j}}{(-x_{12}^2)^{2j+3h-4}} \,.
\end{align}
Expanding the prefactor in $\epsilon$ we find
\begin{align}\label{eqn:diagram-a-final-version-to-expand}
\langle \hat{\op}_j(x_1)  \hat{\bar{\op}}_j(x_2) \rangle^{(\ref{fig:twopointfunctionselfenergy})} &= \frac{g^6 N \delta^{aa} 2^{2j-2}\Gamma(2j+2)}{(2\pi)^{3h}} \left(- \frac{1}{\epsilon} -2 + 3 H_{2j+1}-3\gamma_E \right) \times \\ \nn
& \qquad   \quad  \qquad \qquad \qquad \times \left( N(j,1/2) - \frac{3}{2}\epsilon I_{jj}^{1/2} \right)\frac{(\hat{x}_{12})^{2j}}{(-x_{12}^2)^{2j+3h-4}} \,,
\end{align}
where $N(j,1/2)= 2/(2j+1)$ is given in \eqref{eq:app-orthogonality} and $I_{jj}^{1/2}$ in \eqref{Ijj1/2}.

~\\{\bf Divergent part}\\
Taking into account the normalisation $N(j,1/2)$, see \eqref{eqn:GP-orthogonality} , the leading divergent term reads
\begin{align}\nn
\langle \hat{\op}_j(x_1)  \hat{\bar{\op}}_j(x_2) \rangle^{(\ref{fig:twopointfunctionselfenergy})} &= - g^6 N \delta^{aa} \frac{1}{\epsilon} \frac{2^{2j-7}}{ \pi^6} \Gamma(2j+1)  \frac{(\hat{x}_{12})^{2j}}{(-x_{12}^2)^{2j+2}}\\ \nn
& = - g^2N \frac{1}{\epsilon} \frac{1}{4\pi^2} \langle \hat{\op}_j \hat{\bar{\op}}_j \rangle^{(0)} + \op(\epsilon^0)\,,
\end{align}
where we have taken into account the tree-level correlator \eqref{eqn:two-point-tree-level-N-equal4SYM} in the last equality. Thus the contribution to the anomalous dimension is
\begin{equation}\label{eqn:anomalous-dim-self-energy}
\gamma_j^{\eqref{fig:twopointfunctionselfenergy}}= \frac{g^2N}{4 \pi^2} \,.
\end{equation}

\notocsubsection{Diagram \ref{fig:twopointcourt}}\label{sec:kite-diagram}
Using the Feynman rules for the scalar gluon vertex \eqref{eqn:feynman-rule-scalar-gluon-vertex}, taking into account that there are four possible contractions as well as the colour factors 
$$\tr(T^aT^b) \tr(T^cT^d)f^{ace}f^{bde}=\frac{1}{4} N \delta^{aa}\,,$$ 
letting $l \rightarrow -l$, $k \rightarrow -k$ and introducing two binomial sums, we find
\begin{align}
\langle \hat{\op}_j(p) \bar{\hat{\op}}_j(-p) \rangle^{(\ref{fig:twopointcourt})}\nn
&=   i (i)^{2j}\frac{g^6}{2} N \delta^{aa} \sum_{k,k^\prime=0}^j a_{jk}^{1/2}  a_{jk^\prime}^{1/2} \sum_{n=0}^k \sum_{m=0}^{k^\prime} {k \choose n} {k^\prime \choose m} (-1)^{-k-k^\prime} \\ 
&\phantom{=}  (\hat{p})^{n+m} \int \frac{d^dk,l}{(2\pi)^{2d}} \frac{(\hat{k})^{j-n} (\hat{l})^{j-m}(l+k+2p)\cdot(l+k)}{k^2(p+k)^2(l-k)^2 l^2 (l+p)^2} \,.
\end{align}
Using the basic integral \eqref{eqn:basic-integral} and $(l+k+2p)\cdot(l+k)=-2 p^2-(l-k)^2+k^2+(k+p)^2+l^2+(l+p)^2$ we can rewrite this as
\begin{align}\label{eqn:kite-with-derivatives}
\langle \hat{\op}_j(p) \bar{\hat{\op}}_j(-p) \rangle^{(\ref{fig:twopointcourt})}\nn
=   i (i)^{2j}\frac{g^6}{2}  N \delta^{aa} &\sum_{k,k^\prime=0}^j a_{jk}^{1/2}  a_{jk^\prime}^{1/2} \sum_{n=0}^k \sum_{m=0}^{k^\prime} {k \choose n} {k^\prime \choose m} (-1)^{-k-k^\prime} \\ \nn
\frac{1}{(4\pi)^{d}}\frac{(\hat{p})^{2j}}{(p^2)^{4-d}}   \Big( - 2 & c_{j-n,j-m}(1,1,1,1,1) -   c_{j-n,j-m}(1,1,0,1,1) \\ \nn 
+    & c_{j-n,j-m}(0,1,1,1,1) +   c_{j-n,j-m}(1,0,1,1,1) \\ \nn
 +    & c_{j-n,j-m}(1,1,1,0,1) +   c_{j-n,j-m}(1,1,1,1,0)  \Big)\,.
\end{align}
These integrals have been solved in Appendix \ref{sec:appendix-integrals}. We define\footnote{$c_{nm}(1,1,1,1,1)$ is  $\op(1)$ and can be neglected completely for this analysis, since the Fourier transformation yields another factor of $\epsilon$ such that the x-space expression is $\op(\epsilon)$.}
\begin{align}
&S^{\ref{fig:twopointcourt}}[j]:=\sum_{k,k^\prime=0}^j a_{jk}^{1/2}  a_{jk^\prime}^{1/2} \sum_{n=0}^k \sum_{m=0}^{k^\prime} {k \choose n} {k^\prime \choose m} (-1)^{-k-k^\prime} \\ \nn 
& \qquad \qquad \Big(  -   c_{j-n,j-m}(1,1,0,1,1)  
 +    c_{j-n,j-m}(0,1,1,1,1) +   c_{j-n,j-m}(1,0,1,1,1) \\ 
& \qquad \qquad ~~+     c_{j-n,j-m}(1,1,1,0,1) +   c_{j-n,j-m}(1,1,1,1,0)   \Big)\,.
\end{align}
Extracting only the $1/\epsilon^2$ term, we find (the sum over the first term appears to be zero and would cancel with the four-scalar diagram anyways)
\begin{equation}
S^{\ref{fig:twopointcourt}}[j]= -\frac{1}{\epsilon^2}\frac{2}{(2j+1)(j+1)}+ \op \left(\frac{1}{\epsilon} \right)
\end{equation}
and thus have
\begin{align}
\langle \hat{\op}_j(p) \bar{\hat{\op}}_j(-p) \rangle^{(\ref{fig:twopointcourt})}\nn
&=  - i (i)^{2j} g^6  \frac{2 N \delta^{aa} }{\pi^4}
\frac{(\hat{p})^{2j}}{(p^2)^{4-d}} \frac{1}{\epsilon^2}  \left( \frac{2^{-9}}{(2j+1)(j+1)}  \right)\,.
\end{align}
Fourier transformation \eqref{eqn:result-fourier-trafo-mink} to position space  yields, using $1/\Gamma(4-d)= 2\epsilon + \op(\epsilon^2)$,
\begin{align}
\langle \hat{\op}_j(x_1) \bar{\hat{\op}}_j(x_2) \rangle^{(\ref{fig:twopointcourt})} &= 
g^6 N \delta^{aa}\frac{1}{\epsilon} \frac{2^{2j-7}}{\pi^{6}} \Gamma(3h-4+j) \frac{(\hat{x}_{12})^{2j}}{(x_{12}^2)^{3h-4+j}}  \\ \nn
 &=  g^6 N \delta^{aa} \frac{1}{\epsilon} \frac{\Gamma(2j+1)}{(j+1)} \frac{2^{2j-3}}{(4\pi^2)^2} \frac{1}{\pi^2} \frac{(\hat{x}_{12})^{2j}}{(x_{12}^2)^{2+2j}} \\ \nn
&=\left( \frac{1}{\epsilon} \frac{g^2N}{4\pi^2} \frac{1}{(j+1)}\right)  \langle  \hat{\op}_j  \bar{\hat{\op}}_j \rangle^{(0)} \,,
\end{align}
where the tree-level correlator \eqref{eqn:two-point-tree-level-N-equal4SYM} was used.
Thus, the contribution to the anomalous dimension is
\begin{equation}\label{eqn:anomalous-dim-kite}
\gamma_j^{(\ref{fig:twopointcourt})}=-\frac{g^2N}{4 \pi^2} \frac{1}{(j+1)}\,.
\end{equation}

\notocsubsection{Diagram \ref{fig:twopointfourscalar}}
After introducing a binomial sum, the four-scalar diagram reads
\begin{align}
&\langle \op_{j=0}(p) \bar{\op}_{j=0}(-p) \rangle^{\ref{fig:twopointfourscalar}} \\ \nn
 &=   (g^2)^4 (i)^4  \frac{i}{1!}\left( -\frac{1}{2g^2} f^{gba}f^{gab}\right)  i^{2j} \left( \int d^dl \frac{\hat{p}^k (\hat{p}-\hat{l})^{j-k}}{l^2 (p-l)^2} \right)^2 \\ \nn
& = i g^6 \frac{N \delta^{aa}}{2} \frac{\hat{p}^{2j}}{(p^2)^{4-d}} \frac{1}{(4\pi)^d}  \sum_{k=0}^j a_{jk}a_{jk^\prime} (-1)^{-k-k^\prime} \sum_{n,m}^{k,k^\prime} {k \choose n }{k^\prime \choose m}  c_{j-n,j-m}(1,1,0,1,1) \,.
\end{align} 
%From this it is clear, that the the diagram combines with the kite diagrams, see expression \eqref{eqn:kite-with-derivatives}. We will solve it anyways, for reference and define
%\begin{equation}
%S^{(\ref{fig:twopointfourscalar})}[j]=\sum_{k=0}^j a_{jk}a_{jk^\prime} (-1)^{-k-k^\prime} \sum_{n,m}^{k,k^\prime} {k \choose n }{k^\prime \choose m}  c_{j-n,j-m}(1,1,0,1,1) 
%\end{equation}
The sum can be solved using \eqref{eqn:identity-beta-function}. The diagram does however cancel with the corresponding term in the kite diagram, so no further investigation is necessary.

\notocsubsection{Cancellation of Four-Scalar Interaction Diagram}
Taking into account the cancellation of the four-scalar diagram  and the fact, that the first term in \eqref{eqn:kite-with-derivatives} is of order $\epsilon$ after Fourier transformation and can be completely neglected, we have
\begin{align}
\langle \hat{\op}_j(p) \bar{\hat{\op}}_j(-p) \rangle^{\ref{fig:twopointcourt}+\ref{fig:twopointfourscalar}}\nn
=   i (i)^{2j}\frac{g^6}{2}  N \delta^{aa}
\frac{1}{(4\pi)^{d}}\frac{(\hat{p})^{2j}}{(p^2)^{4-d}} S[j]^{\ref{fig:twopointcourt}+\ref{fig:twopointfourscalar}}\,,
\end{align}
where we defined
\begin{align}
S[j]^{\ref{fig:twopointcourt}+\ref{fig:twopointfourscalar}}:&=\sum_{k,k^\prime=0}^j a_{jk}^{1/2}  a_{jk^\prime}^{1/2} \sum_{n=0}^k \sum_{m=0}^{k^\prime} {k \choose n} {k^\prime \choose m} (-1)^{-k-k^\prime} \\ \nn
\Big(& c_{j-n,j-m}(0,1,1,1,1) +   c_{j-n,j-m}(1,0,1,1,1) \\ \nn 
 +    &  c_{j-n,j-m}(1,1,1,0,1) +   c_{j-n,j-m}(1,1,1,1,0)   \Big)\,.
\end{align}
Fourier transformation yields
\begin{align}\label{eqn:kite+scalar-version-to-expand}
\langle \hat{\op}_j(p) \bar{\hat{\op}}_j(-p) \rangle^{\ref{fig:twopointcourt}+\ref{fig:twopointfourscalar}}
=   -g^6  N \delta^{aa} 2^{2 j-9} 
\frac{1}{\pi^{3h}} \frac{  \Gamma \left(\frac{3 d}{2}+2 j-4\right)}{\Gamma (4-d)} \frac{(\hat{x}_{12})^{2j}}{(-x_{12}^2)^{3d/2-4+2j}} S[j]^{\ref{fig:twopointcourt}+\ref{fig:twopointfourscalar}}\,.
\end{align}

\notocsubsection{Diagram \ref{fig:twopointonecovderivative}}
We have also calculated this diagram in $x$-space as a comparison, which is actually somewhat simpler, since we can perform the integral in $x$-space. In momentum space the diagram reads
\begin{align}\label{eqn:momentum-space-calculaiton-of-diagram-c}
 \langle \hat{\op}_j(p) \bar{\hat{\op}}_j(-p) \rangle^{(\ref{fig:twopointonecovderivative})} &= 2 \langle \hat{\op}_j^A (p) \bar{\hat{\op}}_j^{\text{free}} (-p) \rangle
= g^6 2 i^{2j+1} N \delta^{aa} \frac{\hat{p}^{2j}}{(-p^2)^{4-d}} S^{(\ref{fig:twopointonecovderivative})}[j] \,,
 \end{align}
 where
 \begin{align}
 S^{(\ref{fig:twopointonecovderivative})}[j] &= \sum_{k=1,k^\prime=0}^j \sum_{m=1}^k
 \sum_{n=0}^{m-1} {k \choose m} {m-1 \choose n} a_{jk}^{1/2} a_{jk^\prime}^{1/2} \\ \nn
 & \quad \phantom{-} (-1)^{m+k^\prime} \Big( b_{j-k+n} - b_{j-k+n+1} \Big)\sum_p^{k+k^\prime-m} {k+k^\prime-m \choose p } b_{2j-p} (3-h,1) \\ \nn
 & \quad- (-1)^{m+k^\prime} \Big( b_{k-m+n} - b_{k-m+n+1} \Big)\sum_p^{2j-k-k^\prime-m} {2j-k-k^\prime \choose p } b_{2j-p} (3-h,1)\,.
 \end{align}
We can solve the last binomial sum using \eqref{eqn:identity-beta-function}. Fourier transformation yields
\begin{align}\label{eqn:FT-of-momentum-space-calculaiton-of-diagram-c}
 \langle \hat{\op}_j(p) \bar{\hat{\op}}_j(-p) \rangle^{(\ref{fig:twopointonecovderivative})} = - g^4 \delta^{aa}  2^{2 d+2 j-7}  \frac{g^2N}{\pi^h} \frac{ \Gamma \left(\frac{3 d}{2}+2 j-4\right)}{\Gamma (4-d)}  \frac{(\hat{x}_{12})^{2j}}{(-x_{12}^2)^{3d/2-4+2j}} S^{(\ref{fig:twopointonecovderivative})}[j] \,.
 \end{align}

~\\{\bf Leading divergent term}\\
The leading divergent term in the sum is
 \begin{align}
 S^{(\ref{fig:twopointonecovderivative})}[j] &= \frac{1}{\epsilon^2}\frac{1}{2} \frac{1}{2j+1} \left( 2H_j  + \frac{1}{j+1}-1\right) \frac{1}{(4\pi)^d} +\op(\epsilon^{-1})
 \end{align}
% and thus
 %\begin{align}
 %  \langle \hat{\op}_j(p) \bar{\hat{\op}}_j(-p) \rangle^{(\ref{fig:twopointonecovderivative})} &= \frac{g^6 N \delta^{aa}}{\epsilon^2} \frac{1}{2j+1} \left( 2H_j  + \frac{1}{j+1}-1\right) \frac{i}{(4\pi)^d} \frac{\hat{p}^{2j}}{(-p^2)^{4-d}} +\op(\epsilon)
% \end{align}
and therefore
\begin{align}\nn
 \langle \hat{\op}_j^A (x_1) \bar{\hat{\op}}_j^{\text{free}} (x_2) \rangle^{\ref{fig:twopointonecovderivative}}
 &=   -\frac{g^6 N \delta^{aa}}{\epsilon} \left( 2 H_j + \frac{1}{(j+1)} - 1 \right)
\frac{ 2^{2 j-7} \Gamma (2 j+1)}{\pi ^6}  \frac{(\hat{x}_{12})^{2j}}{(-x_{12}^2)^{2+2j}} + \op(\epsilon^0) \\ \nn
 &=   \left( -\frac{1}{\epsilon} \frac{g^2N}{4 \pi^2} \left( 2 H_j + \frac{1}{(j+1)} - 1 \right) \right)  \langle  \hat{\op}_j  \bar{\hat{\op}}_j \rangle^{(0)} \,.
  \end{align}
Thus the contribution to the anomalous dimension is
\begin{equation}\label{eqn:anomalous-dim-one-gluon-vertex}
\gamma_j^{\ref{fig:twopointonecovderivative}} = \frac{g^2N}{4 \pi^2}  \left(2 H_j + \frac{1}{(j+1)} -1 \right)\,.
\end{equation}

\notocsubsection{Diagram \ref{fig:twopointtwocovderivative}}
 We consider the tree-level contraction of two operators $\hat{\op}^A_j$ as given by \eqref{eqn:twist-operator-one-gauge-field}
\begin{align}\nn
\langle \hat{\op}_j \bar{\hat{\op}}_j\rangle^{\ref{fig:twopointtwocovderivative}} = \langle \hat{\op}_j^{\text{A}} \bar{\hat{\op}}_j^{\text{A}} \rangle 
&=  \frac{1}{4} f^{abc} f^{def} \left(1+(-1)^j\right)^2   \sum_{k,k^\prime=1}^{j}a_{jk}^{1/2} a_{jk^\prime}^{1/2} \sum_{m,m^\prime=1}^{k,k^\prime}{k \choose m} {k^\prime \choose m^\prime} \\ \nn
& \langle \hat{\partial}^{m-1} \hat{A}^a \hat{\partial}^{m^\prime-1} \hat{A}^d \rangle \langle  \hat{\partial}^{k-m} \phi^{12,b}  \hat{\partial}^{j-k} \phi^{12,c}  \hat{\partial}^{k^\prime-m^\prime} \bar{\phi}^{12,e}  \hat{\partial}^{j-k^\prime} \bar{\phi}^{12,f} \rangle\,.
\end{align}
In Feynman-gauge we have
\begin{equation}
\langle \hat{A}^a \hat{A}^d\rangle = - z^\mu z^\nu \langle A^a_\mu A^d_\nu \rangle = z^2 \frac{\Gamma(h-1)\delta^{ad}}{4 \pi^h (-x_{12}^2)^{h-1}} =0\,,
\end{equation}
since $z^2=0$ and thus the diagram is zero in Feynman gauge.

\section{Tree-Level Three-Point Function in the Limit $x_2\to \infty$}\label{sec:three-point-limit-x2-to-infinity}
\begin{figure}[h]
\centering
 \includegraphics[width=.35\textwidth]{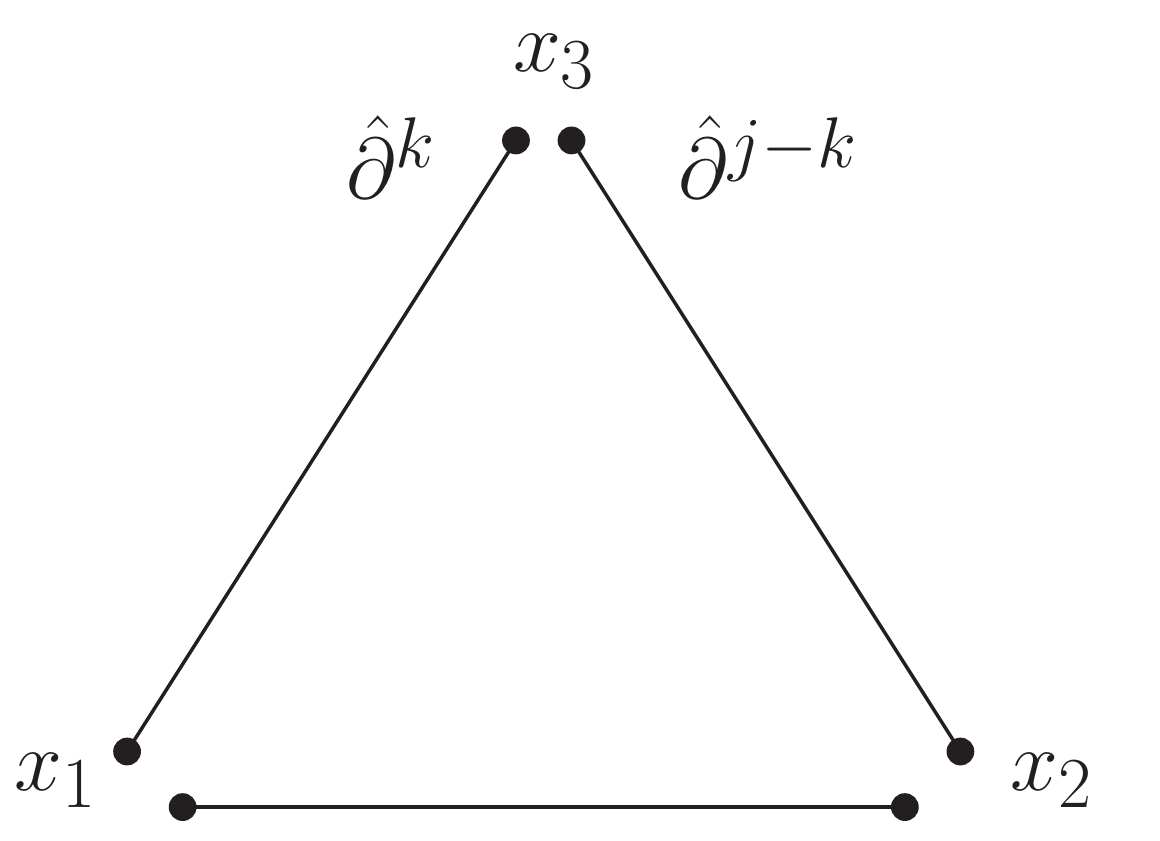}
\caption{3-point function at tree-level}
\label{fig:three-point function at tree}
\end{figure}

The 3-point correlator $\langle \op \tilde \op \hat{\op}_j \rangle$ simplifies considerably in the limit $x_2 \rightarrow \infty$. We will now explicitly evaluate the correlation function in this limit in order to extract the structure constant $C_{\op \tilde\op j}$.  Using the representation \eqref{eqn:ajk-representation-twist-operator} and $a_{jk}=(-1)^{j}a_{j,j-k}$
we have
\begin{align}
\langle \op(x_1) \tilde{\op}(x_2)  \hat{\op}_j(x_3)\rangle &=  \frac{g^6 \delta^{aa}}{8} \frac{\Gamma(h-1)^3}{(4\pi^h)^3}  \frac{1}{(-x_{12}^2)^{h-1}}  \\ \nn
 \sum_{k=0}^{j} &  \, a_{jk}^\nu  \left( \hat{\partial}_3^k \frac{1}{(-x_{13}^2)^{h-1}} \hat{\partial}_3^{j-k}  \frac{1}{(-x_{23}^2)^{h-1}}
+ \hat{\partial}_3^k \frac{1}{(-x_{23}^2)^{h-1}} \hat{\partial}_3^{j-k} \frac{1}{(-x_{13}^2)^{h-1}} \right)\,.
\end{align}
Multiplying this with the factors of $x_{13}^2$ and $x_{23}^2$ as given in \eqref{eqn:3-point limit} and taking the limit $x_2 \rightarrow \infty$ it is easy to see that the right-hand side can only be produced  by the term in which all derivatives act on $x_{13}^2$, i.e.
\begin{align}
 \lim_{x_2 \rightarrow \infty} (-x_{12}^2)^{h-1} (-x_{23}^2)^{h-1} \langle \op \tilde{\op} \hat{\op}_j \rangle =   
\frac{g^6 \delta^{aa}}{8} \frac{\Gamma(h-1)^3}{(4\pi^h)^3}  \left(a_{jj}^\nu+a_{j0}^\nu \right) &\hat{\partial}_3^j \frac{1}{(-x_{13}^2)^{h-1}}\,.
\end{align}
Taking $j$ even and using $a_{jk}^\nu = (-1)^{j}a_{j,j-k}^\nu$ as well as 
\begin{equation}
a_{jj}^\nu=\sum_{m=0}^j c^\nu_{m_j}=  C_j^\nu(1) = {2\nu -1 + j \choose j} ,\qquad \nu = h-3/2,
\end{equation}
performing the derivatives and setting $\nu=1/2$ we find
\begin{align}
& \lim_{x_2 \rightarrow \infty} (-x_{12}^2)^{h-1} (-x_{23}^2)^{h-1} \langle \op \tilde{\op} \hat{\op}_j \rangle \\ \nn
 &\qquad\qquad\qquad\qquad=   g^6 \delta^{aa} 2^{j-2} \frac{\Gamma(h-1)^3}{(4\pi^h)^3} \frac{\Gamma(j+h-1)}{\Gamma(h-1)} {2\nu -1 + j \choose j} 
  \frac{(\hat{x}_{13})^j}{(-x_{13}^2)^{j+h-1}} \,.
\end{align}
%The coefficient $a_{jj}$ is the coefficient where all derivatives act on the first field. In the expression by the Gegenbauer Polynomials this can be seen to be the coefficient of $(\hat{\partial}_a)^j(\hat{\partial}_b)^0$. The Gegenbauer Polynomial can be written as
%\begin{equation}
% C_j^\nu(x) = \sum_{m=0}^j c^\nu_{m_j} x^m
%\end{equation}
%Thus it is clear that the coefficient of $(\hat{\partial_a})^j$  in
%\begin{equation}
%(\hat{\partial_a}+\hat{\partial_b})^j C_j^\nu \left(\frac{\hat{\partial}_a-\hat{\partial}_b}{\hat{\partial}_a+\hat{\partial}_b} \right)
%\end{equation}
%is
Comparing with  \eqref{eqn:3-point limit} 
\begin{equation}
 \lim_{x_2 \rightarrow \infty} (-x_{12}^2)^{h-1} (-x_{23}^2)^{h-1} \langle \op \tilde{\op} \hat{\op}_j \rangle = C_{\op \tilde{\op} j} \frac{(\hat{x}_{13})^j}{(-x_{13}^2)^{j + h-1}}\,,
\end{equation}
we thus get 
\begin{equation}\label{eqn:Co1o2oj}
 C_{\op  \tilde{\op} j}= g^6 \delta^{aa} 2^{j-2} \frac{\Gamma(h-1)^2}{(4 \pi^h)^3} \Gamma(h+j-1)  {2\nu -1 + j \choose j} \,.
\end{equation}
This expression is valid for a scalar theory in general integer dimension $d$, e.g. $(\phi^3)$ - theory  in six dimensions, up to the factors of $\delta^{aa}/8$ that is due to the trace over the gauge group. For $d=4$ we find
\begin{equation}
C_{\op \tilde{\op} j}  = g^6 \delta^{aa} \frac{2^{j-8} }{\pi^6}\Gamma(j+1) 
\end{equation}
in agreement with \eqref{eqn:tree-level-strcuture-constant-N=4}.

\section{Anomalous Dimension from Three-Point Functions}\label{sec:details-anomalous-dimension-from-three-point-function}
As a check of the relative factors and signs between the diagrams we read off the anomalous dimension from the three-point function. After the cancellations between diagrams \ref{fig:3point1loopa}, \ref{fig:3point1loopb}, \ref{fig:3point1loopj}, \ref{fig:3point1looph} resp. the cancellations between diagrams \ref{fig:3point1loopb}, \ref{fig:3point1loopc}, \ref{fig:3point1loopk},  \ref{fig:3point1loopl}  as shown in section \ref{sec:cancellations-three-point function} resp. \ref{sec:cancellations-three-point function mirrored} the only divergent contributions that remain are
\begin{align}\nn
\langle \op \tilde{\op} \hat{\op}_j \rangle^{(1)} &= \frac{1}{2}\langle \op \tilde{\op} \hat{\op}_j \rangle^{\ref{fig:3point1loopa}} + \frac{1}{2}\langle \op \tilde{\op} \hat{\op}_j \rangle^{\ref{fig:3point1loopc}}+\langle \op \tilde{\op} \hat{\op}_j \rangle^{\ref{fig:3point1loope}} + \langle \op \tilde{\op} \hat{\op}_j \rangle^{\ref{fig:3point1loopd}} + \langle \op \tilde{\op} \hat{\op}_j \rangle^{\ref{fig:3point1loopg}+\ref{fig:3point1loopl}}  \\ \nn
&= \langle \op \tilde{\op} \hat{\op}_j \rangle^{\ref{fig:3point1loopa}} + 2 \langle \op \tilde{\op} \hat{\op}_j \rangle^{\ref{fig:3point1loope}}  + \langle \op \tilde{\op} \hat{\op}_j \rangle^{\ref{fig:3point1loopg}+\ref{fig:3point1loopl}}  +\op(\epsilon^0)\,,
\end{align}
where we used the symmetry of the diagrams under $p\to -p$ for $j$ even in the last step. Using 
\begin{equation}\nn
c_{jj}^{1/2} = 2^{1-j} \frac{\Gamma(2j)}{\Gamma(j)\Gamma(j+1)},\qquad b_j(4-d/2,1) = (-1)^j i \frac{\Gamma(j)}{\Gamma(j+1)}\frac{1}{(4\pi)^{2}}  + \op(\epsilon)\,,
\end{equation}
we find that the divergent part of \eqref{eqn:diagram-a-in-limit} is
\begin{equation}\nn
\langle \op \tilde{\op} \hat{\op}_j \rangle^{\ref{fig:3point1loopa}} 
 = i^j  g^8 N \delta^{aa}  \frac{1}{\epsilon} \frac{\Gamma(2j)}{\Gamma(j+1)\Gamma(j+1)}\frac{\hat{p}^j}{(-p^2)^{5-d}}\frac{1}{(4\pi)^{d}}  +\op(\epsilon^0) \,.
\end{equation}
The divergent part of \eqref{eqn:diagram-g-plus-j} is
\begin{equation}\nn
\langle \op \tilde{\op} \hat{\op}_j \rangle^{\ref{fig:3point1loopg}+\ref{fig:3point1loopl}} 
 = - i^j  g^8 N \delta^{aa}  \frac{1}{\epsilon} \frac{\Gamma(2j)}{\Gamma(j+1)\Gamma(j+1)}\frac{1}{(j+1)}\frac{\hat{p}^j}{(-p^2)^{5-d}}\frac{1}{(4\pi)^{d}}  +\op(\epsilon^0) \,.
\end{equation}
Furthermore, the divergent part of diagram \ref{fig:3point1loope} is
\begin{equation}\nn
2\langle \op \tilde{\op} \hat{\op}_j \rangle^{\ref{fig:3point1loope}} 
 =  i^j  g^8 N \delta^{aa}  \frac{1}{\epsilon} \frac{\Gamma(2j)}{\Gamma(j+1)\Gamma(j+1)} \left(H_j+H_{j+1} -1\right)\frac{\hat{p}^j}{(-p^2)^{5-d}}\frac{1}{(4\pi)^{d}}  +\op(\epsilon^0) \,.
\end{equation}
Summing up these contributions we find
\begin{align}\nn
\langle \op \tilde{\op} \hat{\op}_j \rangle^{(1)} &= 
 i^j  g^8 N \delta^{aa}  \frac{\Gamma(2j)}{\Gamma(j+1)\Gamma(j+1)} \left( \frac{H_j }{\epsilon}\right)\frac{\hat{p}^j}{(-p^2)^{5-d}}\frac{1}{(4\pi)^{d}}  +\op(\epsilon^0) \\ \nn
 &=  \left( - \frac{g^2N}{(4\pi)^h} \frac{ 4 H_j}{\epsilon}\right) \langle \op(p) \op(-p) \bar{\op}_j(0) \rangle^{(0)} + \op(\epsilon^0) \\ \nn
 &=  \left( - \frac{Z_{j}^{(1)} }{\epsilon}\right) \langle \op(p) \bar{\op}(-p) \bar{\op}_j(0) \rangle^{(0)} + \op(\epsilon^0)\,,
\end{align}
where we used the tree-level three-point function \eqref{eqn:tree-level-three-point-function-in-momentum-space} and can thus read off the anomalous dimension
\begin{align}\label{eqn:anomalous-dimension-Z-factor}
\gamma_j^{(1)} =   \left(\frac{g^2N}{4\pi^2}\right) 2H_j\,,
\end{align}
which is exactly what we expect and thus serves as a check for the relative factors of the divergent diagrams.

\addcontentsline{toc}{chapter}{Bibliography}
%\bibliographystyle{nb}
%\bibliography{thesis}
%\bibliographystyle{../mybib/nb}
%\bibliography{../mybib/all}

\section*{Software}
\begin{itemize}
\item This thesis was typset in LaTeX.
\item For the calculations Mathematica 8.0 by \emph{Wolfram Research} was used including the package \cite{Czakon:2005rk}.
\item Graphics were compiled using JaxoDraw and Mathematica 8.0.
\end{itemize}

\addcontentsline{toc}{chapter}{Index}
\printindex

%\newpage
%\section*{Selbst\"andigkeitserkl\"arung}
%\addcontentsline{toc}{chapter}{Selbst\"andigkeitserkl\"arung}
%
%Hiermit erkl\"are ich, die vorliegende Arbeit selbst\"andig ohne fremde Hilfe verfasst
%und nur die angegebene Literatur und Hilfsmittel verwendet zu haben.
%
%
%~\\[1cm]
%Konstantin Wiegandt\\
%Berlin, den 3. Mai 2012\\

\end{document}